Università degli Studi di Torino

**Scuola di Dottorato in Scienze e Alta Tecnologia**

**Tesi di Dottorato di Ricerca in Scienze e Alta Tecnologia**
**Indirizzo: Matematica**

# ON THE COMPLEXITY OF STRATIFIED LOGICS

**Luca Vercelli**
**Advisor: prof. Luca Roversi**

**Abstract**

This work deals with Implicit Computational Complexity, a research area that characterizes the Computational Complexity Classes by means of formal tools that are independent from Turing machines, the standard model of reference for such characterizations.

Our primary motivation is the comparison of two different traditions used to characterize the class FPTIME of the polynomial time computable functions. On one side, FPTIME can be captured by Intuitionistic Light Affine Logic (ILAL), a logic derived from Linear Logic, characterized by the structural invariant *Stratification*. On the other side, FPTIME can be captured by Safe Recursion on Notation (SRN), an algebra of functions based on *Predicative Recursion*, a restriction of the standard recursion schema used to define primitive recursive functions. *Stratification* and *Predicative Recursion* seem to share common underlying principles, whose study is the main subject of this work.

The starting point is a known result. Under natural and straightforward assumptions, a relation between Stratification and Predicative Recursion exists. It takes the form of a compositional embedding of a fragment $\mathsf{BC}^-$ of SRN into ILAL.

We propose two different extensions of such an embedding.

The first extension follows from a systematic and uniform analysis of the deductive systems, that we call *subsystems*, we obtain inside MS, a *multimodal and stratified framework* we introduce in this work.

MS generates subsystems. Two features characterize every subsystem $\mathcal{P}$ of MS. (i) Every derivation of $\mathcal{P}$ embodies the same Stratification principle we find in ILAL. (ii) Every $\mathcal{P}$ is defined by fixing an arbitrary set of modal operators, in analogy to the two modalities of ILAL, and choosing deductive rules in a well-defined set of rules that generalize those ones in ILAL. The point of having many modalities is to extend the set of derivations-as-programs available in $\mathcal{P}$. The framework allows a "wild" arbitrariness in the definition of subsystems $\mathcal{P}$, so it is often the case that $\mathcal{P}$ is polynomial time *un*sound, *i.e.* it can compute functions outside FPTIME. We need to develop *polynomial time soundness criteria* to distinguish inside MS those subsystems that only develop polynomial time sound computations.

The starting technical tool to devise criteria is *Context Semantics*. Once given such *semantics flavored* criteria, we supply some purely *syntactic and decidable criteria*, that allow to distinguish if a given subsystem $\mathcal{P}$ of MS is polynomial time sound or polynomial time *un*sound just observing the syntactic form of the rules of $\mathcal{P}$. Among the syntactic criteria there is the one that also determines the *maximal polynomial time sound* subsystems of MS, which are those subsystems that, fixed a set of modalities, contain the largest set of derivations-as-programs we use to control how the normalization proceeds.

We have identified two subsystems of MS worth mentioning.

The first subsystem is $\mathcal{P}^{M}_{\mathbb{LTS}}$. It is as computationally complex as the multiplicative fragment of Linear Logic, but it is a bit more expressive: $\mathcal{P}^{M}_{\mathbb{LTS}}$ contains among its proofs those ones encoding Church Numerals and the first basic operations on them.

The second subsystem is soLAL, a multimodal generalization of ILAL. soLAL allows us to partially fulfill the initial goal of finding a relation between Stratification and Predicative Recursion. Indeed, we can embed into soLAL a fragment $\mathsf{SRN}^-$ of SRN such that $\mathsf{BC}^- \subsetneq \mathsf{SRN}^- \subsetneq \mathsf{SRN}$.

As we said, a second kind of extension of the embedding from BC$^-$ into ILAL is considered, independent from MS. We introduce Light Affine Logic by Levels (LALL), an affine deductive system based on a weaker form of the Stratification principle. While soLAL uses many modalities, LALL preserves the two modalities present in ILAL. Crucially exploiting the weaker form of Stratification of LALL, and the possibility of freely introducing recursive types without altering the complexity cost of the normalization of LALL, we show that *all the finite fragments of* SRN *programs compositionally embed into* LALL.

Summing up, we investigate how to make the relation between Stratification and Predicative Recursion closer. We base our work on two, somewhat orthogonal, approaches. On one side, we generalize the modal aspects of ILAL, while preserving its original notion of Stratification. On the other, we exploit a generalization of the Stratification, while preserving the original set of modal operators. Both the lines of investigation this work develops move the relation between Predicative Recursion and Stratification a step forward. Despite of this, none of them is able to capture the whole SRN, at least not in the standard way.

What we believe we learn thanks to this work is that the very abstract principles FPTIME relies on are still hidden in the syntactic bureaucracy that still contaminates systems like SRN, ILAL, soLAL, and LALL.

## Ringraziamenti

Ringrazio in particolar modo il mio relatore, prof. Luca Roversi, che ha dedicato davvero molto tempo a seguire me e il mio lavoro.

Ringrazio tutto il gruppo "Formal Methods in Computing", ex gruppo $\lambda$, che è stato per me un ambiente molto accogliente.

Ringrazio il Dipartimento di Matematica, che mi ha permesso di lavorare insieme al Dipartimento di Informatica, permettendomi così di sviluppare argomenti che — a mio avviso — interessano entrambi i Dipartimenti.

Tra le tante persone con le quali ho lavorato, un grazie particolare va a Ugo Dal Lago, per le utili discussioni che hanno dato origine ai primi risultati presenti in questa tesi.

Grazie infine a tutte le persone che mi sono state vicine durante la stesura della tesi, un periodo che ricordo come decisamente stressante: Daniela, mia madre, mio fratello, e anche mio padre, il quale sono sicuro mi è stato vicino, pur essendo mancato da ormai 5 anni.

## Acknowledgements

Really many thanks to my advisor, prof. Luca Roversi, who dedicated so much time to follow me up and to supervise this work.

Thanks to the whole "Formal Methods in Computing" group, ex $\lambda$-group, that provided an hospitable working environment to me.

Thanks to the Department of Mathematics, that allowed me to work together with the Department of Computer Science, developing subjects that — at least in my opinion — may interest both the Departments.

In order to write my thesis, I have chatted with several people. Thanks in particular to Ugo Dal Lago for the useful discussions that lead to the first results presented in this work.

At last, thanks to all the people that have been close to me during the stressing period of the writing: Daniela, my mother, my brother, and my father too, who passed away 5 years ago.

# Contents

# Chapter 1

# Introduction

## 1.1 State of the art

**Implicit Computational Complexity.** Our work relates with Implicit Computational Complexity, in the following ICC. We will briefly recall what it is. The theory of Computational Complexity, or CC [Pap94], classifies the computable functions in *computational classes*: for example, FPTIME is the class of all the functions *computable with a Turing machine in a time that is polynomial in the size of the input*, and FPSPACE is the class of all the functions *computable with a Turing machine using an amount of memory that is polynomial in the size of the input*. Now, though the computational model of Turing machines has been a standard for years, it is quite heavy. We mean, for example, it is very difficult to write any *equation* describing a complexity class, and this is because the model of Turing machines is not suitable for that. The ICC tries to characterize the same complexity classes in a way independent from such a model. For this purpose, several different approaches are used. We will deal in particular with two approaches: *algebras of functions* and *light logics*.

In the first approach, a computational class (say, FPTIME) can be described as an *algebra of functions*, i.e. a set of functions containing some basic functions and closed under some operators over functions. This is a style common in recursion theory, where e.g. the computable functions can be represented as the algebra of functions described by Kleene [Kle36, Odi89]. In Sections 2.7 and ff. we will recall two important systems of this kind, SRN [BC92] and the Ramified Recurrence [Lei93], that both characterize the class FPTIME.

On the other side, *light logics* provide a completely different approach to ICC. The so-called *Curry-Howard correspondence* [How80, SrU06] identifies an analogy between *proofs* and *programs*. More precisely, there exists a bijection between proofs of intuitionistic and purely implicative logic, in the formalism of natural deduction, and programs, in the formalism of typed $\lambda$-calculus. The execution of programs corresponds to the normalization of proofs. So, a class of programs also corresponds to a *class of proofs*. This is the idea underlying light logics: a computational class can be represented using a *logic*. Such a logic must be even weaker than the intuitionistic logic, hence the name *light*. Some computational classes have already been characterized that way. Examples of light logics are LLL [Gir98], ILAL [AR02] and SLL [Laf04], all characterizing FPTIME; ELL, characterizing the class of elementary functions [DJ03]; $\mathsf{STA}_B$, characterizing FPSPACE [GMR08]. Most of them are logics derived from the Linear Logic, or LL [Gir87]. Some more details will be found in Chapter 2.

In this work, we will never refer directly to $\lambda$-calculus, nor to the Curry-Howard corre-

$$\frac{}{A \vdash A}\,\text{identity} \qquad \frac{\Gamma \vdash A \quad \Delta, A \vdash B}{\Gamma, \Delta \vdash B}\,\text{cut}$$

$$\frac{\Gamma \vdash A}{\Gamma, B \vdash A}\,\mathbf{W} \qquad \frac{}{\vdash A}\,\mathbf{h}$$

$$\frac{\Gamma, A \vdash B}{\Gamma \vdash A \multimap B}\,\multimap_{\mathcal{R}} \qquad \frac{\Gamma \vdash A \quad \Delta, B \vdash C}{\Gamma, \Delta, A \multimap B \vdash C}\,\multimap_{\mathcal{L}}$$

$$\frac{\Gamma \vdash A \quad \Delta \vdash B}{\Gamma, \Delta \vdash A \otimes B}\,\otimes_{\mathcal{R}} \qquad \frac{\Gamma, A, B \vdash C}{\Gamma, A \otimes B \vdash C}\,\otimes_{\mathcal{L}}$$

$$\frac{\Gamma \vdash A}{\Gamma \vdash \forall \alpha.A}\,\forall_{\mathcal{R}}\,(*) \qquad \frac{\Gamma, A\left[{}^{B}/_{\alpha}\right] \vdash B}{\Gamma, \forall \alpha.A \vdash B}\,\forall_{\mathcal{L}}$$

$$\frac{A \vdash B}{!A \vdash !B}\,! \qquad \frac{\vdash B}{\vdash !B}\,! \qquad \frac{!A, !A, \Gamma \vdash B}{!A, \Gamma \vdash B}\,\text{contraction}$$

Figure 1.1: The rules of ILAL, in a sequent calculus style. (*) Provided $\alpha$ does not occur free in $\Gamma$.

spondence. We will study proofs, (to be precise, proof nets, see Section 2.3 or [Gir96]), and we will study the normalization of proofs, under the assumption that the normalization of proofs is our computational model.

Of course, the two presented approaches to ICC are not the only ones. An overview of other, different, approaches is in the introduction of [Lei94].

**Intuitionistic Linear Logic (LL).** Linear Logic [Gir87], at least in its intuitionistic variant, will be described in Section 2.3; here we just recall that LL is characterized by the presence of two kinds of binary connectives, called *additives* and *multiplicatives*, plus one unary connective "!", or modality, called *exponential*. The traditional rules of *weakening* and *contraction* are restricted to the modal formulæ only:

$$\frac{\Gamma, !A, !A \vdash B}{\Gamma, !A \vdash B}\,\text{contraction} \qquad \frac{\Gamma \vdash B}{\Gamma, !A \vdash B}\,\text{weakening.}$$

In (one of the possible formulations of) LL, the exponentials are controlled by the following rules:

$$\frac{\Gamma, A \vdash B}{\Gamma, !A \vdash B}\,\text{dereliction} \qquad \frac{A_1, \ldots, A_k \vdash B \quad k \geq 0}{!A_1, \ldots, !A_k \vdash !B}\,\text{functorial, or soft, promotion} \qquad \frac{\Gamma, !!A \vdash B}{\Gamma, !A \vdash B}\,\text{digging.}$$

The real source of complexity, in the normalization of proof nets, is generated by the exponentials. So, the "secret" of light logics is a clever control over the exponentials.

**Intuitionistic Light Affine Logic (ILAL).** ILAL is a light logic that was first presented in [Asp98] and [AR02], and its purpose is to characterize FPTIME simplifying the Light Linear

Logic of [Gir98]. The complete list of the rules of ILAL is in Figure 1.1, in a sequent calculus style. We now try to explain how these rules are obtained. They are derived from the rules of LL. In order to lower the complexity of the reductions, w.r.t. LL, ILAL forbids the *dereliction* and *digging* rules, and allows the *functorial promotion* with at most one premise:

$$\frac{A_1, \ldots, A_k \vdash B \quad k \in \{0,1\}}{!A_1, \ldots, !A_k \vdash !B} \; !.$$

At this point, the logic that we have got is *too* weak. Indeed every reduction is performed in polynomial time, but it is not obvious (maybe it is impossible?) to represent the Church numerals inside it, and they are usually required to represent some form of iteration. The solution involves two steps: (i) the introduction of *another modality "§" (paragraph)*, that allows a promotion rule more liberal than the "!", but which does not admit contraction, and (ii) the introduction of the *free weakening* (**W**) rule. **W** is the real difference between ILAL and LLL. The dynamic of the reduction of the free weakening rules allows to *erase* pieces of proofs in a simple way, without any need for additives. As observed in [Asp98]: "*the abstinence from weakening leads to inessential syntactical complications*". So, ILAL just has multiplicative and exponential connectives.

The free weakening has also some negative consequences, that force ILAL to be (i) intuitionistic, and (ii) to include the somewhat weird rule **h** of Figure 1.1. The interested reader is referred to Section 2.4.

**Algebras of Functions vs Light Logics.** Both ILAL and SRN, we already said, are able to characterize the functions in FPTIME. It is reasonable, thus, to compare these two systems.

Both ILAL and SRN allow the representation of all the polytime Turing machines [AR02, Cob65, BC92]. However, *at first sight*, programming in ILAL seems easier than in SRN. The reason is that ILAL handles higher-order types, while SRN only handles first-order functions, from natural numbers to natural numbers. So, for example, [Cas97, Hof03] show that there exist simple algorithms that cannot be written – at least not easily – in the systems based on predicative recurrence. Some of these algorithms can be written in ILAL, instead (but of course not all: the nesting of iteration is limited in ILAL, as well).

On the other side, two facts suggest that SRN could express in a simple way programs that ILAL cannot. First, ILAL is *strongly* polytime, that is polytime under every strategy of reduction [Ter07], while SRN is *weakly* but not strongly polytime, i.e. it is polytime only under some precise strategy [BW96, BDLM06]. And it is reasonable that many more weakly polytime algorithms exist, than strongly polytime. Second, at the moment it is not known a translation that *compositionally* translates programs of SRN into proofs of ILAL. In [MO04], it is shown a simple encoding that translates the programs of $\mathsf{BC}^-$, a proper subset of the programs of SRN, into ILAL, and it is also shown that it is not possible to extend the *same* encoding to the whole SRN (more details in Section 2.9). By the way, at the moment it is not known if $\mathsf{BC}^-$ captures all FPTIME, or not. In the same work [MO04] it is also shown that it is possible to extend $\mathsf{BC}^-$ to some $\mathsf{BC}^\pm$ that captures FPTIME and that compositionally embeds into ILAL; however, $\mathsf{BC}^\pm$ does not help understanding the intrinsic differences between SRN and ILAL, since it contains primitives extraneous to SRN. The situation is summed up in Figure 1.2: it seems necessary to extend ILAL to some system (3) that can contain SRN. Or, at least, we would like to find some extension (2) that captures at least some strict extension (1) of $\mathsf{BC}^-$.

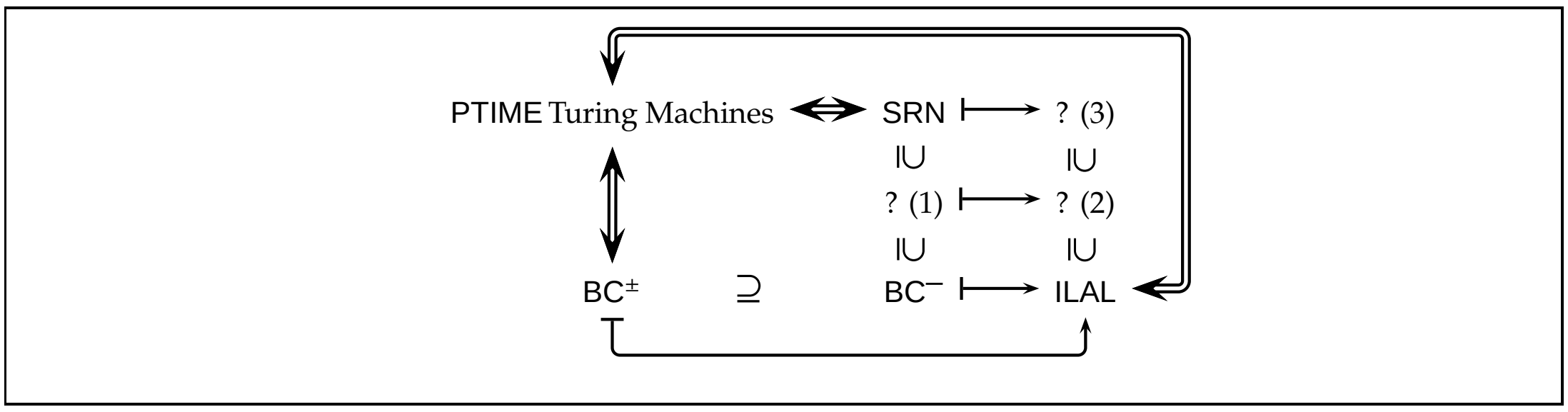

Figure 1.2: Relations between SRN and ILAL.

## 1.2 The problem

**Extending ILAL.** The origin that justifies this work is the search for some extension of ILAL, that nevertheless characterizes the class FPTIME. It should be clear that *many* different extensions are possible, as we can imagine a large variety of rules to add to ILAL; the problem is not to escape from the domain of polynomial time. As an example of how extending ILAL may increase its expressiveness, we will consider two different extensions of ILAL by means of some forms of *dereliction*. Let us consider $\mathsf{ILAL}_!$ as ILAL plus the dereliction rule (already presented above). $\mathsf{ILAL}_!$ is *not* polytime, since it allows to write an hyper-exponential function that we are going to describe. Let us consider the following type for Church numerals:

$$\mathbb{N} = \forall\alpha.!(\alpha \multimap \alpha) \multimap \alpha \multimap \alpha \qquad \text{(the same as in LL)}.$$

Then, one can build the following proof nets (i.e. proofs), with the expected semantics:

$$\begin{aligned} \texttt{succ} &: \mathbb{N} \multimap \mathbb{N} \\ \texttt{add} &: \mathbb{N} \multimap \mathbb{N} \multimap \mathbb{N} \\ \texttt{mul} &: !\,\mathbb{N} \multimap \mathbb{N} \multimap \mathbb{N} \\ \texttt{exp} &: !!\,\mathbb{N} \multimap \mathbb{N} \multimap \mathbb{N} \\ \texttt{hypexp} &: !!!\,\mathbb{N} \multimap \mathbb{N} \multimap \mathbb{N} \end{aligned}$$

This means, for example, that it is possible to build a proof of $\mathbb{N} \multimap \mathbb{N}$ called succ that, whenever cut with a numeral $n$, reduces to $n+1$. Addition in fact is obtained concatenating numerals. Each one of the last three functions is obtained iterating the one that precedes it. For example, multiplication is in Figure 1.3a.

So, $\mathsf{ILAL}_!$ does not work for our purposes. We may then imagine to allow the dereliction on the "§" instead than on the "!":

$$\frac{\Gamma, A \vdash B}{\Gamma, \S A \vdash B}\, D_{\S}.$$

We obtain a system $\mathsf{ILAL}_{\S}$ that does not enjoy cut elimination: there is no way to reduce a §-dereliction in cut with a §-box with some !-premises. We are not frightened by this inconvenience: the proof net containing irreducible cuts will be considered normal forms. We now show that $\mathsf{ILAL}_{\S}$ is not polytime. It is possible to use the following type for the Church numerals:

$$\mathbb{N} = \forall\alpha.!(\alpha \multimap \alpha) \multimap \S(\alpha \multimap \alpha) \qquad \text{(as in ILAL)}.$$

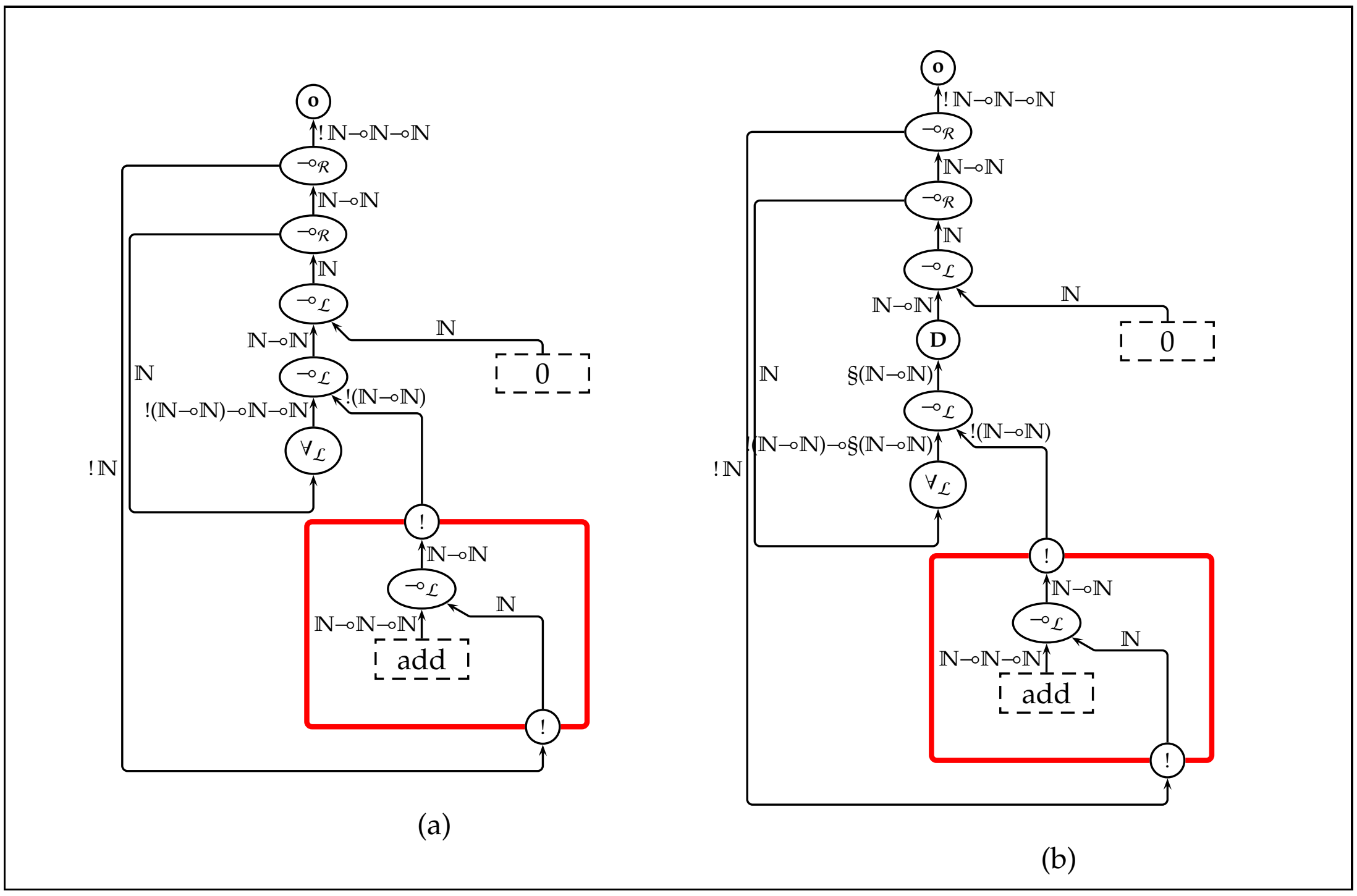

(a) (b)

Figure 1.3: Multiplication proofnet in ILAL plus !-dereliction (a) and in ILAL plus §-dereliction (b). Notice that ℕ denotes a different type in the two cases.

Now we can write, in $\mathsf{ILAL}_\S$, all the functions that we already encoded in $\mathsf{ILAL}_!$, and with the same type. For example, multiplication is in Figure 1.3b. Thus $\mathsf{ILAL}_\S$ is not polytime.

*The previous two examples make clear that, even if a great number of extensions of* ILAL *are possible, we need some way to understand which one are polytime, and which one are not. We need to consider a large number of possible extensions, and to study them in a uniform way.*

## 1.3 Contribution

**The overall picture.** This work has the following main goals.

1. We will present a class of light logics derived from ILAL that *can* be useful for the representation of computational classes. These logics will be called *subsystems* of a framework MS that this thesis introduces and studies.
2. We will compare the two approaches to ICC already described: light logics and algebras of functions. This means that we will try to fill the question marks in Figure 1.2, along two different routes.
   - The first route is to search for an extension of ILAL inside the subsystems of MS, arriving for example to soLAL, a subsystem that we can replace for the question mark (2) in Figure 1.2, for a suitable subsystem $\mathsf{SRN}^-$ of SRN in place of (1).
   - The second route is to turn to a weaker form of stratification that MS, for the moment, does not contain. Such a weaker form is in the logical system LALL (Light Affine Logic by Levels) that we shall present, which, in a precise technical sense, can replace (3).

Rephrasing the *incipit* of [Lei94], *getting to the above two goals will help providing insights into the abstract nature of computing* FPTIME *functions, while offering concepts and methods for generalizing computational complexity to computing over arbitrary structures and to higher type functions.*

We now deepen a bit the description of our contribution.

**Multimodal Stratified framework (MS).** One of the focuses of this work is what we call *Multimodal and Stratified framework for polytime computations* [RV08, RV09], that we will present in Chapter 3. Now we sketch the main features of MS. The framework allows to study in a uniform way a variety of different light logics, all derived from LL, and all described in the formalism of proof nets. Please notice that instead, in this Introduction, we have found easier to deal with the derivation trees of sequent calculus than with the proof nets. The various light logics that we will consider will be called *subsystems* of MS.

Before any formal definition of the framework MS, we prefer to underline what are the key points of all its subsystems. (i) Each subsystem $\mathcal{P}$ is defined as a particular set of *building rules*, i.e. rules that allow the inductive construction of proof nets. We will denote $\mathbf{PN}(\mathcal{P})$ the set of such proof nets. Along the work, we shall refer informally to the elements of $\mathbf{PN}(\mathcal{P})$ as *the proof nets of* $\mathcal{P}$, and we shall say that $\mathcal{P}$ *allows a rule* $R$ whenever $R \in \mathcal{P}$. (ii) All the proof nets that we consider are intuitionistic, that is proof nets with one only conclusion. (iii) Among the possible rules, there is the *linear kernel* of LL, whose rules will be recalled in Section 2.3 and 3.1. This kernel includes the rules for multiplicative connectives and for second-order quantifiers, but *not* the rules for additive connectives. This fact is somewhat traditional in an affine context: the additives can be defined using the free weakening and the second order quantification. (iv) The exponential rules, or modal rules, derived from LL, are generalized in order to handle an arbitrary number of *modalities*, and not only the one of LL or the two ones of ILAL. This is why we say the framework is *multimodal*. This is analogous e.g. to what happens in [Sch94]. From the technical point of view, multimodality is obtained allowing the modalities to vary over a fixed – but arbitrary – set $\mathbb{X}$, that usually will be finite, and defining the framework in a way that $\mathbb{X}$ is used as a free parameter. (v) On the other side, however, the framework puts some restrictions on the exponential rules. Remember that we are generalizing ILAL, so exactly as in ILAL the *dereliction* and *digging* rules are forbidden. These restrictions force the proof nets to reduce in a peculiar way: the promotion rules, inside a proof net, mark off some regions ("boxes") that reduce independently the one from the other, and the number of *nested* boxes around a given node of a proof net do not vary during a reduction. This behavior is called *stratification*, hence the name of the framework.

And now, a *slightly* more formal definition of the framework. The framework is made up of a set of *meta-building rules* and *meta-normalization steps*, that will be defined resp. in Section 3.1 and 3.3. The meta-building rules allow the construction of *meta-proof nets*, i.e. proof nets with some parameters that may vary over a given set of modalities $\mathbb{X}$. For example,

$$\frac{\Gamma, \mathfrak{m}A, \mathfrak{n}A \vdash B}{\Gamma, \mathfrak{q}A \vdash B}\ \mathbf{Y}_{\mathfrak{q}}(\mathfrak{m},\mathfrak{n}) \qquad \mathfrak{m},\mathfrak{n},\mathfrak{q} \in \mathbb{X}$$

is a meta-building rule, that can be instantiated to the "usual" contraction provided $! \in \mathbb{X}$ and $\mathfrak{m} = \mathfrak{n} = \mathfrak{q} =!$. The meta-normalization steps map meta-proof nets into meta-proof nets. So, this implies that, once instantiated the parameters of the meta-normalization steps, we get normalization steps from proof nets to proof nets. For every fixed set $\mathbb{X}$ of modalities, $\mathfrak{B}_{\mathbb{X}}$ is the set of all the *instances* of the meta-building rules, with modalities varying over $\mathbb{X}$. Similarly, $\mathcal{R}(\mathfrak{B}_{\mathbb{X}})$ is the set of all the *instances* of the meta-normalization steps. At last, a *subsystem* of MS is simply a subset $\mathcal{P} \subseteq \mathfrak{B}_{\mathbb{X}}$. Each subsystem **uniquely** identifies (1) a set of proof nets $\mathbf{PN}(\mathcal{P})$, that can be built using all and only the building rules in $\mathcal{P}$, and (2) a set of normalization steps $\mathcal{R}(\mathcal{P}) \subseteq \mathbf{PN}(\mathcal{P})^2$, which contains all and only the normalization steps in $\mathcal{R}(\mathfrak{B}_{\mathbb{X}})$ that correctly map proof nets of $\mathbf{PN}(\mathcal{P})$ into proof nets of $\mathbf{PN}(\mathcal{P})$.

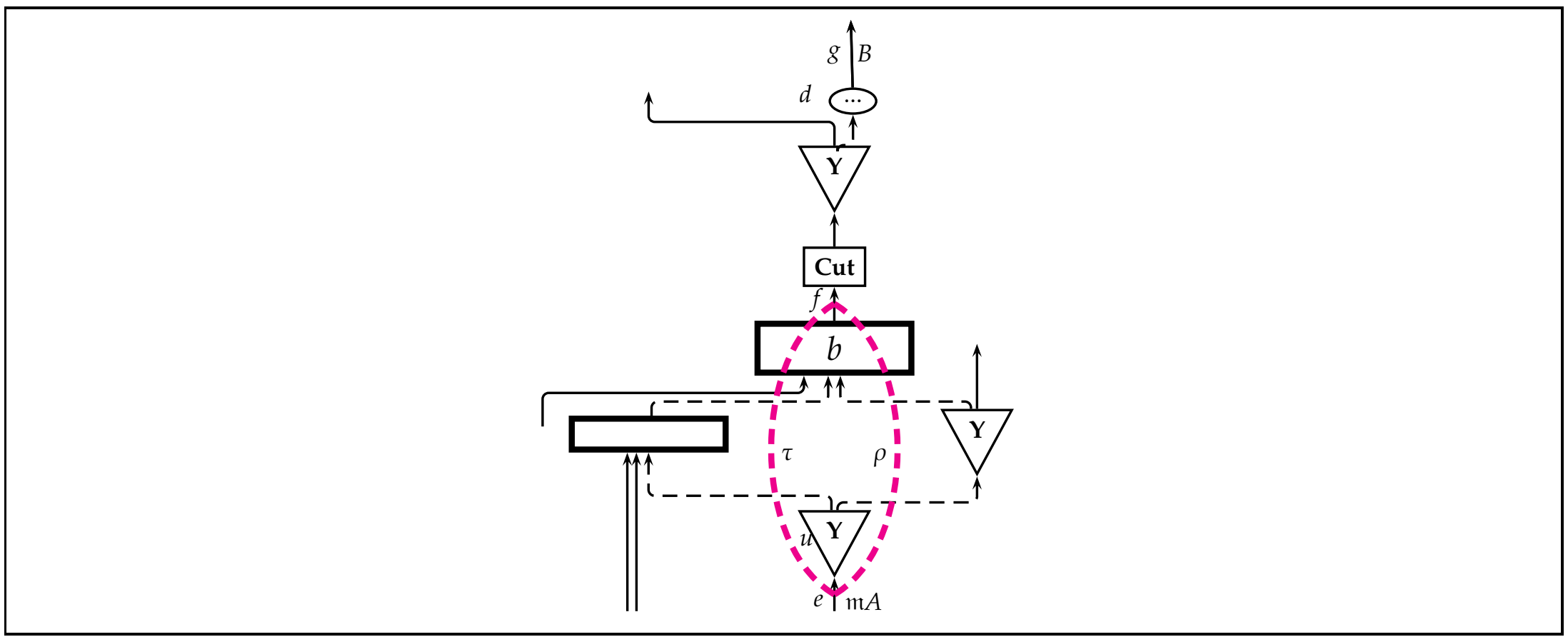

Figure 1.4: The structure between $e$ and $f$ is an example of *spindle*. If the spindle is *dangerous*, it is possible connect it to some $g$ labelled $B = \mathfrak{m}A'$.

**Polytime subsystems.** Up to here, we have described the statical properties of MS. But, as we already said, every subsystem $\mathcal{P}$ of MS identifies in fact a class of programs, and we are interested to the behavior of such programs. As a consequence of the constraints given by *stratification*, fixed a subsystem $\mathcal{P} \subseteq \mathfrak{B}_{\mathbb{X}}$, every reduction in $\mathcal{P}$ is performed in at most an elementary time in the size of the proof (read also: *"every program executes in an elementary time in the size of the arguments"*).

We characterize those subsystems in which the reductions are performed in polytime (i.e. polynomial time). PMS is the class containing such subsystems. We describe how in Chapter 4. There, we will define two basic notions: (a) a *dangerous spindle*, recalled in Figure 1.4, that is a geometrical configuration that can be present inside the proof nets of MS, made up of two and only two paths connecting a contraction to a box; and (b) a *sensible* subsystem, that is a subsystem endowed with a sufficient number of rules, enough to encode a minimum amount of programs. The Polynomiality Criterion (Proposition 4.3.40) tells that:

> A sensible subsystem is polytime if and only if it does not allow the construction of dangerous spindles.

The notions of spindle and dangerous spindle, as well as the proof of the Polynomiality Criterion, rely on the formal technology of *context semantics*, or CS, [DL08]. Notice anyway that the spindles provide an *abstraction* over CS: once defined them, only a small amount of CS is needed to understand if $\mathcal{P}$ can build a dangerous spindle, or not.

Anyway, this *small amount* of CS is still too much. We prove some more properties, more *syntactical*, that still discriminate among polytime and non-polytime subsystems, and nevertheless completely hide CS. For example, Proposition 4.4.15 tells that, given the rules of $\mathcal{P}$, one can understand if $\mathcal{P}$ is polytime or not just looking at the relation $\mathfrak{R}_{\uparrow}$ over $\mathbb{X}$ that the promotion and contraction rules of $\mathcal{P}$ naturally define.

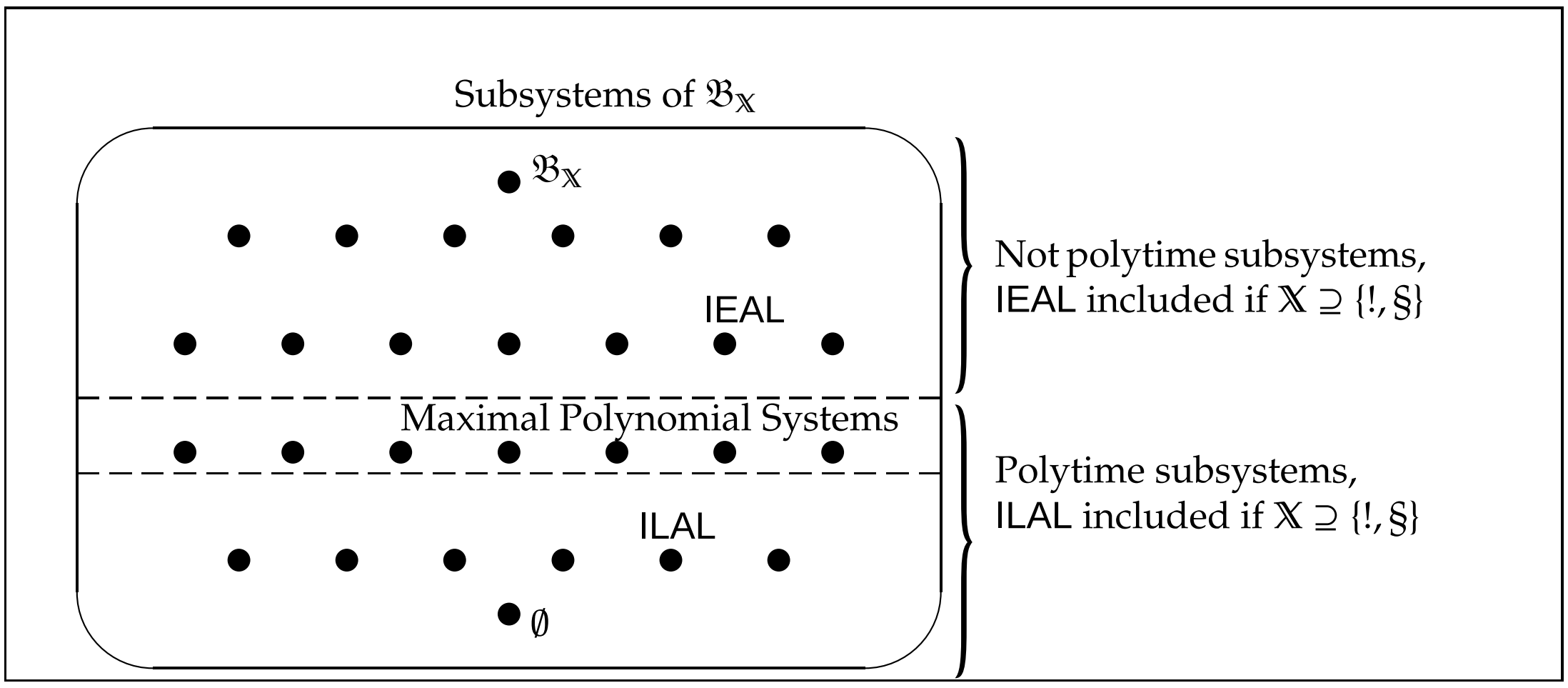

Figure 1.5: A picture of the poytime, non-polytime and maximal subsystems of $\mathfrak{B}_{\mathbb{X}}$.

Last but not least, we study the polynomiality of our subsystems concentrating on the *maximal* subsystems. Intuitively, a subsystem $\mathcal{P}$ of $\mathfrak{B}_{\mathbb{X}}$ is maximal if it is polytime, but any extension $\mathcal{P} \cup \{R\}$, where $R$ is any other rule of $\mathfrak{B}_{\mathbb{X}}$, is no more polytime (see Figure 1.5). So, a generic subsystem $\mathcal{P}$ is polytime iff it can be extended to some polytime subsystem (Proposition 4.4.32). We think that the study of maximal subsystems is important; they are the most interesting subsystems, because they allow the greatest flexibility on composing rules but nevertheless they are polytime. This is the reason that originates the Structure Theorem (Theorem 4.4.25): it describes all and only the maximal subsystems $\mathcal{P}$ just giving conditions on the syntax of the building rules in $\mathcal{P}$.

**Algebras of Functions vs. Subsystems: a restatement of the original problem.** Let us return to the original problem: the embedding of SRN (or some other algebra of functions) into ILAL (or some other light logic). We restate it in the following way: *"find a subsystem of* MS *in which it is possible to encode all the* SRN *programs"*. We face the question inside Chapter 6, but we will not completely succeed in our search. We present a subsystem soLAL (*sorted* ILAL) that can encode some more programs of SRN as compared to ILAL, thus filling the question mark (2) in Figure 1.2 for some suitable super-system (1) of $\mathsf{BC}^{-}$.

soLAL extends ILAL by means of *sorted* modalities: soLAL is made up of many copies of ILAL, each one with its own modalities "$i!$" and "$i\S$", each with its sort $i$. This allows a mechanism forbidden in ILAL: the duplication of a §-box, with the proviso that the sort of the input paragraph box is smaller than the one in output. For example, this means that soLAL proves $i\S\,\mathbb{N} \multimap (i+1)\S(\mathbb{N}\otimes\mathbb{N})$. Of course this duplication is limited in power, because the type of the result is different from the type of the input. As a consequence, we shall see that it is possible to encode in soLAL some programs $g \in \mathsf{SRN}\setminus\mathsf{BC}^{-}$. But, we stress, we are able encoding $g$ under the same very natural assumptions of [MO04]. However, the type of the inputs of $g$ is different from the type of its output, so that $g$ cannot be iterated anymore. We will call $\mathsf{SRN}^-$ the set of functions that can be written in $\mathsf{soLAL}$, and such that $\mathsf{BC}^- \subsetneq \mathsf{SRN}^- \subsetneq \mathsf{SRN}$.

Anyway, we could not find any subsystem filling the question mark (3). This suggests the following considerations.

There is an analogy between the *stratification* typical e.g. of ILAL and the *predicative recurrence* at the basis of SRN. This is the separation of objects – otherwise indistinguishable – into separate levels, or sorts; and such separation is at the basis of the control over complexity. Despite of this similarity, there is really a great difference between the two approaches to ICC. Algebras of functions have a great freedom in both (a) choosing their base functions, and (b) distinguishing different kinds of variables (or, different *types* for numerals). In light logics, all these features can only be obtained as *indirect consequences* of the chosen logical rules and of the chosen representation of the data-types. What we are saying is that the root of the un-relatedness of SRN and light logics lies in the choice of the syntax we start from.

So, for example, the known fact that ILAL is strongly polytime, while SRN is just weakly polytime, is only a consequence of the chosen syntax, and not the origin of such un-relatedness.

To conclude, we anyway think that some subsystem of MS where to embed SRN can still exist, not necessarily stickying to all the natural assumptions in [MO04]. The following observations justify why. The programs of SRN can reduce in exponential time and space, *but* the size of their normal forms is always polynomial in size. This feature is uncommon in subsystems of MS. So, the exponential reductions of SRN appears to us as *degenerate cases of reductions*. Anyway, this is still an on-going work.

**Light Affine Logic by Levels (LALL).** Once "exhausted" the attempts to relate SRN with usual stratified logics, we turn our attention to a weak notion of stratification. We will present in Chapter 7 a polytime system LALL (Light Affine Logic by Levels) that extends ILAL using (i) the so-called "levels" of $\mathsf{L}^4$ [BM10], and (ii) some recursive type, namely the type $\mathscr{S}$ of the Scott numerals [ACP93]. The notion of "level" underpins a different notion of stratification. Thus, LALL is not a subsystem of MS. In Sections 8.2.2 and 8.2.3 we will try to generalize the framework MS in order to capture LALL; but this is still an on-going work.

Inside LALL it is possible to encode *finite initial fragments* of all the SRN programs. We mean, for every SRN program $t$, and for every finite $l \in \mathbb{N}$, we will be able to produce a proof net $\lceil t \rceil^l$ of LALL that represents a program whose output coincides with $t$ for every input of length less than $l$, and returns an approximation of the result on every other input. The encoding $\lceil \cdot \rceil^{\cdot}$ is based on the one given in [MO04]. Scott words and levels are used because of the following properties. (a) For every $l \in \mathbb{N}$, it exists in LALL a proof net $\nabla \mathsf{S}_l[s] \triangleright \mathscr{S} \multimap \mathscr{S} \otimes \mathscr{S}$, that duplicates every number *at most l bits long*. This is not true for Church numerals. (b) Using levels as in $\mathsf{L}^4$, $\S(A \otimes B)$ is provably equivalent to $\S A \otimes \S B$, so that in particular for every $k$ it exists in LALL a proof net $\nabla \mathsf{S}_l[s] \triangleright \S^k\mathscr{S} \multimap \S^k\mathscr{S} \otimes \S^k\mathscr{S}$ with the same semantic of the one in point (a). This is not possible in a stratified context. (a) and (b) together allow a mechanism impossible in ILAL: the duplication of arguments. On the other side, Scott numerals cannot be used to iterate *ad libitum* other functions. This is the reason that obliges us to encode just finite fragments of programs, and not whole programs.

At our knowledge, there are few works in literature that exploit the use of recursive formulæ for calculations [BM04, DLB06].

## 1.4 Guideline for the reader.

Chapter 2 contains basics knowledge about $\lambda$-calculus, proof theory, computational complexity, light logics and predicative recurrence. Most of the Chapter is mainly intended for non-specialists. Section 2.7, on the other side, presents SRN, that will be required later, in Chapters 6 and 7; so probably every reader wants to refer to this Section. The framework MS will be studied from Chapter 3 up to Chapter 6. They are quite technical; we suggest the reader, for a first lecture, to read only the main definitions and theorems. The most important are the Polynomiality Criterion (Proposition 4.3.40) and the Structure Theorem (Theorem 4.4.25); some more propositions are summed up in Figure 4.1. Chapter 7, at last, presents LALL, a light logic that is not a subsystem of MS, but that can be used to encode finite fragments of SRN programs.

# Chapter 2

# Preliminaries

In this Section we recall the basics concepts needed to put this thesis in his correct background.

**The Section is mainly intended for *non-specialists*, who need a picture of the background around this Thesis. So, it should not be surprising that most of details will be skipped, or presented in an informal or naïve way.**

A more detailed picture is e.g. in [Maz02, Sch08].

## 2.1 Proofs as Programs

In this Thesis we shall deal with *proofs*, in some precise formalism to be defined. However, we will see the proofs *as they were programs*: we will say that our proofs *normalize*, thinking that the corresponding program executes; and we will call *results of the computation* the proofs that cannot be normalized any more. Theorem 2.1.2 shall make more precise this correspondence: there exists a bijection between proofs – in the formalism of natural deduction – and programs – in the formalism of $\lambda$-calculus. We are now going to recall what are $\lambda$-calculus and natural deduction, in order to formalize Theorem 2.1.2.

**Untyped $\lambda$-calculus.** $\lambda$-calculus is a model of computation, whose programs are called *$\lambda$-terms* [Bar84]. $\lambda$-terms are syntactical objects defined by the following grammar:

$$t, u ::= x \mid \lambda x.t \mid tu \qquad \forall x \in \mathcal{V}$$

where $\mathcal{V}$ is a countable set of symbols called *variables*. Every $\lambda$-term can be seen as a *program*; the execution of that program is given by the rule of $\beta$-reduction:

$$(\lambda x.t)u \rightarrow_{\beta} u\left[{}^{t}\!/\!_{x}\right].$$

It's possible to prove that every Turing machine can be codified into a $\lambda$-term, and of course there exists a Turing machine that performs the reduction of any given $\lambda$-term. So, this is a model of calculus equivalent to Turing machines.

It is very interesting to study the semantics of these objects [AC98]. Intuitively, $\lambda$-terms are (computable) functions. But of course they cannot be functions according to the usual set-theoretical conventions, as they violate the Axiom of Foundation: $tt$ is a correct $\lambda$-term, while no function can be applied to itself.

Anyway, this is not the matter of our work.

In order to simplify the semantics of $\lambda$-terms, it is often convenient to consider only a *typed* version of the $\lambda$-calculus. This can be done in two different ways.

**Typed $\lambda$-calculus: approach à la Church.** Types and typed $\lambda$-terms are syntactical objects defined as follows:

$$A, B ::= \alpha \mid A \to B \qquad \forall \alpha \in \mathcal{T}$$

$$t, u ::= x^A \mid \left(\lambda x^A . t^B\right)^{A \to B} \mid \left(t^{A \to B} u^A\right)^B \qquad \forall x^A \in \mathcal{V}$$

where $\mathcal{T}$ is a countable set of symbols called *atomic types* and $\mathcal{V}$ is as before. Typed $\lambda$-terms can be seen as a small subset of the $\lambda$-terms, just forgetting the types. Their semantics is clear: a type $A$ is a set, and a term $t^{A \to B}$ is a (computable) function from $A$ to $B$.

**Typed $\lambda$-calculus: approach à la Curry.** It is defined a notion of *typeability* over untyped $\lambda$-terms. So, for example, $\lambda x.x$ is typeable because it can be endowed with a type $A \to A$. Its type it is not unique, because also $B \to B$ does the job. Only typeable $\lambda$-terms are accepted.

**Expressiveness of Typed $\lambda$-calculus.** In the typed settings, only a small set of Turing-computable functions can be represented. In order to restore the completeness, one possible solution is the introduction of some kind of *fix point operator*; but at the moment this is not needed in our work. A different way to restore expressiveness (without however reaching full Turing completeness) is the introduction of the second order quantifier, as in Girard's system F [GTL89].

**Proof theory.** There exist some different formalisms for *proofs*. Each formalism tries to capture what are the elements to make a proof correct. So, each formalism defines one or more axiom to start with, and one or more deduction rules that are allowed to derive theorems.

**Hilbert systems** use only 1 logical rule (*modus ponens*) and an infinite number of different kinds of axioms. Very often, all tautologies are included among these axioms. This kind of systems is quite unnatural and, at the moment, they are not used when studying ICC.

In **natural deduction** [Pra65] there are no axioms at all (well, they are hidden in the syntax), and a lot of rules. Typically this is the formalization of the approach that a mathematician uses to find out a proof. Among these rules there are both *introduction* and *elimination* rules. One usually denotes NJ and NK resp. the intuitionistic and classical version of natural deduction.

In the third formalization, Gentzen's **logistic calculi** or **sequents calculus** [Gen35], there is exactly 1 kind of axiom (*identity*), 1 elimination rule (*cut*, quite similar to *modus ponens*) and a lot of introduction rules. Gentzen denotes LJ and LK resp. the intuitionistic and classical version of logistic calculi. So, most of the axioms of the Hilbert systems are here turned into *logical rules*.

The fourth – and last – formalization that we consider is very different: a **proof net** [Gir96] is a proof represented by a particular graph. Proof nets take their origin in Linear Logic, but can be applied in fact also to classical and intuitionistic logic [Rob03]. The main idea is that in both natural deduction and logistic calculi many proofs of a same statement differ just for the – irrelevant – order of application of the logical rules; using a graph-theoretical representation these differences disappear, and all these proofs are identified in a single proof net.

**Cut-elimination.** The first important result in proof theory is the *Haupsatz*:

THEOREM 2.1.1 (HAUPSATZ (GENTZEN, 1935))
In both LK and LJ, the *cut* rule is not necessary. There exists an algorithm (called *cut-elimination*) that transforms every LJ (resp. LK) proof without non-logical axioms into a *cut-free* LJ (resp. LK) proof. In LJ this algorithm is deterministic.

This means that, if we were just interested in the set of provable sequents, the *cut* rule could be avoided, and logistic calculus would be a system with only *introduction* rules. In a computational perspective, on the contrary, the cut rule will have a very important role.

The algorithm of *cut-elimination* has also an equivalent version in NJ and NK, that we will call *normalization*. Essentially, the normalization removes useless pair of introduction/elimination rules.

**Curry-Howard correspondence.** At last, the already announced Theorem:

THEOREM 2.1.2 (CURRY-HOWARD CORRESPONDENCE)
There exists a bijection between proofs of purely implicative NJ and typed $\lambda$-terms. Under this bijection, the normalization of proofs corresponds to $\beta$-reduction of $\lambda$-terms.

The Curry-Howard correspondence suggests a model of computation based on proofs instead than programs.

In our work we use little natural deduction, and more logistic calculi and proof nets. Every proof net stands for several NJ proofs, and every NJ proof corresponds to several LJ proofs. So we can speak about Curry-Howard correspondence also between typed $\lambda$-terms and purely implicative proof nets, an between typed $\lambda$-terms and purely implicative LJ proofs.

The following syntax is used for making the correspondence of Theorem 2.1.2 effective:

$$\frac{}{x : A \vdash x : A}\, ax \qquad \frac{\Gamma, x : A \vdash t : B}{\Gamma \vdash \lambda x.t : A \multimap B} \multimap I \qquad \frac{\Gamma \vdash t : A \quad \Gamma \vdash s : A \multimap B}{\Gamma \vdash st : B} \multimap E \tag{2.1}$$

Types are omitted, as they are trivially induced by the formulæ. Thanks to deductions (2.1), every proof of $\vdash A$ is associated to a program $t$ such that $\vdash t : A$; $t$ is called *realizer* of the proof, and in some sense $t$ is the program that proves $A$. Now it's clear why the logic is intuitionistic and not classic: $A$ must be proved *constructively*. This vision is called *Heyting semantics* of the proofs; the interested reader can refer to [GTL89].

## 2.2 Computational Complexity

The objects studied by Computational Complexity [Pap94] are both computationally solvable problems, and algorithms that solve them. Informally speaking, CC target is to give each algorithm and each problem a *complexity measure*; for example, how much time a solving program takes to execute, or how many resources does it use? So, the main aim of CC is the identification of a hierarchy of *complexity classes* of problems.

In particular the theory identifies:

- a hierarchy of time classes: PTIME and EXPTIME are the classes of problems that can be solved with a Turing Machine in resp. polynomial and exponential time in the size of the input.
- a hierarchy of space (that is, used memory) classes: L, PSPACE and EXPSPACE are the classes of problems that can be solved with a Turing Machine in resp. logarithmic, polynomial and exponential space in the size of the input.
- for each one of the previous classes, a non-deterministic class: that is, the class of problems that can be solved according to that complexity bound using a *non-deterministic* Turing Machine, which is able to make always the better choice.
- a hierarchy of *computable functions*: e.g. if PTIME identifies the problems solvable in polynomial time, FPTIME identifies the class of the functions computable in polynomial time.

There are hundreds of results actually known about CC. We just remember that:

$$\mathsf{L} = \mathsf{NL} \subseteq \mathsf{PTIME} \subseteq \mathsf{NPTIME} \subseteq \mathsf{PSPACE} = \mathsf{NPSPACE} \subseteq \\ \subseteq \mathsf{EXPTIME} \subseteq \mathsf{NEXPTIME} \subseteq \mathsf{EXPSPACE}$$

and

$$\mathsf{PTIME} \subsetneq \mathsf{EXPTIME} \qquad \mathsf{L} \subsetneq \mathsf{PSPACE} \subsetneq \mathsf{EXPSPACE}$$

but all the other inequalities are not known yet to be proper or not. $\mathsf{PTIME} \stackrel{?}{=} \mathsf{NPTIME}$ is one of the most well-known open problems.

All our work concentrates on the class FPTIME of the functions computable in polynomial time.

## 2.3 Linear Logic, Affine Logic, and Proof nets

**Linear Logic.** The logic that we use in this work is the Linear Logic (LL) [Gir87], mainly in its intuitionistic and affine form [AR02].

Here we sum up the main features of the LL, without any hope to be exhaustive.

- The sequent calculi systems LJ and LK are made up of both *logical* and *structural* rules. Traditionally, structural rules (*contraction* and *weakening*) are accepted by every logician, and the studies concentrate on the logical rules.
- On the contrary LL focuses on the structural rules.
- LL is built around a kernel of rules MALL (Multiplicative and Additive LL) where the rules *contraction* and *weakening* of LK are forbidden. This means, in some sense, that $A \vdash B$ iff $B$ can be proved from $A$ using the formula $A$ *exactly once*. This is the *linearity*.
- MALL is more similar to classical logic than to intuitionistic logic, and nevertheless its cut elimination procedure is deterministic.

$$\dfrac{}{A \vdash A}\ \text{identity} \qquad \dfrac{\Gamma \vdash A \quad \Delta, A \vdash B}{\Gamma, \Delta \vdash B}\ \text{cut}$$

$$\dfrac{\Gamma, A \vdash B}{\Gamma \vdash A \multimap B}\ \multimap_{\mathcal{R}} \qquad \dfrac{\Gamma \vdash A \quad \Delta, B \vdash C}{\Gamma, \Delta, A \multimap B \vdash C}\ \multimap_{\mathcal{L}}$$

$$\dfrac{\Gamma \vdash A \quad \Delta \vdash B}{\Gamma, \Delta \vdash A \otimes B}\ \otimes_{\mathcal{R}} \qquad \dfrac{\Gamma, A, B \vdash C}{\Gamma, A \otimes B \vdash C}\ \otimes_{\mathcal{L}}$$

$$\dfrac{\Gamma \vdash A \quad \Gamma \vdash B}{\Gamma \vdash A \& B}\ \&_{\mathcal{R}} \qquad \dfrac{\Gamma, A \vdash C}{\Gamma, A \& B \vdash C}\ \&_{\mathcal{L}}$$

$$\dfrac{\Gamma \vdash A}{\Gamma \vdash \forall \alpha . A}\ \forall_{\mathcal{R}}\ (*) \qquad \dfrac{\Gamma, A\left[{}^{B}/_{\alpha}\right] \vdash B}{\Gamma, \forall \alpha . A \vdash B}\ \forall_{\mathcal{L}}$$

MALL

$$\dfrac{\Gamma \vdash A}{\Gamma, !B \vdash A}\ \text{weakening} \qquad \dfrac{!A, !A, \Gamma \vdash B}{!A, \Gamma \vdash B}\ \text{contraction}$$

$$\dfrac{!A_1, \ldots, !A_k \vdash B \quad k \geq 0}{!A_1, \ldots, !A_k \vdash !B}\ \text{promotion} \qquad \dfrac{\Gamma, B \vdash A}{\Gamma, !B \vdash A}\ \text{dereliction}$$

Figure 2.1: The rules of Intuitionistic Linear Logic, in a sequent calculus style. (*) Provided $\alpha$ does not occur free in $\Gamma$.

- It appears that the conjunction and disjunction can follow two different behaviours, *additive* or *multiplicative*; MALL uses 4 different connectives instead of the traditional 2.
- MALL is a very weak system. LL extends it using a modality, *bang*, written !. $!A$ means that the formula $A$ can be used as many times as needed in the proofs. This is formally obtained allowing the *contraction* and *weakening* rules restricted to !-formulas.
- *Promotion*, *dereliction*, *digging* are the rules that control the introduction of the modalities in LL.

The rules of (the intuitionistic version of) LL are recalled in Figure 2.1. We privilege the intuitionistic version of such rules because we will never consider the classical ones in our work. An alternative formulation of LL is in Figure 2.2. Such a version is equivalent, in the sense that every formula provable in one system is also provable in the other system. Notice however that, with this second formulation, it is no longer true that "*the cut is the only elimination rule*". Indeed, the *digging* eliminates a symbol of modality. The *soft promotion* is also known as *functorial promotion*; in our work we shall always use this formulation of the promotion rule.

**Affine Logic.** It is a version of LL that allows the use of weakening but not of contraction. Here, every argument has to be used *at most once* instead of *exactly once*. The price to pay is a longer definition of the cut elimination procedure; the more, the cut elimination procedure is not deterministic anymore. This is why one usually deals with the *intuitionistic* version of the affine logic. We have already faced this problem in the Introduction, talking about ILAL.

$$\frac{}{A \vdash A}\ \text{identity} \qquad \frac{\Gamma \vdash A \quad \Delta, A \vdash B}{\Gamma, \Delta \vdash B}\ \text{cut}$$

$$\frac{\Gamma, A \vdash B}{\Gamma \vdash A \multimap B}\ \multimap_{\mathcal{R}} \qquad \frac{\Gamma \vdash A \quad \Delta, B \vdash C}{\Gamma, \Delta, A \multimap B \vdash C}\ \multimap_{\mathcal{L}}$$

$$\frac{\Gamma \vdash A \quad \Delta \vdash B}{\Gamma, \Delta \vdash A \otimes B}\ \otimes_{\mathcal{R}} \qquad \frac{\Gamma, A, B \vdash C}{\Gamma, A \otimes B \vdash C}\ \otimes_{\mathcal{L}}$$

$$\frac{\Gamma \vdash A \quad \Gamma \vdash B}{\Gamma \vdash A \& B}\ \&_{\mathcal{R}} \qquad \frac{\Gamma, A \vdash C}{\Gamma, A \& B \vdash C}\ \&_{\mathcal{L}}$$

$$\frac{\Gamma \vdash A}{\Gamma \vdash \forall \alpha . A}\ \forall_{\mathcal{R}}\ (*) \qquad \frac{\Gamma, A\left[{}^{B}/_{\alpha}\right] \vdash B}{\Gamma, \forall \alpha . A \vdash B}\ \forall_{\mathcal{L}}$$

$$\frac{\Gamma \vdash A}{\Gamma, !B \vdash A}\ \text{weakening} \qquad \frac{!A, !A, \Gamma \vdash B}{!A, \Gamma \vdash B}\ \text{contraction}$$

$$\frac{A_1, \ldots, A_k \vdash B \quad k \geq 0}{!A_1, \ldots, !A_k \vdash !B}\ \text{soft promotion} \qquad \frac{\Gamma, A \vdash B}{\Gamma, !A \vdash B}\ \text{dereliction} \qquad \frac{\Gamma, !!A \vdash B}{\Gamma, !A \vdash B}\ \text{digging}$$

Figure 2.2: Another formulation of Intuitionistic Linear Logic, in a sequent calculus style, *almost* equivalent to the one in Figure 2.1. (*) Provided $\alpha$ does not occur free in $\Gamma$.

**Proof nets.** Together with LL, in [Gir87] are presented proof nets. For example, the (intuitionistic) proof

$$\cfrac{\cfrac{\cfrac{\cfrac{\cfrac{}{A \vdash A}\ \text{identity}}{!A \vdash !A}\ \text{promotion} \qquad \cfrac{}{B \vdash B}\ \text{identity}}{!A, B \vdash !A \otimes B}\ \otimes\text{-right}}{!A, !A, B \vdash !A \otimes B}\ \text{weakening}}{!A, B \vdash !A \otimes B}\ \text{contraction}$$

can be represented by the following (intuitionistic) proof net:

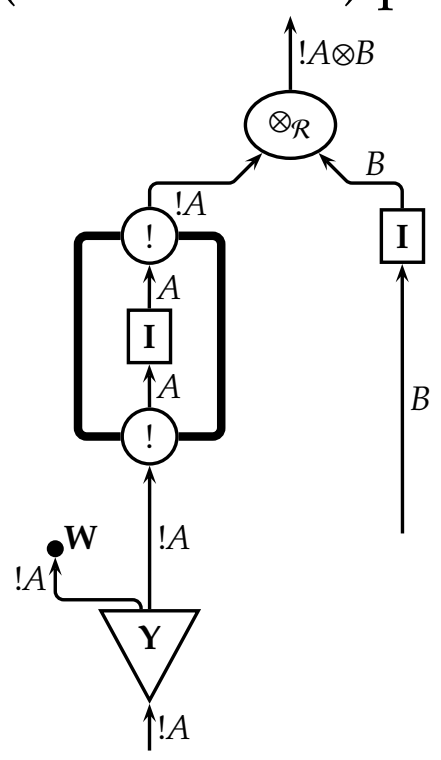

where each rule is represented as a node of the graph. However, not all the rules can be represented as nodes. Some of the rules (in part. the *promotion rule* and the rules for *additives*) involve not only one formula, but the whole context. In order to correctly handle them, one

$$\frac{}{A \vdash A}\text{ identity} \qquad \frac{\Gamma \vdash A \quad \Delta, A \vdash B}{\Gamma, \Delta \vdash B}\text{ cut}$$

$$\frac{\Gamma, A \vdash B}{\Gamma \vdash A \multimap B}\multimap_{\mathcal{R}} \qquad \frac{\Gamma \vdash A \quad \Delta, B \vdash C}{\Gamma, \Delta, A \multimap B \vdash C}\multimap_{\mathcal{L}}$$

$$\frac{\Gamma \vdash A \quad \Delta \vdash B}{\Gamma, \Delta \vdash A \otimes B}\otimes_{\mathcal{R}} \qquad \frac{\Gamma, A, B \vdash C}{\Gamma, A \otimes B \vdash C}\otimes_{\mathcal{L}}$$

$$\frac{\Gamma \vdash A \quad \Gamma \vdash B}{\Gamma \vdash A \& B}\&_{\mathcal{R}} \qquad \frac{\Gamma, A \vdash C}{\Gamma, A \& B \vdash C}\&_{\mathcal{L}}$$

$$\frac{\Gamma \vdash A}{\Gamma \vdash \forall \alpha . A}\forall_{\mathcal{R}}\ (*) \qquad \frac{\Gamma, A\left[{}^{B}/_{\alpha}\right] \vdash B}{\Gamma, \forall \alpha . A \vdash B}\forall_{\mathcal{L}}$$

$$\frac{\Gamma \vdash A}{\Gamma, !B \vdash A}\text{ weakening} \qquad \frac{!A, !A, \Gamma \vdash B}{!A, \Gamma \vdash B}\text{ contraction}$$

$$\frac{A_1, \ldots, A_k \vdash B \quad k \geq 0}{!A_1, \ldots, !A_k \vdash !B}\text{ soft promotion} \qquad \frac{\overbrace{A, \ldots, A}^{n}, \Gamma \vdash B \quad n \geq 0}{!A, \Gamma \vdash B}\text{ multiplexor}$$

Figure 2.3: The rules of Soft Linear Logic, in a sequent calculus style. (*) Provided $\alpha$ does not occur free in $\Gamma$.

has to use additional structures over the graph called *boxes*. They will be presented more in detail in Section 3.1.

In Section 3.1 we will describe exactly the formulæ and the proof nets that we will use. Here, we just anticipate that (i) we will deal with an affine logic, and (ii) we won't deal with additive connectives.

## 2.4 Light logics, ILAL and SLL

Under the Curry-Howard correspondence, the concept of *complexity of programs* is also a *complexity of proofs*, and a *class of complexity* is in fact a *fragment of the (linear) logic*. A *light logic* is a fragment e.g. of LL , and can be used to identify a computational class. This is a relevant change in perspective in the study of CC: no more *Turing Machines* has to be considered, but *proofs*. Of course, we said that there is a strict correspondence between proofs and $\lambda$-terms, so a possible objection is the following: the real *step forward* is done passing from Turing machines to $\lambda$-terms, and there is no real further advantage passing from $\lambda$-terms to proofs. However, using light logics it is easy to characterize classes of proofs, while using $\lambda$-calculus it is less easy to characterize classes of programs.

A certain number of logical systems have already been introduced to this purpose; we recall in particular that both Intuitionistic Light Affine Logic (ILAL) [AR02] and Soft Linear Logic (SLL) [Laf04] characterize the class of polynomial time-calculable functions.

Two words are worth writing about these two systems. The rules of ILAL have already been reported in Figure 1.1; SLL is in Figure 2.3. The two systems are deeply different. ILAL is said *stratified*, while the *multiplexor* rule of SLL is inherently *unstratified*. A formal definition of stratification will be find in Section 3.3, and we don't want anticipate it here. Intuitively, this means that the multiplexor rule introduces a modality ("!") outside a *single* formula of a sequent, but not outside all the other formulas of the sequent. This is forbidden in ILAL, whose proofs are rigidly organized in *levels* of modalities. On the other side, ILAL poses some restrictions in the number of premises present in the Promotion rule. A comparison between ILAL and SLL can be found, e.g., in [GRV].

Now, we want underline some more technical aspect of ILAL. As we anticipated in the Introduction, the free weakening has at least two negative consequences. (a) The classical version of this system is non-confluent, and this is the reason why ILAL is intuitionistic, while LLL is not. (b) In order to normalize the proofs, a long list of reduction rules is needed, and the more, ILAL must contain the dæmon rule (**h**), which makes ILAL inconsistent [AR02]. We underline that the rule **h** is only required during the reductions, and *not* in the real proofs. **h** is a bit like the *imaginary unit*: it may be needed during the computations, but it does not appear in the input data, nor in *most of* the results.

## 2.5 Implicit Computational Complexity

Light logics are part of a larger area of research called Implicit Computational Complexity (ICC), that tries, using different strategies, to characterize complexity classes without referring to Turing Machines. For example, a different approach relates to the more traditional *recursion theory*. In this perspective, a computational class must correspond to some subclass of the algebra of functions of Kleene. We will deal about this approach in the next sections.

## 2.6 Algebras of functions, and Kleene functions

As we already pointed out, we focus on two different traditions for representing complexity classes. One, is the one of light logics; the other one, is the one of algebras of functions, derived from the functions of Kleene. Here, we recall this second approach.

An **algebra of functions** is a set of functions $\mathcal{A}$ containing some *base functions* and closed under some *operators over functions*. The elements of $\mathcal{A}$ may be seen as *functions*, but also as *programs* that calculate them. So, we shall both talk about functions in $\mathcal{A}$ and programs in $\mathcal{A}$.

All this functions are *first-order*, in the sense that they are defined over a common *ground set*; for example, $\mathbb{N}$, or $\mathbb{N} \times \ldots \times \mathbb{N}$. We do not handle functions of functions and such things. Specifically, we deal with algebras of functions *over the binary words* $\mathbf{W}$, assuming the natural correspondence between $\mathbf{W}$ and the natural numbers $\mathbb{N}$.

The Kleene functions are the most well-known algebra of functions. This is defined as the smallest set of partial and first-order functions $\mathscr{F}$ from $\mathbb{N} \times \ldots \times \mathbb{N}$ to $\mathbb{N}$ such that:

- $\mathscr{F}$ contains the constant function 0, the successor, the projections;
- $\mathscr{F}$ is closed under Composition, Recursion, and Minimization, as described in Figure 2.4. There, $\overrightarrow{x}^{\mathsf{n}}$ denotes a vector with $\mathsf{n}$ components.

As we already pointed out, the Kleene functions are in fact *algorithms* that calculate functions. So we may refer to them also as Kleene programs. Second observation, the *Minimalization* is the only operator that allows the construction of *non-total* functions. At last, the reason that make important such functions:

THEOREM 2.6.1
The Kleene functions correspond exactly to the Turing-computable functions.

In the following two Sections we will study two restrictions of the Kleene algebra that capture exactly the FPTIME class. The keypoints will be (i) to forbid the minimization schema; and (ii) to *restrict* the power of the recurrence schema. It will be useful the following terminology. Inside a recursion schema of Figure 2.4, the last argument of $h$ is called **critical argument**. The first argument of $f$, that is also the first argument of $h$, is the **recurrence argument**, and we say that *it drives the unfolding of* $f$.

## 2.7 Safe Recursion on Notation

Safe Recursion on Notation, or SRN, also called BC, is the algebra of total and first-order functions over binary words recalled in Figure 2.5. The base functions are the successors, the predecessor, the projections and the conditional. SRN is the smallest set of functions containing the base functions and closed under Safe Recursion and Safe Composition. The key feature of SRN is the presence of two kinds of variables, called resp. *normal* and *safe*. Each variable can be used only in a specific way, depending on its kind. In particular, safe variables cannot be used as recurrence arguments.

From an extensional point of view, the functions in SRN are exactly the functions in FPTIME [BC92]. On the intentional side, SRN can be seen as a programming language; every SRN program executes in polynomial time under a certain strategy [BW96], and every FPTIME function has an SRN program computing it. The system SRN is inspired to the one of [Cob65], but it really improves this latter: the system presented by Cobham explicitly puts a polynomial upper bound on the recursion scheme, while in SRN such a bound is a consequence of the structure of the system.

We shall often write $\mathsf{SRN}^{\mathsf{n};\mathsf{s}}$ to denote the SRN functions with $\mathsf{n}$ normal arguments and $\mathsf{s}$ safe arguments.

## 2.8 Ramified Recurrence

Leivant's Ramified Recurrence system [Lei93] is an algebra of total and first-order functions with ground the *tiered words* $\mathbf{W}_i$, $i \in \mathbb{N}$. All the sets $\mathbf{W}_i$ are equal to the sets of the binary words; however, they must be kept disjoint, as they are used in different ways. $\mathbf{W}_i$ is the set of words of *sort* $i$. The base functions are the two word successors and all the projections; the algebra is the smallest set of functions containing the base functions and closed by composition and *predicative recurrence*:

$$f = \mathrm{rec}[g_\epsilon, g_0, g_1] \quad \text{defined by} \quad \begin{cases} f(\epsilon, \vec{x}) & = g_\epsilon(\vec{x}) \\ f(w0, \vec{x}) & = g_0(w, \vec{x}, f(w, \vec{x})) \\ f(w1, \vec{x}) & = g_1(w, \vec{x}, f(w, \vec{x})) \end{cases}$$

$$f'\left(\overrightarrow{x}^{\mathsf{n}}\right) = f\left(g_1\left(\overrightarrow{x}^{\mathsf{n}}\right), \dots, g_{\mathsf{n}'}\left(\overrightarrow{x}^{\mathsf{n}}\right)\right) \tag{2.2}$$

$$\begin{cases} f\left(0, \overrightarrow{y}^{\mathsf{n}}\right) = g\left(\overrightarrow{y}^{\mathsf{n}}\right) \\ f\left(x+1, \overrightarrow{y}^{\mathsf{n}}\right) = h\left(x, \overrightarrow{y}^{\mathsf{n}}, f\left(x, \overrightarrow{y}^{\mathsf{n}}\right)\right) \end{cases} \tag{2.3}$$

$$\mu(f)\left(\overrightarrow{y}^{\mathsf{n}}\right) = \begin{cases} \min\left\{z \in \mathbb{N} \mid f\left(z, \overrightarrow{y}^{\mathsf{n}}\right) = 0\right\} & \text{if it exists,} \\ \text{undefined} & \text{else.} \end{cases} \tag{2.4}$$

Figure 2.4: The Kleene functions: Composition (2.2), Recursion (2.3) and Minimization (2.4).

$$\mathsf{s}_i(\,;x) = 2x + i \qquad i \in \{0,1\} \tag{2.5}$$

$$\mathtt{pred}(\,;2x+i) = x \tag{2.6}$$

$$\pi_k^{\mathsf{n},\mathsf{s}}\left(\overrightarrow{x}^{\mathsf{n}}; \overrightarrow{y}^{\mathsf{s}}\right) = \begin{cases} x_k & \text{if } x \leq \mathsf{n} \\ y_k & \text{if } x > \mathsf{n} \end{cases} \tag{2.7}$$

$$\mathtt{cond}(\,;a,b,c) = \begin{cases} b & \text{if } a \text{ odd} \\ c & \text{if } a \text{ even} \end{cases} \tag{2.8}$$

$$f'\left(\overrightarrow{x}^{\mathsf{n}}; \overrightarrow{y}^{\mathsf{s}}\right) = f\Big(g_1\left(\overrightarrow{x}^{\mathsf{n}};\right), \dots, g_{\mathsf{n}'}\left(\overrightarrow{x}^{\mathsf{n}};\right); h_1\left(\overrightarrow{x}^{\mathsf{n}}; \overrightarrow{y}^{\mathsf{s}}\right), \dots, h_{\mathsf{s}'}\left(\overrightarrow{x}^{\mathsf{n}}; \overrightarrow{y}^{\mathsf{s}}\right)\Big) \tag{2.9}$$

$$\begin{cases} f\left(\mathbb{0}, \overrightarrow{x}^{\mathsf{n}}; \overrightarrow{y}^{\mathsf{s}}\right) = g\left(\overrightarrow{x}^{\mathsf{n}}; \overrightarrow{y}^{\mathsf{s}}\right) \\ f\left(\mathsf{s}_i(w), \overrightarrow{x}^{\mathsf{n}}; \overrightarrow{y}^{\mathsf{s}}\right) = h_i\left(w, \overrightarrow{x}^{\mathsf{n}}; \overrightarrow{y}^{\mathsf{s}}, f\left(w, \overrightarrow{x}^{\mathsf{n}}; \overrightarrow{y}^{\mathsf{s}}\right)\right) & i \in \{0,1\} \end{cases} \tag{2.10}$$

---

$$f'\left(\overrightarrow{x}^{\mathsf{n}}; \overrightarrow{y_1}^{\mathsf{s}_1}, \dots, \overrightarrow{y_{\mathsf{s}'}}^{\mathsf{s}_{\mathsf{s}'}}\right) = f\Big(g_1\left(\overrightarrow{x}^{\mathsf{n}};\right), \dots, g_{\mathsf{n}'}\left(\overrightarrow{x}^{\mathsf{n}};\right); h_1\left(\overrightarrow{x}^{\mathsf{n}}; \overrightarrow{y_1}^{\mathsf{s}_1}\right), \dots, h_{\mathsf{s}'}\left(\overrightarrow{x}^{\mathsf{n}}; \overrightarrow{y_{\mathsf{s}'}}^{\mathsf{s}_{\mathsf{s}'}}\right)\Big) \tag{2.11}$$

$$\begin{cases} f\left(\mathbb{0}, \overrightarrow{x}^{\mathsf{n}}; \overrightarrow{y}^{\mathsf{s}}\right) = g\left(\overrightarrow{x}^{\mathsf{n}}; \overrightarrow{y}^{\mathsf{s}}\right) \\ f\left(\mathsf{s}_i(w), \overrightarrow{x}^{\mathsf{n}}; \overrightarrow{y}^{\mathsf{s}}\right) = h_i\left(w, \overrightarrow{x}^{\mathsf{n}}; f\left(w, \overrightarrow{x}^{\mathsf{n}}; \overrightarrow{y}^{\mathsf{s}}\right)\right) & i \in \{0,1\} \end{cases} \tag{2.12}$$

Figure 2.5: SRN: (2.5) Successors, (2.6) Predecessor, (2.7) Projections, (2.8) Conditional, (2.9) Safe Composition, (2.10) Safe Recursion. (2.11) and (2.12): Linear variants of Recursion and Composition.

provided that:

$$\begin{array}{rcl} g_\epsilon & : & \mathcal{W} \rightarrow \mathbf{W}_n \\ g_0 & : & \mathbf{W}_m \times \mathcal{W} \times \mathbf{W}_n \rightarrow \mathbf{W}_n \\ g_1 & : & \mathbf{W}_m \times \mathcal{W} \times \mathbf{W}_n \rightarrow \mathbf{W}_n \\ f & : & \mathbf{W}_m \times \mathcal{W} \rightarrow \mathbf{W}_n \end{array}$$

for some $\mathcal{W} = \mathbf{W}_{i_1} \times \ldots \times \mathbf{W}_{i_k}$ and $m > n$. The *ramified recurrence* hence states that the sort of the recurrence argument ($m$) must be strictly greater than the sort of the critical argument ($n$).

These functions are exactly the functions in FPTIME.

## 2.9 Relationship between light logics and algebras of functions

At a first glance, there are similarities between ILAL and systems based on predicative recurrence. They all rely on a notion of *stratification*, i.e. distinction of objects in classes that can only communicate each other under particular restrictions. Moreover, in ILAL it is easy to encode an *iteration schema* similar to the recursion schema: whenever $h \triangleright A \vdash A$ is (a proof net calculating) a function from $A$ to itself, it is possible to find a proof net $f \triangleright \mathbb{W}, A \vdash \S A$ calculating the iterates of $h$. $\mathbb{W} = \forall\alpha.!(\alpha \multimap \alpha) \multimap !(\alpha \multimap \alpha) \multimap \S(\alpha \multimap \alpha)$ is a type that encodes Church words in ILAL. Notice that the output of $f$ is not of type $A$, but $\S A$, implying that $f$ cannot be iterated any more.

Following such an idea, it has been proved [MO04] that it is possible to embed a small fragment $\mathsf{BC}^-$ of SRN into ILAL. This means that it exists a map that *compositionally* translates a program $t\left(\overrightarrow{x}^{\mathsf{n}}; \overrightarrow{y}^{\mathsf{s}}\right) \in \mathsf{BC}^{-\mathsf{n};\mathsf{s}}$ into some proof net of ILAL, whose type is $\underbrace{\mathbb{W}, \ldots, \mathbb{W}}_{\mathsf{n}}, \underbrace{\S^m \mathbb{W}, \ldots, \S^m \mathbb{W}}_{\mathsf{s}} \vdash \S^m \mathbb{W}$ for some $m \geq 0$, being $\mathbb{W}$ as before. $\mathsf{BC}^-$ contains the base functions and it is closed under Linear Safe Recursion (2.12) and Linear Safe Composition (2.11). At the moment it is not known if $\mathsf{BC}^-$ contains all the FPTIME functions, or not. Surely, it does not contain all the SRN programs.

However, it is not possible to extend this result, using the same encoding, to the whole SRN. This last observation, just sketched in [MO04], is quite interesting; we can generalize it in the following way. Let us imagine to add $\texttt{concat}(\,; x, y)$ among the base functions of SRN, with the intended semantics. We get a system SRN′ that is no more polytime, because it contains the following function exp:

$$\begin{array}{rcl} \texttt{dup}(w; x) & = & \texttt{concat}(\,; \pi_2^{1,1}(w; x), \pi_2^{1,1}(w; x)) \\ \left\{ \begin{array}{r} \texttt{exp}(0;\,) \\ \texttt{exp}(sw;\,) \end{array} \right. & \begin{array}{c} = \\ = \end{array} & \begin{array}{l} 1 \\ \texttt{dup}(w; \texttt{exp}(w;)) \end{array} \end{array}$$

Observe, in particular, that (i) in ILAL there exists a proof net representing $\texttt{concat}(\,; x, y)$; (ii) exp cannot be defined with Linear Safe Composition / Recursion, according to the fact that ILAL is polytime. In particular, the contraction on the safe variables needed to build $\texttt{dup}(\,; x)$ does not exist.

We can conclude observing that ILAL is somehow *orthogonal* to SRN, with regard to the encoding of Murawski and Ong. As a consequence, if we want to embed SRN inside some extension of ILAL, it will be necessary to slightly change the chosen encoding, or the system ILAL. This is exactly what we shall do.

## 2.10 Relations, Orders, Preorders.

In some Sections of Chapter 4 we will need the following, standard, notions. A **binary relation** $R$ over elements of a set $X$ is $R \subseteq X \times X$. The relation is **reflexive** if $\forall x \in X(xRx)$; **antireflexive** if $\forall x \in X \neg(xRx)$; **symmetric** if $\forall x, y \in X(xRy \to yRx)$; **antisymmetric** if $\forall x, y \in X(xRy \wedge yRx \to x = y)$; **connected** or **total** if $\forall x, y \in X(xRy \vee yRx)$.

A **(partial) preorder** is a reflexive and transitive relation. A **(partial) order** is a reflexive, antisymmetric and transitive relation. A **linear order**, or **total order**, is an order that is also connected. Similarly one defines **linear (or total) preorders**. Totality implies reflexivity, so that a linear order is just a transitive, antisymmetric and total relation. A **strict order** is an antireflexive, antisymmetric and transitive relation. Please notice that (i) a strict order is *not* an order; (ii) it is possible to define a **linear strict order**, however this is necessarily not connected, so to define it it is necessary to state that $\forall x, y \in X(x \neq y \to xRy \vee yRx)$. Every order naturally induces a strict order.

An **equivalence** is a reflexive, symmetric and transitive relation. If $\sim$ is an equivalence, $\overline{x} = \{y \in X \mid x \sim y\}$ is the **equivalence class** of $x$, and $X' = X/\sim\, = \{\overline{x} \mid x \in X\}$ is the **quotient set**.

- Every transitive relation $R$ can be completed to a partial preorder. Just add all and only the couples $xRx$. This is the smallest partial preorder extending $R$. We shall use this fact in Section 4.4.2.
- Given a partial preorder $R$ over $X$, it induces an equivalence over $X$: $x \sim y$ in $X$ iff $xRy$ and $yRx$. Then, $R$ naturally induces a partial order $R'$ over the quotient set $X'$:

  $$\overline{x}R'\overline{y} \qquad \text{iff} \qquad xRy$$

  We shall use this fact in Section 4.4.3.
- Every partial order can be extended to a linear order, but not in a unique way; in literature there are several algorithms for doing that. Probably we shall never use this fact.

## 2.11 Set-theoretical notations

In this Section we write down some of the notational conventions that are used in this work.

- $\omega$ is (essentially) the set of natural numbers. $2^\omega$ is (essentially) the set of real numbers.
- Let $A, B$ are sets. $B^A$ is the set of functions from $A$ to $B$. It is also denoted $A \to B$.
- If $f$ is a function, $f''(A) \stackrel{\text{def}}{=} \{f(x) \mid x \in A\}$ is the **image** of $A$ under $f$.
- $\mathscr{P}(A)$ is the **power set** of $A$.

- $\mathscr{P}_{<\omega}(A)$ is the set of finite subsets of $A$.
- $\mathscr{M}(A)$ is the set of **multisets** on $A$:

$$B \in \mathscr{M}(A) \longleftrightarrow (B \text{ is a function}, B : A \to \mathbb{N})\,.$$

  We will call **support** of $B \in \mathscr{M}(A)$ the set $\{a \in A \mid B(a) \neq 0\} \subseteq A$. If $a \in A$, $B(a)$ is the **multiplicity** of $a$ in $B$.

- $\mathscr{M}_{<\omega}(A)$ is the set of finite multisets on $A$:

$$B \in \mathscr{M}_{<\omega}(A) \longleftrightarrow (B \in \mathscr{M}(A) \wedge B \text{ has finite support})\,.$$

- $A^{<\omega}$ is the set of finite sequences of elements of $A$ (or **strings**):

$$A^{<\omega} \stackrel{\text{def}}{=} \{s : \{0, 1, \ldots, n-1\} \to A \mid n \in \omega\}\,.$$

  We write its element this way: $\vec{x} \in A^{<\omega}$, $(x_0, \ldots, x_k) \in A^{<\omega}$; $\varepsilon$ is the empty string. We use also the convention $A^* \stackrel{\text{def}}{=} A^{<\omega}$, $A^+ \stackrel{\text{def}}{=} A^{<\omega} \setminus \{\varepsilon\}$.

# Chapter 3

# The framework MS

This Chapter is devoted to the presentation of the Multimodal Stratified framework (MS). As we already said, MS is a framework that allows to study a variety of different logics, all derived from LL and ILAL, in a uniform way; such logics will be called *subsystems* (Section 3.2). MS is described in Section 3.1, in the language of proof nets. MS, as well as all its subsystems, is a *rewriting system* [KBV01]: in order to perform computations, it is necessary to describe a *rewriting relation* that transforms a proof net into another one. This relation is the *normalization of proof nets*, i.e., the procedure that eliminates as many cut nodes as possible inside a given proof net. It will be described in Section 3.3.

The last Section 3.4 contains some more technical considerations, and the reader can freely skip it. In Section 3.4.1 we will compare the framework MS with the *very* more general theory of *Abstract Rewriting Systems (ARS)*; every subsystem of MS is, in fact, an ARS. So, we can deduce some properties of MS from the theory of ARS. Section 3.4.2 discusses the definition of *size of a proof net*, comparing it with other possible definitions. We will find out that the definition we use is reasonable. At last, Section 3.4.3 shows that SLL cannot be compositionally embedded into any subsystem of MS.

## 3.1 Meta-Proof nets and Proof nets

We are going to define what is a proof net; we follow, essentially, the definition in [DL08]. However, in order to handle an arbitrary set of modalities $\mathbb{X}$, we distinguish formulæ and meta-formulæ, as well as proof nets and meta-proof nets.

Definition 3.1.1 (Meta-Formulæ and Formulæ) *Let us fix a countable set $\mathcal{V}$, whose elements will be called* **propositional variables***, and a countable set $\mathcal{X}$, whose elements will be called* **meta-modalities***. We will use $\alpha, \beta, \gamma, \ldots$ to indicate variables and $\mathfrak{m}, \mathfrak{n}, \mathfrak{p}, \ldots$ to denote meta-modalities. Let us consider the meta-formulæ generated by the following grammar:*

$$\begin{aligned} F &::= L \mid M \\ L &::= \alpha \mid F \otimes F \mid F \multimap F \mid \forall \alpha.F \\ M &::= \mathfrak{m}F \end{aligned}$$

*The set of* **meta-formulæ** *with start symbol F is denoted $\mathcal{F}$; meta-formulæ with start symbol M are called* **modal** *and their set is denoted $\mathcal{M}$; and meta-formulæ with start symbol L are* **linear or non-modal***, and their set is $\mathcal{L}$.*

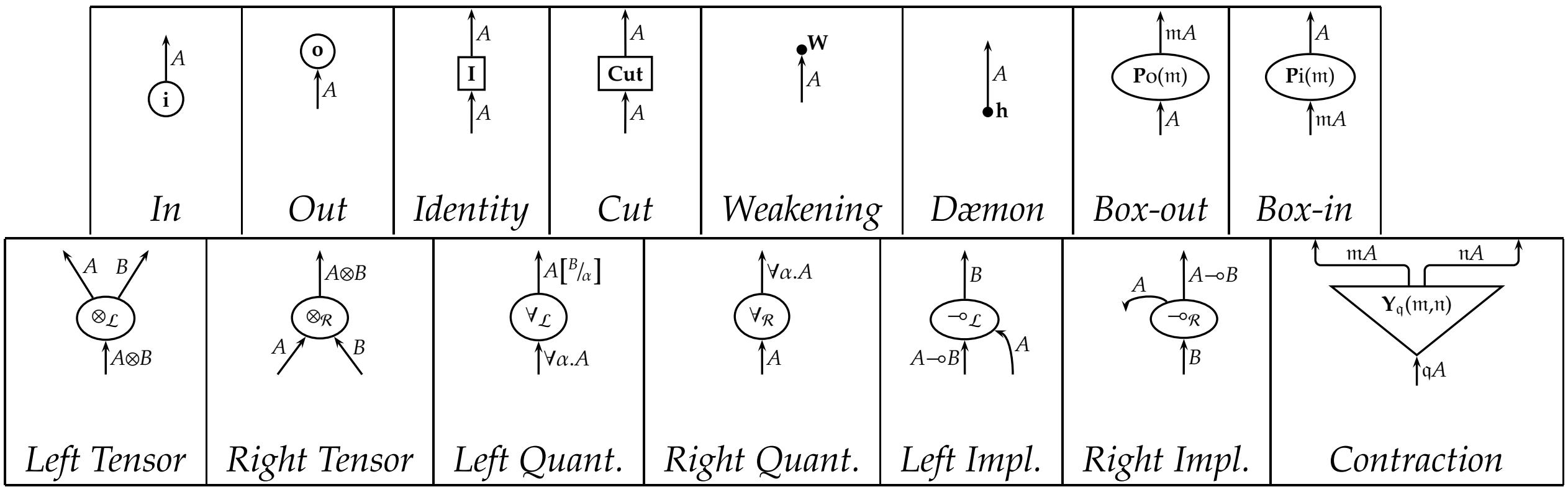

Figure 3.1: The nodes of the meta-proof nets, with $\mathfrak{m}, \mathfrak{n}, \mathfrak{q} \in \mathcal{X}$, $A, B \in \mathcal{F}$.

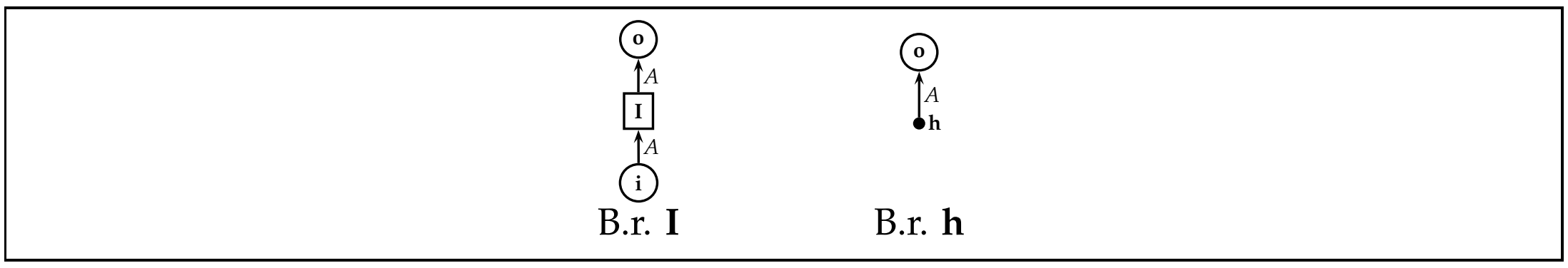

Figure 3.2: The meta-building rules (b.r.) of MS: base cases. $A \in \mathcal{F}$.

*For every fixed set $\mathbb{X}$, the* **formulæ (with modalities in $\mathbb{X}$)** *are obtained instantiating the parameters $\mathfrak{m} \in \mathcal{X}$ with modalities $\mathfrak{m} \in \mathbb{X}$. The set of formulæ with modalities in $\mathbb{X}$ is $\mathcal{F}_{\mathbb{X}}$. Formulæ obtained from the modal meta-formulæ are* **modal formulæ***, and their set is $\mathcal{M}_{\mathbb{X}}$. Formulæ obtained from the linear meta-formulæ are* **linear formulæ***, and their set is $\mathcal{L}_{\mathbb{X}}$.*

Notice in particular that a modal formula can be hidden inside a non-modal formula.

We shall use letters $A, B, C, \ldots$ to range over formulæ, or meta-formulæ; and Greek letters $\Gamma, \Delta, \Phi, \Psi$ to range over multisets of formulæ.

We shall use the notation $A\left[{}^{B}/_{y}\right]$ to denote substitution of $y$ with $B$ in the (meta-)formula $A$. Unless explicitly stated we shall make confusion between the concepts of *formula* and *occurrence of formula*. There are no restrictions, in principle, on the set $\mathbb{X}$; but in most of the cases we shall use only finite $\mathbb{X}$'s.

Definition 3.1.2 (Boxed-graph) *A* **boxed graph** *is a pair $(\mathcal{G}, \mathsf{B})$, where $\mathcal{G}$ is a graph, and $\mathsf{B}$ is a set of subgraphs of $\mathcal{G}$, that are called* **boxes***.*

A boxed graph can be drawn just as a graph, with the addition of the border of the boxes. In this way, we will seldom use the formal definition of boxed graph: on the contrary, we will draw it.

Definition 3.1.3 (Meta-proof nets and Proof nets) *A* **meta-proof net** *is a boxed graph $(\mathcal{G}, \mathsf{B})$, where: (i) $\mathcal{G}$ is a finite directed graph, whose nodes are instances of the nodes in Figure 3.1 and whose edges are labelled with meta-formulæ in $\mathcal{F}$; (ii) $(\mathcal{G}, \mathsf{B})$ can be built inductively using the* **meta-building rules** *in Figure 3.2 and 3.3, in the following way. The graphs in Figure 3.2 are meta-proof nets, with no boxes. Let us assume that the following are meta-proof nets:*

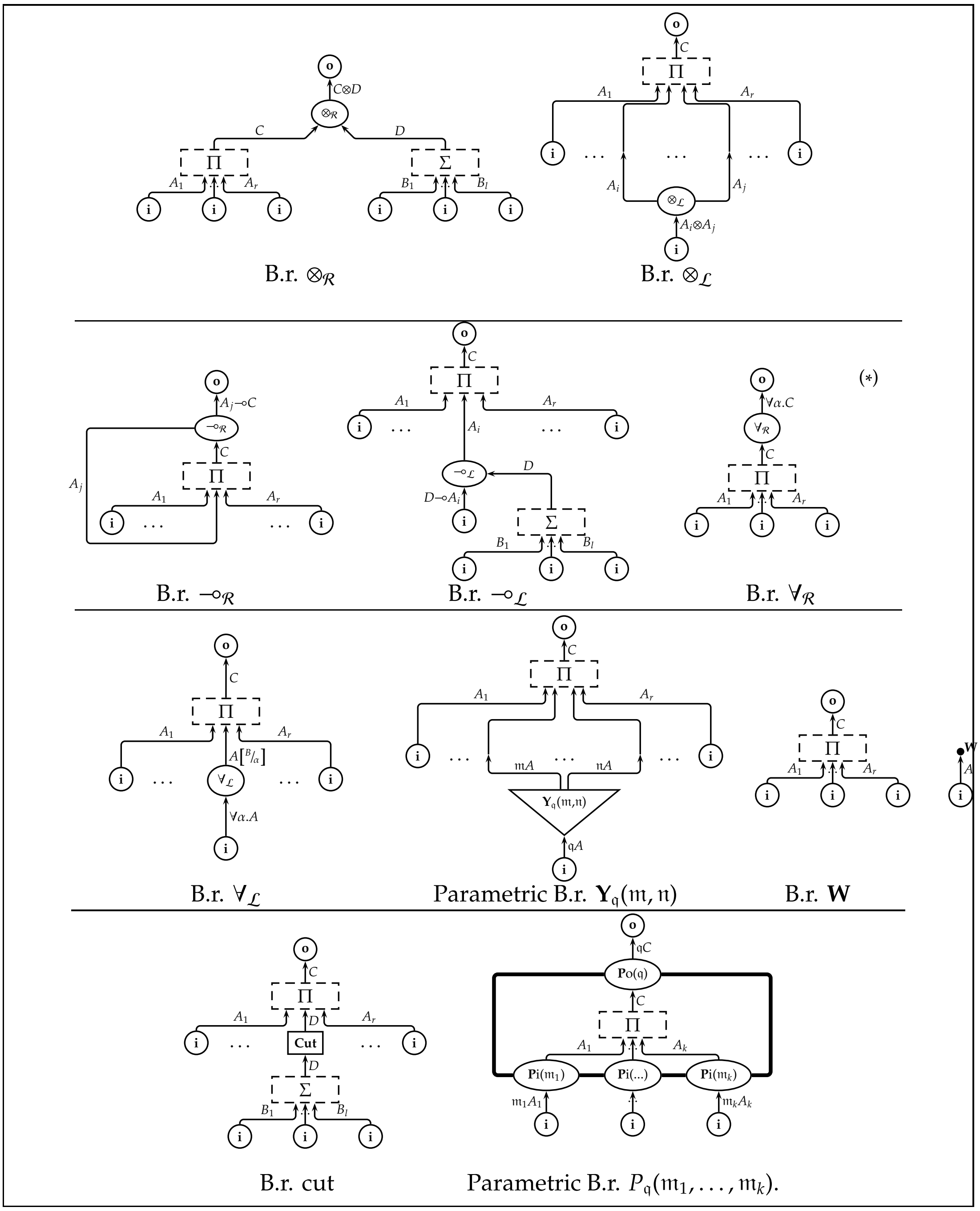

Figure 3.3: The meta-building rules (b.r.) of MS: inductive cases. $\mathfrak{m}, \mathfrak{n}, \mathfrak{q}, \mathfrak{m}_1, \ldots, \mathfrak{m}_k \in \mathcal{X}$, $A, B \in \mathcal{F}$. (*) $\alpha$ is not free in $A_1, \ldots, A_r$. Please notice that the promotion rule is in fact a *whole set* of rules, one for each $k$.

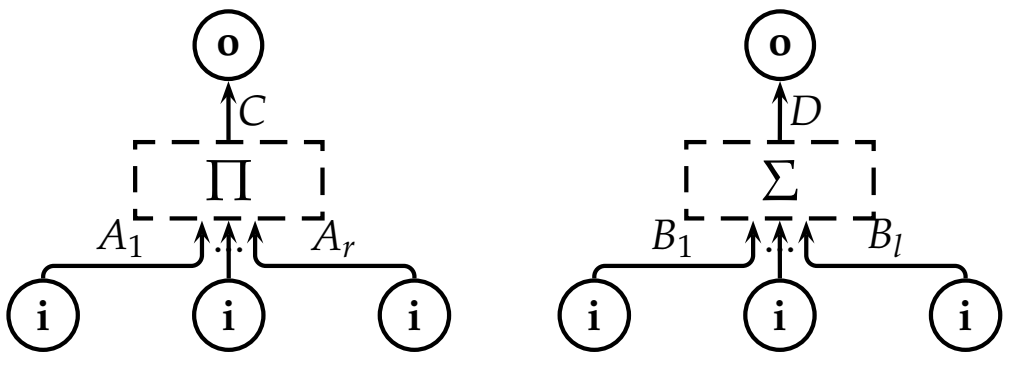

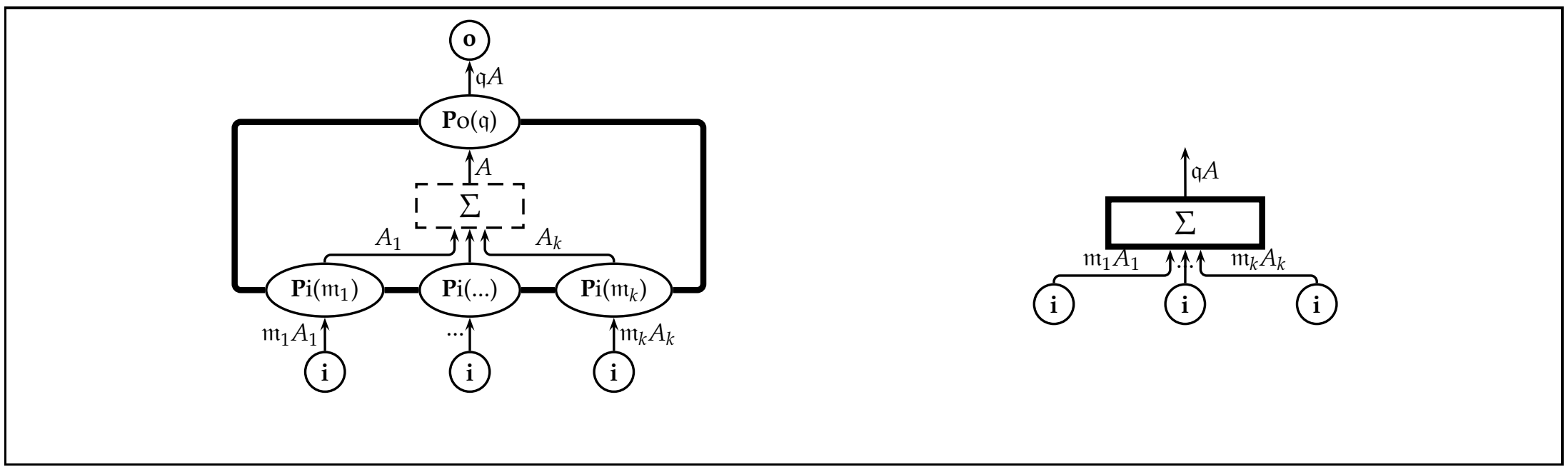

Figure 3.4: On the left, an example of box. On the right, the same box in a more compact notation.

*Then, also the graphs in Figure 3.3 are meta-proof nets.*

*For every fixed set of modalities $\mathbb{X}$, the meta-proof nets with formulæ instantiated in $\mathcal{F}_{\mathbb{X}}$ are called* **proof nets***.*

Some comments are needed about our *building rules*. Some B.r. require that the meta-formulæ labelling the edges have some precise form. For example, in the cut B.r., $A_i = D$ for some $i$, and in the contraction $A_i = \mathfrak{m}A$ and $A_j = \mathfrak{n}A$ for some $i, j$. The $\forall_{\mathcal{R}}$ B.r. satisfies the usual requirement that $x$ is not free in $A_1, \ldots, A_r$. All the meta-B.r. depend on $\mathbb{X}$, because once fixed $\mathbb{X}$, the formulæ labelling the edges must be in $\mathcal{F}_{\mathbb{X}}$. But, two meta-B.r., the *contraction* and the *promotion*, depend on $\mathbb{X}$ in a deeper way. The contraction is just *one* meta-B.r. for meta-proof nets; only when instantiated, it can give origin to many B.r. for proof nets. The case of the promotion is slightly different; there are *countably many* different meta-B.r., one for each possible $k$. Once instantiated, each of them can give origin to many B.r. for proof nets. The premises are unordered: we consider, e.g., $P_{\mathfrak{m}}(\mathfrak{n}, \mathfrak{n}, \mathfrak{q})$, $P_{\mathfrak{m}}(\mathfrak{n}, \mathfrak{q}, \mathfrak{n})$ and $P_{\mathfrak{m}}(\mathfrak{q}, \mathfrak{n}, \mathfrak{n})$ as the same B.r.. Sometimes we will draw the boxes omitting their *Po* and *Pi* nodes (see Figure 3.4).

In the case of the cut B.r., we shall use the compact notation $\Pi \overset{A_i}{\bowtie} \Sigma$ to talk about the resulting proof net. When no confusion is possible, we will simply write $\Pi \bowtie \Sigma$.

The **terminal nodes** in a proof net $\Pi$ are its **o** node and all its **i** nodes.

We shall use letters $\Pi, \Sigma, \Lambda, \Theta$ to range over proof nets.

Definition 3.1.4 (Basic Proof nets Definitions) *If $\Pi$ is a proof net or a meta-proof net:*

- *$V_\Pi$ denotes the set of its nodes;*
- *$E_\Pi$ is the set of its edges;*
- *$B_\Pi$ is the set of its $Po$ nodes (we shall often refer to the $Po$ nodes as the* **boxes** *of $\Pi$);*
- *$\alpha_\Pi : V_\Pi \to \{\mathbf{i}, \mathbf{o}, Pi, Po, \otimes_{\mathcal{R}}, \otimes_{\mathcal{L}}, \ldots\}$ is the function associating each node with its kind;*
- *$\beta_\Pi : E_\Pi \to \mathcal{F}_{\mathbb{X}}$ is the function labelling edges of $\Pi$ with formulas;*
- *if $x$ is a $Pi$ node, it lies on the border of a box $\mathrm{bo}(x) \in B_\Pi$;*
- *$P_\Pi : B_\Pi \to \mathbb{N}$ is the function telling how many premises has a box;*
- $\mathrm{boundary}(\Pi) \overset{\mathrm{def}}{=} \{v \in V_\Pi \mid \alpha_\Pi(v) \in \{\mathbf{i}, \mathbf{o}\}\}$.

Definition 3.1.5 (Principal and Secondary Inputs and Outputs) *Let $v$ be a node $\multimap_{\mathcal{L}}$ and $u$ be a node $\multimap_{\mathcal{R}}$ as in figures:*

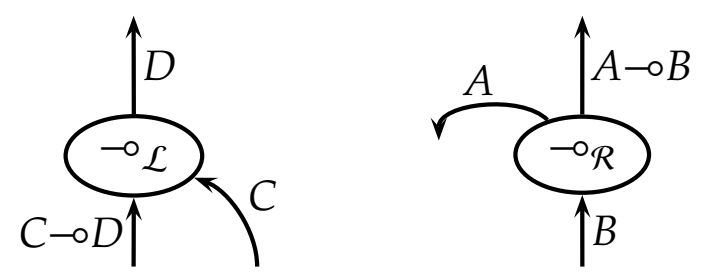

*The edge labelled $C \multimap D$ is the* **principal input** *of $v$ and the edge labelled $C$ is the* **secondary (or auxiliary) input** *of $v$. The edge labelled $A \multimap B$ is the* **principal output** *of $u$, the edge labelled $A$ is the* **secondary (or auxiliary) output** *of $u$.*

*We call also* **reverse edge of** *$u$ its secondary output.*

Remark 3.1.6 Every proof net has a single conclusion: the system is *intuitionistic*.

Remark 3.1.7 Inside a proof net $\Pi$, boxes are in bijection with *Po* nodes.

Definition 3.1.8 (Sequent Associated to a Proof net) *If $\Pi$ is a proof net with premises $\Gamma$ and conclusion $C$, we write*

$$\Pi \triangleright \Gamma \vdash C$$

*and we say $\Gamma \vdash C$ is the* **sequent associated to** *$\Pi$, or that $\Pi$* proves.

The latter definition is suggested by the usual correspondence between proof nets and sequent calculus.

**h** nodes and the **h** B.r. come directly from ILAL. They are necessary in handling reductions of free weakenings, but they have not a meaningful corresponding rule in sequent calculus.

Finally:

Definition 3.1.9 MS *is the set of all the meta-building rules. We will use labels that recall the names of the B.r.:*

$$\mathsf{MS} = \{\mathbf{I}, \mathrm{cut}, \mathbf{h}, \mathbf{W}, \otimes_{\mathcal{L}}, \otimes_{\mathcal{R}}, \multimap_{\mathcal{L}}, \multimap_{\mathcal{R}}, \forall_{\mathcal{L}}, \forall_{\mathcal{R}}, \mathbf{Y}_{\mathfrak{q}}(\mathfrak{m}, \mathfrak{n})\} \cup \{P_{\mathfrak{q}}(\mathfrak{m}_1, \ldots, \mathfrak{m}_k) \mid k \in \mathbb{N}\}$$

*For each $\mathbb{X}$, $\mathfrak{B}_{\mathbb{X}}$ is the set of the building rules of* MS instantiated *with modalities in $\mathbb{X}$.* **PN**$(\mathfrak{B}_{\mathbb{X}})$ *is the set of the proof nets built with the B.r. in $\mathfrak{B}_{\mathbb{X}}$.*

$$\mathfrak{B}_{\mathbb{X}} = \{\mathbf{I}, \mathrm{cut}, \mathbf{h}, \mathbf{W}, \otimes_{\mathcal{L}}, \otimes_{\mathcal{R}}, \multimap_{\mathcal{L}}, \multimap_{\mathcal{R}}, \forall_{\mathcal{L}}, \forall_{\mathcal{R}}\} \cup \{\mathbf{Y}_{\mathfrak{q}}(\mathfrak{m}, \mathfrak{n}) \mid \mathfrak{m}, \mathfrak{n}, \mathfrak{q} \in \mathbb{X}\} \cup \{P_{\mathfrak{q}}(\mathfrak{m}_1, \ldots, \mathfrak{m}_k) \mid k \in \mathbb{N}, \mathfrak{q}, \mathfrak{m}_1, \ldots, \mathfrak{m}_k \in \mathbb{X}\}$$

*We will call* **linear kernel** *the set of all the non-modal B.r. in $\mathfrak{B}_{\mathbb{X}}$.*

$$K = \{\mathbf{I}, \mathrm{cut}, \mathbf{h}, \mathbf{W}, \otimes_{\mathcal{L}}, \otimes_{\mathcal{R}}, \multimap_{\mathcal{L}}, \multimap_{\mathcal{R}}, \forall_{\mathcal{L}}, \forall_{\mathcal{R}}\}$$

**Measures.** We can now define the following measures on the proof nets.

We shall use letters $x, y, z, \ldots$ to range over elements of $V_\Pi \cup E_\Pi$.

Definition 3.1.10 *Let $\Pi$ be a proof net. For every $x \in V_\Pi \cup E_\Pi$ we define $\partial(x)$, the* **depth (or level) of** *$x$:*

- *If $x \in V_\Pi$, $\partial(x)$ is the maximum number of (nested) boxes that contain $x$.*
- *If $x = (u, v) \in E_\Pi$, $\partial(x) \stackrel{\text{def}}{=} \max\{\partial(u), \partial(v)\}$.*

*The* **depth of** *$\Pi$ is $\partial(\Pi) \stackrel{\text{def}}{=} \max_{w \in V_\Pi} \partial(w)$.*

In particular, if $u$ is a $Pi$ or $Po$ node, its depth is computed as $u$ were *outside* the box it is on the border of.

DEFINITION 3.1.11 *We denote* $V^d_\Pi$ *the set of nodes at dept* $d$ *of* $\Pi$*. Similarly for* $E^d_\Pi$ *and* $B^d_\Pi$.

REMARK 3.1.12 The terminal nodes only occur at level 0 of $\Pi$ and $P$-nodes only at level $d < \partial(\Pi)$.

DEFINITION 3.1.13 *Let* $\Pi$ *be a proof net.*

- *Its* **size** *is*

$$|\Pi| = \#\{\text{nodes in } \Pi\text{, at every depth}\}.$$

- *If* $0 \leq d \leq \partial(\Pi)$*, the* **size at level** $d$ *of* $\Pi$ *is*

$$|\Pi|_d = \#\{\text{nodes in } \Pi\text{, at depth } d\}$$

This way,

$$|\Pi| = \sum_{0 \leq d \leq \partial(\Pi)} |\Pi|_d.$$

We count every terminal node, every $Pi$-node and every $Po$-node as part of the size, for they are part of our syntax. Lemma 3.4.4 and remark 3.4.5 will show that we could, in fact, harmlessly ignore them.

DEFINITION 3.1.14 *If* $\Pi$ *is a proof net and* $0 \leq d \leq \partial(\Pi)$*,* $b_d(\Pi)$ *is the* **number of boxes at level** $d$*.*

Namely, this is the number of $Po$ nodes. In particular $b_{\partial(\Pi)}(\Pi) = 0$.

LEMMA 3.1.15 (BASIC RELATIONS BETWEEN MEASURES)
For every proof net $\Pi$ and for every $0 \leq d \leq \partial(\Pi)$, $b_d(\Pi) \leq |\Pi|_d$ and $b_d(\Pi) \leq |\Pi|_{d+1}$.

**Proof.** The first relation is trivial. For the second one, notice that every box contains at least one node. □

Figure 3.5 gives an example. In that proof net, $\partial(\Pi) = 2$, $|\Pi|_0 = 13$, $|\Pi|_1 = 5$ (remember that $Pi$ and $Po$ nodes are considered *outside* their box), $|\Pi|_2 = 1$, and so $|\Pi| = 22$. The boxes are: $b_0(\Pi) = 3$, $b_1(\Pi) = 1$, $b_2(\Pi) = 0$.

**Structure of proof nets.** In order to prove Lemma 3.1.18, about the structure of the proof nets, we need to introduce the notion of *serialization* of a proof net, i.e. the sequence of B.r. that make a proof net.

DEFINITION 3.1.16 (SERIALIZATION) *Let* $\Pi$ *be a proof net. By definition,* $\Pi$ *is built by a sequence of building rules* $(R_0, \ldots, R_k)$*. We say that* $(R_0, \ldots, R_k)$ *is a* **serialization** *of* $\Pi$.

Every proof net may have more than one serialization.

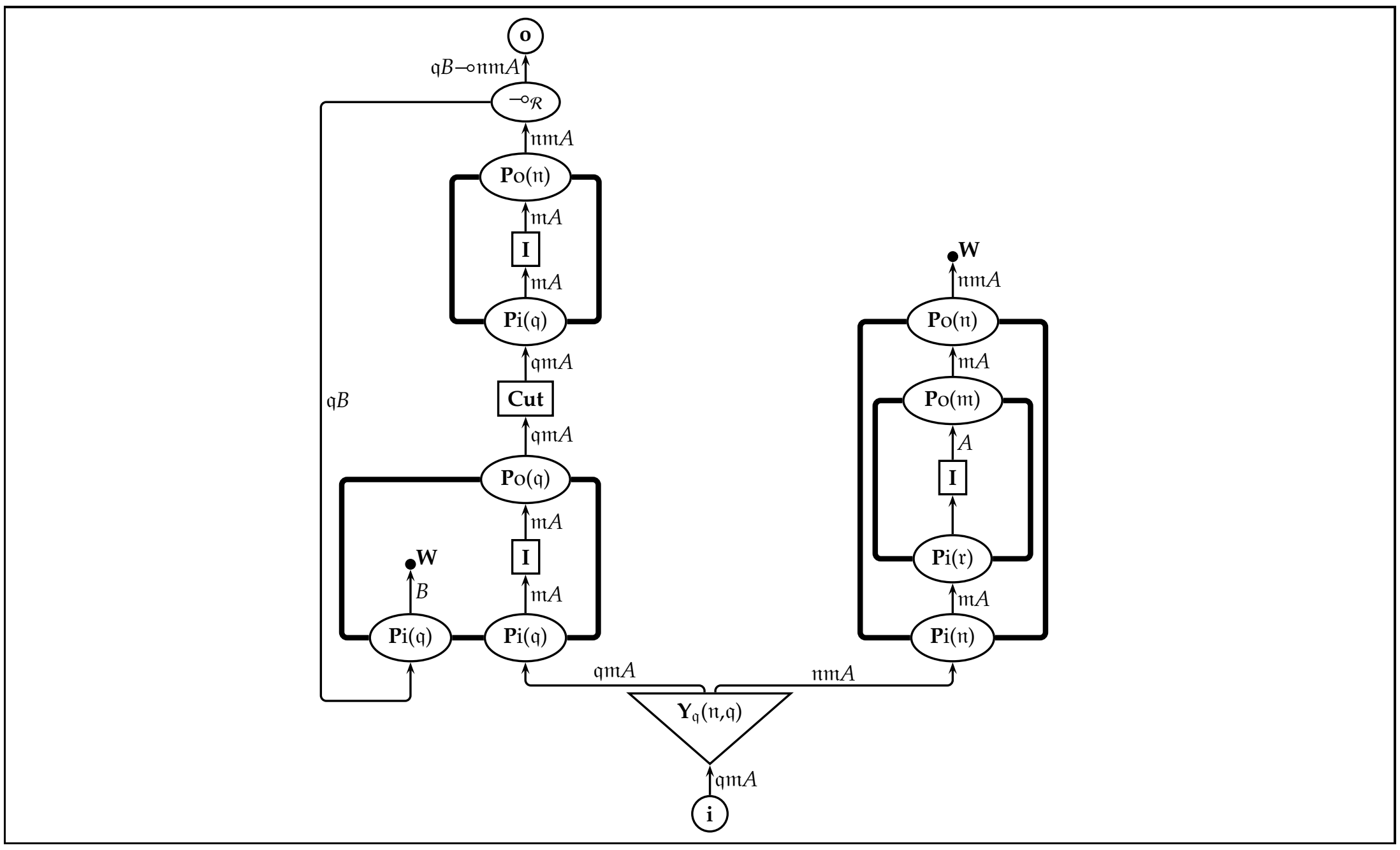

Figure 3.5: An example of proof net.

DEFINITION 3.1.17 (LEFT AND RIGHT NODES) *We call* **right nodes** *the nodes* $\otimes_{\mathcal{R}}$*,* $\multimap_{\mathcal{R}}$*,* $\forall_{\mathcal{R}}$*;* **left nodes** *the nodes* $\otimes_{\mathcal{L}}$*,* $\multimap_{\mathcal{L}}$*,* $\forall_{\mathcal{L}}$*,* **W***,* **Y***.*

LEMMA 3.1.18 (STRUCTURE OF CUT-FREE PROOF NETS)
Let $\Pi$ a proof net of MS without any cut node at level 0. We can draw the level 0 of $\Pi$ in a way such that:

1. all identity, **h** and **W** node and every box lie on a single horizontal line;
2. all right nodes lie *over* that line;
3. all left nodes lie *below* that line;
4. the output node is on the top;
5. all the input nodes are on the bottom;
6. all edges, except for reverse edges and for secondary outputs of $\multimap_{\mathcal{L}}$ nodes, are directed upward.

However, this is not the way we usually draw proof nets: we find easier to draw the secondary inputs of $\multimap_{\mathcal{L}}$ nodes from bottom to top too.

**Proof.** $\Pi$ is built up thanks to a serialization $(R_0, R_1, \ldots, R_n)$; it's easy to prove the statement by induction on $n$, just looking back at the building rules of definition 3.1.3. □

**Subnets and Modules.** In our work we shall use often the notion of *module*, i.e. a piece of a proof net.

DEFINITION 3.1.19 (MODULES GENERATED) [1] *Let* $\Pi$ *be a proof net. Let us consider a subset*

[1]Terminology comes from [DR89]

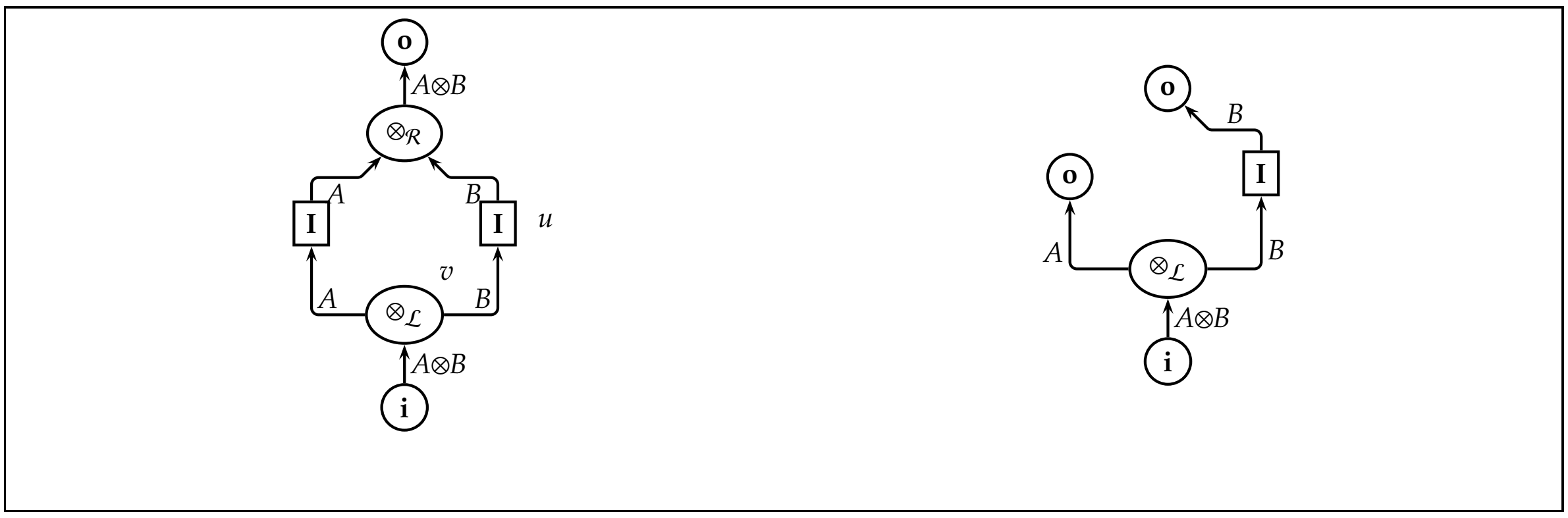

Figure 3.6: On the left, a proof net $\Pi$. On the right, the module generated by $\{u, v\}$.

$V \subseteq V_\Pi$. *The* **module** $\Sigma$ **of** $\Pi$ **generated by** $V$ *is a boxed graph defined as follows.*

- *All nodes in* $V$ *are in* $V_\Sigma$*.*
- *If* $v \in V$, $\alpha(v) \in \{Po, Pi\}$*, then* $\Sigma$ *is the whole box* $\mathrm{bo}(v)$*.*
- *If* $u, v \in V$ *and* $(u, v) \in E_\Pi$*, then* $(u, v) \in E_\Sigma$*.*
- *If* $u \in V$, $v \notin V$ *and* $(u, v) \in E_\Pi$*, then there exists an output node* $w \in V_\Sigma$*, and an edge* $(u, w) \in E_\Sigma$*.*
- *If* $u \in V$, $v \notin V$ *and* $(v, u) \in E_\Pi$*, then there exists an input node* $w \in V_\Sigma$*, and an edge* $(w, u) \in E_\Sigma$*.*

Definition 3.1.20 (Modules) *We call* **module** *a boxed graph* $\Sigma$ *if there exist* $\Pi$ *and* $V \subseteq V_\Pi$ *such that* $\Sigma$ *is the module of* $\Pi$ *generated by* $V$*.*

An example is given in Figure 3.6. For each module $\Sigma$ of $\Pi$ there exists a **canonical injection** $i_\Sigma : (V_\Sigma \setminus \mathrm{boundary}(\Sigma)) \hookrightarrow V_\Pi$. When no confusion arises, if $v \in V_\Sigma$ we will write $v \in V_\Pi$ meaning that $i_\Sigma(v) \in V_\Pi$. Notice that a module is not necessarily connected.

Remark 3.1.21 (Of Notational Nature) We shall use the notation $\Pi \triangleright \Gamma \vdash \Delta$ meaning that the module $\Pi$ has premises labelled with $\Gamma$ and conclusions labelled with $\Delta$. If $\Pi \triangleright \Gamma, \Phi \vdash \Delta$ and $\Sigma \triangleright \Gamma' \vdash \Phi, \Delta'$ are two modules, $\Pi \bowtie \Sigma \triangleright \Gamma, \Gamma' \vdash \Delta, \Delta'$ is the module obtained connecting $\Pi$ and $\Sigma$ through cut nodes, provided it is still a module. On the contrary, $\Pi \circledast \Sigma \triangleright \Gamma, \Gamma' \vdash \Delta, \Delta'$ is the module obtained connecting $\Pi$ and $\Sigma$ *just by edges*, provided it is still a module.

Definition 3.1.22 (Subnets) *A* **subnet** *of* $\Pi$ *is a module of* $\Pi$ *that is also a proof net.*

**Environment subnets, surface graphs, paths.** In this work, two notions of *path* will be defined. The first one, here below, is *similar to* the standard notion of graph-theoretical path. The other one relates with a dynamical interpretation of the proof nets, and will be presented in Section 4.3.1. *Environment Subnets* and *Surface Graphs* are objects required to define graph-theoretical paths in a simple way.

Definition 3.1.23 *Let us consider a proof net* $\Pi$ *and let* $\Sigma$ *be a module of* $\Pi$*. The* **environment subnet** *of* $\Sigma$ *is the largest proof net contained in the innermost box containing* $\Sigma$*, if it exists; it is* $\Pi$ *otherwise.*

This is also the largest subnet containing $\Sigma$ and containing only nodes of depth $\partial(x)$.

In the following, if $G$ is a graph, we use notation $V_G$ (resp. $E_G$) for the set of nodes (resp. edges) of $G$.

Definition 3.1.24 *Let $\Pi$ be a proof net and $\Sigma$ be a module of $\Pi$. Let $\Theta$ be the environment subnet of $\Sigma$. The* **surface graph** $G_\Sigma$ **generated by** $\Sigma$ *is a directed graph such that*
1. *The nodes of $G_\Sigma$ are the nodes at least depth in $\Theta$ different from Pi.*
2. *For $(u,v) \in E_\Theta$ with least depth, and $\alpha(v) \neq Pi$, $(u,v) \in E_{G_\Sigma}$.*
3. *For $(u,v) \in E_\Theta$ with least depth, and $\alpha(v) = Pi$, $(u, \mathrm{bo}(v)) \in E_{G_\Sigma}$.*
4. *No other edges are in $G_\Sigma$.*

*The* **surface graph of** $\Pi$ *is $G_\Pi \stackrel{\text{def}}{=} G_{\{x\}}$ where $x$ is any node at level 0 in $\Pi$.*

Definition 3.1.25 *Let $\Pi$ be a proof net, and $x, y \in V_\Pi \cup E_\Pi$ within the same environment graph. A* **(directed) graph-theoretic path** *$\tau$ from $x$ to $y$ is a (directed) path from $x$ to $y$ in $G_{\{x\}}$, that is a finite sequence*

$$\tau \stackrel{\text{def}}{=} (x_0, x_1, \ldots, x_n) \in {V_\Pi}^*$$

*such that $(x_i, x_{i+1}) \in E_\Pi$, $x_0 = x$ and $x_n = y$.*

**Undirected graph-theoretic paths** *are defined in similar way.*

## 3.2 Subsystems

We have defined what are MS and $\mathfrak{B}_\mathbb{X}$ for every $\mathbb{X}$. As anticipated, we are going to define a family of light logics that are called *subsystems* of MS.

Definition 3.2.1 *For every fixed set $\mathbb{X}$, a* **subsystem** *$\mathcal{P}$ of $\mathfrak{B}_\mathbb{X}$ is a subset of $\mathfrak{B}_\mathbb{X}$.* **PN**$(\mathcal{P})$ *is the set of proof nets definable with the rules in $\mathcal{P}$.*

*A subsystem of* MS *is a subsystem of some $\mathfrak{B}_\mathbb{X}$.*

Remark 3.2.2 (On the Used Terminology) We shall often refer to the elements of **PN**$(\mathcal{P})$ as **the proof nets of** $\mathcal{P}$, and we shall say that $\mathcal{P}$ **allows a rule** $R$ whenever $R \in \mathcal{P}$.

E.g. $\mathcal{P} = \{\otimes_\mathcal{L}, \mathbf{I}, \mathbf{Y}_\mathfrak{q}(\mathfrak{m},\mathfrak{n}), P_\mathfrak{m}(\mathfrak{m},\mathfrak{m})\}$ and $\mathcal{S} = \{\otimes_\mathcal{L}, \mathbf{I}, \mathbf{Y}_\mathfrak{m}(\mathfrak{m},\mathfrak{n}), P_\mathfrak{m}(\mathfrak{m}), P_\mathfrak{m}(\mathfrak{n},\mathfrak{q})\}$ are two different sets of construction rules. Please notice that $\mathfrak{B}_\mathbb{X}$ is, in fact, the largest subsystem of $\mathfrak{B}_\mathbb{X}$. We adopt the following **convention**. We shall write $P_\mathfrak{m}(\mathfrak{r}^?, \vec{\mathfrak{n}})$ to denote the set of the 2 rules $P_\mathfrak{m}(\mathfrak{r}, \vec{\mathfrak{n}})$ and $P_\mathfrak{m}(\vec{\mathfrak{n}})$; and $P_\mathfrak{m}(\mathfrak{r}^*, \vec{\mathfrak{n}})$ to denote the infinite set of rules $\{P_\mathfrak{m}(\underbrace{\mathfrak{r}, \ldots, \mathfrak{r}}_{k}, \vec{\mathfrak{n}}) \mid k \geq 0\}$.

Definition 3.2.3 *Let $\mathcal{P}$ be a subsystem. We say it is* **downward closed** *iff whenever $\mathcal{P}$ has $P_\mathfrak{m}(\mathfrak{m}_1, \ldots, \mathfrak{m}_k)$, it has also all the rules $P_\mathfrak{m}(\mathfrak{m}_{i_1}, \ldots, \mathfrak{m}_{i_l})$ for every sublist $(\mathfrak{m}_{i_1}, \ldots, \mathfrak{m}_{i_l})$ of $(\mathfrak{m}_1, \ldots, \mathfrak{m}_k)$.*

For example, a subsystem without promotion rules is trivially downward closed.

Definition 3.2.4 *Let $\mathbb{X}$ be the smallest set such that $\mathcal{P} \subseteq \mathfrak{B}_\mathbb{X}$. We say that the* **number of modalities** *of $\mathcal{P}$ is $\#\mathbb{X}$.*

In particular, whenever $\mathcal{P} \subseteq \mathfrak{B}_{\mathbb{X}}$, $\mathcal{P}$ has *at most* $\#\mathbb{X}$ modalities.

We will be interested in a particular class of subsystems:

Definition 3.2.5 *$\mathcal{P} \subseteq \mathfrak{B}_{\mathbb{X}}$ is a* **sensible subsystem** *if:*

1. *$\mathbb{X}$ is finite.*
2. *$\mathcal{P}$ contains identity, cut and weakening.*
3. *Either $\mathbb{X} = \emptyset$, or $\mathcal{P}$ contains at least one among $\multimap_{\mathcal{R}}$ and* **h***.*
4. *For every modality $\mathfrak{n} \in \mathbb{X}$, $\mathcal{P}$ contains the* closed box $P_{\mathfrak{n}}()$.
5. *$\mathcal{P}$ is* downward closed.

"Sensible" shortens *computationally sensible* subsystems, where we can erase/duplicate all the structure we might need when writing proof nets-as-programs. Sensible subsystems will be relevant thanks to Proposition 4.3.40. Notice that, even if a subsystem $\mathcal{P}$ allows a rule $R$, in general we are not sure that $R$ can be really used inside some proof net. Think for example to $\mathcal{P} = \{\mathbf{I}, P_{\mathfrak{m}}(\mathfrak{n}, \mathfrak{q})\}$: a promotion rule with 2 premises cannot be used in any proof net of $\mathbf{PN}(\mathcal{P})$. If $\mathcal{P}$ is sensible, at least all its closed promotion rules can be really used:

Fact 3.2.6 (About the Definition of Subsystem)
Point 3. in the Definition 3.2.5 can be equivalently replaced by the following:
*3bis.* "For every $\mathfrak{n}$, the promotion rule $P_{\mathfrak{n}}()$ can be used inside some proof net of $\mathbf{PN}(\mathcal{P})$."

## 3.3 Normalization

Following the tradition of light logics, we are going to define a procedure of *normalization* for the proof nets of MS. In Section 3.1 we have distinguished meta-proof nets and proof nets. Here, similarly, we distinguish between *meta-normalization steps* and *normalization steps*. The meta-*ns* reduce meta-proof nets, while the *ns* reduce proof nets of every $\mathfrak{B}_{\mathbb{X}}$. Then, we will define also a normalization procedure for all the subsystems $\mathcal{P}$ of $\mathfrak{B}_{\mathbb{X}}$, restricting *ad hoc* the normalization of $\mathfrak{B}_{\mathbb{X}}$.

The normalization tries to erase as many cuts as possible. For every $\mathbb{X}$, we shall see that $\mathfrak{B}_{\mathbb{X}}$ (recall: $\mathfrak{B}_{\mathbb{X}}$ is a subsystem of $\mathfrak{B}_{\mathbb{X}!}$) enjoys the *cut-elimination property*: it is possible to normalize a proof net arriving to another proof net cut-free (Corollary 3.3.13). This property does not hold, in general, for the subsystems of $\mathfrak{B}_{\mathbb{X}}$. This is why we adopt the term *normalization* instead of the more traditional term *cut-elimination*.

Definition 3.3.1 *We will call* **meta-normalization steps (or meta-*ns*)** *of* MS *all the rewriting rules in Figures 3.7, 3.8, 3.9, 3.10. The left side of a rule is the* **redex***, the right-hand side is the* **reduct***.*

*For every fixed set $\mathbb{X}$, a* **normalization step (or *ns*)** *is a meta-*ns *with formulæ instantiated in $\mathcal{F}_{\mathbb{X}}$. $\mathcal{R}(\mathfrak{B}_{\mathbb{X}})$ is the set of such* ns.

*Every* ns *has a* **name** *(e.g. $[\forall_{\mathcal{R}}/\forall_{\mathcal{L}}]$, $[\mathbf{W}/P]$, …).*

Definition 3.3.2 *For every fixed $\mathbb{X}$, a* **normalization strategy** *$\sigma$ is a function that takes a proof net of $\mathfrak{B}_{\mathbb{X}}$ and tells which cut eliminating. Formally, $\sigma : \mathbf{PN}(\mathfrak{B}_{\mathbb{X}}) \to \mathbf{PN}(\mathfrak{B}_{\mathbb{X}})$, such that (i) $\Pi \to \sigma(\Pi)$ or $\sigma(\Pi) = \Pi$, and (ii) if there is a reducible cut in $\Pi$ then $\sigma(\Pi) \neq \Pi$.*

Notice we do not require this function to be computable. We shall use letters $\sigma, \tau$ for strategies.

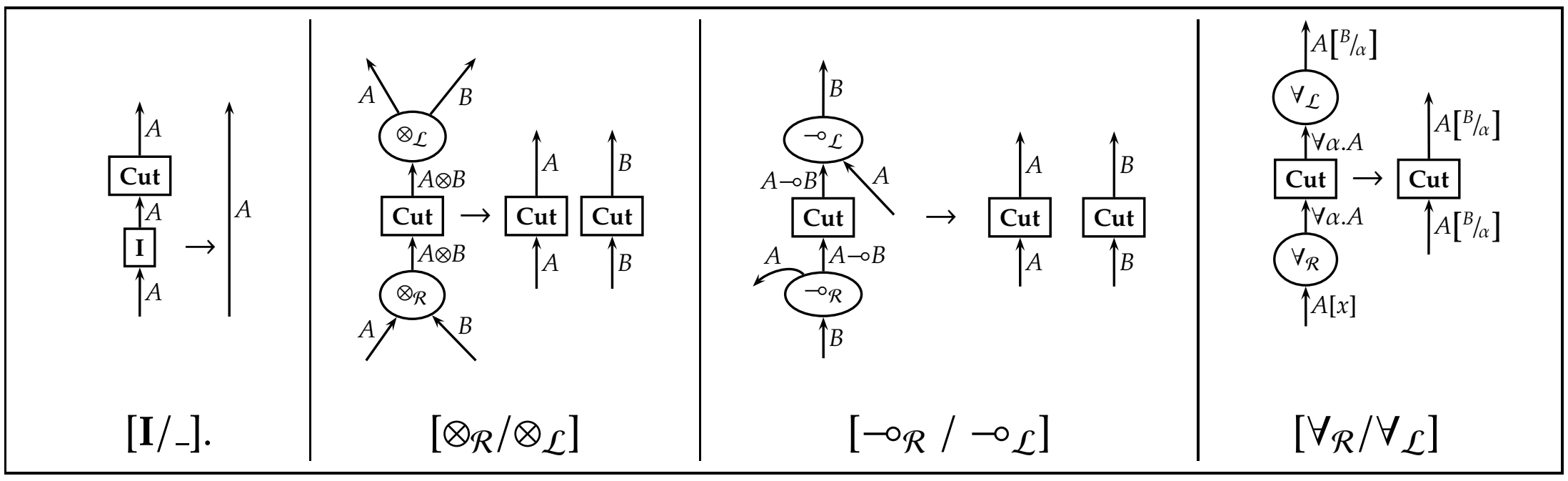

Figure 3.7: *Linear* meta-*ns*. $[\forall_{\mathcal{R}}/\forall_{\mathcal{L}}]$ substitutes $B$ for $\alpha$ in the whole proof net rooted at $\forall_{\mathcal{R}}$. We do not explicitly write the sceheme $[_/\mathbf{I}]$ because analogous to $[\mathbf{I}/_]$.

In particular:

DEFINITION 3.3.3 *The* **closed normalization strategy** *requires that an instance of the modal cuts ($[P/P]$ and $[P/\mathbf{Y}]$) can be reduced only if the lowermost box they involve is closed.*

DEFINITION 3.3.4 (NOTATIONS)
- *If the proof net $\Pi$ rewrites to $\Sigma$ in a single* ns, *we write* $\Pi \to \Sigma$.
- *The transitive closure of $\to$ will be denoted $\to^*$.*
- $\Pi \to^n \Sigma$ *iff* $\Pi \to^* \Sigma$ *in exactly $n$ steps.*
- $\Pi \to_\sigma \Sigma$ *iff* $\Pi \to^* \Sigma$ *using the strategy $\sigma$.*
- $\Pi \to_d \Sigma$ *iff* $\Pi \to \Sigma$ *reducing a* cut *node at level $d$.*
- $\Pi \to_S \Sigma$ *iff* $\Pi \to \Sigma$ *is a* ns *with name $S$.*
- *Combinations of the previous symbols shall be used.*

DEFINITION 3.3.5 *A proof net $\Pi$ is in* **normal form according to a strategy** $\sigma$ *iff* $\sigma(\Pi) = \Pi$. *A proof net is in* **normal form** *if it is in formal form according every possible strategy.*

DEFINITION 3.3.6 *We will call* **normalization (resp. $\sigma$-normalization) of a proof net** $\Pi$ *a sequence of proof nets*

$$\langle \Pi = \Pi_0, \Pi_1, \ldots, \Pi_n \rangle$$

*where $\Pi_i \to \Pi_{i+1}$ (resp. $\Pi_i \to_\sigma \Pi_{i+1}$), and such that $\Pi_n$ is in normal form.*

LEMMA 3.3.7 (STABILITY UNDER REDUCTION)
Applying a *ns* to a meta-proof net leads to another meta-proof net. Applying a *ns* to a proof net of $\mathfrak{B}_{\mathbb{X}}$ leads to another proof net of $\mathfrak{B}_{\mathbb{X}}$.

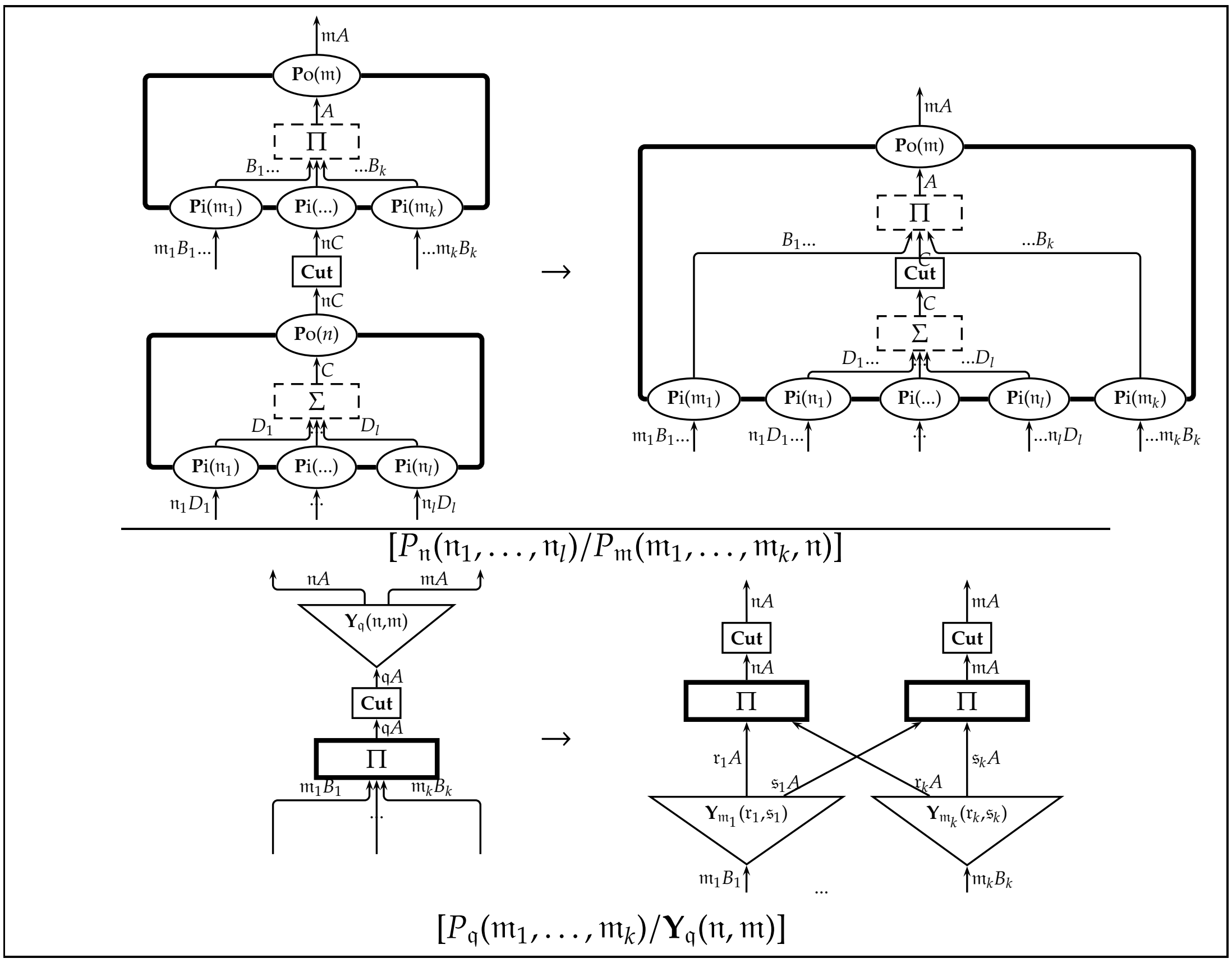

Figure 3.8: Meta-*ns* involving boxes. $[P_{\mathfrak{q}}(\vec{\mathfrak{r}})/\mathbf{Y}_{\mathfrak{q}}(\mathfrak{n},\mathfrak{m})]$ can generate a whole set of *ns* according to instances of the parameters $\vec{\mathfrak{r}}$ and $\vec{\mathfrak{s}}$.

**Proof.** The second statement is a consequence of the first one. So we prove just the first one. Given a meta-proof net $\Pi$ and one of its cuts $u$, the elimination of $u$ using a single *ns* leads to another boxed graph $\Pi'$. We want to show $\Pi'$ is a meta-proof net.

The following is *not* a proof by induction; however it relies on the inductive definition of proof net.

We will not prove the thesis for all possible cuts, just a relevant one.

Let us consider $u$ a cut $[\multimap_{\mathcal{R}}\ /\ \multimap_{\mathcal{L}}]$. $\Pi$ has been built, *by definition*, using some sequence of construction rules $R^i_j$, as in Figure 3.11.

Note that we don't know exactly the order of application of these rules; however e.g. $R^4_3$ needs to come after $R^4_1$ and $R^4_2$, because $\Sigma_1$ and $\Sigma_2$ need to exist.

Now, let us apply our *ns* to the cut $c$. We get a new boxed graph $\Pi'$, which is essentially the same as $\Pi$: in particular, proof nets $\Lambda_1$, $\Lambda_2$, $\Lambda_3$ are unchanged. However, rules $R^4_1$, $R^4_2$, $R^4_3$ are not applied: they are replaced by two *cut* rules, that create a proof net $\Sigma'$ with the same inputs and outputs of $\Sigma$. At last, all the nodes added by rules $R^5_i$ are unchanged. So, $\Pi'$ can be built by the rules in Figure 3.12. This shows that also $\Pi'$ in built using construction rules, and so it is a proof net.

All other *ns* are proved similarly. □

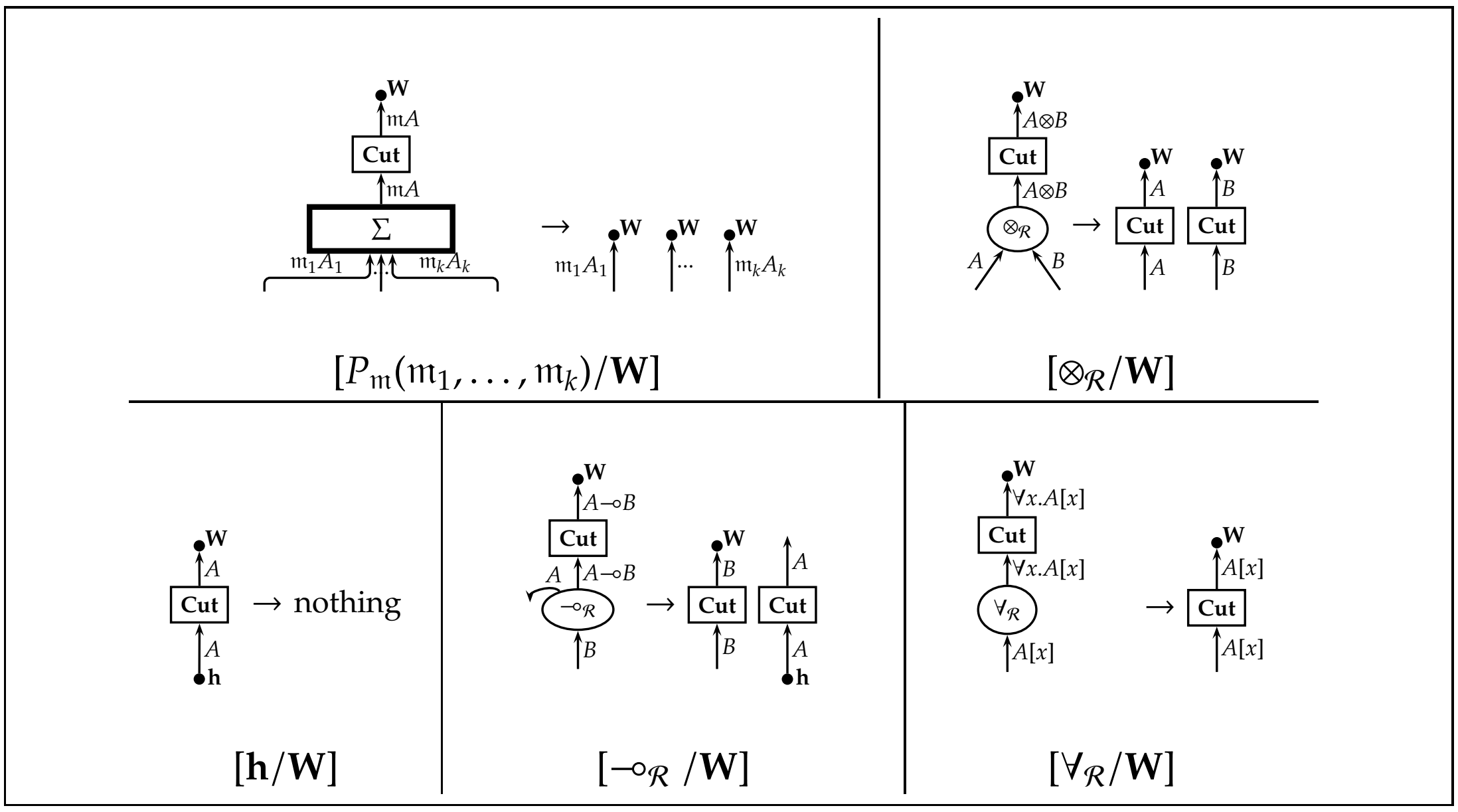

Figure 3.9: Meta-*ns* involving **W**.

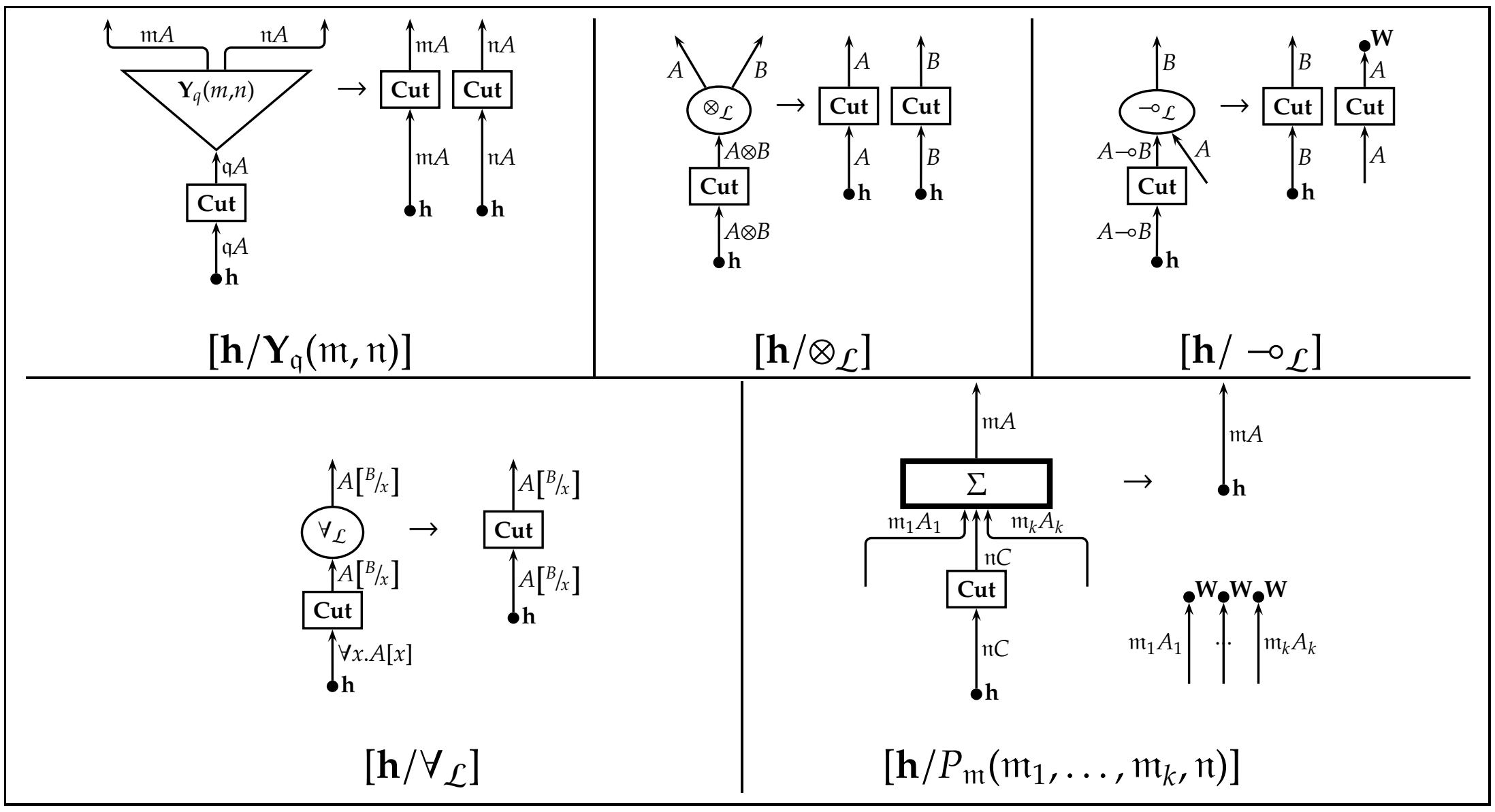

Figure 3.10: Meta-*ns* involving **h**.

Note in particular that this proof is, essentially, a proof of stability of the *sequentialization* of a given proof net Π. This different from the standard, and more elegant, proof that uses some correcteness criterion for proof nets. The reason is that, at the moment, we do not have any working correctness criterion for our proof nets.

$R^1_1$
$\dots$
$R^1_k$ is a rule that creates $\Lambda_1 \triangleright \Gamma \vdash A$

$R^2_1$
$\dots$
$R^2_l$ is a rule that creates $\Lambda_2 \triangleright \Delta, B \vdash C$

$R^3_1$
$\dots$
$R^3_m$ is a rule that creates $\Lambda_3 \triangleright \Theta, A \vdash B$

$R^4_1$ is a rule $\multimap_{\mathcal{L}}$ that creates $\Sigma_1 \triangleright \Gamma, \Delta, A \multimap B \vdash C$ from $\Lambda_1$ and $\Lambda_2$.

$R^4_2$ is a rule $\multimap_{\mathcal{R}}$ that creates $\Sigma_2 \triangleright \Theta \vdash A \multimap B$ from $\Lambda_3$.

$R^4_3$ is a rule *cut* that creates $\Sigma \triangleright \Gamma, \Delta, \Theta \vdash C$ from $\Sigma_1$ and $\Sigma_2$ introducing $c$.

$R^5_1$
$\dots$
$R^5_z$ creates $\Pi$

Figure 3.11: Proof of Stability Lemma.

$R^1_1$
$\dots$
$R^1_k$ creates $\Lambda_1 \triangleright \Gamma \vdash A$

$R^2_1$
$\dots$
$R^2_l$ creates $\Lambda_2 \triangleright \Delta, B \vdash C$

$R^3_1$
$\dots$
$R^3_m$ creates $\Lambda_3 \triangleright \Theta, A \vdash B$

$\tilde{R}^4_1$ is a rule *cut* that creates $\Sigma'_1 \triangleright \Gamma, \Theta \vdash B$ from $\Lambda_1$ and $\Lambda_3$.

$\tilde{R}^4_2$ is a rule *cut* that creates $\Sigma' \triangleright \Gamma, \Delta, \Theta \vdash C$ from $\Sigma_1$ and $\Lambda_2$.

$\tilde{R}^5_1$
$\dots$
$\tilde{R}^5_z$ creates $\Pi'$

where rules $\tilde{R}^5_i$ are the same rules $R^5_i$ applied to a different proof net.

Figure 3.12: Proof of Stability Lemma.

**Subsystems and normalization.** We have already seen that $\mathfrak{B}_{\mathbb{X}}$ is a rewriting system endowed with all *ns* in $\mathcal{R}(\mathfrak{B}_{\mathbb{X}})$, that are the *instances* of the meta-*ns*. Now, what about the subsystems? Notice that it is possible that $\Pi \in \mathbf{PN}(\mathcal{P}) \subseteq \mathbf{PN}(\mathfrak{B}_{\mathbb{X}})$, $\Sigma \in \mathbf{PN}(\mathfrak{B}_{\mathbb{X}})$, $\Sigma \notin \mathbf{PN}(\mathcal{P})$, and $\Pi \rightarrow \Sigma$ in $\mathfrak{B}_{\mathbb{X}}$. So, we state that *each subsystem $\mathcal{P} \subseteq \mathfrak{B}_{\mathbb{X}}$ is a rewriting system endowed with all and only the* ns *of $\mathcal{R}(\mathfrak{B}_{\mathbb{X}})$ that map proof nets of $\mathcal{P}$ to proof nets of $\mathcal{P}$*. We shall call $\mathcal{R}(\mathcal{P}) \subseteq \mathcal{R}(\mathfrak{B}_{\mathbb{X}})$ the set of such rules; please notice that $\mathcal{R}(\mathcal{P})$ is well-defined (i.e. unique).

Every subsystem is in fact a rewriting systems, see Section 3.4.1.

In this Section we will prove that each $\mathcal{P}$: (i) is strongly normalizing (Proposition 3.3.11), but (ii) in general, is not Church-Rosser (Lemma 3.3.14).

**Lifts and Residues.**

Definition 3.3.8 (Lifts and Residues of Nodes) *Let $\Pi \rightarrow \Pi'$. There is a natural map from $V_{\Pi'}$ to $V_{\Pi}$. We call it* **lift**, *denoted* Lift. *Observe that more than one node of $\Pi'$ can be mapped to the same node of $\Pi$. We call* **residue** *of $u \in V_{\Pi}$ all the nodes whose lift is $u$:*

$$\mathrm{Res}(u) = \mathrm{Lift}^{-1}(\{u\}) \subseteq V_{\Pi'}.$$

The definition can be generalized to $\Pi \rightarrow^* \Pi'$; in this case we denote $\mathrm{Res}_{\Pi}^{\Pi'}(u)$ the set of residues of $u$ in $\Pi'$.

Note in particular that a node can have no residue, if it disappears after reduction; on the contrary a single cut node can have more that one residue.

Note moreover that an **h** node can be a residue of a **W** node, and vice versa.

**Stratification.**

Definition 3.3.9 *A subsystem of* MS *is* **stratified** *if the rewriting rules assure that the depth of each node $u \in V_{\Pi}$ different from a cut and the depth of every node $v \in \mathrm{Res}_{\Pi}^{\Pi'}(u)$ coincide.*

Of course, it is still possible that a node disappears during reduction.

Note in particular that the depth of a stratified proof net cannot increase.

Lemma 3.3.10 (Stratification in MS)
Every subsystem $\mathcal{P} \subseteq \mathfrak{B}_{\mathbb{X}}$ is stratified.

**Proof.** Just look at the rewriting rules in Figures 3.7, 3.8, 3.9, 3.10. □

By the way, if we had considered also the cut nodes in the Definition 3.3.9, this result would be false: consider e.g. the $[P/P]$ reduction.

**Basic properties of subsystems.**

Proposition 3.3.11 (Elementary Upper Bound)
For each $d \in \mathbb{N}$ there exists an elementary function $e_d(n)$ that bounds the reduction time of all the proof nets $\Pi \in \mathbf{PN}(\mathfrak{B}_{\mathbb{X}})$ with depth $d$ and size $n$.

Vice versa, for every elementary function $e(n)$ there exists a depth $d$ and sequence of proof nets $\Pi_n \in \mathbf{PN}(\mathfrak{B}_{\mathbb{X}})$ such that every $\Pi_n$ reduces in time $\geq e(n)$.

**Proof.** We can embed $\mathfrak{B}_{\mathbb{X}}$ in IEAL just collapsing all the modalities $\mathfrak{n}$ into !. It is well-known that IEAL reduces in elementary time.

Vice versa, it's clear that every IEAL proof can be encoded in $\mathfrak{B}_{\mathbb{X}}$, using a particular modality $\mathfrak{m}$ instead of !. So $\mathfrak{B}_{\mathbb{X}}$ is at least as expressive as IEAL. □

COROLLARY 3.3.12 (TERMINATION)
Every normalization terminates in a finite number of steps.

COROLLARY 3.3.13 (CUT-ELIMINATION)
The subsystem $\mathfrak{B}_{\mathbb{X}}$enjoys the cut-elimination property.

**Proof.** Every cut can be reduced, using the *ns* of $\mathfrak{B}_{\mathbb{X}}$. Corollary 3.3.12 tells that every normalization eventually terminates; hence the normal form must be cut-free. □

On the contrary, Church-Rosser confluence property does not hold:

LEMMA 3.3.14 (NON-DETERMINISM IN NORMALIZATION)
There are proof nets $\Sigma$, $\Pi_1$, $\Pi_2$ such that $\Sigma \to^* \Pi_1$, $\Sigma \to^* \Pi_2$, and $\Pi_1$ and $\Pi_2$ are two different proof nets in normal form.

This is due to non-determinism in $[\mathbf{Y}/P]$ *ns*.

## 3.4 Technical considerations

### 3.4.1 Abstract Rewriting Systems

Here we will recall the definition of *Abstract Rewriting System (ARS)* from [Klo92, KBV01]. Each subsystem $\mathcal{P}$ is in fact an instance of ARS. So, some (few) theorems holding for a generic ARS also hold for $\mathcal{P}$. We underline that, in literature, *terms rewriting systems (TRS)* are more widely used than ARS. TRS are a particular case of ARS, whose objects are *terms*; but each $\mathcal{P}$ is *not* a TRS, because its objects are proof nets instead of terms.

DEFINITION 3.4.1 *An* **Abstract Rewriting System (or ARS)** *is an algebraic structure* $\mathscr{A} = \langle A, \to \rangle$ *where* $A$ *is a set and* $\to$ *is a binary relation over* $A$*.* $\twoheadrightarrow$ *is the transitive and reflexive closure of* $\to$*.* $\leftarrow$ *is the inverse relation of* $\leftarrow$*.* $\twoheadleftarrow$ *the transitive and reflexive closure of* $\leftarrow$*.* $\equiv$ *is the* **convertibility** *relation: it is the smallest equivalence relation containing* $\twoheadrightarrow$*.* $a \in A$ *is a* **normal form (n.f.)** *if there is no* $b$ *such that* $a \to b$*.* $a$ *is a n.f. of* $b$ *if* $a$ *is a n.f. and* $b \twoheadrightarrow a$*.*

**(WCR)** $\mathscr{A}$ *is* **weakly Church Rosser** *if* $\forall a, b, c \in A\, (c \leftarrow a \to b \Rightarrow \exists d \in A\, (c \twoheadrightarrow d \twoheadleftarrow b))$ .

**(CR)** $\mathscr{A}$ *is* **Church Rosser** *if* $\forall a, b, c \in A\, (c \twoheadleftarrow a \twoheadrightarrow b \Rightarrow \exists d \in A\, (c \twoheadrightarrow d \twoheadleftarrow b))$ .

**(WN)** $\mathscr{A}$ *is* **weakly normalizing** *if every* $a \in A$ *has a n.f.*

**(SN)** $\mathscr{A}$ *is* **strongly normalizing** *if every reduction* $a_0 \to a_1 \to \ldots$ *eventually terminates.*

- $\mathscr{A}$ *is* **complete** *if it is SN and CR.*

**(NF)** $\mathscr{A}$ *has the* **normal form property** *if* $\forall a, b \in A$*,* $a$ *is a n.f. and* $a \equiv b$ *implies* $b \twoheadrightarrow a$*.*

**(UN)** *$\mathscr{A}$ has the* **unique normal form property** *if $\forall a, b \in A$, $a, b$ are n.f. and $a \equiv b$ implies $a = b$.*

**(FB)** *$\mathscr{A}$ is* **finitely branching** *if $\forall a \in A$ there exist at most a finite number of $b \in A$ such that $a \to b$.*

*If $\mathscr{A} = \langle A, \to_A \rangle$ and $\mathscr{B} = \langle B, \to_B \rangle$ are two ARS such that $A \subseteq B$, then:*

- *$\mathscr{B}$ is a* **conservative extension** *of $\mathscr{A}$ if $\forall a, a' \in A\ (a \to_A a' \Leftrightarrow a \to_B a')$.*
- *$\mathscr{A}$ is a* **substructure** *of $\mathscr{B}$ if, moreover, $\forall a \in A\ \forall b \in B\ (a \to_B b \Rightarrow b \in A)$.*

The following is proved as Theorem 1.0.7 in [Klo92]:

Theorem 3.4.2 (Basic Properties of ARS)
1. CR $\Rightarrow$ WCR.
2. SN $\Rightarrow$ WN.
3. CR $\Rightarrow$ NF $\Rightarrow$ UN.
4. SN + WCR $\Rightarrow$ CR.
5. UN + WN $\Rightarrow$ CR.

Now, we apply the previous definitions and theorems to our subsystems. For every fixed $\mathbb{X}$, the structure $\mathscr{A} = \langle \mathbf{PN}(\mathfrak{B}_{\mathbb{X}}), \to \rangle$ is an ARS. If $\mathbb{X}$ is finite, such ARS is FB; indeed, every proof net has a finite number of cuts, and each of them can reduce in a finite number of ways, being $\mathbb{X}$ finite. Moreover, $\mathscr{A}$ is SN (Corollary 3.3.12), but it is not WCR (Lemma 3.3.14), so neither CR.

For every $\mathcal{P} \subseteq \mathfrak{B}_{\mathbb{X}}$, the structure $\mathscr{B} = \langle \mathbf{PN}(\mathcal{P}), \to \rangle$ is an ARS, too. $\mathscr{A}$ is a conservative extension of $\mathscr{B}$; however, $\mathscr{B}$ is *not* in general a substructure of $\mathscr{A}$, because it is possible that $\Pi \in \mathbf{PN}(\mathcal{P})$ reduces – in $\mathfrak{B}_{\mathbb{X}}$ – to some $\Pi' \notin \mathbf{PN}(\mathcal{P})$. $\mathscr{B}$ is FB and SN (properties inherited by $\mathfrak{B}_{\mathbb{X}}$), but not in general WCR. Theorem 3.4.2 implies that whenever $\mathscr{B}$ is WCR it is also CR, and so it is also NF and UN. Summing up:

Corollary 3.4.3 (Basic Properties of Subsystems)
For every $\mathcal{P} \subseteq \mathfrak{B}_{\mathbb{X}}$, the structure $\mathscr{B} = \langle \mathbf{PN}(\mathcal{P}), \to \rangle$ is an ARS. $\mathscr{B}$ is SN. $\mathscr{B}$ is WCR if and only if it is CR; and if it is WCR it is also NF and UN. If $\mathbb{X}$ is finite, $\mathscr{B}$ is FB.

This property will be used in Section 5.2.

### 3.4.2 The definition of Size

In Section 3.1 we have defined the *size of a proof net* $\Pi$ as the number $|\Pi|$ of nodes appearing in it. This definition, apparently trivial, hides in fact some important issues. (i) All the premises of a promotion box contribute to $|\Pi|$; but in fact the promotion is one single rule, both in the sequent calculus or in the natural deduction formalisms. Similarly, the size includes also the *terminal nodes*, i.e. premises and conclusion of the proof net, that do not correspond to any logical rule. We will prove that the premises of the box contribute to square the size of $\Pi$, that is: if $N$ is the number of nodes in $\Pi$ different from a *Pi* node or a terminal node, then $|\Pi| = O\left(N^2\right)$ (Lemma 3.4.4). As a consequence, since we are only interested in the polynomial time, the two definitions are equivalently good. (ii) $|\Pi|$ does not consider the size of the formulæ appearing as label of $\Pi$. This is quite common in literature [AR02, Laf04]; and indeed there is a deep reason in this choice. For example, in either ILAL or SLL, if we count also the time required to change the formulæ of $\Pi$, the reduction of $\Pi$ cannot be performed in polytime. Anyway we shall see that the size of the formulæ has an impact only over subsystems that allow the second order quantification (Lemma 3.4.7 and 3.4.8). (iii) The Computational Complexity is traditionally formulated in terms of Turing machines. What is the relationship between $|\Pi|$ and the number of bits required to encode $\Pi$ on the tape of a Turing machine? We will see that this last number is polynomial in $|\Pi|$.

Lemma 3.4.4 (Size and Terminal Nodes)
Given a net $\Pi$, let $s$ be its size, $\partial$ its depth, $\widehat{s}$ its size without counting the terminal nodes, and $\widetilde{s}$ its size without any terminal node and any *Pi* node at any depth. Then

$$\begin{aligned} s &= O(\widehat{s}); \\ s &= O(\partial \cdot \widetilde{s}) = O\left(\widetilde{s}^2\right). \end{aligned}$$

**Proof.** Let us call $n(\Pi)$ the number of non-terminal nodes (including boxes) of $\Pi$ at level 0; $t(\Pi)$ the number of terminal nodes (of course at level 0); $N(\Pi) = \widetilde{s}$ the number of non-terminal nodes at any depth; and $T(\Pi)$ the number of terminal nodes and *Pi* nodes, at any depth. $n(\Pi) + t(\Pi) = |\Pi|_0$ while $N(\Pi) + T(\Pi) = |\Pi|$, and $\widehat{s} = |\Pi| - t(\Pi)$.

We show by induction on $\partial$ that $t(\Pi) \leq 3N(\Pi)$.

(i) If $\partial = 0$, there are no boxes; so each non-terminal node is connected to at most 3 nodes; and each terminal node is connected to some non-terminal node; so $t(\Pi) \leq 3N(\Pi)$.

(ii) Let $\partial > 0$, and $B = b_0(\Pi)$. Inside the $B$ boxes we find $B$ subnets, which we can call $\Pi_1, \ldots, \Pi_B$, whose depth is less than $\partial$. We can consider the *Pi* nodes at level 1 as the terminal nodes of $\Pi_1, \ldots, \Pi_B$. Now, $N(\Pi) = N(\Pi_1) + \ldots + N(\Pi_B) + n(\Pi)$. Following the argument for the base-case,

$$\begin{aligned} t(\Pi) &\leq 3(n(\Pi)) + t(\Pi_1) + \ldots + t(\Pi_B) \\ &\leq 3n(\Pi) + 3N(\Pi_1) + \ldots + 3N(\Pi_B) \qquad \text{by inductive hypothesis} \\ &\leq 3N(\Pi). \end{aligned}$$

This shows that $t(\Pi) \leq 3N(\Pi) = 3\widehat{s}$, and $s = \widehat{s} + t(\Pi) = O(\widehat{s})$. Now, we have to study $T$ instead of $t$. Again, by induction:

(i) If $\partial = 0$, $T(\Pi) = t(\Pi) \leq 3N(\Pi) = 3 \cdot 1 \cdot N(\Pi)$.

(ii) If $\partial > 0$,

$$\begin{aligned} T(\Pi) &= T(\Pi_1) + \ldots + T(\Pi_B) + t(\Pi) \\ &\leq 3 \cdot (\partial - 2) \cdot N(\Pi_1) + \ldots + 3 \cdot (\partial - 2) \cdot N(\Pi_B) + 3N(\Pi) \\ &\leq 3 \cdot (\partial - 1) \cdot N(\Pi) \\ &= 3 \cdot (\partial - 1) \cdot \widetilde{s}. \end{aligned}$$

And so $s = N(\Pi) + T(\Pi) = O(\partial \cdot \widetilde{s})$. □

REMARK 3.4.5 (ABOUT OUR THEOREMS) The reason why we have proved Lemma 3.4.4 is that we were initially worried whether the theorems that we will prove in Chapter 4, and in particular the Polynomiality Criterion 4.3.40, could not hold if we change the definition of size according to Lemma 3.4.4. The answer is yes, they still hold; the point is to use larger polynomials.

**On the size of the formulæ.** Let us move to the second question: what about the size of the formulæ?

DEFINITION 3.4.6 *We suppose that a binary coding for formulas has been declared, and the lengths of the formulæ are calculated in bits. Given a proof net* $\Pi$ *of* MS*, we will call* $\gamma(\Pi)$ *the* **maximum length of formulas** *appearing on edges of* $\Pi$.

LEMMA 3.4.7
If $\Pi$ is a net of MS, **without any** $\forall$ **nodes**. Then $\gamma(\Pi)$ cannot increase during reduction.

**Proof.** Just look back at the *ns*: sometimes formulas disappear, sometimes $\mathfrak{m}A$ is replaced by $\mathfrak{n}A$, but the size of formulas does not increase. □

LEMMA 3.4.8
Let $\mathbb{X}$ be fixed, and $\Pi, \Pi' \in \mathbf{PN}(\mathfrak{B}_{\mathbb{X}})$, such that $\Pi \to \Pi'$. Then $\gamma(\Pi') \leq \gamma(\Pi)^2$.

**Proof.** The *heavier* case is that of $[\forall_{\mathcal{L}}/\forall_{\mathcal{R}}]$. Let us imagine we are eliminating the following cut of $\Pi$:

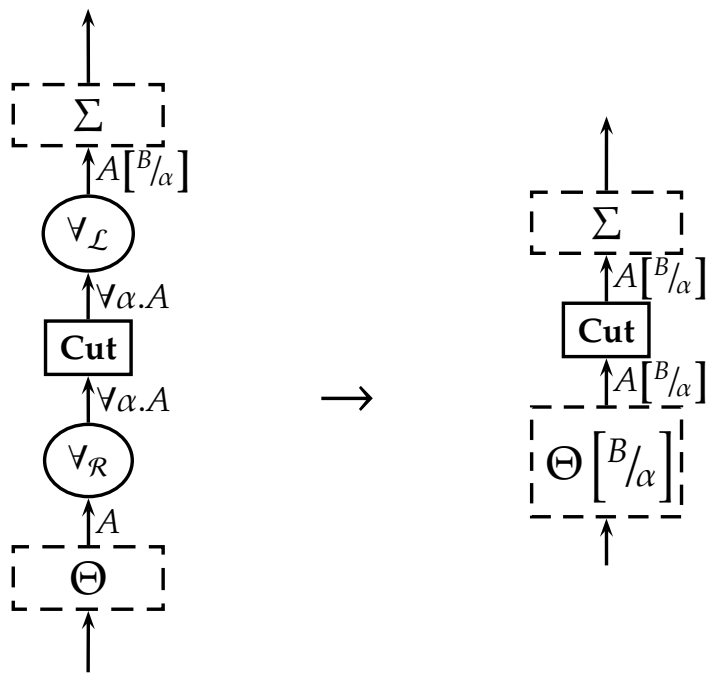

In this case we have to change all the labels $x$ in $\Pi$ with another formula $B$, which could be larger. $B$ is at most $\gamma(\Theta) \leq \gamma(\Pi)$ long; and all other formulæ in $\Theta$ can contain at most $\gamma(\Theta) \leq \gamma(\Pi)$ occurrences of $x$; so $\gamma(\Pi') \leq \gamma(\Pi)^2$. □

So, if $\Pi$ has $O(n)$ cuts of kind $[\forall_{\mathcal{L}}/\forall_{\mathcal{R}}]$, and they all are eliminable, after the reduction

$$\gamma(\Pi') \leq \gamma(\Pi)^{2^{O(n)}}$$

and $\Pi'$ will require an exponential number of bits to be completely encoded. This is why we will accept the tradition to *ignore* the length of the formulæ when encoding a proof net.

**Relationship with Turing machines.** In this last paragraph, we compare the size of a proof net with the size of an *encoding* of the same proof net on the tape of a Turing machine. A proof net $\Pi$ is compounded of:

- A graph $V_\Pi$ of $|\Pi|$ nodes and at most $3|\Pi|$ edges: it can be encoded using an adiacency matrix ($|\Pi|^2$ bits) or an incidence matrix ($\leq 3|\Pi|^2$ bits), or a list of edges as couple of nodes ($\leq 3|\Pi| \cdot 2\log(|\Pi|)$ bits).
- Some additional *boxes*, which are identified by their nodes: $O(|\Pi|)$ nodes per each box, so $O(|\Pi| \cdot \log(|\Pi|))$ bits.
- A label for each node, chosen among 15 different shapes: $4 \cdot |\Pi|$ bits.
- A label for each edge, which is an arbitrarily long formula. We can decide to ignore them, *or* to encode them using at most $O(\gamma(\Pi))$ bits each, and $O(|\Pi| \cdot \gamma(\Pi))$ bits all.

So, without labels, the proof net can be codified by $O(|\Pi| \cdot \log(|\Pi|))$ bits, and with labels $O(|\Pi| \cdot \log(|\Pi|) + |\Pi| \cdot \gamma(\Pi))$ bits. It follows that $|\Pi|$ is a good choice for the complexity of the graph.

### 3.4.3 Why SLL is not a subsystem of MS

SLL, the Soft Linear Logic of [Laf04], is a polytime logic derived from LL. It has been briefly recalled in Section 2.4; it has the additive and multiplicative rules of LL, plus the exponential rules reported in Figure 2.3, in a sequent calculus style. Our original target was the immersion of SLL into MS. That is, we wanted to find some subsystem $\mathcal{P} \subseteq$ MS and a compositional embedding from proof nets of SLL into proof nets of $\mathcal{P}$, such that every reduction in SLL corresponds to a reduction in $\mathcal{P}$, too. Notice that this target would be interesting, because SLL is *not* a stratified system, as it has the *multiplexor* rule.

The idea was the following. Take a proof net $\Pi$ of SLL. Some new boxes have to be added to $\Pi$; each multiplexor in $\Pi$ with $n = 2$ premises has to be transformed in a contraction, plus two !-premises of some new box; if $n > 2$ the multiplexor will be replaced with an *ad hoc* number of contractions; if $n = 1$ the multiplexor (or *dereliction*) has to be transformed into a !-premise of some new box.

However this is not possible, as the example in Figure 3.13 shows. The reader can try to add some new boxes to this proof net; he will see that, if any box passes trough the lower dereliction, then the two formulas labelling the premises of the lower multiplexor will be different.

On the other side, it is possible to write SLL as a stratified system, allowing a more liberalized use of the modalities; see e.g. [GRV]. We will return on this topic in Section 8.2.2.

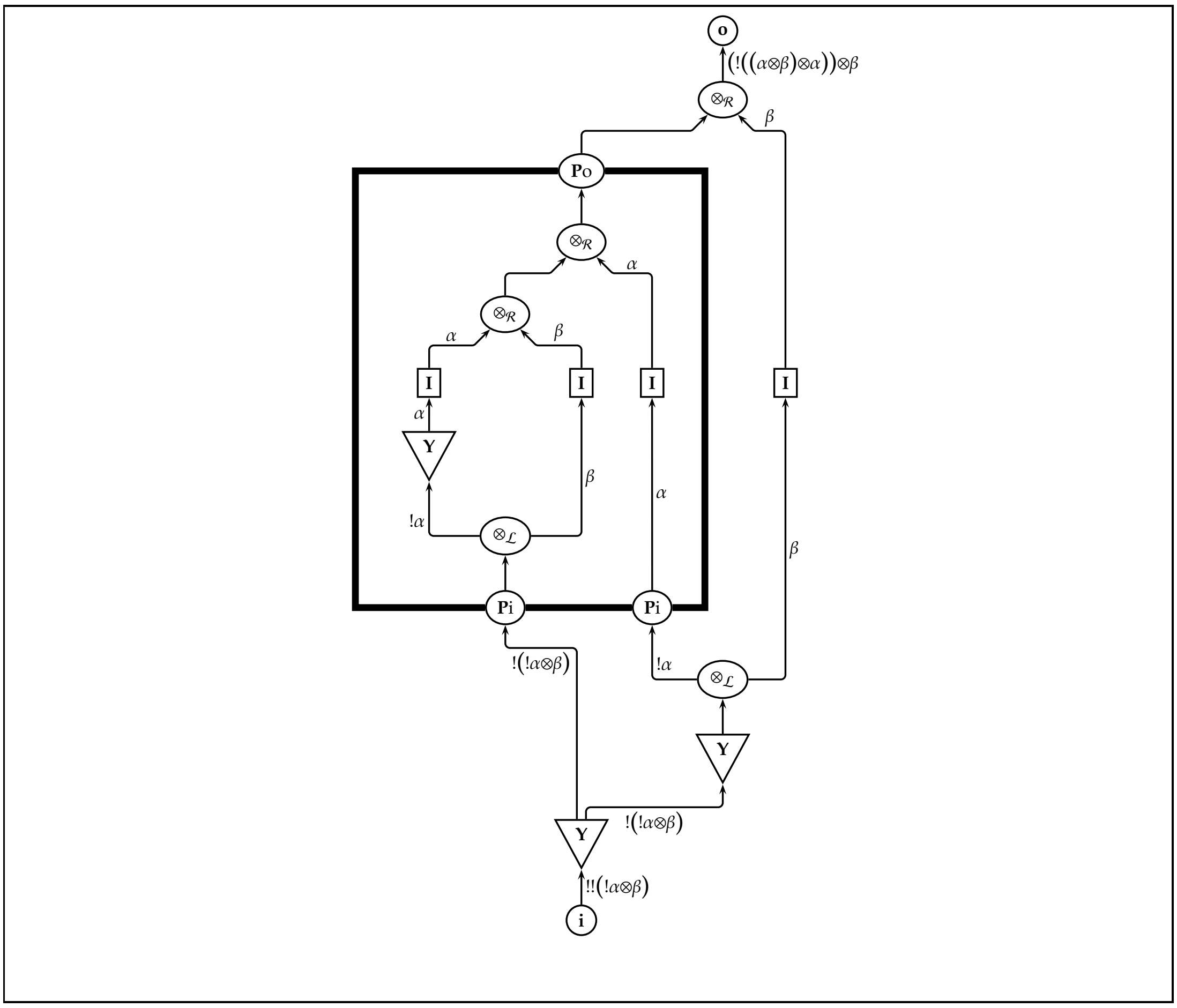

Figure 3.13: The counterexample described in Section 3.4.3.

# Chapter 4

# Polynomiality in MS

In this Chapter we study PMS, the class of the subsystems of MS whose reductions are performed in polynomial time. A formal definition of polynomiality is given in Section 4.1. All the subsystems of MS are stratified; thus, we will begin our study generalizing the proof of strong polynomial time of ILAL (Section 4.2).

Then, we will provide some conditions that make a subsystem polytime or not polytime. These theorems are divided in two Sections: 4.3 and 4.4. The first one contains some "*semantical*" criteria for polynomial time, while the second one moves towards the more "*syntactical*" criteria. "Semantical" here means that they rely on the context semantics of the proof nets; while the "syntactical" criteria recognize the polynomiality of a subsystem "*just looking at its rules*".

The first criterion of the Chapter is Proposition 4.3.40. It tells that a *sensible* subsystem $\mathcal{P}$ is polytime if and only if its rules do not allow the construction of a *dangerous spindle*. The statement requires some definitions. We recall that a subsystem is sensible if it contains a minimum amount of rules, needed for basic programming (Definition 3.2.5). A spindle and a dangerous spindle are geometrical configurations that may occur inside proof nets (Definition 4.3.26 and Figure 4.4a). The notion of spindle relies on context semantics (CS); so in Section 4.3.1 and 4.3.2 we will recall the basics of CS. The Criterion is then proved in Section 4.3.3.

After that, we will find out some more properties equivalent to the Criterion. In particular, Proposition 4.4.5 tells that a subsystem $\mathcal{P}$ is polytime if and only if its *augmented subsystem* $\mathcal{S}(\mathcal{P})$ contains some specific rules. $\mathcal{S}(\mathcal{P})$ is defined as an extension of $\mathcal{P}$ containing certain rules derivable in $\mathcal{P}$. Similarly, Proposition 4.4.15 assures that a subsystem is polytime if it does not contain certain rules, described by means of a relation among modalities $\mathfrak{R}_{\uparrow}$. Another interesting result (Lemma 4.4.13) says that the question "*Is $\mathcal{P}$ polytime?*" is decidable, at least in most of the practical cases. It should be clear that these last theorems do not mention at all the CS. Figure 4.1 here below summarizes the main results of Sections 4.3 and 4.4. Beware that the results recalled in the Figure require a number of definitions that the reader can find in the appropriate Sections.

At last, in Section 4.5 we adapt (some of) the previous theorem to some non-sensible system. We are interested in particular in subsystems that use an infinite number of modalities, as we think that they are very natural in a theoretical perspective.

| Proposition | At page | Syntactic? | Easy to use? | Statement |
|---|---|---|---|---|
| 4.3.32 | 62 | | | Let $\mathcal{P} \subseteq \mathfrak{B}_{\mathbb{X}}$. Let us assume that every *chain of spindles* that can be built in $\mathcal{P}$ cannot have more than $L$ spindles, for some constant $L \in \mathbb{N}$. Then $\mathcal{P}$ is strongly polytime. |
| 4.3.40 | 66 | | (sometimes) | Let $\mathcal{P} \subseteq \mathfrak{B}_{\mathbb{X}}$ sensible, with $M$ modalities. The following are equivalent:<br>1. $\mathcal{P}$ is polytime.<br>2. $\mathcal{P}$ cannot build any dangerous chain of spindles.<br>3. $\mathcal{P}$ cannot build any *dangerous spindle*.<br>4. $\mathcal{P}$ cannot build any chain (neither if not dangerous!) of more than $M$ spindles. |
| 4.3.43 | 67 | | | If $\mathcal{P} \subseteq \mathfrak{B}_{\mathbb{X}}$ enjoys *strong determinacy*, then it is polytime. |
| 4.4.5 | 69 | ✓ | ✓ | Let $\mathcal{P} \subseteq \mathfrak{B}_{\mathbb{X}}$ be a sensible system. Then $\mathcal{P}$ is not polytime iff its *augmented system* $\mathcal{S}(\mathcal{P})$ has the two rules $P_{\mathfrak{q}}(\mathfrak{r}, \mathfrak{s})$ and $\mathbf{Y}_{\mathfrak{q}}(\mathfrak{r}, \mathfrak{s})$, for some value of the parameters. |
| 4.4.8 | 70 | ✓ | | Let $\mathcal{P}$ be a sensible subsystem with $M$ modalities. Then $\mathcal{P} \in \mathsf{PMS}$ iff $\mathfrak{R}_\diamond$ is a strict partial order. |
| 4.4.12 | 72 | ✓ | | Let $\mathcal{P}$ be a sensible subsystem. $\mathcal{P}$ has a dangerous chain of spindles iff there exist $\mathfrak{m}, \mathfrak{n}, \mathfrak{q}$ such that $\langle q\rangle \overline{\mathfrak{R}_\triangle} \langle \mathfrak{m}, \mathfrak{n}\rangle \overline{\mathfrak{R}_\triangle} \langle \mathfrak{q}\rangle$. |
| 4.4.15 | 73 | ✓ | ✓ | Let $\mathcal{P} \subseteq \mathfrak{B}_{\mathbb{X}}$ be sensible. $\mathcal{P}$ is polytime if and only if, for every 6 elements $\mathfrak{q}_1 \equiv \mathfrak{q}_2 \equiv \mathfrak{r}_1 \equiv \mathfrak{r}_2 \equiv \mathfrak{s}_1 \equiv \mathfrak{s}_2$, i.e. all equivalent under $\widehat{\mathfrak{R}_\uparrow}$, $\mathcal{P}$ does not contain both $\mathbf{Y}_{\mathfrak{q}_1}(\mathfrak{r}_1, \mathfrak{s}_1)$ and $P_{\mathfrak{q}_2}(\mathfrak{r}_2, \mathfrak{s}_2)$. |
| 4.4.32 | 80 | ✓ | | Let $\mathcal{P} \subseteq \mathfrak{B}_{\mathbb{X}}$ be any sensible subsystem. Then, $\mathcal{P}$ is polytime iff $\mathcal{P} \subseteq \mathcal{Q}$ for some $\mathcal{Q} \subseteq \mathfrak{B}_{\mathbb{X}}$ maximal. |

Figure 4.1: Summa of the results contained in this Chapter.

## 4.1 Polynomial time and size

DEFINITION 4.1.1 (REDUCTION TIME AND USED SIZE OF PROOF NET.) *Let $\Pi$ be a proof net.*

- *Its* **reduction time** *is* $[\Pi] \stackrel{\text{def}}{=} \max\{n \mid \Pi \to^n \Sigma\}$. *The notion can be made relative to a strategy* $\sigma$: $[\Pi]_\sigma \stackrel{\text{def}}{=} \max\{n \mid \Pi \to^n_\sigma \Sigma\}$, *and to every level* $d \le \partial(\Pi)$: $[\Pi]^d \stackrel{\text{def}}{=} \max\{n \mid \Pi \to^n_d \Sigma\}$.
- *Its* **used size** *is* $||\Pi|| \stackrel{\text{def}}{=} \max\{|\Sigma| \mid \Pi \to^* \Sigma\}$. *The notion can be made relative to a strategy* $\sigma$: $||\Pi||_\sigma \stackrel{\text{def}}{=} \max\{|\Sigma| \mid \Pi \to^n_\sigma \Sigma\}$, *to a level* $d \le \partial(\Pi)$: $||\Pi||^d \stackrel{\text{def}}{=} \max\{|\Sigma| \mid \Pi \to^n_d \Sigma\}$, *and to the combination of two levels* $i, j$ *such that* $i < j \le \partial(\Pi)$: $||\Pi||^j_i \stackrel{\text{def}}{=} \max\{|\Sigma|_i \mid \Pi \to^*_j \Sigma\}$. *Specifically,* $||\Pi||^j_i$ *is the maximal used size at level* $i$*, when every reduction step occurs at level* $j$.

DEFINITION 4.1.2 (WEAK AND STRONG POLYSTEP SYSTEMS) $\mathcal{P}$ *is* **weakly polystep** *iff for every* $d \in \mathbb{N}$ *there exists a polynomial* $p_d(n)$ *and a strategy* $\sigma$ *such that for every* $\Pi \in \mathbf{PN}(\mathcal{P})$,

$[\Pi]_\sigma \leq p_d(|\Pi|)$.
*$\mathcal{P}$ is* **(strongly) polystep** *iff for every $d \in \mathbb{N}$ there exists a polynomial $p_d(n)$ such that for every $\Pi \in$ **PN**$(\mathcal{P})$, $[\Pi] \leq p_d(|\Pi|)$.*

Definition 4.1.3 (Weak and Strong Polysize Systems) *$\mathcal{P}$ is* **weakly polysize** *iff for every $d \in \mathbb{N}$ there exists a polynomial $p_d(n)$ and a strategy $\sigma$ such that for every $\Pi \in$ **PN**$(\mathcal{P})$, $||\Pi||_\sigma \leq p_d(|\Pi|)$.*
*$\mathcal{P}$ is* **(strongly) polysize** *iff for every $d \in \mathbb{N}$ there exists a polynomial $p_d(n)$ such that for every $\Pi \in$ **PN**$(\mathcal{P})$, $||\Pi|| \leq p_d(|\Pi|))$.*

Definition 4.1.4 (Weak and Strong Polytime Systems) *$\mathcal{P}$ is* **(strongly) polytime** *iff it is both strongly polystep and strongly polysize.*
*$\mathcal{P}$ is* **weakly polytime** *iff it is both weakly polystep and weakly polysize, under the same strategy $\sigma$.*

Remark 4.1.5 (About Definition of Polytime Systems) Our definition of *polytime system* needs some comments. Why is it necessary to consider different polynomials $p_d(n)$ when depth changes? Let us consider two alternative definitions, each one based on the attempt to use one only polynomial:

1. $\mathcal{P}$ is weakly polynomial iff there exists a strategy $\sigma$ and a polynomial $p(n)$ such that every $\sigma$-normalization of a proof net $\Pi$ takes at most $p(|\Pi|)$ steps.

   This is not so interesting, because $\mathcal{P}$ cannot codify all polytime functions.

2. $\mathcal{P}$ is weakly polynomial iff there exists a strategy $\sigma$ such that for every proof $\Pi$ of $\mathcal{P}$ there exists a polynomial $p_\Pi(n)$ such that the $\sigma$-normalization of $\Pi$ takes at most $p_\Pi(|\Pi|)$ steps.

   This is uninteresting, as well, because every system would be weakly polynomial according to every strategy $\sigma$, just choose $p_\Pi(n)$ large enough.

Now, let us arrive to the main problem.

Problem 4.1.6 Given a system $\mathcal{P} \subseteq$ MS, is it (strongly) polynomial? or weakly polynomial?

Of course, for each $\mathbb{X}$, $\mathfrak{B}_\mathbb{X}$ is not polynomial, because of Proposition 3.3.11 at page 39.

Definition 4.1.7 PMS ***is the class of the subsystems of some $\mathfrak{B}_\mathbb{X}$ that are polytime.***

The following Lemma is a trivial property that we will use often:

Lemma 4.1.8 (Monotonicity)
Let $\mathcal{P} \subseteq \mathcal{Q}$ be two subsystems, and $\mathcal{Q} \in$ PMS. Then, $\mathcal{P} \in$ PMS, too.

**Proof.** All the reductions $\Pi \to^* \Sigma$ that can be performed in $\mathcal{P}$ may be performed in $\mathcal{Q}$, too. Now, in $\mathcal{Q}$ there exists a polynomial $p(x)$ such that $[\Pi], ||\Pi|| \leq p\,(|\Pi|)$, and the same polynomial bounds the reductions in $\mathcal{P}$. □

Remark 4.1.9 (Cost of Box Duplication) Some authors (e.g. [NM02]) underline that the rewriting rules have not all the same *cost* (where the *cost* of a single *ns* is the time needed to perform it): box duplication should take a time proportional to the content of the box itself. Actually we agree with them, but for sake of simplicity we considered box-duplication as constant-time.

## 4.2 Following ILAL

First of all, we generalize the proof of polynomial soundness for ILAL [AR02]. We classify the cuts in Figures 3.7, 3.8, 3.9, 3.10 in the following way:

Definition 4.2.1 (Classification of Cuts) *We call* **modal cuts** *the cut nodes named* $[P/P]$, $[P/\mathbf{Y}]$ *and* $[\mathbf{I}/_]$; **linear cuts** *the cut nodes named* $[\otimes/\otimes]$, $[\forall/\forall]$, $[\multimap / \multimap]$ *and* $[\mathbf{I}/_]$; *and* **garbage collector cuts (*gc*)** *the cuts* $[\mathbf{W}/_]$, $[_/\mathbf{h}]$ *and* $[\mathbf{I}/_]$.

We notice that:

1. The $[\mathbf{I}/_]$ cuts belong to all the categories.
2. The linear cuts can always be eliminated, in every system $\mathcal{P}$ where they exist; eliminating them at some level $i$ can lead to some new cuts at the same level $i$, of unknown name.
3. The elimination of a (reducible) modal cut can lead to some new modal cut or *gc* cut at the same level (plus some other cuts at the higher levels).
4. The elimination of a (reducible) *gc* cut can only lead to some new *gc* cut at the same level.

These observations allow the following definition:

Definition 4.2.2 (Canonical Strategies) *A reduction of* $\Pi$ *is said* **canonical** *if it has the following shape:*

$$\Pi = \Pi_0 \to^* \Pi'_0 \to^* \Pi_1 \to^* \Pi'_1 \to^* \ldots \to^* \Pi'_\partial \to^* \Pi_{\partial+1} \to^* \Sigma$$

*where* $\partial = \partial(\Pi)$, $\Pi'_i$ *is obtained reducing all the linear cuts at level i in* $\Pi_i$, $\Pi_{i+1}$ *is obtained reducing all the modal cuts at level i in* $\Pi'_i$ *and* $\Sigma$ *is obtained reducing all the gc cuts in* $\Pi'_{\partial+1}$.

Lemma 4.2.3 (Canonical Strategies Are The Worst Ones)
Let $\Pi$ be a proof net. For every reduction $\sigma : \Pi = \Pi_0 \to \Pi_1 \to \ldots \to \Pi_l = \Sigma$ there exists a canonical reduction $\tau : \Pi = \Pi'_0 \to \Pi'_1 \to \ldots \to \Pi'_{l'} = \Sigma$ such that (i) $l' \geq l$ and (ii) $\max\{|\Pi'_i| \mid i \leq l'\} \geq \max\{|\Pi_j| \mid j \leq l\}$.

Moreover, if $\sigma$ is a reduction performed in some $\mathcal{P} \subseteq$ MS, then also $\tau$ is performed in $\mathcal{P}$.

**Proof.** We proceed by induction on $l$. If $l \leq 1$, of course the reduction is canonical. Otherwise, we can assume that $\Pi_0 \to \ldots \to \Pi_{l-1}$ can be transformed into some canonical reduction $\Pi = \Pi'_0 \to \Pi'_1 \to \ldots \to \Pi'_k = \Pi_{l-1}$, while $\Pi_{l-1} \to_\alpha \Pi_l$ does not follow this canonical reduction. The reduction of $\alpha$ does not follow a canonical reduction because there is another (reducible) cut $\beta$ that should be reduced before $\alpha$. We claim that we can reduce $\beta$ before $\alpha$, with some further minor modifications in the reduction sequence. Let us distinguish some cases:

1. $\alpha$ is a modal cut and $\beta$ a linear cut at the same level. This case is easy: $\alpha$ and $\beta$ do not interact at all with each other (they are not a critical pair), so there is no difference in the order of their reduction. We exchange them, and we build a new sequence, with the same length $l' = l$ and with the same greatest size.

2. $\alpha$ is a *gc* cut and $\beta$ a modal or linear cut. We can exchange the order of the 2 cuts. Now, it's possible that some new reducible cuts appear, because according to $\sigma$ it was erased before reduction; we just reduce them and we get a new sequence with $l' \geq l$. The new sequence has the same greatest size than the previous one.

3. $\alpha$ has depth $d'$ and $\beta$ has depth $d < d'$. Let us call $b$ the box at level $d$ in which $\alpha$ is contained. Let us consider the possible $\beta$'s.
   - Of course, if $\beta$ is not a cut with $b$, we can just exchange $\alpha$ and $\beta$.
   - Similarly if $\beta$ is $[P/P]$ with the box $b$.
   - If $\beta$ is $[P/\mathbf{Y}]$ with the box $b$, we can reduce $\beta$ before $\alpha$: in this way $b$ is duplicated. The new normalization sequence is strictly longer than before, $l' > l$. Moreover the greatest size of this new sequence can be greater than the greatest size of the previous one.

□

Lemma 4.2.4 (A Condition for Polynomiality)
Let $\mathcal{P} \subseteq$ MS. Let us assume that for every $\partial$ there exist two polynomials $p(x), q(x)$ such that for every proof net $\Pi$ of $\mathcal{P}$ with $\partial(\Pi) \leq \partial$ and for every $d \leq \partial(\Pi)$:

$$[\Pi]^d \leq p(|\Pi|), \qquad ||\Pi||^d \leq q(|\Pi|).$$

Then $\mathcal{P}$ is strongly polytime.

**Proof.** We prove that the result holds for every *canonical* reduction; Lemma 4.2.3 tell us that in fact it holds for every reduction.

Let us consider a canonical reduction $\sigma$:

$$\Pi = \Pi_0 \to^* \Pi'_0 \to^* \Pi_1 \to^* \Pi'_1 \to^* \ldots \to^* \Pi_{\partial+1} \to^* \Sigma.$$

$\Pi'_i$ is obtained from $\Pi_i$ in linear time $|\Pi_i|$, and the size is not increased: $|\Pi'_i| \leq |\Pi_i|$.
$\Pi_{i+1}$ is obtained from $\Pi'_i$ in time $p(|\Pi'_i|)$, and the size is at most $|\Pi_{i+1}| \leq q(|\Pi'_i|)$.
$\Sigma$ is obtained from $\Pi_{\partial+1}$ again in linear time $|\Pi_{\partial+1}|$, and the size does not increase: $|\Sigma| \leq |\Pi_{\partial+1}|$. So

$$|\Pi_i| \leq q\left(|\Pi'_{i-1}|\right) \leq q\left(|\Pi_{i-1}|\right) \leq q\left(q\left(|\Pi_{i-2}|\right)\right) \leq \ldots \leq q^{(i)}\left(|\Pi|\right) \leq q^{(\partial)}\left(|\Pi|\right).$$

This means that $\mathcal{P}$ is polysize. Moreover, the reduction takes at most

$$\begin{aligned} & |\Pi_0| + p\left(|\Pi'_0|\right) + |\Pi_1| + p\left(|\Pi'_1|\right) + \ldots + p\left(|\Pi_\partial|\right) + |\Pi_{\partial+1}| \\ \leq\ & |\Pi_0| + |\Pi_1| + \ldots + |\Pi_{\partial+1}| + p\left(|\Pi'_0|\right) + p\left(|\Pi'_1|\right) + \ldots + p\left(|\Pi_\partial|\right) \\ \leq\ & (\partial+2)\, q^{(\partial)}\left(|\Pi|\right) + (\partial+1)\, p\left(q^{(\partial)}\left(|\Pi|\right)\right) \end{aligned}$$

steps, and so $\mathcal{P}$ is also polystep. □

This proof of weak polynomiality is a generalization of (part of) the proof of weak polynomial soundness for ILAL in [AR02]. The proof of strong polynomiality is a variation of the proof of strong polynomiality for ILAL in [Ter07], but with a more combinatorial flavour.

## 4.3 Semantical Criteria for Polytime Soundness

The first propositions that we shall prove to distinguish polytime and not polytime subsystems $\mathcal{P}$ are *semantical*, in the sense that they are based on the context semantics of the proof nets of $\mathcal{P}$.

### 4.3.1 Context semantics: definitions

We adapt the context semantics technology of [DL08]. Context semantics essentially defines a notion of CS-*path* over the nodes of a proof net. A CS-path between two nodes tells that they can *communicate* each other. For example, let us imagine that a proof net $\Pi$ contains two nodes $u, v$, one $\otimes_{\mathcal{L}}$ and the other one $\otimes_{\mathcal{R}}$, and that during some reduction $\Pi \to^* \Pi'$ the residues $u', v'$ of $u, v$ become connected through a cut, so that they can finally annihilate. In this case, the context semantics says that, in $\Pi$, there is a CS-*path* between $u$ and $v$. So, CS-paths can be used to predict the future reductions of a proof net. Technically, CS-paths are obtained using *contexts*, that is quadruples $(e, \mathcal{U}, V, b)$, that record the nodes already encountered over a CS-path. We will describe them in a moment.

We just stress that our framework is simpler, since *digging* and *dereliction* are missing in MS. The first consequence is that we can use contexts $(e, \mathcal{U}, b)$ with just three components. Though, our version of context semantics must supply a quantitative computational model for systems where (i) the full cut-elimination does not hold, and (ii) the *weakening* is on arbitrary formulæ.

Definition 4.3.1 (Exponential Signatures) *An* **exponential signature** *is a formula generated by the following grammar:*

$$t \stackrel{\text{def}}{=} \mathtt{e} \mid \mathtt{r}(t) \mid \mathtt{l}(t).$$

*$\mathcal{E}$ is the set of exponential signatures.*

Definition 4.3.2 (Stacks) *A* **stack element** *is an element of $S \stackrel{\text{def}}{=} \{\mathtt{a}, \mathtt{o}, \mathtt{f}, \mathtt{s}, \mathtt{x}\} \cup \mathcal{E}$. A* **stack** *is a finite non empty sequence of stack elements, that is an element $s \in S^+$.*

Definition 4.3.3 (Polarities) *A* **polarity** *is an element of $\mathcal{B} \stackrel{\text{def}}{=} \{+, -\}$. If $c$ is a polarity, we denote $c \downarrow$ the other one. When needed, we can consider the two polarities as $\pm 1$.*

Definition 4.3.4 (Contexts) *Let $\Pi$ a proof net. Then a* **context** *of $\Pi$ is an element of*

$$\mathrm{C}_{\Pi} \stackrel{\text{def}}{=} E_{\Pi} \times S^+ \times \mathcal{B}.$$

The nodes of MS induce rewriting rules over contexts:

Definition 4.3.5 (Rewriting Relation Among Contexts) *Let $\Pi$ be a proof net, $C, C' \in \mathrm{C}_{\Pi}$. We write $C \mapsto_{\Pi} C'$, if $C'$ can be obtained from $C$ through the rewriting relation described in Figure 4.2.*

| | | | |
|---|---|---|---|
| 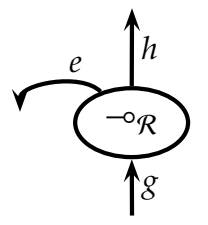 | $(g, U, +) \mapsto_\Pi (h, U \cdot \mathtt{o}, +)$<br>$(e, U, -) \mapsto_\Pi (h, U \cdot \mathtt{a}, +)$<br>$(h, U \cdot \mathtt{o}, -) \mapsto_\Pi (g, U, -)$<br>$(h, U \cdot \mathtt{a}, -) \mapsto_\Pi (e, U, +)$ | 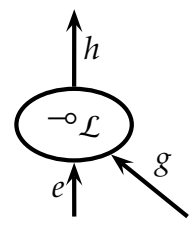 | $(g, U, +) \mapsto_\Pi (e, U \cdot \mathtt{a}, -)$<br>$(h, U, -) \mapsto_\Pi (e, U \cdot \mathtt{o}, -)$<br>$(e, U \cdot \mathtt{a}, +) \mapsto_\Pi (g, U, -)$<br>$(e, U \cdot \mathtt{o}, +) \mapsto_\Pi (h, U, +)$ |
| 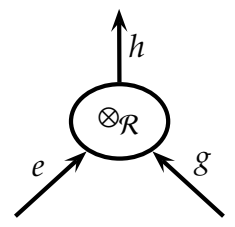 | $(e, U, +) \mapsto_\Pi (h, U \cdot \mathtt{f}, +)$<br>$(g, U, +) \mapsto_\Pi (h, U \cdot \mathtt{x}, +)$<br>$(h, U \cdot \mathtt{f}, -) \mapsto_\Pi (e, U, -)$<br>$(h, U \cdot \mathtt{x}, -) \mapsto_\Pi (g, U, -)$ | 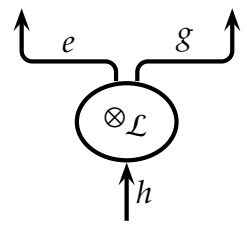 | $(h, U \cdot \mathtt{f}, +) \mapsto_\Pi (e, U, +)$<br>$(h, U \cdot \mathtt{x}, +) \mapsto_\Pi (g, U, +)$<br>$(e, U, -) \mapsto_\Pi (h, U \cdot \mathtt{f}, -)$<br>$(g, U, -) \mapsto_\Pi (h, U \cdot \mathtt{x}, -)$ |
| 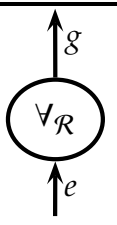 | $(e, U, +) \mapsto_\Pi (g, U \cdot \mathtt{s}, +)$<br>$(g, U \cdot \mathtt{s}, -) \mapsto_\Pi (e, U, -)$ | 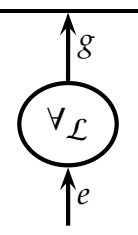 | $(e, U \cdot \mathtt{s}, +) \mapsto_\Pi (g, U, +)$<br>$(g, U, -) \mapsto_\Pi (e, U \cdot \mathtt{s}, -)$ |
| 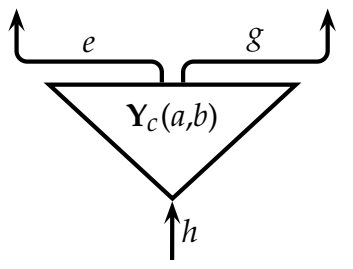 | $(h, U \cdot \mathtt{l}(t), +) \mapsto_\Pi (e, U \cdot t, +)$<br>$(h, U \cdot \mathtt{r}(t), +) \mapsto_\Pi (g, U \cdot t, +)$<br>$(e, U \cdot t, -) \mapsto_\Pi (h, U \cdot \mathtt{l}(t), -)$<br>$(g, U \cdot t, -) \mapsto_\Pi (h, U \cdot \mathtt{r}(t), -)$ | 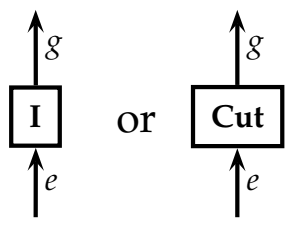 | $(e, U, +) \mapsto_\Pi (g, U, +)$<br>$(g, U, -) \mapsto_\Pi (e, U, -)$ |
| 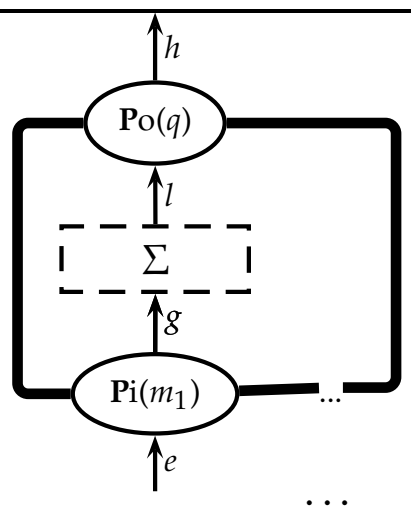 | $(e, U, +) \mapsto_\Pi (h, U, +)$<br>$(h, U, -) \mapsto_\Pi (e, U, -)$ | | |

Figure 4.2: Rewriting Relation among contexts. Notice that if $(e, U, b) \mapsto_\Pi (e', U', b')$ then also $(e', U', b' \downarrow) \mapsto_\Pi (e, U, b \downarrow)$.

Definition 4.3.6 (Canonical, Initial, Final Contexts) $(e, t \cdot U, b)$ *is* **canonical** *if $t$ is an exponential signature, $U$ does not contain exponential signatures, and if $U$ contains an even number of* $\mathtt{a}$*'s, then $b$ is $+$, otherwise, if $U$ contains an odd number of* $\mathtt{a}$*'s, then $b$ is $-$.*
$(e, t, +)$ *is* **initial** *if it is canonical and $t$ is exponential.*
$((u, v), \mathtt{e} \cdot U, b)$ *is* **final** *if it is canonical, and either*

1. *$b = +$ and $\alpha(v) \in \{\mathbf{W}, \mathbf{o}\}$, or*

2. *$b = -$ and $\alpha(v) = \mathbf{i}$.*

Canonical, initial and final contexts are the atomic steps that we use to compose the *correct* CS-paths to travel along the edges of a proof net. "Correct" means that a CS-path highlights the pairs of nodes that, eventually, will interact by means of a cut-elimination step, in the course of the normalization. The ultimate goal of defining CS-paths is to use them for walking through a net from the root of any box to either a weakening node that will erase it, or a terminal node of the whole net or of a proof net inside a box. The existence of many canonical CS-paths from the same box root means the existence of contraction nodes that, in principle, will duplicate the box. In particular, any initial context is defined to go upward — thanks to its positive polarity — in the proof net, traversing contraction nodes — thanks to its exponential signature —. Any final context says that, starting from the root of a box, we have walked through a net "consuming" all the directions settled by an exponential signature, getting to the empty exponential signature $\mathsf{e}$. Figure 4.3 is an example of complete CS-path from the root of the rightmost box to the output of the net, only highlighting the stack and the polarity. The leftmost box is "skipped" by the sequence of contexts associated to the net, because we are in a stratified setting.

Recall that we have defined graph-theoretical paths on page 33. As anticipated, the following is a *different* notion of path:

Definition 4.3.7 (CS-Paths and Maximal CS-Paths) *Let $\Pi \in \mathbf{PN}(\mathfrak{B}_{\mathbb{X}})$, $b \in V_\Pi$ a $Po$ node or a contraction, $v \in V_\Pi$. A* **CS-path** *from $b$ to $v$ is a finite sequence*

$$\tau \stackrel{\text{def}}{=} (u_0, u_1 \ldots, u_{n-1}, u_n) \in {V_\Pi}^*$$

*of nodes in $\Pi$, where $b = u_0$ and $v = u_n$, for which there is a sequence of contexts*

$$T \stackrel{\text{def}}{=} (((b, u_1), U_0, b_0), \ldots, ((u_{n-1}, v), U_{n-1}, b_{n-1})) \in {C_\Pi}^*$$

*such that (i) $((b, u_1), U_0, b_0)$ is initial, and (ii) $(e_i, U_i, b_i) \mapsto_\Pi (e_{i+1}, U_{i+1}, b_{i+1})$, for every $0 \leq i \leq n-1$.*

*Moreover, $\tau$ is maximal if $((u_{n-1}, v), U_{n-1}, b_{n-1})$ is final.*

*Finally, we call $T$ the* **sequence of contexts associated to** *$\tau$.*

Please notice that, differently from [DL08], a CS-path is made up of nodes, and *not* of contexts. This implies that two CS-paths are equal when they pass through the same nodes, regardless their contexts.

The definition of maximal CS-paths confirms the intuition given before. Every of them starts from the root of a box $b$ and shows where the root of a copy of $b$ will eventually occur, after every instance of the rules only in Figures 3.7, and 3.8 have been applied.

Notice that an initial context is canonical, as well as all the contexts in a CS-path are canonical.

A CS-path is *not*, in general, a graph-theoretical CS-path, because of the orientation of edges.

Definition 4.3.8 (CS-Path Length) *If $\tau = (u_0, u_1, \ldots, u_n)$ is a CS-path, the* **length** *of $\tau$ is $\text{len}(\tau) \stackrel{\text{def}}{=} n$.*

Definition 4.3.9 (Cycles) *A* **cycle** *is a CS-path $\tau$ such that, if $((e_i, U_i, b_i))_{i \leq n}$ is its associated sequence of contexts, then $e_0 = e_n$ and $b_0 = b_n$. A proof net is* **acyclic** *if it has no cycles.*

Definition 4.3.10 (CS-Paths, Number of CS-Paths) *Let $\Pi \in \mathbf{PN}(\mathfrak{B}_{\mathbb{X}})$, $b \in B_\Pi$. Then*

$$O_\Pi(b) \stackrel{\text{def}}{=} \{\tau \mid \tau \text{ is a maximal CS-path from } b \text{ to some } v \in V_\Pi\}.$$

$$R_\Pi(b) \stackrel{\text{def}}{=} \# O_\Pi(b).$$

Like in [DL08] we call $R_\Pi(b)$ the **number of copies** of $b$. The expression "number of copies" anticipates the content of Lemma 4.3.16: $R_\Pi(b)$ shall give a bound on the number of copies of $b$ that can be generated during a reduction. This is the content of DEFINITION 4.3.11 (PROOF NET WEIGHT) *Let* $\Pi$ *be a proof net. Its* **weight at depth** $d \leq \partial(\Pi)$ *is*

$$\mathrm{W}_d(\Pi) \stackrel{\text{def}}{=} \sum_{b \in B^d_\Pi} R_\Pi(b).$$

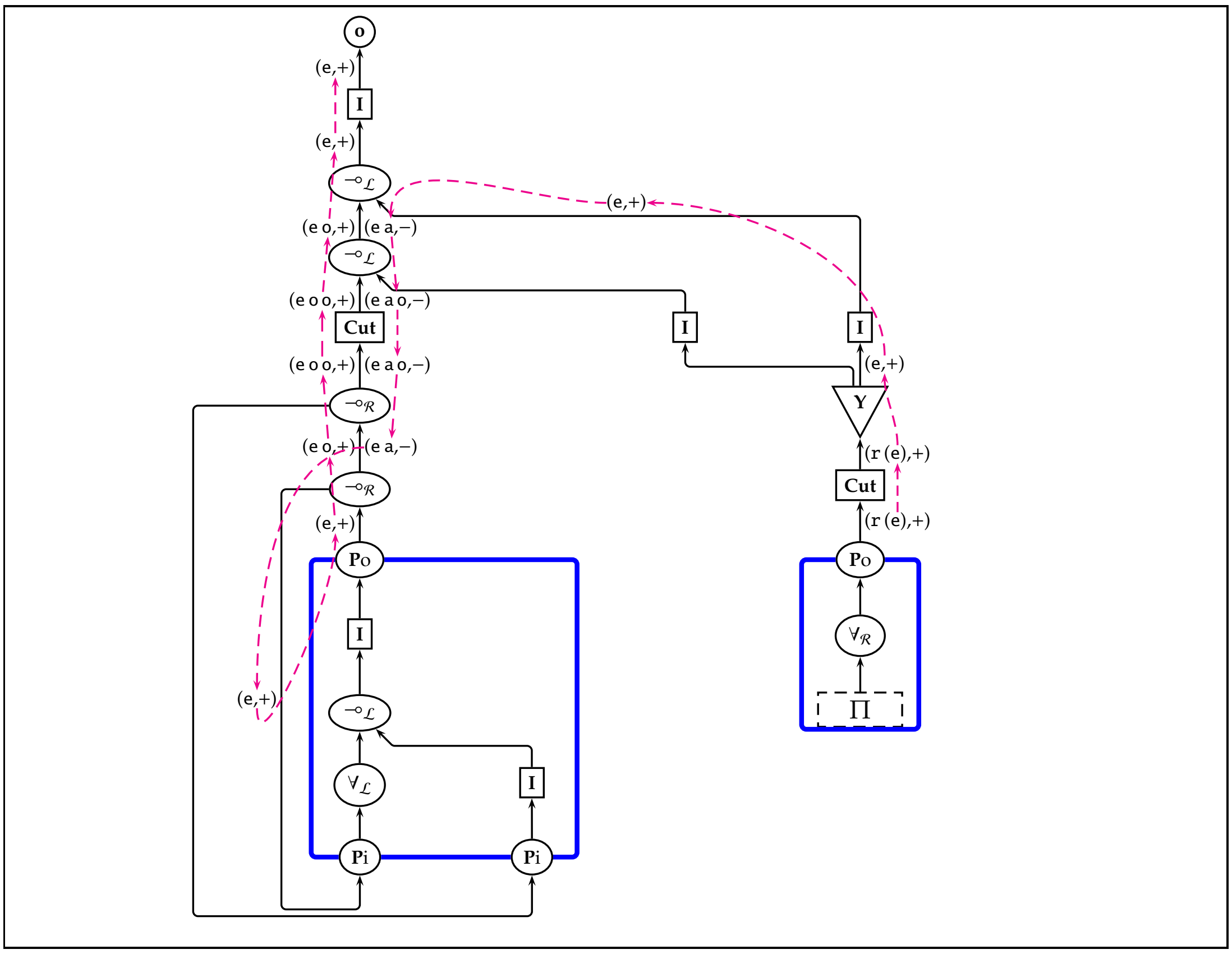

Figure 4.3: An example of maximal path.

DEFINITION 4.3.12 (PROOF NET MODIFIED WEIGHT) *Let* $\Pi$ *be a proof net. Its* **modified weight at depth** $d \leq \partial(\Pi)$ *is*

$$\mathrm{T}_d(\Pi) = \sum_{u \in V^d_\Pi} \mathrm{T}_d(\Pi, u)$$

$$\mathrm{T}_d(\Pi, u) \stackrel{\text{def}}{=} \begin{cases} 1 & \alpha(u) \in \{\mathbf{I}, \mathrm{cut}, \mathbf{W}, \mathbf{h}, Pi(\mathfrak{m}), \mathbf{i}, \mathbf{o}\} \\ 3 & \alpha(u) \in \{\otimes_\mathcal{L}, \otimes_\mathcal{R}, \forall_\mathcal{L}, \forall_\mathcal{R}, \multimap_\mathcal{L}, \multimap_\mathcal{R}, \mathbf{Y}_\mathfrak{q}(\mathfrak{m}, \mathfrak{n})\} \\ 2 \cdot (P_\Pi(u) + 1) \cdot R_\Pi(u)^2 & \alpha(u) \in \{Po(\mathfrak{m})\}. \end{cases}$$

($P_\Pi$ was defined on page 28). Notice that this weight is slightly simpler than the one defined in [DL08]. Again, the simplification is possible thanks to the *stratification* of our setting.

### 4.3.2 Properties of the Context semantics

We show that the main properties of the context semantics, as proved in [DL08], keep holding in our setting, in which, we recall, we have unconstrained weakening and, for any generic $\mathcal{P}$ of MS, we cannot assure full cut elimination. Recall, indeed, that cut elimination holds for each full $\mathfrak{B}_{\mathbb{X}}$; so, in all the following proofs, we will imagine that the reductions are performed in $\mathfrak{B}_{\mathbb{X}}$, instead than in particular $\mathcal{P} \subseteq \mathfrak{B}_{\mathbb{X}}$.

In order to simplify our proofs, we firstly print some properties (from Remark 4.3.13 up to Lemma 4.3.15) that do not hold in the general CS settings, but only in our specific stratified setting.

Remark 4.3.13 (Relating contexts in a CS-path and the top of a stack.) If $(e, t \cdot U, b)$ is a canonical context *in a* CS*-path* $\tau$, the top of the stack $t \cdot U$ relates to the formula $\beta(e)$ that labels $e$. If $\beta(e)$ is $A \otimes B$, then the top is $\mathtt{f}$ or $\mathtt{x}$. Similarly, $\beta(e) = \forall\alpha.A$ corresponds to the $\mathtt{s}$, while $\beta(e) = A \multimap B$ to one between $\mathtt{o}$ and $\mathtt{a}$. Finally, if $\beta(e) = \natural A$, the top element must be the exponential signature $t$.

Remark 4.3.14 (Some Rules are Useless) We remark that, thanks to what we have just underlined, the canonical contexts *inside a* CS*-path* $\tau$ can never reach neither a $Po$ node, nor a $\mathbf{Y}$ node with negative polarity. Indeed, a negative polarity implies that at least one $\mathtt{a}$ is present on the stack, while the context of an edge incident in a contraction or $Po$ node has $t$ as stack. As a consequence, 3 of the the rewriting rules from contexts to contexts are useless: they are the three rules that arrive to nodes $Po, \mathbf{Y}$ with negative polarity. Also, a CS-path only consumes an (initial) exponential signature, never adding $\mathtt{l}$ or $\mathtt{r}$ symbols.

In traditional CS, composition of CS-paths is not in general a CS-path. Instead, with our limited notion of CS, the property holds:

Lemma 4.3.15 (Compositions of CS-Paths is a CS-Path)
Let $\tau$ be a CS-path from $u$ to $v$ inside some proof net $\Pi$, where $v$ is a $Po$ or a contraction; and $\sigma$ be a CS-path from $v$ to $w$. Then, the path obtained joining $\tau$ and $\sigma$ is still a CS-path.

**Proof.** The CS-path $\tau$ starts in $u$ with context $(e_1, t_1, +)$ and arrives in $v$ with context $(e_2, t_2, +)$. The CS-path $\sigma$ starts in $v$ with context $(e_3, t_3, +)$ and arrives in $w$ with context $(e_4, t_4, +)$. We pointed out that $\tau$ cannot add new exponential signatures to $t_1$; so $t_1 = T(t_2)$. So, $t_2$ is untouched and we can freely replace it with $t_3$. We have found another different sequence of contexts for $\tau$: it starts in $u$ with context $(e_1, T(t_3), +)$ and arrives in $v$ with context $(e_2, t_3, +)$. This sequence of context may go on up to $w$ following the contexts for $\sigma$. So we found a sequence of CS-paths for the composition of $\tau$ and $\sigma$. □

The following Lemma explains the meaning of the expression "number of copies".

Lemma 4.3.16 (Number of Copies)
Let $\Pi \to_d^* \Sigma$, $b \in B_\Pi^d$ and $\Theta$ the proof net inside $b$ at depth $d+1$. Then, in $\Sigma$, there are at most $R_\Pi(b)$ equal residues ("copies") of $\Theta$.

Definition 4.3.17 *Let $\Pi$ reduce to $\Sigma$ in a single* ns*, by firing a cut $c$. Also, let $\tau = (u_0, \ldots, u_{i-1}, c, u_{i+1}, \ldots, u_n)$ be a* CS*-path in $\Pi$ that passes by $c$. The* **residue** $\mathrm{Res}(\tau)$ **of** *$\tau$ is the set of the residues of the nodes of $\tau$ in $\Sigma$.*

LEMMA 4.3.18 (THE RESIDUE OF A CS-PATH IS A CS-PATH)
Let $\Pi \to \Sigma$ through the reduction of a cut $c$. Let $\tau = (u_0, \ldots, u_{i-1}, c, u_{i+1} \ldots, u_n)$ be a CS-path in $\Pi$ that passes by $c$. The residue $\mathrm{Res}(\tau)$ of $\tau$, ordered in the natural way, is still a CS-path.

**Proof.** The proof is by cases on the possible *ns*. □

One may imagine to relax the Definition 4.3.17 allowing the path $\tau$ not passing by $c$. However in this case Lemma 4.3.18 would be false: think to the $[\mathbf{h}/P]$ *ns*, where the residue of a path may be disconnected.

On the other side, every CS-path in $\Sigma$ is the residue of – at least – one CS-path in $\Pi$; we call $\Phi$ the relation that associates a CS-path in $\Sigma$ with all the corresponding CS-paths in $\Pi$. $\Phi$ is not, in general, a function.

LEMMA 4.3.19
Every proof net $\Pi$ is acyclic.

**Proof.** Let us consider a canonical reduction for $\Pi$:

$$\Pi = \Pi_0 \to \Pi_1 \to \ldots \to \Pi_n \to^* \Sigma$$

where $n$ has been chosen least (please see on page 50 the definition of canonical strategy). We proceed by induction on $n$.

**Base case.** $n = 0$ means $\Pi$ contains only *gc* cuts. Straightforward to prove, by induction on the structure of $\Pi$, that if $(e, U, b) \mapsto^+_\Pi (e, V, c)$ then $b \neq c$.

**Inductive case.** If $\Pi \to_S \Pi_1$, by inductive hypothesis we can assume $\Pi_1$ is acyclic. Now, every CS-path $\tau$ in $\Pi$ corresponds to (at least) CS-path $\Phi(\tau)$ in $\Pi_1$, as we have already said; and the correspondence $\Phi$ preserves acyclicity. [1]

□

LEMMA 4.3.20
For each $\Pi$ and $b \in B_\Pi$, $R_\Pi(b) \geq 1$.

**Proof.** Let us consider a sequence of contexts starting from a box $b$ with a context $(e, t, +)$, where $t$ is an exponential signature. Provided $t$ is long enough, there is no way to *stop* such a sequence, unless (i) we arrive in a $\mathbf{W}, \mathbf{i}$ or $\mathbf{o}$ node, or (ii) the CS-path is a cycle and returns at the starting point. But this second situation is forbidden by the previous lemma. So there is at least a canonical sequence of contexts starting from $b$ and arriving in a $\mathbf{W}, \mathbf{i}$ or $\mathbf{o}$ node. Choosing $t$ of the correct size, we can make the last context a final context, and this is a maximal CS-path. □

LEMMA 4.3.21
If $\Pi$ at level $d$ has only *gc* and $[P/P]$ cuts, then for every $b \in B_\Pi$, $R_\Pi(b) = 1$.

[1]This latter observation would not be obvious if $S$ were a *gc* step.

**Proof.** By structural induction on $\Pi$. □

LEMMA 4.3.22 ($\mathrm{W}_d$ AND $\mathrm{T}_d$ DECREASE)
Given a proof net $\Pi$, let us suppose $\Pi \to_S \Sigma$. Then

1. $\mathrm{W}_d(\Pi) = \mathrm{W}_d(\Sigma)$, if $S \in \{[\multimap / \multimap], [\otimes/\otimes], [\forall/\forall], [P/\mathbf{Y}]\}$;
2. $\mathrm{W}_d(\Pi) = \mathrm{W}_d(\Sigma) + R_\Pi(b)$, if $S = [P/P]$ and $b$ is one of the merged boxes;
3. $\mathrm{W}_d(\Pi) \geq \mathrm{W}_d(\Sigma)$, if $S$ is a *gc* step;
4. $\mathrm{T}_d(\Pi) > \mathrm{T}_d(\Sigma)$;

**Proof.** In the first case the CS-paths and the boxes are untouched, so for every $b$, $R_\Pi(b) = R_\Sigma(\mathrm{Res}(b))$. In the second case, a box $b$ disappears, so the sum that defines $\mathrm{W}_d$ decreases of exactly $R_\Pi(b)$. In the third case, also, a box may disappear depending on what kind of node is cut with the *weakening* node.

For the last point, the reader can check that $T_d(\Pi)$ strictly decreases for every possible reduction step; we only prove the most difficult case. Let us assume $\Pi \to_S \Sigma$ by a $[P/\mathbf{Y}]$ normalization step; a box $b$ is duplicated in $b_1$, $b_2$. Let $P = P_\Pi(b) = P_\Pi(b_i)$ and $R_i = R_\Pi(b_i)$; it holds $R_\Pi(b) = R_1 + R_2$. Now, if we call $x$ the weight of a generic contraction, $y$ the weight of a cut, and $z$ the weight of a Pi node,

$$\begin{aligned} \mathrm{T}_d(\Pi) &= A + x + y + \mathrm{T}_d(b) + P \cdot z \\ &= A + 3 + 1 + 2(P+1)(R_1 + R_2)^2 + P \end{aligned}$$

$$\begin{aligned} \mathrm{T}_d(\Sigma) &= A + 2y + \mathrm{T}_d(b_1) + \mathrm{T}_d(b_2) + 2P \cdot \mathrm{T}_d(\text{a } Pi) + P \cdot x \\ &= A + 2 + 2(P+1)R_1^2 + 2(P+1)R_2^2 + 2P + 3P. \end{aligned}$$

$$T_d(\Pi) - T_d(\Sigma) = 2 + 2(P+1) \cdot 2R_1R_2 - 4P \geq 2 + 4(P+1) - 4P = 1 > 0$$

remembering that every $R_i$ is $\geq 1$. □

LEMMA 4.3.23 (WEIGHT AND MODIFIED WEIGHT ARE POLYNOMIALLY RELATED)
There is a polynomial $p(x, y)$ such that for every $\Pi$ and every $d$

$$\mathrm{T}_d(\Pi) \leq p\left(\mathrm{W}_d(\Pi), |\Pi|_d\right); \qquad \mathrm{W}_d(\Pi) \leq \mathrm{T}_d(\Pi).$$

**Proof.**

$$\mathrm{T}_d(\Pi) \leq 3|\Pi|_d + 2|\Pi|_d \sum_{b \in B^d_\Pi} R^2_\Pi(b) \leq 3|\Pi|_d + 2|\Pi|_d \left(\mathrm{W}_d(\Pi)\right)^2,$$

$$\mathrm{W}_d(\Pi) = \sum_{b \in B^d_\Pi} R_\Pi(b) \leq \sum_{b \in B^d_\Pi} R^2_\Pi(b) \leq \mathrm{T}_d(\Pi).$$

□

Lemma 4.3.24 (The Modified Weight bounds Time and Space)
Let $\Pi$ be a proof net. Then, for every $i > d$,

$$[\Pi]^d, ||\Pi||_d^d \leq T_d(\Pi), \qquad ||\Pi||_i^d \leq |\Pi|_i \cdot T_d(\Pi), \qquad ||\Pi||^d \leq |\Pi| \cdot T_d(\Pi).$$

**Proof.**

(i) Let us consider a reduction $\Pi \to_d^n \Sigma$. We already know that $T_d(\cdot)$ strictly decreases during reduction; so $n \leq T_d(\Pi)$. Since every node has weight at least 1, $|\Pi|_d \leq T_d(\Pi)$. So $||\Pi||_d^d \leq T_d(\Pi)$.

(ii) Let us consider a reduction $\Pi \to_d^* \Sigma$ and fix a level $i > d$. For every $u$ node at level $i$, inside a box $b \in B_\Pi^d$, at most $R_\Pi(b)$ copies of $u$ will appear in $\Sigma$ (Lemma 4.3.16). Let $b_1, \ldots, b_k$ be all the boxes at level $d$, and $S_1, \ldots, S_k$ their sizes at level $i$. Then, $||\Pi||_i^d \leq S_1 \cdot R_\Pi(b_1) + \ldots + S_k \cdot R_\Pi(b_k) \leq |\Pi|_i \cdot T_d(\Pi)$, since $R_\Pi(b_i) \leq \mathrm{T}_d(\Pi, b_i)$.

(iii) Immediate corollary of (ii).

□

Corollary 4.3.25
Let $\mathcal{P} \subseteq \mathfrak{B}_\mathbb{X}$. Let us assume there exists a polynomial $r(x)$ such that for every proof net $\Pi$ of $\mathcal{P}$ and for every $d$, $\mathrm{W}_d(\Pi) \leq r(|\Pi|)$. Then $\mathcal{P}$ is strongly polytime.

**Proof.** Lemma 4.3.23 and 4.3.24 and hypothesis imply that $[\Pi]^d, ||\Pi||^d, ||\Pi||_i^d \leq P(|\Pi|)$ for some polynomial $P(x)$. Then Lemma 4.2.4 implies the thesis. □

### 4.3.3 Strong Polytime and Spindles

This is one of our most important sections. Up to this point, we have found a condition (Corollary 4.3.25) that allows a subsystem to be polytime. However, that condition is quite general, we need something more precise. Here, we will define a geometrical condition, based on the geometrical structure called *spindle*, that can be satisfied by a subsystem $\mathcal{P}$. We will show that, if such a condition is satisfied, then also the hypothesis of Corollary 4.3.25 are satisfied, so that the subsystem is polytime. The main result is in Proposition 4.3.40.

Definition 4.3.26 *Let $\mathcal{P} \subseteq \mathfrak{B}_\mathbb{X}$, $\Pi \in$ **PN**$(\mathcal{P})$; $e \in E_\Pi$ an edge labelled $\mathfrak{m}A$ entering a contraction $u$; $f \in E_\Pi$ an edge labelled $\mathfrak{n}B$ outgoing a P∘ node $b$; $\partial(e) = \partial(f)$. A* **spindle** *$\mathfrak{m}A : \Sigma : \mathfrak{n}B$ between $e$ and $f$ (or also between $u$ and $b$) is a pair of* CS*-paths: $\tau$ from $e$ to $f$ passing through the left conclusion of $u$; $\rho$ from $e$ to $f$ passing through the right conclusion of $u$; and such that $\tau$ and $\rho$ are the only* CS*-paths connecting $e$ with $f$. $e$, $f$, $u$ are resp. the* **principal premise**, **principal conclusion** *and* **principal contraction** *of $\Sigma$. An edge that is premise (resp. conclusion) of a node of $\Sigma$, but that is not part of $\Sigma$, is said* **non-principal** *premise (resp. conclusion). We shorten $\mathfrak{m}A : \Sigma : \mathfrak{n}B$ with $\mathfrak{m} : \Sigma : \mathfrak{n}$.*

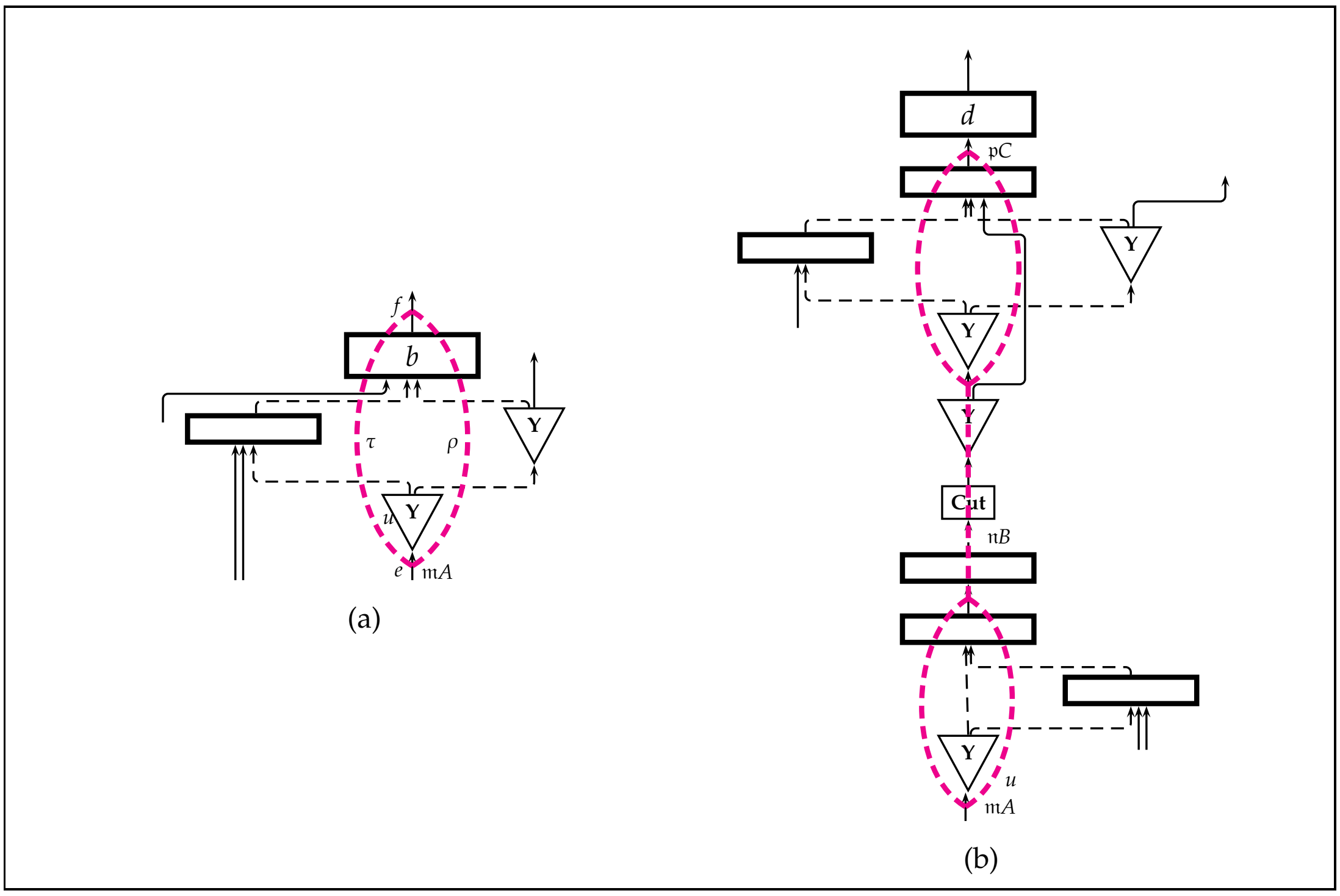

Figure 4.4: (a) A spindle, (b) a chain of two spindles.

Definition 4.3.27 *Let $\mathcal{P} \subseteq \mathfrak{B}_{\mathbb{X}}$, $\Pi \in$ **PN**$(\mathcal{P})$, $\Pi$ containing $r \geq 1$ spindles $\mathfrak{m}_1 A_1 : \Sigma_1 : \mathfrak{n}_1 B_1, \ldots, \mathfrak{m}_r A_r : \Sigma_r : \mathfrak{n}_r B_r$. Let us suppose that each spindle $\Sigma_i$ is between nodes $u_i$ and $b_i$, and for each $i < r$, $\Pi$ contains a* CS*-path between $b_i$ and $u_{i+1}$. The graph composed by the $r$ spindles and the $r-1$* CS*-paths is called a* **chain of** *$r$* **spindles***, and we shall write $\mathfrak{m}_1 A_1 : \Theta : \mathfrak{n}_r B_r$ for it. $|\Theta| = r$ denotes the number of spindles it contain.*

Notice that a spindle is a chain of 1 spindle.

Definition 4.3.28 *Let $\mathcal{P} \subseteq \mathfrak{B}_{\mathbb{X}}$, $\Pi \in$ **PN**$(\mathcal{P})$, $\mathfrak{m}A : \Theta : \mathfrak{n}B$ a chain of $r \geq 1$ spindles inside $\Pi$, between $u$ and $b$. Let us assume that $\Pi$ contains another node $d$ and a* CS*-path $\chi$ such that $\chi$ connects $b$ to $d$, and the last edge of $\chi$ is labelled $\mathfrak{m}A'$ (*possibly $d = b$ and $\chi = \emptyset$*). Then, the graph made up of $\Theta$ and $\chi$ is a* **dangerous chain of spindles***; we write $\mathfrak{m}A : (\Theta \cup \chi) : \mathfrak{m}A'$ for it. A* **dangerous spindle** *is a dangerous chain with $r = 1$.*

Note that if $(\Theta \cup \chi)$ is a dangerous chain of spindles starting and ending with the same formula $\mathfrak{m}A$, we can build other chains $\Theta_n \stackrel{\text{def}}{=} \underbrace{(\Theta \cup \chi) \bowtie \ldots \bowtie (\Theta \cup \chi)}_{n}$ as long as we want. We call **exponential chain** such an object $\Theta_n$, for each $n$. Please notice that, *a priori*, we are not sure that an exponential chain can reduce; but, if it reduces, the reduction takes an exponential time. We will prove that in Lemma 4.3.35.

Remark 4.3.29 (About the Definition of Spindle) Consider the following, alternative, definition. A chain of spindles is a single spindle in which we relax the request: "*the two* CS*-paths*

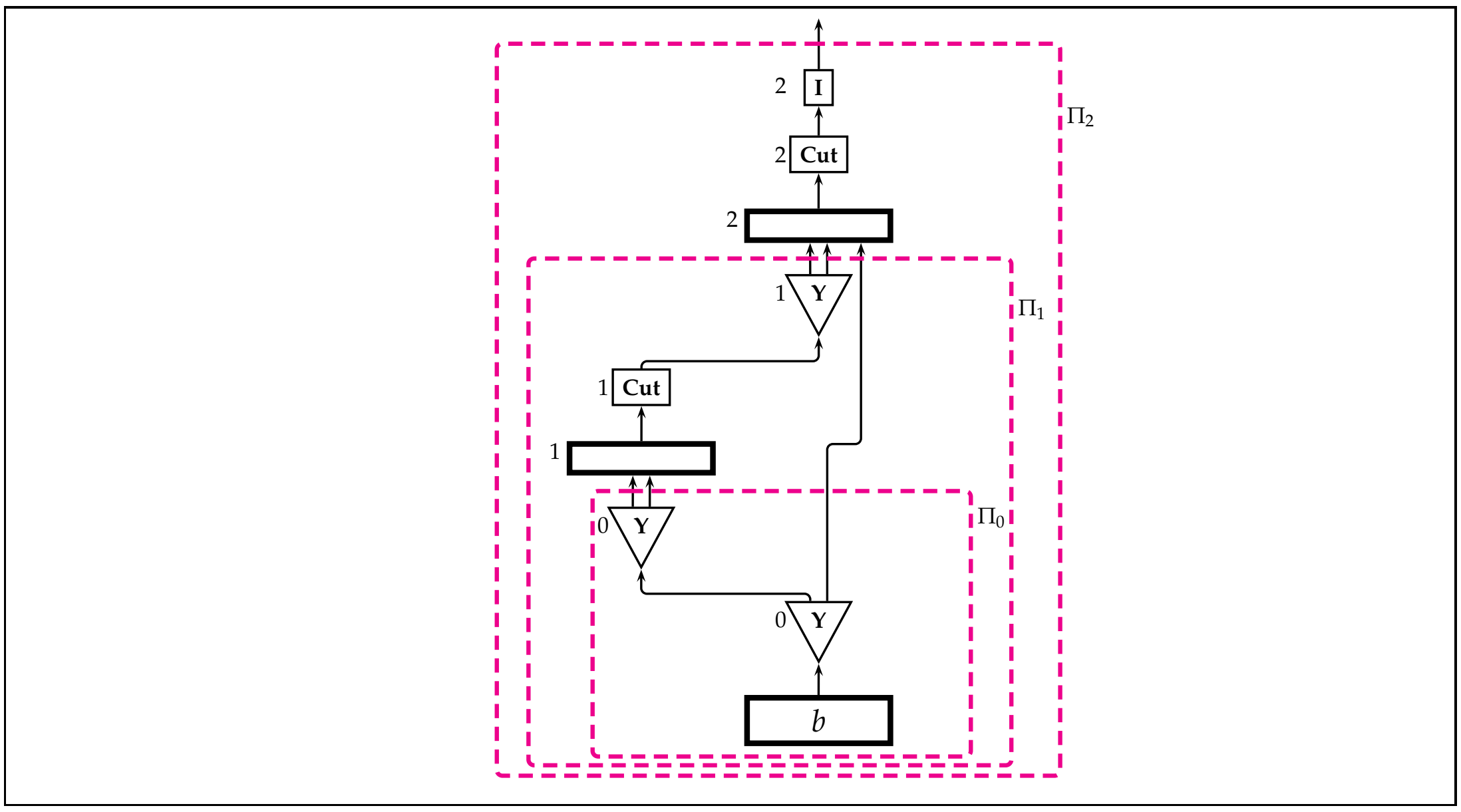

Figure 4.5: An example of proof net, whose nodes have been classified as in proof of Lemma 4.3.31.

*$\rho$ and $\tau$ are disjoint*". Such a definition is not equivalent to the previous one; but, more important, it is not good for us because we want to control the number $r$ of the spindles in the chain.

**Lemma 4.3.30** (One-Ring-Long Chains)
Let $\mathcal{P} \subseteq \mathfrak{B}_{\mathbb{X}}$. If $\mathcal{P}$ has a (dangerous) chain of spindles then it also has a (dangerous) spindle.

**Proof.** Let us take a chain of $r \geq 1$ spindles $\Theta$. We build a module $\Theta'$ containing: (a) the *lower* spindle of $\Theta$; (b) the $r-1$ paths that connect the $r$ spindles; (c) for each one of the other $r-1$ spindles, we chose exactly one among the two CS-paths composing it. We get a module $\Theta'$ with exactly one spindle. If $\Theta \cup \chi$ was dangerous, then $\Theta' \cup \chi$ is dangerous, too. □

**Lemma 4.3.31** (Short Chains and Number of CS-Paths)
Let $\mathcal{P} \subseteq \mathfrak{B}_{\mathbb{X}}$. Let us assume that every chain of spindles that can be built in $\mathcal{P}$ cannot have more than $L$ spindles, for some constant $L \in \mathbb{N}$. Then there exists a polynomial $p(x)$ such that for every $\Pi \in \mathbf{PN}(\mathcal{P})$, for every $d \leq \partial(\Pi)$ and for every $b \in B^d_\Pi$,

$$R_\Pi(b) \leq p\left(|\Pi|_d\right).$$

**Proof.** Let $b$ be a fixed box at depth $d$ in $\Pi$.

1. We want to classify the links $u \in V^d_\Pi$ in classes: class-0, class-1, class-2, ... according to *how many consecutive spindles there are between $b$ and $u$*. Let us consider all the possible CS-paths $\tau$ between $b$ and $u$. Let us call $\Sigma^u_b$ the module made up of all the nodes of all the $\tau$'s. Let us consider all the possible chains of spindles $\Phi^u_b, \Psi^u_b, \ldots, \Omega^u_b$ whose nodes are among the nodes of $\Sigma^u_b$. The node $u$ is said *class-$i$* if $i = \max\left\{\left|\Phi^u_b\right|, \left|\Psi^u_b\right|, \ldots, \left|\Omega^u_b\right|\right\}$. Figure 4.5 may help the comprehension. Notice that there are at most $L+1$ classes $(0, \ldots, L)$ by hypothesis.

2. Let $m_i$ be the *number of Contractions* of class-$i$ (they are all at depth $d$), and $m = m_0 + \ldots + m_L$. We observe that $m \leq |\Pi|_d$.
3. For $0 \leq i \leq L$, let $\Pi_i$ be the *module containing all and only the class-$j$ nodes*, for every $j \leq i$.
4. We define *maximal relatively to* $\Pi_i$ a CS-path $\tau$ starting from $b$ such that there exists a maximal CS-path $\tau'$ starting from $b$ whose intersection with $\Pi_i$ is exactly $\tau$. So, for example, the CS-paths maximal relatively to $\Pi_L$ are exactly the maximal CS-paths in $\Pi$.
5. We call $R_i$, for $0 \leq i \leq L$, the number of CS-paths starting from $b$ and maximal relatively to $\Pi_i$. The definitions imply $R_\Pi(b) = R_L$.

We prove that $R_i \leq \prod_{j=0}^{i} \left(m_j + 1\right)$ by induction on $i$. Two CS-paths separate only in a contraction node, but *not* in a $\otimes_{\mathcal{L}}$ node, which, instead, forces to go in one specific direction. In $\Pi_0$ there are no spindles, meaning $R_0 \leq m_0 + 1$. By induction, let $R_{i-1} \leq \prod_{j=0}^{i-1} \left(m_j + 1\right)$. Then, the CS-paths maximal relatively to $\Pi_i$ are at most $R_i \leq R_{i-1} \cdot (m_i + 1) \leq \prod_{j=0}^{i} \left(m_j + 1\right)$. So, $R_\Pi(b) = R_L \leq \prod_{j=0}^{L}(m_i + 1) \leq {|\Pi|_d}^{L+1}$. □

Lemma 4.3.31 and Corollary 4.3.25 directly implies:

Proposition 4.3.32 (Short Chains and Polynomiality)
Let $\mathcal{P} \subseteq \mathfrak{B}_{\mathbb{X}}$. Let us assume that every chain of spindles that can be built in $\mathcal{P}$ cannot have more than $L$ spindles, for some constant $L \in \mathbb{N}$. Then $\mathcal{P}$ is strongly polytime.

**Some examples.** Some useful didactic examples now follow (from Lemma 4.3.33 up to Example 4.3.36). The goal is to begin to understand, given a subsystem $\mathcal{P}$, if it is polytime or not. The following two Lemmas, in particular, are easy applications of Proposition 4.3.32; but, they can be easily proved also as corollaries of Proposition 4.3.40.

Lemma 4.3.33 (Some Polytime Systems)
Let $\mathcal{P} \subseteq \mathfrak{B}_{\mathbb{X}}$, with $\mathbb{X} = \{1, 2, \ldots, M\}$, ordered in the usual way, and whose rules satisfy the following restrictions:

1. if $\mathcal{P}$ satisfies $P_q(m_1, \ldots, m_k)$, then $q$ is greater or equal than every $m_i$, and equal to at most one among them.
2. if $\mathcal{P}$ satisfies $\mathbf{Y}_c(a, b)$, then $c$ is less or equal than both $a$ and $b$.

Then $\mathcal{P}$ is strongly polytime.

**Proof.** Let us consider a chain of $r$ spindles. Let us consider two among the 'main' $r$ contractions of the chain. Let us assume the upper one, $v$, has label $\mathbf{Y}_c(a, b)$, and the lower one, $u$, has label $\mathbf{Y}_d(e, f)$. There are two CS-paths starting from $u$ with modalities $e, f$ and arriving in $v$ with modality $c$. The CS-paths can contain only boxes and contractions: so, by hypothesis, $e \leq c$ and $f \leq c$. But the two CS-paths must join in a box, let us call it $B$; and at least one of the CS-paths, by hypothesis, strictly decreases its modality passing through $B$. We can assume $e \leq c$ and $f < c$. Now remember that $c \leq a, b$, and we realize that the two labels must be different: $(c, a, b) \neq (d, e, f)$. And this holds for every couple of contractions $u, v$ in the chain: so $r \leq M$. We can now apply Lemma 4.3.31 and Proposition 4.3.32 to conclude that the system is strongly polynomial. □

Similarly:

Lemma 4.3.34 (Some Polytime Systems)
Let $\mathcal{P} \subseteq \mathfrak{B}_{\mathbb{X}}$, with $\mathbb{X} = \{1, 2, \ldots, M\}$, ordered in the usual way, and whose rules satisfy the following restrictions:

1. if $\mathcal{P}$ satisfies $P_q(m_1, \ldots, m_k)$, then $p$ is greater than every $m_i$.
2. if $\mathcal{P}$ satisfies $\mathbf{Y}_c(a, b)$, then $c \leq a$ and $c < b$.

Then $\mathcal{P}$ is strongly polytime.

Lemma 4.3.35 (Some Non-Polytime Systems)
Let $\mathcal{P} \subseteq \mathfrak{B}_{\mathbb{X}}$. Let us assume that: (i) $\mathcal{P}$ can build a dangerous chain $\mathfrak{m}A : (\Sigma \cup \chi) : \mathfrak{m}A$ of spindles (in part. a single dangerous spindle); (ii) $\mathfrak{m}A$ is a tautology, that is $\mathcal{P}$ can prove $\vdash \mathfrak{m}A$; (iii) $\Sigma \cup \chi$ is a proof net; and (iv) $\mathcal{P}$ enjoys cut-elimination. Then, $\mathcal{P}$ is neither polysize, nor polystep.

**Proof.** Remember that $\Sigma \cup \chi$ is a chain of spindles between 2 edges labelled with the same modality. In this case, moreover the two edges are also labelled with the same formula $\mathfrak{m}A$. Using (ii) there exists a closed box $b \triangleright\ \vdash \mathfrak{m}A$ in $\mathcal{P}$. Using $b$ and (iii), we can build the proof net $\Theta_n = \underbrace{\Sigma \bowtie \ldots \bowtie \Sigma}_{n} \bowtie b$. Now, let us try to reduce $\Theta_n$. In particular, thanks to (iv), we can decide to reduce the whole level 0, but not the upper levels. In this way, the reduction is not performed in polysize: the content of $b$ is duplicated $2^n$ times. Now, at level 1 there are a least $2^n$ new cuts; reducing them, we understand that this reduction is not polystep, too. □

Example 4.3.36 We must be careful in the identification of spindles. Let us consider the following system with 4 modalities:

$$\mathcal{P} = \{\mathbf{I}, \multimap_{\mathcal{R}}, \multimap_{\mathcal{L}}, P_{\mathfrak{m}}(\mathfrak{n}, \mathfrak{q}), P_{\mathfrak{n}}(\emptyset), P_{\mathfrak{q}}(\emptyset), P_{\mathfrak{r}}(\mathfrak{n}, \mathfrak{n}, \mathfrak{q}, \mathfrak{q}), \mathbf{Y}_{\mathfrak{m}}(\mathfrak{n}, \mathfrak{n}), \mathbf{Y}_{\mathfrak{n}}(\mathfrak{q}, \mathfrak{q})\}\,.$$

If we only look at the rules, we may think $\mathcal{P}$ can build a spindle, the more, a dangerous spindle:

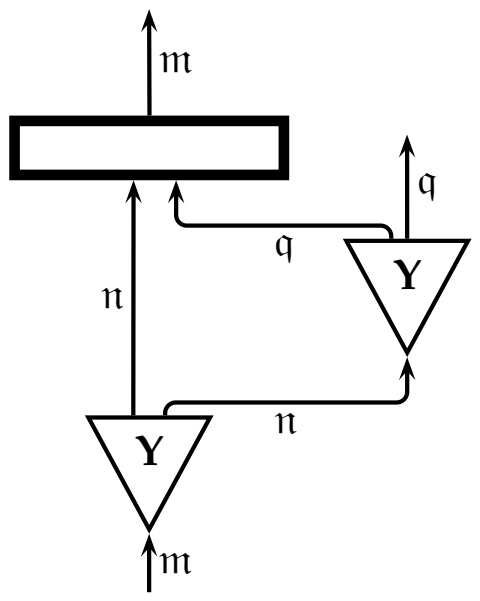

However, there is no way to place this spindle inside some proof net. Indeed, by induction on the construction rules, the reader can check that: *(i) for every $\Pi \in \mathbf{PN}(\mathcal{P})$ containing the rule $P_{\mathfrak{m}}(\mathfrak{n}, \mathfrak{q})$ at level 0, there are at most two modal inputs, one with modality $\mathfrak{n}$ and the other one with modality $\mathfrak{q}$; (ii) for every $\Pi$ containing the rule $P_{\mathfrak{r}}(\mathfrak{n}, \mathfrak{n}, \mathfrak{q}, \mathfrak{q})$ there are at most 4 modal inputs, with modality $\mathfrak{m}$, $\mathfrak{n}$ or $\mathfrak{q}$; and in particular, (iii) $\mathcal{P}$ can only build chains of at most 1 spindle.* So this system is polytime by Proposition 4.3.32.

The system $\mathcal{P}'$ obtained adding e.g. a **W** rule is not polytime:

$$\mathcal{P}' = \{\mathbf{I}, \multimap_{\mathcal{R}}, \multimap_{\mathcal{L}}, P_{\mathfrak{m}}(\mathfrak{n}, \mathfrak{q}), P_{\mathfrak{n}}(\emptyset), P_{\mathfrak{q}}(\emptyset), P_{\mathfrak{r}}(\mathfrak{n}, \mathfrak{n}, \mathfrak{q}, \mathfrak{q}), \mathbf{Y}_{\mathfrak{m}}(\mathfrak{n}, \mathfrak{n}), \mathbf{Y}_{\mathfrak{n}}(\mathfrak{q}, \mathfrak{q}), \mathbf{W}\}\,.$$

Of course the strange behaviour of $\mathcal{P}$ cannot occur in presence of the rules $\otimes_{\mathcal{R}}$ or **W**, nor in presence of only *standard* $\mathbf{Y}^i$ contractions. Indeed the first two rules always allow to create a proof net from a dangerous chain; and the standard contraction is almost useless inside a spindle.

### 4.3.4 A first Criterion for Polynomial Time Soundness.

The previous results (Proposition 4.3.32, Lemma 4.3.33, and Lemma 4.3.35) give some necessary and sufficient conditions for the polynomial time soundness of subsystems. Here, *we restrict our attention to sensible subsystems*, that have been defined on page 34. In this case we improve those theorems coming in particular to Lemma 4.3.39, which simplifies Lemma 4.3.35, and Proposition 4.3.40, that gives a criterion, i.e. a both sufficient and necessary condition, for polytime soundness.

We need the following two technical Lemmas.

Fact 4.3.37 (Structural Properties of Sensible Subsystems)
Let $\mathcal{P}$ be a sensible subsystem.

1. For every $l \geq 0$, $\mathcal{P}$ contains $\Pi_l \triangleright A, \ldots, A \vdash A$, with $A = \gamma \multimap \gamma$, $l$ occurrences of $A$ as assumptions, and $\partial(\Pi_l) = 0$.
2. Moreover, if $\mathbf{h}$ belongs to $\mathcal{P}$, then the formula $A$ can be any one in $\mathcal{F}_{\mathbb{X}}$.
3. If $\mathcal{P}$ proves $A, \ldots, A \vdash A$, with $l$ occurrences of $A$, then it proves also $\mathfrak{m}_1 A, \ldots, \mathfrak{m}_k A \vdash \mathfrak{q} A$ for $k \leq l$, for every $P_{\mathfrak{q}}(\mathfrak{m}_1, \ldots, \mathfrak{m}_k) \in \mathcal{P}$.

**Proof.** $\Pi_l$ is built using one $\multimap_{\mathcal{R}}$ or $\mathbf{h}$ rules, plus $l$ occurrences of the free weakening. □

Lemma 4.3.38 (Building Good Spindles)
Let $\mathcal{P} \subseteq \mathfrak{B}_{\mathbb{X}}$ be a sensible subsystem that can build a (dangerous) spindle $\Sigma$. In $\mathcal{P}$ there is another (dangerous) spindle $\Sigma''$ such that: (a) $\Sigma''$ contains only Contractions, Box-out and Cut nodes at level 0; (b) Every edge $e$ at level 0 has label $\mathfrak{q}A$, for some fixed $A$; (c) $\partial(\Sigma'') = 1$; and (d) The only premise of $\Sigma''$ is the principal one.

**Proof.** We begin the proof considering $\Sigma$ to be a standard, non-dangerous, spindle. The goal is to transform, step by step, $\Sigma$ into $\Sigma''$.

Let $\Pi \in \mathbf{PN}(\mathcal{P})$ be the proof net containing $\Sigma$, an example of which is in Figure 4.6a. We reduce all the linear cuts in $\Pi$ at level 0. We get to $\Pi' \in \mathbf{PN}(\mathcal{P})$ with $\Sigma'$, the reduct of $\Sigma$, in it. (See Figure 4.6b.) $\Sigma'$ is still a (dangerous) spindle, as well as the residuals of the two CS-paths $\rho, \tau$ of $\Sigma$ are still CS-paths in $\Pi'$. The three CS-paths cannot contain linear nodes. Indeed, we just observe that, e.g., $\tau$ must begin in a contraction, with label $\mathfrak{n}A$, and must stop in a box, with label $\mathfrak{m}B$. So it cannot cross neither any right-node, otherwise it would add a non-modal symbol to $\mathfrak{n}A$, and so the last formula could not be $\mathfrak{m}B$, nor any left-node, otherwise it would remove a non-modal symbol from $\mathfrak{n}A$.

We transform $\Sigma'$, just generated, into another graph $\Sigma^\star$ by replacing every proof net enclosed in a box $P_{\mathfrak{q}}(\mathfrak{m}_1, \ldots, \mathfrak{m}_k)$ of $\Sigma'$ by $\Pi_k \triangleright A, \ldots, A \vdash A$ as in point 1 of Fact 4.3.37 Consequently, every $\mathfrak{q}B$ at level 0 in $\Sigma'$ becomes $\mathfrak{q}A$ in $\Sigma^\star$, for some $\mathfrak{q}$. This replacement implies $\partial(\Sigma^\star) = 1$ (Figure 4.6c).

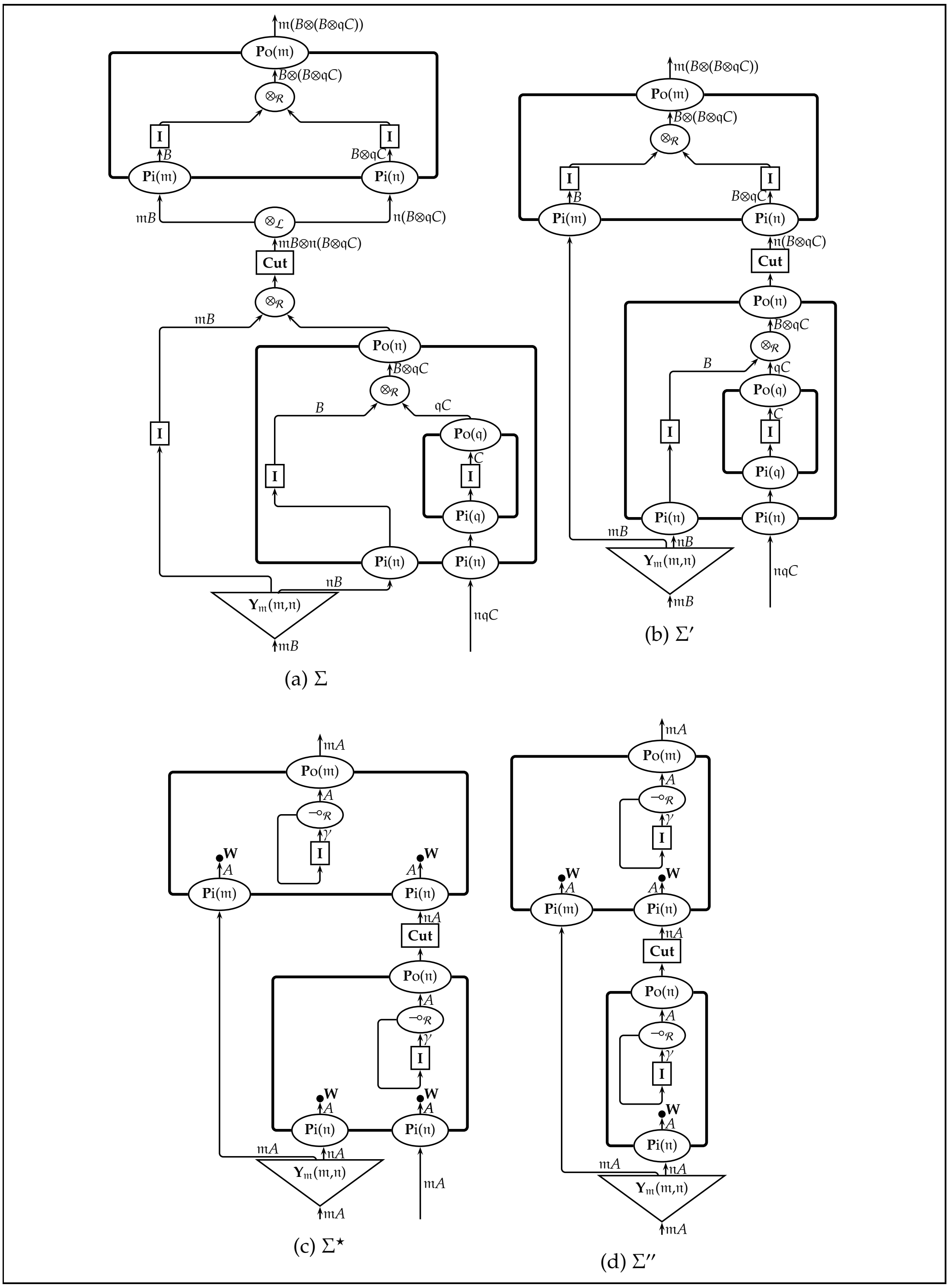

Figure 4.6: Examples relative the proof of Lemma 4.3.38.

$\Sigma^{\star}$ is a graph satisfying the requirements (a)-(b)-(c) in the statement. However in general

it does not satisfy (d). Moreover $\Sigma^\star$ is not necessarily a spindle, because it may be not contained in a proof net of $\mathcal{P}$. We can modify $\Sigma^\star$ to get a $\Sigma''$ that satisfies the point (d). If $e$ is a non-principal premise of $\Sigma^\star$, entering some box $b$ of $\Sigma^\star$, then we remove the edge $e$ and we reduce the number of premises of $b$, according to point 3 of Fact 4.3.37. Now, we can plug every non-principal conclusion of $\Sigma''$, exiting upward from some contraction $u$ of $\Sigma''$, with a weakening node: we get a proof net $\Pi''$ containing $\Sigma''$, thus showing that $\Sigma''$ is a spindle.

At last, if $\Sigma \cup \chi$ was dangerous, we can apply the same transformation to the CS-path $\chi$, thus arriving to $\chi''$; $\Sigma'' \cup \chi''$ is a dangerous spindle. □

Lemma 4.3.39 (Some Non-Polytime Sensible Systems)
Let $\mathcal{P} \subseteq \mathfrak{B}_{\mathbb{X}}$ be sensible. Let us assume that $\mathcal{P}$ can build a dangerous spindle. Then $\mathcal{P}$ is neither polysize, nor polystep.

**Proof.** The proof is similar to the one of Lemma 4.3.35. However, here we have more relaxed hypothesis on $\Sigma$. Firstly, we must transform the given dangerous spindle $\mathfrak{n}:(\Sigma \cup \chi):\mathfrak{n}$ into $\mathfrak{n}A:(\Sigma'' \cup \chi''):\mathfrak{n}A$, as described in Lemma 4.3.38. Fact 4.3.37 assures the existence of a proof net $b \triangleright \vdash \mathfrak{n}A$. Now, we call $\Theta_n = \underbrace{(\Sigma'' \cup \chi'') \bowtie \ldots \bowtie (\Sigma'' \cup \chi'')}_{n} \bowtie b$. Remember that we called *exponential chain* such an object. $\Theta_n$ is not a proof net because it can have many conclusions, but it can be placed easily inside a proof net $\Pi_n$ closing all its open conclusions with free weakening nodes. Please notice that $|\Pi_n| = O(n)$. At last, we reduce the first level of $\Pi_n$. This is possible, thanks to the downward closure of $\mathcal{P}$: the reduction is performed *bottom-up*, starting from the lower box $b$. As we already notice when proving Lemma 4.3.35, such a reduction is neither polysize, nor polystep. □

Proposition 4.3.40 (A Polynomiality Criterion)
Let $\mathcal{P} \subseteq \mathfrak{B}_{\mathbb{X}}$ sensible, with $M$ modalities. The following are equivalent:
1. $\mathcal{P}$ is polytime.
2. $\mathcal{P}$ cannot build any dangerous chain of spindles.
3. $\mathcal{P}$ cannot build any dangerous spindle.
4. $\mathcal{P}$ cannot build any chain (*neither if not dangerous!*) of more than $M$ spindles.

**Proof.**

4. ⇒ 1. We can apply Proposition 4.3.32 with $L = M$ and we get that $\mathcal{P}$ is polytime.

¬2. ⇒ ¬1. If $\mathcal{P}$ can build a dangerous chain of spindles, Lemma 4.3.39 says that $\mathcal{P}$ is neither polystep nor polysize.

¬4. ⇒ ¬2. Let us assume that $\mathcal{P}$ can build a chain $\Theta$ of $r > M$ spindles. Let $w_1, \ldots, w_r$ the principal contractions of the spindles in $\Theta$. By the pigeons-hole principle, at least two principal contractions share the same input label; let $w_i$ and $w_j$ be the two incriminated nodes, and $\Sigma$ be the chain of $j - i \leq M$ spindles between them. $\Sigma$ is dangerous.

3. ⇒ 2. Obvious by definition.

2. ⇒ 3. This is implied by Lemma 4.3.30. □

If the system is not sensible, the implication 1. ⇒ 2. may fail; however the implications 2. ⇔ 3. ⇔ 4. ⇒ 1. still hold.

COROLLARY 4.3.41 (POLYSIZE IMPLIES POLYSTEP)
Let $\mathcal{P} \subseteq \mathfrak{B}_{\mathbb{X}}$ a polysize sensible system. Then it is also polystep.

**Proof.** Because if $\mathcal{P}$ reduces in polysize, then the same proof of $\neg$2. $\Rightarrow \neg$1. holds: $\mathcal{P}$ cannot build any dangerous chains of spindles. So, the criterion tells us that, in fact, $\mathcal{P}$ is polytime. □

This means, essentially, that no sensible subsystem $\mathcal{P}$ of MS can represent PSPACE, unless, of course, PSPACE=PTIME.

We shall apply in Chapter 6 the propositions proved in this Section.

### 4.3.5 Strong Determinacy

We generalize the definition of "strong determinacy" present in [DL08]. In that work the strong determinacy of ILAL is used to prove that ILAL is polytime.

DEFINITION 4.3.42 (STRONG DETERMINACY) *A subsystem $\mathcal{P} \subseteq \mathfrak{B}_{\mathbb{X}}$ enjoys* **strong determinacy** *iff for every $\Pi \in \mathbf{PN}(\mathcal{P})$ and every $C \in C_{\Pi}$ there is at most 1 context $D$ such that $D \mapsto_{\Pi} C$.*

LEMMA 4.3.43 (STRONG DETERMINACY AND POLYNOMIALITY)
If $\mathcal{P} \subseteq \mathfrak{B}_{\mathbb{X}}$ enjoys strong determinacy, then it is polytime.

**Proof.** Strong determinacy is equivalent to the absence of spindles in every $\Pi$. □

So, this is a condition sufficient but not necessary for polynomiality.

## 4.4 Syntactical Criteria for Polytime Soundness

Here, we move towards a different kind of Criteria. We want to avoid the use of context semantics, and we would like to understand if a given subsystem $\mathcal{P}$ is polytime "*just looking at the building rules it contains*".

### 4.4.1 Augmented subsystems

Here we will present a class of subsystems that by definition are closed w.r.t. their *admissible* rules. Recall that, traditionally, a rule is "admissible" inside a logic if it can be derived inside that logic, so that it can be freely added to the subsystem without changing the set of provable statements. Here, we will use the term "admissible" in a slightly different way: only *some of* the derivable rules are *admissible*, and the admissible rules are really derivable *only for sensible subsystems*. We shall see that if a subsystem $\mathcal{P}$ is polytime, the we can freely add all such admissible rules and we get another polytime system (Proposition 4.4.4). We will call *augmented* this larger subsystem. Moreover, we shall see that checking if the augmented subsystem is polytime or not is quite easy (Proposition 4.4.5).

DEFINITION 4.4.1 *Given a system $\mathcal{P} \subseteq \mathfrak{B}_{\mathbb{X}}$, we define inductively its* **admissible (promotion and contraction) rules**.

- *If $\mathcal{P}$ has the rule $\mathrm{P}_{\mathfrak{m}}(\mathfrak{m}_1, \ldots, \mathfrak{m}_k)$ or $\mathbf{Y}_{\mathfrak{q}}(\mathfrak{m}, \mathfrak{n})$, then it is admissible for $\mathcal{P}$;*

- *If rules* $P_{\mathfrak{m}}(\mathfrak{r}, \mathfrak{m}_1, \ldots, \mathfrak{m}_k)$ *and* $P_{\mathfrak{r}}(\mathfrak{n}_1, \ldots, \mathfrak{n}_l)$ *are admissible for* $\mathcal{P}$*, so it is the rule* $P_{\mathfrak{m}}(\mathfrak{n}_1, \ldots, \mathfrak{n}_l, \mathfrak{m}_1, \ldots, \mathfrak{m}_k)$.
- *If rules* $P_{\mathfrak{m}}(\mathfrak{r}, \mathfrak{m}_1, \ldots, \mathfrak{m}_k)$ *and* $\mathbf{Y}_{\mathfrak{q}}(\mathfrak{r}, \mathfrak{s})$ *are admissible for* $\mathcal{P}$*, then also the rule* $P_{\mathfrak{m}}(\mathfrak{q}, \mathfrak{m}_1, \ldots, \mathfrak{m}_k)$ *is admissible for* $\mathcal{P}$.
- *If rules* $P_{\mathfrak{m}}(\mathfrak{m}_1, \ldots, \mathfrak{m}_k)$ *and* $\mathbf{Y}_{\mathfrak{m}}(\mathfrak{r}, \mathfrak{s})$ *are admissible for* $\mathcal{P}$*, then also the rules* $P_{\mathfrak{r}}(\mathfrak{m}_1, \ldots, \mathfrak{m}_k)$ *and* $P_{\mathfrak{s}}(\mathfrak{m}_1, \ldots, \mathfrak{m}_k)$ *are admissible for* $\mathcal{P}$.
- *If rules* $\mathbf{Y}_{\mathfrak{m}}(\mathfrak{r}, \mathfrak{s})$ *and* $\mathbf{Y}_{\mathfrak{r}}(\mathfrak{q}, \mathfrak{p})$ *are admissible for* $\mathcal{P}$*, then also the rules* $\mathbf{Y}_{\mathfrak{m}}(\mathfrak{q}, \mathfrak{s})$ *and* $\mathbf{Y}_{\mathfrak{m}}(\mathfrak{q}, \mathfrak{p})$ *are admissible for* $\mathcal{P}$.
- *If rules* $\mathbf{Y}_{\mathfrak{m}}(\mathfrak{r}, \mathfrak{s})$ *and* $P_{\mathfrak{m}}(\mathfrak{q})$ *are admissible for* $\mathcal{P}$*, then also the rule* $\mathbf{Y}_{\mathfrak{q}}(\mathfrak{r}, \mathfrak{s})$ *is admissible for* $\mathcal{P}$.
- *If rules* $\mathbf{Y}_{\mathfrak{m}}(\mathfrak{r}, \mathfrak{s})$ *and* $P_{\mathfrak{q}}(\mathfrak{s})$ *are admissible for* $\mathcal{P}$*, then also the rule* $\mathbf{Y}_{\mathfrak{m}}(\mathfrak{r}, \mathfrak{q})$ *is admissible for* $\mathcal{P}$.

If $\mathcal{P}$ is sensible, then all its admissible rules can be derived in $\mathcal{P}$. This can be proved by induction. For example, if $\mathcal{P}$ has the two rules $P_{\mathfrak{m}}(\mathfrak{r}, \mathfrak{m}_1, \ldots, \mathfrak{m}_k)$ and $\mathbf{Y}_{\mathfrak{q}}(\mathfrak{r}, \mathfrak{s})$, so that $P_{\mathfrak{m}}(\mathfrak{q}, \mathfrak{m}_1, \ldots, \mathfrak{m}_k)$ is admissible for $\mathcal{P}$, than this latter rule can be proved using **W**.

Definition 4.4.2 (Augmented and Saturated Systems) *Let* $\mathcal{P} \subseteq \mathfrak{B}_{\mathbb{X}}$*. The* **augmented system** $\mathcal{S}(\mathcal{P})$ *of* $\mathcal{P}$ *is* $\mathcal{P}$ *plus every admissible rule for* $\mathcal{P}$*.*

*If* $\mathcal{P} = \mathcal{S}(\mathcal{P})$ *the system is said* **saturated**.

Notice that $\mathcal{S}(\mathcal{P})$ contains the closed boxes and it is downward closed. By the definition it follows that every augmented system $\mathcal{S}(\mathcal{P})$ is saturated, i.e. $\mathcal{S}(\mathcal{S}(\mathcal{P})) = \mathcal{S}(\mathcal{P})$. By definition, the augmented system $\mathcal{S}(\mathcal{P})$ of $\mathcal{P}$ contains one promotion rule for each module of $\mathcal{P}$ that can be built with cuts, contractions and boxes but without spindles.

Every proof net of $\mathcal{S}(\mathcal{P})$, in fact, can use rules that are not in $\mathcal{P}$; anyway, *if* $\mathcal{P}$ *is sensible*, each one of this rules can be decomposed in a number of *more primitive* rules that are in $\mathcal{P}$:

Definition 4.4.3 (Primitive Proof nets) *Let* $\mathcal{P} \subseteq \mathfrak{B}_{\mathbb{X}}$ *be sensible. Consider a proof net* $\Pi$ *in* $\mathcal{S}(\mathcal{P})$*: all the rules of* $\Pi$ *can be substituted by a finite number of rules of* $\mathcal{P}$*; call* $\Pi'$ *the proof net of* $\mathcal{P}$ *obtained by the rewriting. We call* $\Pi'$ *the* **primitive proof net** *of* $\Pi$*.*

An example is in Figure 4.7.

Proposition 4.4.4 (Polynomiality of $\mathcal{P}$ and Polynomiality of $\mathcal{S}(\mathcal{P})$)
Let us consider a sensible system $\mathcal{P} \subseteq \mathfrak{B}_{\mathbb{X}}$. $\mathcal{S}(\mathcal{P}) \in$ PMS iff $\mathcal{P} \in$ PMS.

**Proof.**

**⇒)** This is just because $\mathcal{P} \subseteq \mathcal{S}(\mathcal{P})$, and the Monotonicity Lemma 4.1.8 holds.

**⇐)** If $\mathcal{P}$ is polytime, let us assume by contradiction that $\mathcal{S}(\mathcal{P})$ is not polytime. Using Proposition 4.3.40, we know that $\mathcal{S}(\mathcal{P})$ can build a dangerous spindle $\Sigma \cup \chi$ between two occurrences of $\mathbf{Y}_{\mathfrak{q}}(\mathfrak{r}, \mathfrak{s})$, while $\mathcal{P}$ cannot. However, $\mathcal{P}$ and $\mathcal{S}(\mathcal{P})$ may differ only for the admissible rules: $\Sigma$ contains an admissible promotion or contraction rule of $\mathcal{P}$. But, every admissible rule may be split in a larger number of promotion and contraction rules of $\mathcal{P}$: so the primitive module $\Sigma' \cup \chi'$ of $\Sigma \cup \chi$ is still a dangerous spindle, and it exists both in $\mathcal{S}(\mathcal{P})$ and $\mathcal{P}$. This means that $\mathcal{P}$ is not polytime, a contradiction.

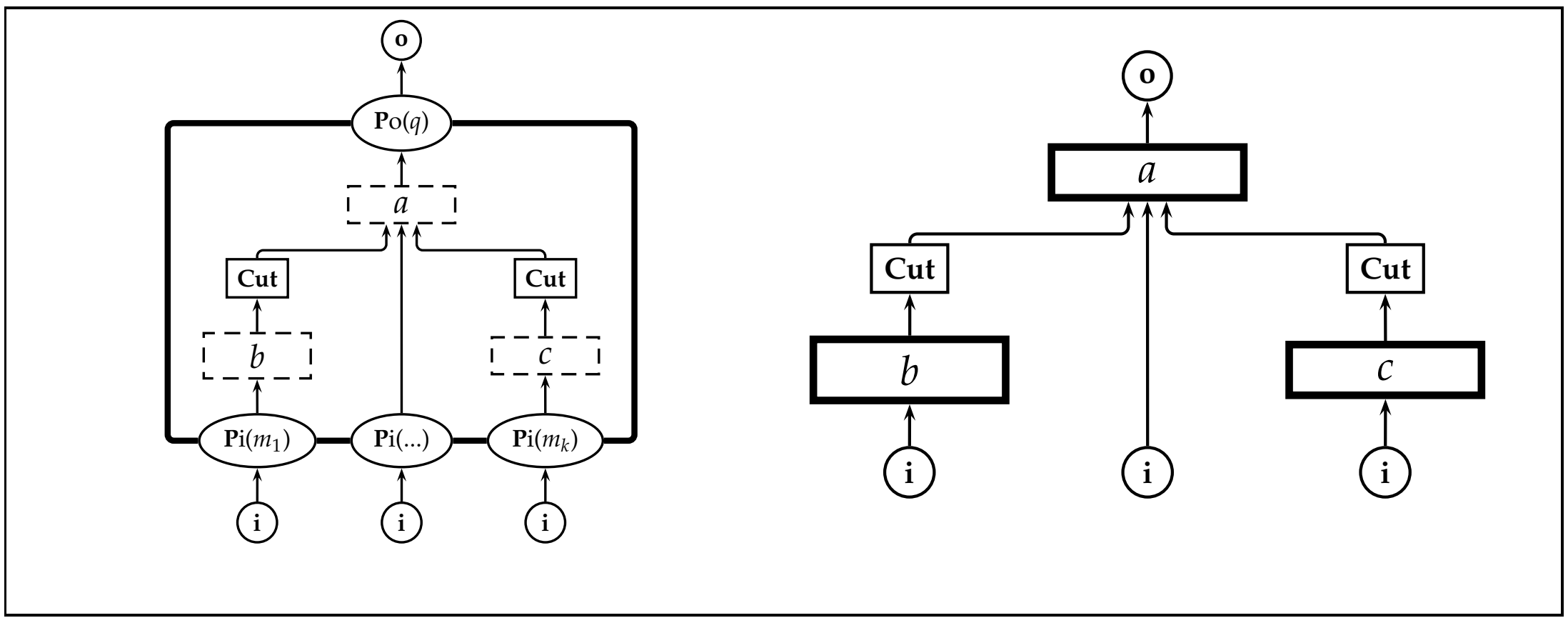

Figure 4.7: An example: a possible proof net in an augmented system $\mathcal{S}(\mathcal{P})$, and its primitive proof net in the original system $\mathcal{P}$.

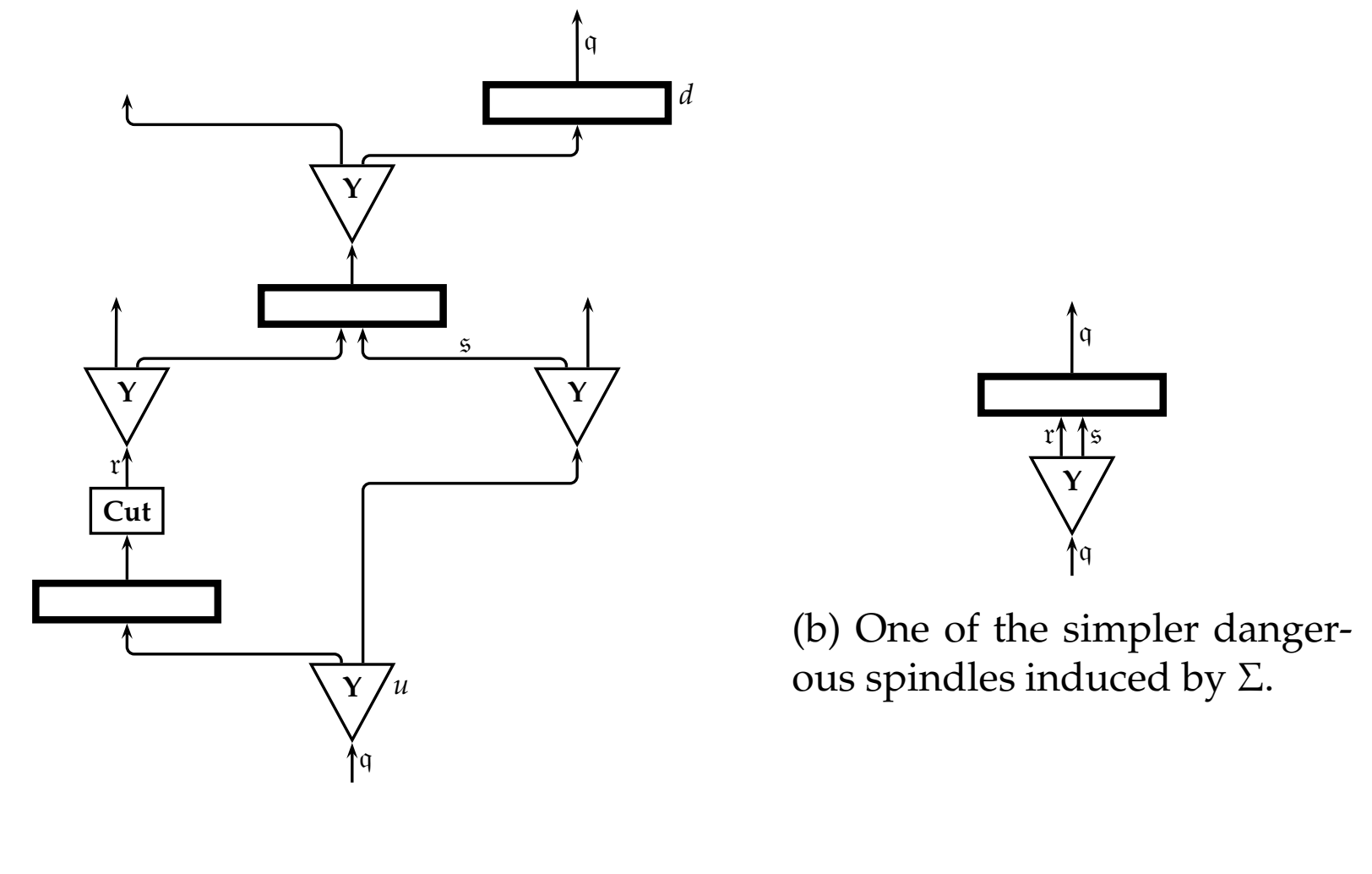

(a) $\Sigma$

(b) One of the simpler dangerous spindles induced by $\Sigma$.

Figure 4.8: An example of dangerous spindle $\Sigma$, as in proof of Proposition 4.4.5.

□

**Proposition 4.4.5 (Polynomiality Criterion Based On Saturation)**
Let $\mathcal{P} \subseteq \mathfrak{B}_{\mathbb{X}}$ be a sensible system. Then $\mathcal{P}$ is not polytime iff $\mathcal{S}(\mathcal{P})$ has the two rules $P_{\mathfrak{q}}(\mathfrak{r}, \mathfrak{s})$ and $\mathbf{Y}_{\mathfrak{q}}(\mathfrak{r}, \mathfrak{s})$, for some value of the parameters.

**Proof.** Using Proposition 4.4.4, $\mathcal{P}$ is polytime iff $\mathcal{S}(\mathcal{P})$ is polytime. "$\Leftarrow$" case. The presence of these two rules leads to a dangerous chain of spindles and so $\mathcal{S}(\mathcal{P})$ is not polytime. "$\Rightarrow$" case. If $\mathcal{S}(\mathcal{P})$ is not polytime it must have a dangerous spindle $\Sigma \cup \chi$ between a contraction $u$ and a promotion or contraction $d$, where the modality $\mathfrak{s}$ that labels the conclusion of $d$ labels also the premise of $u$. An example is in Figure 4.8. But, thanks to saturated-ness, we can merge all the nodes and boxes between $u$ and $w$ in a single box, and so we build a new dangerous spindle $\Sigma' \cup \chi'$ made up of only 1 box and 1 contraction, as the thesis requires. □

### 4.4.2 Relations among Modalities

Throughout this Section, we consider $\mathcal{P} \subseteq \mathfrak{B}_{\mathbb{X}}$ be a fixed subsystem with $M < \infty$ modalities (for example a sensible subsystem). We define a number of different relations among the $M$ modalities. These relations will be used to define some syntactical criteria for the polytime soundness of $\mathcal{P}$.

The notions of *order*, *preorder* and sim. have been recalled in Section 2.10.

Definition 4.4.6 $\mathfrak{m}\,\mathfrak{R}_{\uparrow}\,\mathfrak{n}$ *whenever* $\mathcal{P}$ *can build a* CS*-path whose edges are all labelled with modal formulæ, the first one with modality* $\mathfrak{m}$ *and the last one with* $\mathfrak{n}$.

This relation is transitive. It can be built considering all the rules $P_{\mathfrak{n}}(\mathfrak{m})$ and $\mathbf{Y}_{\mathfrak{m}}(\mathfrak{n}, \mathfrak{q})$, for some $\mathfrak{q}$, that can be effectively *used* in some proof net of $\mathcal{P}$, and then taking the transitive closure. Please recall that it is not obvious that a rule is *used* inside some proof net of $\mathcal{P}$, unless it is sensible (cfr. Fact 3.2.6).

Definition 4.4.7 $\mathfrak{m}\,\mathfrak{R}_{\diamond}\,\mathfrak{n}$ *whenever* $\mathcal{P}$ *can build a spindle* $\mathfrak{m} : \Sigma : \mathfrak{n}$.

This relation is transitive, and it is a sub-relation of $\mathfrak{R}_{\uparrow}$. For the moment we neglect the algorithmic construction of $\mathfrak{R}_{\diamond}$, even if we think that such algorithm must exist. However this relation is the most important from a theoretical point of view:

Lemma 4.4.8 (Polynomiality Criterion Based On $\mathfrak{R}_{\diamond}$)
Let $\mathcal{P}$ be a sensible subsystem, and $\mathfrak{R}_{\diamond}$ as before. Then $\mathcal{P} \in$ PMS iff $\mathfrak{R}_{\diamond}$ is a strict partial ordering.

**Proof.** We already noticed that $\mathfrak{R}_{\diamond}$ is transitive. $\mathcal{P}$ has a dangerous chain of spindles iff $\exists\mathfrak{n}(\mathfrak{n}\,\mathfrak{R}_{\diamond}\,\mathfrak{n})$, that is if $\mathfrak{R}_{\diamond}$ is not anti-reflexive. Now, if $\mathfrak{R}_{\diamond}$ is a partial strict ordering then it is anti-reflexive, so $\mathcal{P}$ has no dangerous chains; conversely, if $\mathcal{P}$ has no dangerous chains, $\forall\mathfrak{n}(\neg\mathfrak{n}\,\mathfrak{R}_{\diamond}\,\mathfrak{n})$, that is $\mathfrak{R}_{\diamond}$ is anti-reflexive, and thanks to transitivity we conclude also that $\mathfrak{R}_{\diamond}$ is anti-symmetric, and so a strict partial ordering. □

We are going to describe a further relation $\overline{\mathfrak{R}_{\triangle}}$. Differently from $\mathfrak{R}_{\diamond}$, $\overline{\mathfrak{R}_{\triangle}}$ is quite complex, however it is easy to show an algorithm that builds it.

Let $\mathcal{P}$ be a **sensible** subsystem of $\mathfrak{B}_{\mathbb{X}}$ with $M < \infty$ modalities. Let

$$D_M = \{\langle\mathfrak{m}\rangle \mid \mathfrak{m} \in \mathbb{X}\} \cup \{\langle\mathfrak{m}, \mathfrak{n}\rangle \mid \mathfrak{m}, \mathfrak{n} \in \mathbb{X}\}\,;$$

$D_M$ contains the multisets with domain $\{1, \ldots, M\}$ with at least one and most two elements. Letters $A, B, C, \ldots$ will temporarily range over elements of $D_M$. $|A| \in \{0, 1\}$ is the cardinality of $A$ (please notice that $|\langle\mathfrak{m}, \mathfrak{m}\rangle| = 2$ !!!).

By the way, one can show that $D_M$ has $\frac{1}{2}\,M(M+3)$ elements.

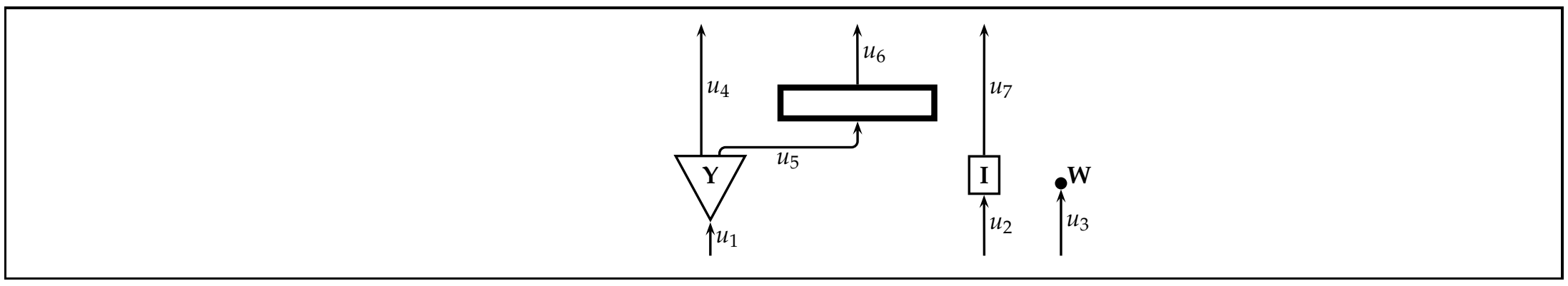

Figure 4.9: An example of quasi-connected module, with principal premises $\{u_1, u_2\}$ and principal conclusions $\{u_4, u_6, u_7\}$. Notice that $u_3$ cannot be a principal premise.

Definition 4.4.9 *We define the binary relation $\mathfrak{R}_\triangle$ on $D_M$ by:*

- *if $\mathcal{P}$ has $P_\mathfrak{q}(\mathfrak{m})$, then $\langle\mathfrak{m}\rangle\,\mathfrak{R}_\triangle\langle\mathfrak{q}\rangle$ and $\forall\mathfrak{s}\langle\mathfrak{m},\mathfrak{s}\rangle\,\mathfrak{R}_\triangle\langle\mathfrak{q},\mathfrak{s}\rangle$;*
- *if $\mathcal{P}$ has $P_\mathfrak{q}(\mathfrak{m},\mathfrak{n})$, then $\langle\mathfrak{m},\mathfrak{n}\rangle\,\mathfrak{R}_\triangle\,\langle\mathfrak{q}\rangle$;*
- *if $\mathcal{P}$ has $\mathbf{Y}_\mathfrak{q}(\mathfrak{m},\mathfrak{n})$, then $\langle\mathfrak{q}\rangle\,\mathfrak{R}_\triangle\langle\mathfrak{m},\mathfrak{n}\rangle$, $\langle\mathfrak{q}\rangle\,\mathfrak{R}_\triangle\langle\mathfrak{m}\rangle$, $\forall\mathfrak{s}\langle\mathfrak{q},\mathfrak{s}\rangle\,\mathfrak{R}_\triangle\langle\mathfrak{m},\mathfrak{s}\rangle$, $\langle\mathfrak{q}\rangle\,\mathfrak{R}_\triangle\langle\mathfrak{n}\rangle$, and $\forall\mathfrak{s}\langle\mathfrak{q},\mathfrak{s}\rangle\,\mathfrak{R}_\triangle\langle\mathfrak{n},\mathfrak{s}\rangle$,*

*$\overline{\mathfrak{R}_\triangle}$ is the transitive and reflexive closure of $\mathfrak{R}_\triangle$.*

$\overline{\mathfrak{R}_\triangle}$ is not, in general, a partial order, even restricting to polytime subsystems.

Lemma 4.4.10
$\mathcal{P}$ is saturated iff $\mathfrak{R}_\triangle = \overline{\mathfrak{R}_\triangle}$.

**Proof.** Look back at the definition of $\mathfrak{R}_\triangle$ and at the definition of $\mathcal{S}(\mathcal{P})$. Their definitions are step-by-step analogous. So, $\mathcal{P}$ is saturated iff its $\mathfrak{R}_\triangle$ relation is the greatest possible, i.e. $\mathfrak{R}_\triangle$ is closed. □

If $\Sigma$ is a module (not necessarily connected!) of $\mathcal{P}$ and $A, B \in D_M$, we write $A : \Sigma : B$ meaning that $\Sigma$ has (at least) $|A|$ premises labelled with modal formulas with modality $\mathfrak{s}$ for every element $\mathfrak{s}$ of $A$, and (at least) $|B|$ conclusions labelled with modal formulas with modality $\mathfrak{s}$ for every element $\mathfrak{s}$ of $B$. We call **principal premises** and **principal conclusions** these $|A|$ and $|B|$ edges. We say that $\Sigma$ is **quasi-connected** if for every principal premise there is a CS-path connecting it to a principal conclusion, and for every principal conclusion there is a CS-path connecting it to a principal premise. In particular every connected module is also quasi-connected. An example is given in Figure 4.9

Lemma 4.4.11
If $A\,\overline{\mathfrak{R}_\triangle}\,B$ then there exists a module $A : \Sigma : B$ quasi-connected.

**Proof.** If $A\,\overline{\mathfrak{R}_\triangle}\,B$, then either $A = B$, or $(A, B)$ is in the reflexive closure of $\mathfrak{R}_\triangle$. Firstly, we consider the case $A = B$. If $A = \langle\mathfrak{m}\rangle$, the thesis can be proved taking as $A : \Sigma : A$ the proof net that contains just one identity node. If $A = \langle\mathfrak{m},\mathfrak{n}\rangle$, the thesis can be proved taking as $A : \Sigma : A$ the disconnected module containing two identity nodes, one introducing $\mathfrak{m}$ and the other one introducing $\mathfrak{n}$. In both cases $\Sigma$ is quasi-connected.

Now let $A \neq B$. In this case $(A, B)$ is in the reflexive closure of $\mathfrak{R}_\triangle$. Being $D_M$ a finite set, $A\,\overline{\mathfrak{R}_\triangle}\,B$ implies that we can find a finite number of elements $A_i$ such that $A = A_0\,\mathfrak{R}_\triangle\,A_1\,\mathfrak{R}_\triangle\,A_2\,\mathfrak{R}_\triangle \ldots \mathfrak{R}_\triangle\,A_k = B$. We proceed by induction on $k$.

**If** $k = 1$, that is $A \,\mathfrak{R}_\triangle\, B$, we know that one of the following holds:

1. $A = \langle \mathfrak{m}, \mathfrak{n} \rangle \,\mathfrak{R}_\triangle\, \langle \mathfrak{q} \rangle = B$ and $\mathcal{P}$ has $P_\mathfrak{q}(\mathfrak{m}, \mathfrak{n})$;
2. $A = \langle \mathfrak{q} \rangle \,\mathfrak{R}_\triangle \langle \mathfrak{m}, \mathfrak{n} \rangle = B$ iff $\mathcal{P}$ has $\mathbf{Y}_\mathfrak{q}(\mathfrak{m}, \mathfrak{n})$;
3. $A = \langle \mathfrak{q} \rangle \,\mathfrak{R}_\triangle \langle \mathfrak{q}' \rangle = B$ and $\mathcal{P}$ has $P_{\mathfrak{q}'}(\mathfrak{q})$ or $\mathbf{Y}_\mathfrak{q}(\mathfrak{q}', \mathfrak{s})$ for some $\mathfrak{s}$.
4. $A = \langle \mathfrak{q}, \mathfrak{n} \rangle \,\mathfrak{R}_\triangle \langle \mathfrak{q}', \mathfrak{n} \rangle = B$ and $\mathcal{P}$ has $P_{\mathfrak{q}'}(\mathfrak{q})$ or $\mathbf{Y}_\mathfrak{q}(\mathfrak{q}', \mathfrak{s})$ for some $\mathfrak{s}$.

In each case we can build a module $A : \Sigma : B$ quasi-connected:

1. one single $\mathcal{P}$ has $P_\mathfrak{q}(\mathfrak{m}, \mathfrak{n})$;
2. one single $\mathbf{Y}_\mathfrak{q}(\mathfrak{m}, \mathfrak{n})$;
3. one among $P_{\mathfrak{q}'}(\mathfrak{q})$ and $\mathbf{Y}_\mathfrak{q}(\mathfrak{q}', \mathfrak{s})$.
4. two connected components, the first one having one among $P_{\mathfrak{q}'}(\mathfrak{q})$ and $\mathbf{Y}_\mathfrak{q}(\mathfrak{q}', \mathfrak{s})$, the second one having a single identity node.

**If** $k > 1$, let us consider $A \,\overline{\mathfrak{R}_\triangle}\, A_{k-1} \,\mathfrak{R}_\triangle\, B$. By inductive hypothesis there exists a quasi-connected $A : \Sigma_1 : A_{k-1}$. Then, $A_{k-1} \,\mathfrak{R}_\triangle\, B$ tells us there exists a quasi-connected $A_{k-1} : \Sigma_2 : B$, as previously showed. Plugging the principal conclusions of $\Sigma_1$ in the principal premises of $\Sigma_2$ we obtain a quasi-connected module $A : \Sigma : B$ as required.

□

Lemma 4.4.12 (Polynomiality Criterion Based On $\overline{\mathfrak{R}_\triangle}$)
Let $\mathcal{P}$ be a sensible subsystem. $\mathcal{P}$ has a dangerous chain of spindles iff there exist $\mathfrak{m}, \mathfrak{n}, \mathfrak{q}$ such that $\langle q \rangle \,\overline{\mathfrak{R}_\triangle} \langle \mathfrak{m}, \mathfrak{n} \rangle \,\overline{\mathfrak{R}_\triangle} \langle \mathfrak{q} \rangle$.

In other words, if $\mathcal{P}$ is polytime, the graph of $\mathfrak{R}_\triangle$ may contain cycles; however it cannot contain cycles with both elements of cardinality 1 and cardinality 2.

**Proof.** (1) Let us assume $\mathcal{P}$ has a dangerous chain $\Theta \cup \chi$. $\Theta \cup \chi$ is composed of a contraction $u$, let us say $\mathbf{Y}_\mathfrak{p}(\mathfrak{r}, \mathfrak{s})$, a box $b$ with 2 premises, and three CS-paths $\tau, \rho, \chi$ of (possibly many) promotion boxes and contraction nodes, where $\tau$ and $\rho$ connect $u$ to $b$ and $\chi$ exits from $b$ and terminates with modality $\mathfrak{p}$. By construction, $\langle \mathfrak{p} \rangle \,\overline{\mathfrak{R}_\triangle} \langle \mathfrak{r}, \mathfrak{s} \rangle \,\overline{\mathfrak{R}_\triangle} \langle \mathfrak{p} \rangle$.

(2) Let us assume $\langle \mathfrak{q} \rangle \,\overline{\mathfrak{R}_\triangle} \langle \mathfrak{m}, \mathfrak{n} \rangle \,\overline{\mathfrak{R}_\triangle} \langle \mathfrak{q} \rangle$. Using Lemma 4.4.11 we can build the quasi-connected modules $\langle \mathfrak{q} \rangle : \Sigma_1 : \langle \mathfrak{m}, \mathfrak{n} \rangle$ and $\langle \mathfrak{m}, \mathfrak{n} \rangle : \Sigma_2 : \langle \mathfrak{q} \rangle$. Plugging the principal conclusions of $\Sigma_1$ with the principal premises of $\Sigma_2$ we get a quasi-connected module $\langle \mathfrak{q} \rangle : \Sigma : \langle \mathfrak{q} \rangle$. But notice that, thanks to the definition of quasi-connectedness, $\Sigma$ is also a dangerous chain of spindles. □

Corollary 4.4.13 (Polynomiality of Sensible Subsystems is Decidable)
Let $\mathcal{P}$ be a sensible subsystem, and let us assume that the promotion rules of $\mathcal{P}$ are described by means of regular expressions. Then, the question "Is $\mathcal{P}$ polytime?" is decidable.

**Proof.** Notice that $M$, the number of modalities in $\mathcal{P}$, is finite, so also the number of possible promotion rules "described by means of regular expressions" is finite. So, the algorithm is the following.

1. Make a list of all the rules $\mathbf{Y}_\mathfrak{q}(\mathfrak{m}, \mathfrak{n})$, $P_\mathfrak{q}(\mathfrak{m})$ and $P_\mathfrak{q}(\mathfrak{m}, \mathfrak{n})$ belonging to $\mathcal{P}$. This is possible because (i) $\mathcal{P}$ has $M < \infty$ modalities, and (ii) $\mathcal{P}$ is finitely described. So this list is finite.

2. Build the relation $\mathfrak{R}_\triangle$. Each rule found in (i) gives at most a constant number of couples in $\mathfrak{R}_\triangle$.
3. Build the relation $\overline{\mathfrak{R}_\triangle}$, the transitive and reflexive closure of $\mathfrak{R}_\triangle$.
4. Exhaustively search for $\mathfrak{q}, \mathfrak{m}, \mathfrak{n}$ such that $\langle\mathfrak{q}\rangle \overline{\mathfrak{R}_\triangle} \langle\mathfrak{m}, \mathfrak{n}\rangle \overline{\mathfrak{R}_\triangle} \langle\mathfrak{q}\rangle$.

□

The forthcoming Proposition 4.4.15 is one of our most useful results. It is interesting both in a theoretical and in an applicative perspective, thanks to the simplicity of the definition of $\mathfrak{R}_\uparrow$. The reader can compare Proposition 4.4.15 and Proposition 4.4.5; he will find some analogies, due to the fact that $\mathcal{P}$ and its augmented subsystem $\mathcal{S}(\mathcal{P})$ share the same relation $\mathfrak{R}_\uparrow$. Recall that every transitive relation can be completed to a partial preorder (Section 2.10).

Definition 4.4.14 *$\widehat{\mathfrak{R}_\uparrow}$ is the smallest partial preorder containing $\mathfrak{R}_\uparrow$.*

In the following statement, $\equiv$ denotes the equivalence under $\widehat{\mathfrak{R}_\uparrow}$.

Proposition 4.4.15 (Criterion of Polynomiality based on $\mathfrak{R}_\uparrow$)
Let $\mathcal{P} \subseteq \mathfrak{B}_\mathbb{X}$ be sensible. $\mathcal{P}$ is polytime if and only if, for every 6 elements $\mathfrak{q}_1 \equiv \mathfrak{q}_2 \equiv \mathfrak{r}_1 \equiv \mathfrak{r}_2 \equiv \mathfrak{s}_1 \equiv \mathfrak{s}_2$, i.e. all equivalent under $\widehat{\mathfrak{R}_\uparrow}$, $\mathcal{P}$ does not contain both $\mathbf{Y}_{\mathfrak{q}_1}(\mathfrak{r}_1, \mathfrak{s}_1)$ and $P_{\mathfrak{q}_2}(\mathfrak{r}_2, \mathfrak{s}_2)$.

**Proof.** $\mathcal{P}$ is not polytime iff $\mathcal{P}$ builds a dangerous spindle $a:(\Sigma \cup \chi):a$. By definitions, this is equivalent to say that $\mathcal{P}$ has:
- a contraction $\mathbf{Y}_a(b, c)$,
- a promotion $P_d(e, f)$,
- a CS-path beginning with modality $b$ and ending with modality $e$,
- a CS-path beginning with modality $c$ and ending with modality $f$,
- a CS-path beginning with modality $d$ and ending with modality $a$ (this is $\chi$).

And this is equivalent to say that $\mathcal{P}$ has a contraction $\mathbf{Y}_a(b, c)$ and a promotion $P_d(e, f)$ such that $a \equiv b \equiv c \equiv d \equiv e \equiv f$. □

### 4.4.3 Maximality in MS

We now consider the following question: is there a notion of *largest* subsystem of MS which is polytime? In this Chapter we will show that there is not a single *greatest* subsystem, but a whole class of *maximal* subsystems. The word *maximal* is here used in a technical way, according to algebra and set theory (think to *maximal ideals* [Vin03], or to *maximal elements* of a partial order [DP02]). Here we will define formally what is a maximal subsystem, and we will give some examples of them. Later, we will prove a Structure Theorem 4.4.25 that characterizes, among the sensible subsystems, all and only the maximal ones. The Structure Theorem is completely syntactical, it describes exactly which rules a maximal subsystem must have.

The study of maximal subsystems is important for at least two reasons:

(a) The maximal subsystems are useful in themselves, as they supply the greatest sets of rules, and nevertheless they are polytime. This has impact on the questions that we posed in the Introduction: if we search for a subsystem able to fill the question mark (3) in Figure 1.2, it is a good idea to begin the search from the maximal subsystems.

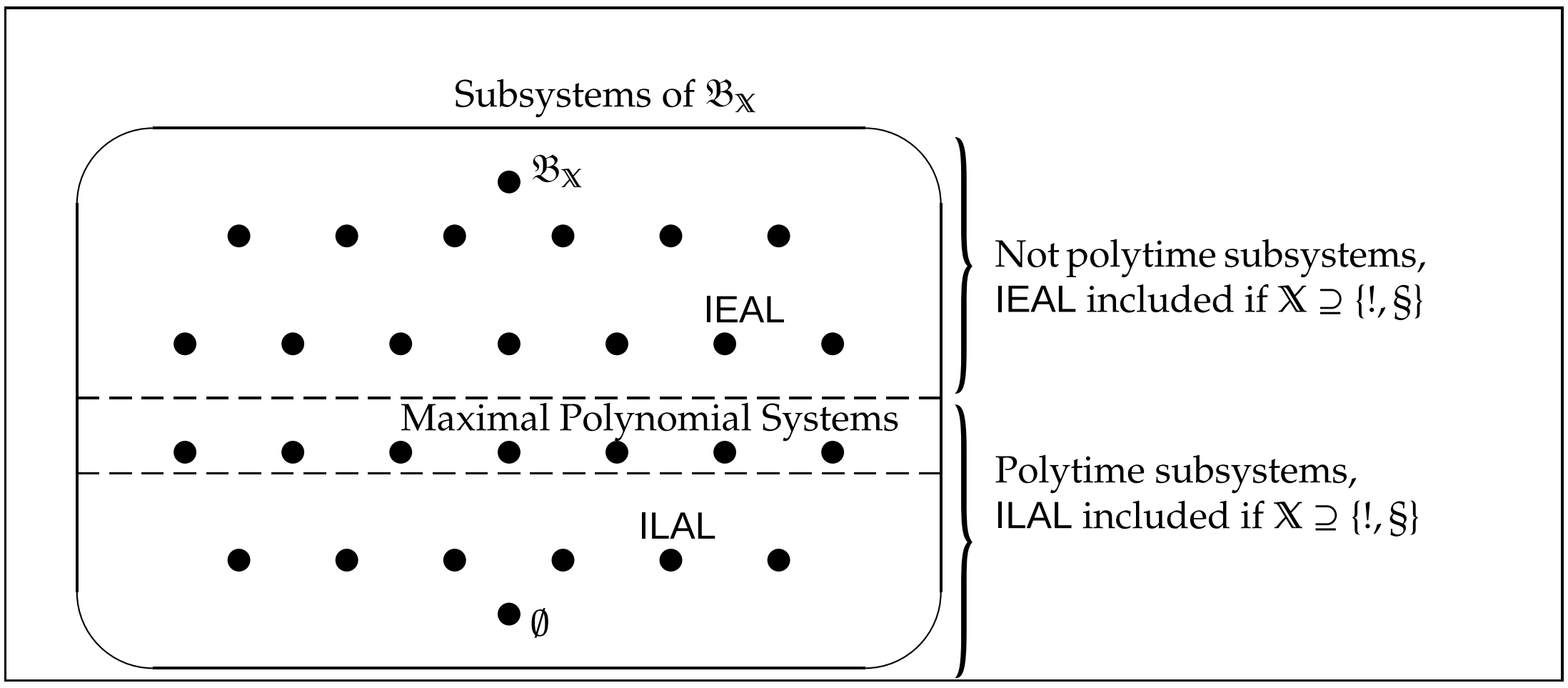

Figure 4.10: A picture of the poytime, non-polytime and maximal subsystems of $\mathfrak{B}_{\mathbb{X}}$.

(b) As we said, the Structure Theorem is completely syntactical; as a corollary, we shall find another syntactical characterization of the polytime subsystems based on maximality (Proposition 4.4.32).

On the other side, we shall see in Chapter 5 that the maximal subsystems do not have good computational properties. Only few of them enjoy the *cut-elimination property*; and almost none of them is *confluent*. We will recall in Chapter 5 what this means exactly. Here, we just point out that the two properties are desirable if we want to use a certain subsystem as a programming language, and maximal subsystems *do not* enjoy them in general. We think that this problem reduces only partially the importance of maximal subsystems. The problem affects the above point (a), but not point (b), so that maximal subsystems are still useful on a practical viewpoint. For example, we can imagine to have a non-maximal subsystem $\mathcal{P}$ that can be extended *in a unique way* to a maximal subsystem $\mathcal{Q}$, thus proving that $\mathcal{P}$ is polytime, and nevertheless $\mathcal{P}$ is confluent. This is the situation, for example, of soLAL (Section 6.2).

Definition 4.4.16 (Maximal Subsystems) *A subsystem $\mathcal{P} \subseteq \mathfrak{B}_{\mathbb{X}}$ is maximal (for polytime) if $\mathcal{P} \in$ PMS, and whenever $\mathcal{P} \subsetneq \mathcal{P}' \subseteq \mathfrak{B}_{\mathbb{X}}$, it holds $\mathcal{P}' \notin$ PMS.*

The picture in Figure 4.10 should help understanding. A weaker notion of maximality is:

Definition 4.4.17 (Relatively Maximal Subsystems) *A subsystem $\mathcal{P} \subseteq \mathfrak{B}_{\mathbb{X}}$ of $\mathfrak{B}_{\mathbb{X}}$ is maximal (for polytime) relatively to $\mathcal{T} \subseteq \mathfrak{B}_{\mathbb{X}}$ if $\mathcal{P} \in$ PMS, and whenever $\mathcal{P} \subsetneq \mathcal{P}' \subseteq \mathcal{T}$, it holds $\mathcal{P}' \notin$ PMS.*

In particular, it is interesting to consider maximality relatively to $\mathcal{T} = \mathsf{MS}^{\mathsf{Y}}$, i.e. the restriction of MS to standard contractions only.

Definition 4.4.18 (Incompatibility) *We will call* **not compatible** *a rule $R$ w.r.t. a subsystem $\mathcal{P}$ if $\mathcal{P} \cup \{R\} \notin$ PMS.*

**Some examples.** For didactic purposes, we are going to describe some maximal subsystems. Most of these subsystems contains the linear kernel $K$, defined on page 29.

LEMMA 4.4.19 (TRIVIALLY MAXIMAL SUBSYSTEMS)
The following subsystems are maximal for polytime:

(a) The subsystem containing every $\mathbf{Y}_{\mathfrak{p}}(\mathfrak{r},\mathfrak{s})$ for every $\mathfrak{p},\mathfrak{r},\mathfrak{s}$, every $P$ rule, $\otimes_{\mathcal{L}}$, $\multimap_{\mathcal{R}}$ and all the rules introducing $\forall$, but no $\mathbf{W}$'s, nor $\otimes_{\mathcal{R}}$, nor $\multimap_{\mathcal{L}}$.
(b) The subsystem containing $K$, every $\mathbf{Y}_{\mathfrak{p}}(\mathfrak{r},\mathfrak{s})$ and every $P_{\mathfrak{r}}(\mathfrak{s})$ for every $\mathfrak{p},\mathfrak{r},\mathfrak{s}$.

The following subsystem is maximal for polytime, relatively to $\mathsf{MS}^{\mathbf{Y}}$:

(c) The subsystem containing $K$ and every $P$ rule, but no contractions.

**Proof.** We show these subsystems reduce in linear time and they are moreover maximal.

(a) Consider a subsystem without $\mathbf{W}$, $\otimes_{\mathcal{R}}$, $\multimap_{\mathcal{L}}$. In this case we can add all the rules we want, including $P$ and $\mathbf{Y}$ rules... indeed we will never use them! That's because there is no way to have more than one premise in a proof net. The subsystem reduces in linear time. This subsystem is maximal: for example, let us try adding $\mathbf{W}$; of course it is not compatible e.g. with $\mathbf{Y}_{\mathfrak{m}}(\mathfrak{m},\mathfrak{m})$ and $P_{\mathfrak{m}}(\mathfrak{m},\mathfrak{m})$.
(b) Second case, consider a subsystem made up of all and only rules $P_{\mathfrak{r}}(\mathfrak{s})$, for all $\mathfrak{r},\mathfrak{s}$. Here, there is no possible duplication, so we can add every $\mathbf{Y}_{\mathfrak{p}}(\mathfrak{r},\mathfrak{s})$ and $\mathbf{W}$. This last subsystem is maximal. Indeed, add some rule $P_{\mathfrak{r}}(\mathfrak{s},\mathfrak{q})$; then cut it with $P_{\mathfrak{s}}(\mathfrak{r})$, $P_{\mathfrak{q}}(\mathfrak{r})$ and $\mathbf{Y}_{\mathfrak{r}}(\mathfrak{q},\mathfrak{q})$, and you have built an exponential chain.
(c) At last, we could have $\mathbf{W}$ and all $P$'s, but no contractions. Also this subsystem trivially reduces in linear time. This is a maximal subsystem too: if we add $\mathbf{Y}^{\mathfrak{m}}$, that rule is not compatible with $P_{\mathfrak{m}}(\mathfrak{m},\mathfrak{m})$. (However note that some other $\mathbf{Y}_{\mathfrak{p}}(\mathfrak{r},\mathfrak{s})$ could be compatible).

□

LEMMA 4.4.20 (ONE MODALITY MAXIMAL SUBSYSTEMS)
Let $\mathbb{X} = \{1\}$. The following two subsystems $\mathcal{P} \subseteq \mathfrak{B}_{\mathbb{X}}$ are the only maximal subsystems relatively to $\mathfrak{B}_{\mathbb{X}}$:

(a) The subsystem containing $\mathbf{Y}^1$, every $P$ rule, $\otimes_{\mathcal{L}}$, $\multimap_{\mathcal{R}}$ and the rules introducing $\forall$, but neither $\mathbf{W}$, nor $\otimes_{\mathcal{R}}$, nor $\multimap_{\mathcal{L}}$.
(b) The subsystem containing $K$, $\mathbf{Y}^1$ and $P_1(1^?)$;
(c) The subsystem containing $K$ and $P_1(1^*)$ rules, but no contractions.

**Proof.** Note that in one-modality subsystems the only possible contraction is $\mathbf{Y}^1$, so $\mathfrak{B}_{\mathbb{X}} = \mathfrak{B}_{\mathbb{X}}{}^{\mathbf{Y}}$. First, we show these subsystems are maximal:

(a),(b) Analogous to Lemma 4.4.19 (a) and (b).
(c) There is no way of getting a spindle, so this is polytime. Of course we cannot add $\mathbf{Y}_1(1,1)$ because it's not compatible with $P_1(1,1)$.

Now, the point is, are there some more maximal subsystems? We prove there are not. If $\mathcal{P}$ is maximal and does not contain $\mathbf{Y}^1$, it must be (c). If $\mathcal{P}$ is maximal and contains $\mathbf{Y}^1$, it can contain $P_1(1^*)$ (so it must be (b)) or it doesn't (so it must be (a)).

What about $\forall, \otimes, \multimap$? Except for the trivial case (a), they are not dangerous, and so they are needed in every maximal subsystem. □

**Basic properties of maximal subsystems.** From now on, we will consider *sensible and maximal subsystems*. Our main result is the Structure Theorem 4.4.25. In order to prove it, we shall study the relations $\mathfrak{R}_{\uparrow}, \mathfrak{R}_{\triangle}, \overline{\mathfrak{R}_{\triangle}}$ among the modalities, that were defined in Section 4.4.2. We are interested in particular to understand whether these relations are linear or partial orders or preorders. The definitions of linear/partial (pre-)orders were given in Section 2.10.

In this paragraph we give some basic properties about the relations $\mathfrak{R}_{\uparrow}, \mathfrak{R}_{\triangle}, \overline{\mathfrak{R}_{\triangle}}$; they will be required to prove the Structure Theorem. We give also some other, unrelated, properties. In particular, Lemma 4.4.21 reveals the relationship between maximal subsystems and *saturated* subsystems, that were studied in Section 4.4.1, and Corollary 4.4.23 shows that the question: *"is $\mathcal{P}$ maximal?"* is in most cases decidable.

LEMMA 4.4.21 (MAXIMAL VS. SATURATED SUBSYSTEMS)
All the sensible maximal subsystems are saturated.

**Proof.** By contradiction, let $\mathcal{P}$ be maximal but not saturated. Then, $\mathcal{S}(\mathcal{P})$ would be polytime for Lemma 4.4.4; and so $\mathcal{P}$ would not be maximal, a contradiction. □

However the converse is false; for example, a sensible subsystem with all and only the *closed* boxes, and without contractions, is trivially saturated, but it is not maximal.

LEMMA 4.4.22 (RULES THAT A MAXIMAL SUBSYSTEM MUST CONTAIN)
Let $\mathcal{P}$ be a sensible maximal subsystem.

1. If $\mathcal{P}$ has $P_{\mathfrak{q}}(\overrightarrow{\mathfrak{m}}, \mathfrak{n}, \mathfrak{n})$ it also has $P_{\mathfrak{q}}(\overrightarrow{\mathfrak{m}}, \mathfrak{n}^*)$.
2. If $\mathcal{P}$ has $P_{\mathfrak{q}}(\mathfrak{m}, \mathfrak{n})$ and $P_{\mathfrak{q}}(\mathfrak{n}, \mathfrak{r})$ it also has $P_{\mathfrak{q}}(\mathfrak{m}, \mathfrak{n}, \mathfrak{r})$.
3. $\forall \mathfrak{q} \in \mathbb{X}$, $\mathcal{P}$ has at least $P_{\mathfrak{q}}(\mathfrak{q}^?)$. That is, $\mathfrak{R}_{\uparrow}$ is reflexive.
4. $\mathfrak{R}_{\triangle} = \overline{\mathfrak{R}_{\triangle}}$.

**Proof.** The justification to the first point develops as follows. Sensibility is needed by Proposition 4.3.40. $\mathcal{P}$ cannot have $P_{\mathfrak{q}}(\overrightarrow{\mathfrak{m}}, \mathfrak{n}^*)$ if such a rule generates a spindle. In that case, the same spindle exists thanks to $P_{\mathfrak{q}}(\overrightarrow{\mathfrak{m}}, \mathfrak{n}, \mathfrak{n})$ against the maximality of $\mathcal{P}$. An analogous argument can be used to justify the second point. The third one is obvious because $P_{\mathfrak{q}}(\mathfrak{q}^?)$ has at most an assumption. □

COROLLARY 4.4.23 (MAXIMALITY IS DECIDABLE)
Let $\mathcal{P}$ be a sensible subsystem, and let us assume that the promotion rules of $\mathcal{P}$ are described by means of regular expressions. Then, the problem "Is $\mathcal{P}$ maximal w.r.t. polynomial time?" is decidable.

**Proof.** We present a very naïve algorithm that solves the problem. Notice that $\mathcal{P}$ has $M < \infty$ modalities, so the possible rules $\mathbf{Y}_{\mathfrak{q}}(\mathfrak{n}, \mathfrak{m})$, $P_{\mathfrak{q}}(\mathfrak{m})$ and $P_{\mathfrak{q}}(\mathfrak{m}, \mathfrak{n})$ are finite. So, (1) for each such rule $R$, check if the subsystem $\mathcal{P} \cup \{R\}$ is polytime. This is possible thanks to Lemma 4.4.13. If so, we are sure that $\mathcal{P}$ is not maximal. Otherwise, (2) we must check that the rules described by Fact 4.4.22 are in $\mathcal{P}$. No other rules can be considered, so the subsystem is maximal iff the answer to (2) is positive. □

LEMMA 4.4.24 ($\Re_\uparrow$ A LINEAR PREORDER)
Let $\mathcal{P}$ be a sensible maximal subsystem. Then $\Re_\uparrow$ is linear preorder, i.e. transitive, reflexive and connected.

**Proof.** $\Re_\uparrow$ is always transitive. Here it is also reflexive, because of the presence of $P_\mathfrak{q}(\mathfrak{q})$. We have to prove that it is connected, that is: $\forall \mathfrak{m}, \mathfrak{n}\ (\mathfrak{m} \Re_\uparrow \mathfrak{n} \vee \mathfrak{n} \Re_\uparrow \mathfrak{m})$. Let us assume that $\mathcal{P}$ is maximal and $\neg\left(\mathfrak{m} \Re_\uparrow \mathfrak{n}\right)$; we will show that $\mathfrak{n} \Re_\uparrow \mathfrak{m}$. This implies that adding $P_\mathfrak{n}(\mathfrak{m})$ to $\mathcal{P}$ gives a non-polytime subsystem $\mathcal{P}'$. So we can find a dangerous spindle $\mathfrak{q} : (\Sigma \cup \chi) : \mathfrak{q}$ in $\mathcal{P}'$ containing $P_\mathfrak{n}(\mathfrak{m})$. This spindle is not in $\mathcal{P}$, however in $\mathcal{P}$ we can find a module $\Sigma' \cup \chi'$ such that, adding a correct number of instances of $P_\mathfrak{n}(\mathfrak{m})$ boxes, gives $\Sigma$. In particular we find in $\Sigma'$ (and so in $\mathcal{P}$) two paths, one from modality $\mathfrak{q}$ to $\mathfrak{m}$ and the other one from $\mathfrak{n}$ to $\mathfrak{q}$, that can be composed to create a path from $\mathfrak{n}$ to $\mathfrak{m}$: so $\mathfrak{n} \Re_\uparrow \mathfrak{m}$. □

So, from now on, we shall write also $\preceq$ in place of $\Re_\uparrow$, $\prec$ for the corresponding strict preorder, and $\mathfrak{m} \equiv \mathfrak{n}$ whenever both $\mathfrak{m} \preceq \mathfrak{n}$ and $\mathfrak{n} \preceq \mathfrak{m}$.

Differently from $\Re_\uparrow$, $\overline{\Re_\triangle}$ is not a preorder; in general it is not even connected. For example, consider $\mathbb{X} = \{1, 2, 3, 4, 5\}$,

$$\mathcal{P} = \{P_4(3, 4), \mathbf{Y}_4(1, 4), P_5(5, 1), \mathbf{Y}_5(5, 2)\}.$$

$\mathcal{P}$ is polytime because $\Re_\uparrow$ is a linear order:

$$3 \prec 4 \prec 1 \prec 5 \prec 2$$

and satisfies the hypothesis of Lemma 4.3.33. But it is not maximal: it can be embedded in some other maximal subsystem $\mathcal{P}'$. We are not interested in the exact form of $\mathcal{P}'$; we just want to observe that $\mathcal{P}'$ cannot have neither the rule $\mathbf{Y}_1(2, 3)$, nor $P_1(2, 3)$, because both of them permit the construction of some dangerous chain. So both $\neg\left(\langle 1\rangle \overline{\Re_\triangle} \langle 2, 3\rangle\right)$ and $\neg\left(\langle 2, 3\rangle \overline{\Re_\triangle} \langle 1\rangle\right)$. That is, $\overline{\Re_\triangle}$ is not connected.

So, it turns out that the relation $\preceq$ has a great importance in the study of maximal systems, and we shall use it (instead of $\overline{\Re_\triangle}$, or other) in proving the Structure Theorem.

**Structure Theorem for Maximal Subsystems.** We are going to restrict our attention to the subsystem in which $\preceq$ is a **linear order** (i.e. a connected partial order, see Section 2.10). Later, we shall see that this assumption is fully justified, i.e. maximal subsystems whose $\preceq$ is a linear order are more interesting that the other ones.

THEOREM 4.4.25 (STRUCTURE OF MAXIMAL SENSIBLE SUBSYSTEMS)
Let $\mathcal{P}$ be a sensible subsystem with a linear $\preceq$ and with modalities in $\mathbb{X}$. $\mathcal{P}$ is maximal iff $\mathcal{P}$ contains exactly the following rules:

1. all the rules $\mathbf{Y}_\mathfrak{q}(\mathfrak{m}, \mathfrak{n})$ for every choice of $\mathfrak{q} \prec \mathfrak{m}, \mathfrak{n}$ in $\mathbb{X}$;
2. all the rules $\mathbf{Y}_\mathfrak{q}(\mathfrak{q}, \mathfrak{n})$ for every choice of $\mathfrak{q} \prec \mathfrak{n}$ in $\mathbb{X}$;
3. all the rules $P_\mathfrak{q}(\mathfrak{q}^?, \mathfrak{m}_1^*, \ldots, \mathfrak{m}_k^*)$ for every choice of $\mathfrak{m}_1, \ldots, \mathfrak{m}_k \prec \mathfrak{q}$ in $\mathbb{X}$;
4. only one among $\mathbf{Y}_\mathfrak{q}(\mathfrak{q}, \mathfrak{q})$ and $P_\mathfrak{q}(\mathfrak{q}^*, \mathfrak{m}_1^*, \ldots, \mathfrak{m}_k^*)$, for every $\mathfrak{m}_1 \ldots, \mathfrak{m}_k \prec \mathfrak{q}$ in $\mathbb{X}$.

**Proof.** *"Only if" direction.* We want to prove that every system $\mathcal{P}$ containing the rules described in points 1, 2, 3, 4 is maximal. The order $\preceq$ prevents to construct dangerous spindles thanks to its linearity. So, Proposition 4.3.40 implies $\mathcal{P}$ is polytime. Moreover, if $\mathbf{Y}_{\mathfrak{q}}(\mathfrak{q},\mathfrak{q})$ belongs to $\mathcal{P}$, adding $P_{\mathfrak{q}}(\mathfrak{q}^*,\mathfrak{m}_1^*,\ldots,\mathfrak{m}_k^*)$, against point 4, would allow to construct a dangerous spindle. The same would be by starting with $P_{\mathfrak{q}}(\mathfrak{q}^*,\mathfrak{m}_1^*,\ldots,\mathfrak{m}_k^*)$ in $\mathcal{P}$ and, then, adding $\mathbf{Y}_{\mathfrak{q}}(\mathfrak{q},\mathfrak{q})$.

*"If" direction.* We assume $\mathcal{P}$ is maximal and we want show that it must contain at least all the rules described in the statement. We prove it by contradiction first relatively to points 1, 2, and 3, then to point 4. A contradiction relative to points 1, 2, and 3 requires to assume that $\mathcal{P}$ has not a rule $R$ described in 1, 2, 3. Let $\mathcal{P}' = \mathcal{P} \cup \{R\}$. The maximality of $\mathcal{P}$ implies that $\mathcal{P}'$ is not polytime. Now, thanks to Proposition 4.3.40, $\mathcal{P}' \notin$ PMS implies that $\mathcal{P}'$ can build a dangerous spindle $\mathfrak{r}:(\Sigma \cup \chi):\mathfrak{r}$. We are going to see that this causes various kinds of contradictions.

Let $u$ be an instance of $R = \mathbf{Y}_{\mathfrak{q}}(\mathfrak{m},\mathfrak{n})$. For example, we can assume that one of the two distinct CS-paths in $\Sigma$ crosses $u$ from $\mathfrak{q}A$ to $\mathfrak{m}A$. By definition of $\preceq$, $\mathfrak{r} \preceq \mathfrak{q} \prec \mathfrak{m} \preceq \mathfrak{r}$, so contradicting the linearity of $\preceq$.

Let $u$ be an instance of $R = \mathbf{Y}_{\mathfrak{q}}(\mathfrak{q},\mathfrak{n})$. The previous case excludes that one of the two distinct CS-paths in $\Sigma$ crosses $u$ from $\mathfrak{q}A$ to $\mathfrak{n}A$. So, it must cross $u$ from its premise, labeled $\mathfrak{q}A$ to its conclusion, labeled $\mathfrak{q}A$. This means that $\mathfrak{r}:\Sigma' \cup \chi':\mathfrak{r}$, obtained from $\Sigma \cup \chi$ just removing $u$, exists in $\mathcal{P}$, which could not be polytime.

The case $R = P_{\mathfrak{q}}(\mathfrak{q},\mathfrak{m}_1,\ldots,\mathfrak{m}_k)$ combines the two previous ones.

Now we move to prove point 4. Without loss of generality, we can prove the thesis for $P_{\mathfrak{q}}(\mathfrak{q},\mathfrak{q})$ instead of $P_{\mathfrak{q}}(\mathfrak{q}^*,\mathfrak{m}_1^*,\ldots,\mathfrak{m}_k^*)$. By contradiction, let $\mathcal{P}$ be maximal and let the thesis be false. So, there are $R_1 = \mathbf{Y}_{\mathfrak{q}}(\mathfrak{q},\mathfrak{q})$ and $R_2 = P_{\mathfrak{q}}(\mathfrak{q},\mathfrak{q})$ such that either $\mathcal{P}$ has both of them, *or* $\mathcal{P}$ has none of them. In the first case the two rules would build a spindle, making $\mathcal{P}$ not polytime. In the second case, neither $R_1$ nor $R_2$ belong to $\mathcal{P}$. This means that neither $\mathcal{P}_1 = \mathcal{P} \cup \{R_1\}$, nor $\mathcal{P}_2 = \mathcal{P} \cup \{R_2\}$ are polytime, because, recall, $\mathcal{P}$ is maximal. So, there have to exist both $\mathfrak{r}_1:(\Sigma_1 \cup \chi_1):\mathfrak{r}_1$ in $\mathcal{P}_1$ that involves an instance $u_1$ of $R_1$ (Figure 4.11a), and $\mathfrak{r}_2:(\Sigma_2 \cup \chi_2):\mathfrak{r}_2$ in $\mathcal{P}_2$ involving the instance $b_2$ of $R_2$ (Figure 4.11b). Thanks to Lemma 4.3.38 we can assume that all the labels in $\Sigma_1 \cup \chi_1$ and $\Sigma_2 \cup \chi_2$ are of the form $\mathfrak{n}A$, for a fixed $A$ and some $\mathfrak{n}$'s. Then, $u_1$ must be the *principal* contraction of $\Sigma_1$. If not, we could eliminate it as we did for $u$ in point 2, proving that $\mathcal{P}$ is not polytime. This is why $\mathfrak{r}_1 = \mathfrak{q}$ in $\Sigma_1 \cup \chi_1$ of Figure 4.11a. Now, let $\Theta$ be the module obtained from $\Sigma_1 \cup \chi_1$ removing $u_1$; we can build $\Theta$ in $\mathcal{P}$ (Figure 4.11c). $\Theta$ has two premises $\mathfrak{q}A$ and one conclusion $\mathfrak{q}A$, exactly as the box $b_2$ in $\Sigma_2$. So, we can replace $b_2$ in $\Sigma_2 \cup \chi_2$ with $\Theta$, getting a new dangerous spindle $\mathfrak{r}_2:(\Sigma \cup \chi):\mathfrak{r}_2$ in $\mathcal{P}$. But $(\Sigma \cup \chi) \in \mathbf{PN}(\mathcal{P})$, so that $\mathcal{P}$ is not polytime. □

Thanks to the Structure Theorem, we can count the maximal subsystems:

**Corollary 4.4.26 (Number of Maximal Subsystems)**
Let $\#\mathbb{X} = M$. There are $2^M$ maximal and sensible subsystems of $\mathfrak{B}_{\mathbb{X}}$, whose $\preceq$ relation is a linear order.

**Proof.** $2^M$ because, for each of the $M$ modalities, one has to choose among the rule $P_i(i^*)$ or $\mathbf{Y}^i$. □

It follows an observation about the *spindles relation* $\mathfrak{R}_\diamond$, defined in Section 4.4.2:

**Corollary 4.4.27 ($\mathfrak{R}_\diamond$ coincides with $\prec$)**
Let $\mathcal{P}$ be a maximal sensible subsystem whose $\mathfrak{R}_\uparrow$ (or $\preceq$) relation is a linear order. Then its $\mathfrak{R}_\diamond$ is a strict linear order, and so it is equal to $\prec$.

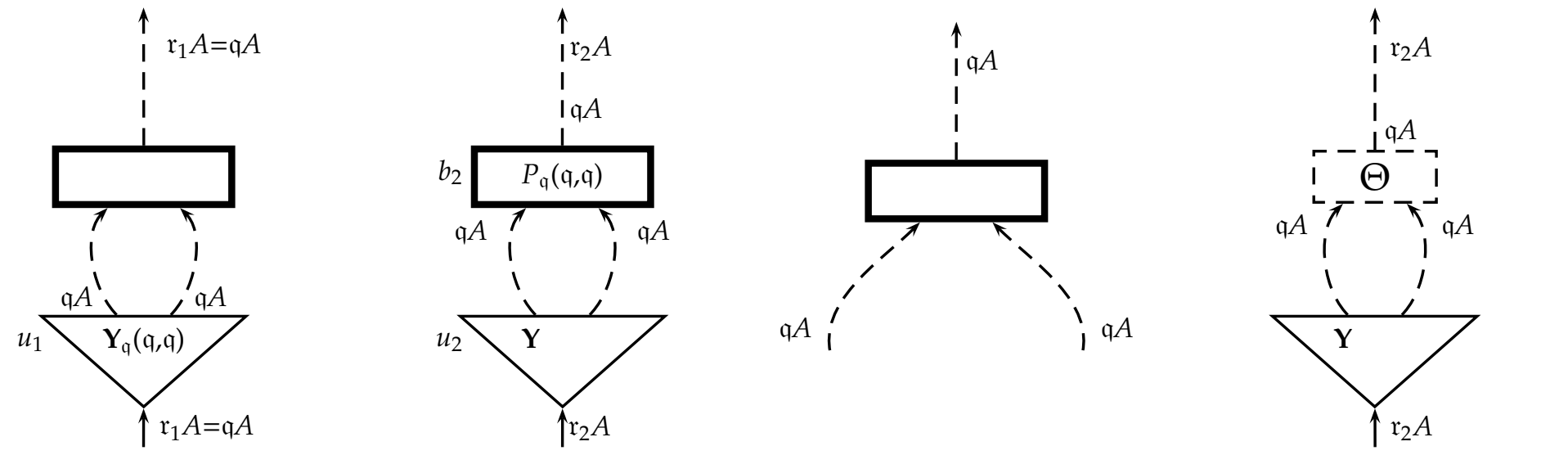

(a) $\mathfrak{r}_1 : (\Sigma_1 \cup \chi_1) : \mathfrak{r}_1$ in $\mathcal{P}_1$ (b) $\mathfrak{r}_2 : (\Sigma_2 \cup \chi_2) : \mathfrak{r}_2$ in $\mathcal{P}_2$ (c) $\Theta$ from $\Sigma_1$ (d) $\mathfrak{r}_2 : (\Sigma \cup \chi) : \mathfrak{r}_2$ in $\mathcal{P}$

Figure 4.11: The proof nets used in the proof of Theorem 4.4.25

**Proof.** Thanks to Structure Theorem, for every $\mathfrak{m} \prec \mathfrak{n}$ we can build a spindle using $\mathbf{Y}_{\mathfrak{m}}(\mathfrak{m}, \mathfrak{n})$ and $P_{\mathfrak{n}}(\mathfrak{m}, \mathfrak{n})$, and so $\mathfrak{m}\,\mathfrak{R}_\diamond\,\mathfrak{n}$. □

If $\mathfrak{R}_\uparrow$ is not an order, the Corollary 4.4.27 may fail. Consider for example $\mathbb{X} = \{1, 2, 3\}$,

$$\mathcal{P} = \{P_1(1^?), P_2(1^*, 2^*, 3^*), P_3(1^*, 2^*, 3^*), \mathbf{Y}_1(x, y) \text{ for } x, y \leq 3\}.$$

This is maximal, $1\,\mathfrak{R}_\diamond\,2$, $1\,\mathfrak{R}_\diamond\,3$, but 2 and 3 incomparable.

**What if $\preceq$ is not an Order.** We shall see that this case can be brought back to the previous one. If $\preceq$ is a linear preorder over $\mathbb{X}$, then it naturally induces a linear order over the quotient $\mathbb{X}' = \mathbb{X}/\!\equiv$, where $\mathfrak{m} \equiv \mathfrak{n}$ iff $\mathfrak{m} \preceq \mathfrak{n}$ and $\mathfrak{n} \preceq \mathfrak{m}$ (cfr. Section 2.10). $\overline{\mathfrak{m}}$ denotes the equivalence class of $\mathfrak{m}$.

Definition 4.4.28 (Natural Quotient of $\mathcal{P}$) *Let $\mathcal{P} \subseteq \mathfrak{B}_{\mathbb{X}}$ sensible. Recall that $\preceq$ is a preorder, and let $\mathbb{X}'$ the quotient of $\mathbb{X}$ as just described. Let $\mathcal{P}' \subseteq \mathfrak{B}_{\mathbb{X}'}$ be the smallest subsystem of $\mathfrak{B}_{\mathbb{X}'}$ such that:*

$$P_{\mathfrak{q}}\left(\overrightarrow{\mathfrak{m}}\right) \in \mathcal{P} \;\Rightarrow\; P_{\overline{\mathfrak{q}}}\left(\overrightarrow{\overline{\mathfrak{m}}}\right) \in \mathcal{P}',$$
$$\mathbf{Y}_{\mathfrak{q}}(\mathfrak{m}, \mathfrak{n}) \in \mathcal{P} \;\Rightarrow\; \mathbf{Y}_{\overline{\mathfrak{q}}}(\overline{\mathfrak{m}}, \overline{\mathfrak{n}}) \in \mathcal{P}'.$$

*$\mathcal{P}'$ is called* **natural quotient** *of $\mathcal{P}$.*

Fact 4.4.29
For every $\mathcal{P}$, the relation $\preceq$ of $\mathcal{P}'$ is a linear order.

The Fact directly follows from the construction.

Lemma 4.4.30 (Linear Orders are Interesting)
Let $\mathcal{P} \subseteq \mathfrak{B}_{\mathbb{X}}$ sensible, and $\mathcal{P}' \subseteq \mathfrak{B}_{\mathbb{X}'}$ be its natural quotient. Then, $\mathcal{P}$ is polytime iff $\mathcal{P}'$ is polytime.

**Proof.** "⇒". If $\mathcal{P}$ can build a dangerous chain of spindles, then by construction $\mathcal{P}'$ can build the same chain, too.

"$\Leftarrow$". If $\mathcal{P}'$ can build a dangerous chain of spindles $\Theta \cup \chi$, then we are not sure that $\mathcal{P}$ can build the same chain. Indeed, it is possible that (1) $\mathfrak{m}, \mathfrak{n} \in \mathbb{X}$ have been collapsed into $\overline{\mathfrak{m}} \in \mathbb{X}'$, (2) $\Sigma'$ contains two nodes e.g. $P_{\overline{\mathfrak{m}}}(\ldots)$ and $\mathbf{Y}_{\overline{\mathfrak{m}}}(\ldots)$ connected through a cut, but (3) such rules arrive from the rules $P_{\mathfrak{m}}(\ldots)$ and $\mathbf{Y}_{\mathfrak{n}}(\ldots)$ of $\mathcal{P}$, that cannot be connected through a cut. But, by construction, in that case $\mathfrak{m} \preceq \mathfrak{n}$, so $\mathcal{P}$ can build a CS-path $\tau$ whose edges are all labelled with modal formulæ, the first one with modality $\mathfrak{m}$ and the last one with $\mathfrak{n}$. $\tau$ can be used to connect the rules $P_{\mathfrak{m}}(\ldots)$ and $\mathbf{Y}_{\mathfrak{n}}(\ldots)$, thus allowing the construction of $\Sigma' \cup \chi'$. □

We have seen that the Structure Theorem 4.4.25 holds only for sensible and maximal subsystems whose $\preceq$ relation is a linear order. Here we generalize it. If $\preceq$ is a preorder, but not an order, the equivalent modalities must be considered all equal:

Corollary 4.4.31 (to the Structure Theorem of Maximal Sensible Subsystems)
Let $\mathcal{P} \subseteq \mathfrak{B}_{\mathbb{X}}$ be a sensible subsystem. $\mathcal{P}$ is maximal iff $\mathcal{P}$ contains exactly the following rules:

1. all the rules $\mathbf{Y}_{\mathfrak{q}}(\mathfrak{m}, \mathfrak{n})$ for every choice of $\mathfrak{q} \prec \mathfrak{m}, \mathfrak{n}$ in $\mathbb{X}$;
2. all the rules $\mathbf{Y}_{\mathfrak{q}}(\mathfrak{q}', \mathfrak{n})$ for every choice of $\mathfrak{q} \prec \mathfrak{n}$ and $\mathfrak{q} \equiv \mathfrak{q}'$ in $\mathbb{X}$;
3. all the rules $P_{\mathfrak{q}}(\mathfrak{q}'^{?}, \mathfrak{m}_1^*, \ldots, \mathfrak{m}_k^*)$ for every choice of $\mathfrak{m}_1, \ldots, \mathfrak{m}_k \prec \mathfrak{q}$ and $\mathfrak{q} \equiv \mathfrak{q}'$ in $\mathbb{X}$;
4. only one among $\mathbf{Y}_{\mathfrak{q}}(\mathfrak{q}, \mathfrak{q})$ and $P_{\mathfrak{q}}(\mathfrak{q}^*, \mathfrak{m}_1^*, \ldots, \mathfrak{m}_k^*)$, for every $\mathfrak{m}_1 \ldots, \mathfrak{m}_k \prec \mathfrak{q}$ in $\mathbb{X}$;
5. whenever $\mathcal{P}$ has the rule $\mathbf{Y}_{\mathfrak{q}}(\mathfrak{q}, \mathfrak{q})$, and $\mathfrak{q}_1 \equiv \mathfrak{q}_2 \equiv \mathfrak{q}_3 \equiv \mathfrak{q}$, then $\mathcal{P}$ also has $\mathbf{Y}_{\mathfrak{q}_1}(\mathfrak{q}_2, \mathfrak{q}_3)$.
6. similarly, whenever $\mathcal{P}$ has the rule $P_{\mathfrak{q}}(\mathfrak{q}^*, \mathfrak{m}_1^*, \ldots, \mathfrak{m}_k^*)$, and $\mathfrak{q}_0 \equiv \mathfrak{q}_1 \equiv \ldots \equiv \mathfrak{q}_t \equiv \mathfrak{q}$, then $\mathcal{P}$ also has $P_{\mathfrak{q}_0}(\mathfrak{q}_1, \ldots, \mathfrak{q}_t, \mathfrak{m}_1^*, \ldots, \mathfrak{m}_k^*)$.

**Proof.** Every two equivalent modalities $\mathfrak{q} \equiv \mathfrak{q}'$ must be considered the same, as it is possible to build a CS-path connecting them. □

**A syntactical criterion for polytime soundness.** At last, we can derive a criterion for polynomial soundness based on maximality:

Proposition 4.4.32 (Criterion of Polynomiality based on Maximality)
Let $\mathcal{P} \subseteq \mathfrak{B}_{\mathbb{X}}$ be any sensible subsystem, and $\mathcal{P}' \subseteq \mathfrak{B}_{\mathbb{X}'}$ be its natural quotient (Definition 4.4.28). Then, the following are equivalent:

1. $\mathcal{P}$ is polytime.
2. $\mathcal{P} \subseteq \mathcal{Q}$ for some $\mathcal{Q} \subseteq \mathfrak{B}_{\mathbb{X}}$ maximal.
3. $\mathcal{P}'$ is polytime.
4. $\mathcal{P}' \subseteq \mathcal{Q}'$ for some $\mathcal{Q}' \subseteq \mathfrak{B}_{\mathbb{X}'}$ maximal.

For a given $\mathcal{P}$, it is easy to verify the condition of point 2., using the Corollary 4.4.31 to the Structure Theorem, or the condition of point 4., using directly the Structure Theorem 4.4.25.
**Proof.**

1. $\Leftrightarrow$ 3. Already proved in Lemma 4.4.30.

2. $\Rightarrow$ 1. If $\mathcal{P} \subseteq \mathcal{Q}$ and $\mathcal{Q}$ is maximal, then $\mathcal{Q} \in$ PMS, so $\mathcal{P} \in$ PMS because of Lemma 4.1.8.

4. $\Rightarrow$ 3. Exactly as 2. $\Rightarrow$ 1.

3. ⇒ 4. We have shown that the relation $\mathfrak{R}_\uparrow$ for $\mathcal{P}'$ is a linear order. If $\mathcal{P}'$ is polytime, then it cannot build dangerous spindles. As a consequence, all the $\mathbf{Y}$ and $P$ rules in $\mathcal{P}'$ respect the constraints on $\mathbb{X}$ requested by Theorem 4.4.25. On the other side, $\mathcal{P}'$ is polytime and maximal, so it cannot build dangerous spindles (Proposition 4.3.40). As a consequence, also the constrains to the rules $\mathbf{Y}$ and $P$ requested by points 3. and 4. of Theorem 4.4.25 hold. We have shown that all the rules of $\mathcal{P}'$ are among the rules of some larger maximal subsystem $\mathcal{Q}'$.

1. ⇒ 2. We have two proofs of this. On one side, it is possible to adapt the proof of 3. ⇒ 4, using Corollary 4.4.31 instead of Theorem 4.4.25. We prefer to follow a different proof. We already know that $\mathcal{P}'$ is polytime, contained into some maximal $\mathcal{Q}' \subseteq \mathfrak{B}_{\mathbb{X}'}$. So, $\mathcal{Q}$ can be build as follows. $\mathcal{Q}$ is a maximal system described in Corollary 4.4.31, with modalities in $\mathbb{X}$ and with the following partial order over $\mathbb{X}$: $\mathfrak{m} \leq \mathfrak{n}$ iff $\overline{\mathfrak{m}} \leq \overline{\mathfrak{n}}$. Notice that this is *not* the partial order $\mathfrak{R}_\uparrow$ induced by $\mathcal{P}$, but an extension of it.

□

## 4.5 Infinite modalities

Subsystems with infinite modalities are not sensible by definition. Moreover, they do not have any practical application, because no calculator will handle them. And nevertheless, they are very natural in a theoretical perspective: why 2, or 3, or 43 modalities, and not *all* the possible natural numbers as modalities?

If the subsystem $\mathcal{P}$ has infinite modalities, the Polynomiality Criterion 4.3.40 fails, as we see in the following example. Consider $\mathbb{X} = \mathbb{N}$, and the subsystem

$$\mathcal{P} = \{\text{linear rules}\} \cup \{P_n(n,n) \mid n \in \mathbb{N}\} \cup \{\mathbf{Y}_n(n+1, n+1) \mid n \in \mathbb{N}\}.$$

$\mathcal{P}$ does not build any dangerous spindles, however it can build arbitrarily long chains of spindles, using always different modalities, so that $\mathcal{P}$ is not polytime.

Now, we will prove some conditions for the polynomiality of some subsystems with infinite modalities. Recall that $i \,\mathfrak{R}_\diamond\, j$ means that $\mathcal{P}$ can build a spindle, starting with modality $i$ and arriving to a modality $j$ (see page 70).

Lemma 4.5.1 (A Sufficient Condition for Polynomiality)
Let $\mathcal{P} \subseteq \mathrm{MS}_{\mathbb{N}}$ be a system with infinite modalities. If $\exists n \in \mathbb{N}\ \nexists i_1, \ldots, i_n \in \mathbb{N}\ (i_1 \,\mathfrak{R}_\diamond\, i_2 \,\mathfrak{R}_\diamond \ldots \mathfrak{R}_\diamond\, i_n)$, i.e. if $\mathcal{P}$ cannot build any chain of $n$ spindles, then $\mathcal{P}$ is polytime.

**Proof.** Observe that Proposition 4.3.32 still holds, as it has not hypothesis on the number of the modalities. Our hypothesis says that there is a bound $n$ to the number of consecutive spindles that compose a chain of spindles; so Lemma 4.3.32 implies the system is polytime. □

We call **sensible**$^\infty$ a system with infinite modalities that, apart from this, respects the definition of sensible subsystem.

**Lemma 4.5.2 (A Sufficient Condition for Non-Polynomiality)**
Let $\mathcal{P}$ be a sensible$^\infty$ system. If $\forall n \in \mathbb{N}\ \exists i_1, \ldots, i_n \in \mathbb{N}\ (i_1 \mathfrak{R}_\diamond i_2 \mathfrak{R}_\diamond \ldots \mathfrak{R}_\diamond i_n)$, i.e. if $\mathcal{P}$ can build chains of spindles $i_1 : \Theta_n : i_n$ arbitrarily long, and moreover the size of these chains is $|\Theta_n| = O(n)$, then $\mathcal{P}$ is not polytime.

**Proof.** For every $n$ we find a chain of $n$ spindles. This chain can be chosen with size $O(n)$ by hypothesis, and depth 1, thanks to Lemma 4.3.38. So, following the proofs of Lemma 4.3.35 and 4.3.39, we understand that $\mathcal{P}$ is not polytime. □

We can see that there is a third, possible case, not contemplated by the two previous Lemmas. This is the case in which $\mathcal{P}$ can build chains of spindles arbitrarily long, but however their size is very big, e.g. $O(2^n)$. In this case, actually we don't know if $\mathcal{P}$ is polytime or not. If $\mathcal{P}$ is saturated and sensible$^\infty$, the previous Lemmas can be extended in order to include all the three cases:

**Corollary 4.5.3 (A Polynomiality Criterion for Saturated Subsystems)**
Let $\mathcal{P}$ be a subsystem sensible$^\infty$ and saturated. Then $\mathcal{P}$ is polytime if and only if $\exists n \in \mathbb{N}\ \nexists i_1, \ldots, i_n \in \mathbb{N}\ (i_1 \mathfrak{R}_\diamond i_2 \mathfrak{R}_\diamond \ldots \mathfrak{R}_\diamond i_n)$.

**Proof.** "$\Leftarrow$". If $\exists n \in \mathbb{N}\ \nexists i_1, \ldots, i_n \in \mathbb{N}\ (i_1 \mathfrak{R}_\diamond i_2 \mathfrak{R}_\diamond \ldots \mathfrak{R}_\diamond i_n)$, then $\mathcal{P}$ is polytime for Lemma 4.5.1.
"$\Rightarrow$". Let $\forall n \in \mathbb{N}\ \exists i_1, \ldots, i_n \in \mathbb{N}\ (i_1 \mathfrak{R}_\diamond i_2 \mathfrak{R}_\diamond \ldots \mathfrak{R}_\diamond i_n)$. This means that, for every $n$, a chain of spindles $\Theta_n$ can be built. Notice that, whenever $\mathfrak{m} : \Sigma : \mathfrak{n}$ is a spindle, the hypothesis of saturation let us to create another spindle $\mathfrak{m} : \Sigma' : \mathfrak{n}$ with just one box and one contraction. As a consequence, from $\Theta_n$ it is possible to build another chain of spindles $\Theta'_n$ with size $O(n)$. Lemma 4.5.2 can now be applied. □

Corollary 4.5.3 highlights the difference between sensible and sensible$^\infty$ subsystems. For sensible$^\infty$ subsystems, Proposition 4.4.4 fails: it is possible that a sensible$^\infty$ subsystem $\mathcal{P}$ is polytime, while its augmented system is not.

A different (but similar) kind of polynomiality can be proved, in analogy to Lemma 8.1.5:

**Lemma 4.5.4 (A Sufficient Condition for a Weak Polynomiality)**
Let $\mathcal{P} \subseteq \mathsf{MS}_{\mathbb{X}}$ be a system with infinite modalities. Let us suppose that $\mathcal{P}$ cannot build dangerous spindles. Let us suppose moreover that (i) there is a linear ordering among the modalities of $\mathbb{X}$, (ii) this linear order is a well-ordering with order type at most $\omega$, and (iii) the rules of $\mathcal{P}$ respect this ordering. Then, there exists a family of polynomials $\{p_{\partial,M}(x) \mid \partial, M \in \mathbb{N}\}$ such that for every $\Pi \in \mathbf{PN}(\mathcal{P})$ with depth at most $\partial$ and whose modalities have order-type at most $M$, $[\Pi]$ and $||\Pi||$ are bounded by $p_{\partial,M}(|\Pi|)$.

**Proof.** The long hypothesis about the well-ordering is needed to obtain that, whenever $\Pi \to^* \Sigma$, then also the modalities of $\Sigma$ have order-type at most $M$. Now, if $\mathcal{P}$ is a subsystem that uses only the first $M$ modalities of the ordering of $\mathbb{X}$, we apply Proposition 4.3.40 and we find that $\mathcal{P}$ is polytime. Finally, we put the obtained polynomials all together. □

# Chapter 5

# Computational Properties in MS

In this Chapter we will face two properties that, traditionally, are considered indefeasible in light logics: *cut-elimination* and *confluence*. A logic enjoys the cut-elimination property if *every* cut inside a proof can always be reduced. A logic enjoys confluence (or: it is Church-Rosser) if whenever a proof can reduce in two different ways, then the two reducts may still reduce to a common form. Such two properties are not guaranteed for subsystems of MS.

In Section 5.1 we shall characterize, among the maximal subsystems, those ones enjoying the cut-elimination property. In Section 5.2 we will see that almost every maximal subsystem is not confluent. We shall also see some properties that force a subsystem to be confluent. For example, the subsystems that have cut-elimination and that use only standard contractions (i.e. of kind $\mathbf{Y}_{\mathfrak{m}}(\mathfrak{m},\mathfrak{m})$) are also confluent (Lemma 5.2.6).

## 5.1 About Cut-elimination

FACT 5.1.1 (CUTS ELIMINABLE IN THE LINEAR KERNEL)
Let $\mathcal{P} \subseteq \mathfrak{B}_{\mathbb{X}}$. All the linear cuts can always be eliminated. If $\mathcal{P}$ contains the linear kernel, also the *gc* cuts can be eliminated.

**Proof.** The linear cuts annihilate, without any need of new nodes. The *gc* cuts may create new nodes, but all these are among the nodes of the linear kernel. □

FACT 5.1.2 (CUTS ELIMINABLE IN SATURATED SUBSYSTEMS)
Let $\mathcal{P} \subseteq \mathfrak{B}_{\mathbb{X}}$ be saturated. Then, every $[P/P]$ cut can be reduced.

**Proof.** By definition of saturation. □

FACT 5.1.3 (CUTS ELIMINABLE IN SENSIBLE SUBSYSTEM)
Let $\mathcal{P} \subseteq \mathfrak{B}_{\mathbb{X}}$ sensible. Every cut involving a closed box is reducible.

**Proof.** The definition of sensible subsystem requires both the presence of all the closed boxes and the downward closure. □

PROPOSITION 5.1.4 (CHARACTERIZATION OF SENSIBLE MAXIMAL SUBSYSTEMS WITH CUT-ELIMINATION)
Let us consider $\mathcal{P} \subseteq$ MS sensible and maximal w.r.t. polytime soundness, such that its $\Re_{\uparrow}$ relation is a linear order. Then, $\mathcal{P}$ enjoys cut elimination iff it exists $\mathfrak{q} \in \mathbb{X}$ such that
(i) for all the modalities $\mathfrak{m}$, with $\mathfrak{m} \leq \mathfrak{q}$, $\mathcal{P}$ has $\mathbf{Y}^{\mathfrak{m}}$, and so not $P_{\mathfrak{m}}(\mathfrak{m}^*)$;
(ii) for all the modalities $\mathfrak{m}$, with $\mathfrak{q} < \mathfrak{m}$, $\mathcal{P}$ has $P_{\mathfrak{m}}(\mathfrak{m}^*)$, and so not $\mathbf{Y}^{\mathfrak{m}}$.

**Proof.** Let us assume there exists such $\mathfrak{q}$; we want to prove that $\mathcal{P}$ enjoys cut-elimination. The only problematic *ns* are $[P/P]$ and $[P/\mathbf{Y}]$. If a box $P_{\mathfrak{m}}(\mathfrak{n}_1, \ldots, \mathfrak{n}_k, \mathfrak{r})$ is cut with a box $P_{\mathfrak{r}}(\mathfrak{s}_1, \ldots, \mathfrak{s}_l)$, the linear order says that $\forall i, j,\ \mathfrak{s}_i \leq \mathfrak{r}$ and $\mathfrak{n}_j, \mathfrak{r} \leq \mathfrak{m}$. The cut must reduce to $P_{\mathfrak{m}}(\vec{\mathfrak{n}}, \vec{\mathfrak{q}})$. There are three cases. (1) If at most one among the $\mathfrak{n}_i$'s and $\mathfrak{s}_j$'s is $\mathfrak{m}$, then the rule $P_{\mathfrak{m}}(\vec{\mathfrak{n}}, \vec{\mathfrak{q}})$ is in $\mathcal{P}$ for maximality. (2) If 2 or more among the $\mathfrak{n}_i$'s are $\mathfrak{m}$, then the $\mathcal{P}$ has the rule $P_{\mathfrak{m}}(\mathfrak{m}^*)$ by maximality, and so it has also $P_{\mathfrak{m}}(\vec{\mathfrak{n}}, \vec{\mathfrak{s}})$. (3) If 2 or more among the $\mathfrak{n}_i$'s and $\mathfrak{s}_j$'s are $\mathfrak{m}$, and at least one among the $\mathfrak{s}_j$'s is $\mathfrak{m}$, then $\mathfrak{r} = \mathfrak{m}$ and one of the boxes is in fact $P_{\mathfrak{m}}(\mathfrak{m}, \mathfrak{m}, \ldots)$; again, for maximality $\mathcal{P}$ must have the rule $P_{\mathfrak{m}}(\vec{\mathfrak{n}}, \vec{\mathfrak{s}})$. So, let us consider a cut between a contraction and a box. There are several different cases. (1) If the box is closed, the reduction can be performed for downward closure. (2) $\mathbf{Y}_{\mathfrak{s}}(\mathfrak{m}, \mathfrak{n})$, with $\mathfrak{s} < \mathfrak{m}$ and $\mathfrak{s} \leq n$, is cut with $P_{\mathfrak{s}}(\mathfrak{r}_1, \ldots, \mathfrak{r}_k, t)$ with $\mathfrak{r}_i < \mathfrak{s}$ and $t \leq \mathfrak{s}$. This can be reduced by maximality, using rules $P_{\mathfrak{m}}(\mathfrak{s}, \ldots, \mathfrak{s}, \mathfrak{m})$, $P_n(r_1, \ldots, r_k, \mathfrak{s})$, $\mathbf{Y}_{\mathfrak{r}_i}(\mathfrak{s}, \mathfrak{r}_i)$ and $\mathbf{Y}_t(\mathfrak{m}, \mathfrak{s})$. (3) $\mathbf{Y}_{\mathfrak{s}}(\mathfrak{s}, \mathfrak{s})$ is cut with $P_{\mathfrak{s}}(\mathfrak{r}_1, \ldots, \mathfrak{r}_k, t)$ with $r_i < \mathfrak{s}$ and $t \leq \mathfrak{s}$. Of course $\mathfrak{s} \leq \mathfrak{q}$, so we can reduce it using rules $P_{\mathfrak{s}}(\mathfrak{r}_1, \ldots, \mathfrak{r}_k, t)$, $\mathbf{Y}^{\mathfrak{r}_i}$ and $\mathbf{Y}^t$. (4) Symmetric: $\mathbf{Y}_{\mathfrak{s}}(m, n)$, with $\mathfrak{s} < \mathfrak{m}$ and $\mathfrak{s} \leq \mathfrak{n}$, is cut with $P_{\mathfrak{s}}(\mathfrak{r}_1, \ldots, \mathfrak{r}_k, \mathfrak{s}^*)$ with $\mathfrak{r}_i < \mathfrak{s}$. Now $\mathfrak{q} < \mathfrak{s}$, so we can reduce the cut using rules $P_{\mathfrak{m}}(\mathfrak{m}^*)$, $P_{\mathfrak{n}}(\mathfrak{n}^*, \mathfrak{s}^*)$, $\mathbf{Y}_{\mathfrak{r}_i}(\mathfrak{m}, \mathfrak{n})$, $\mathbf{Y}_{\mathfrak{s}}(\mathfrak{m}, \mathfrak{s})$.

Conversely, let us assume that $\mathcal{P}$ does not satisfy this condition. We can find $\mathfrak{m}, \mathfrak{n}$, where $\mathfrak{m} < \mathfrak{n}$ and $\nexists \mathfrak{r}(\mathfrak{m} < \mathfrak{r} < \mathfrak{n})$, such that $\mathcal{P}$ has the rules $P_{\mathfrak{m}}(\mathfrak{m}^*)$ and $\mathbf{Y}^{\mathfrak{n}}$. As an example, the reader can try to reduce a cut between $\mathbf{Y}^{\mathfrak{n}}$ and $P_{\mathfrak{n}}(\mathfrak{m}, \mathfrak{m})$; it is not possible. □

The following improves Corollary 4.4.26:

COROLLARY 5.1.5 (NUMBER OF MAXIMAL SUBSYSTEMS WITH CUT-ELIMINATION)
Let $\#\mathbb{X} = M$. There are $M$ maximal and sensible subsystems of $\mathfrak{B}_{\mathbb{X}}$, whose $\preceq$ relation is a linear order and that satisfy cut-elimination.

**Proof.** $M$ because $\mathfrak{q}$ of Proposition 5.1.4 can be chosen in just $M$ ways. □

## 5.2 About Confluence

Intuitively, a rewriting system is confluent whenever each computation leads to a unique result. This has been formally stated in Section 3.4.1, where we have recalled moreover that different kinds of confluence (called CR, WCR, UN) exist for ARS. In a first moment, when our aim was just to study polynomial soundness of our subsystems, confluence was not an issue. But, when we are interested also to programming inside a subsystem $\mathcal{P}$, confluence becomes a useful property: each program comes to a unique result. In this Section we will provide some condition that allows a subsystem to be confluent (Corollary 5.2.2); but, more importantly, we show that most of the maximal sensible subsystems are *not* CR (Lemma 5.2.4).

We rephrase the properties CR, WCR, UN in terms of proof nets. In that way, we will never need to refer to ARS.

- A subsystem $\mathcal{P}$ is WCR if, whenever $\Pi, \Sigma_1, \Sigma_2 \in \mathbf{PN}(\mathcal{P})$, $\Pi \to \Sigma_1$ and $\Pi \to \Sigma_2$, there exists $\Theta \in \mathbf{PN}(\mathcal{P})$ such that $\Sigma_1 \to^* \Theta$ and $\Sigma_2 \to^* \Theta$.
- A subsystem $\mathcal{P}$ is CR if, whenever $\Pi, \Sigma_1, \Sigma_2 \in \mathbf{PN}(\mathcal{P})$, $\Pi \to^* \Sigma_1$ and $\Pi \to^* \Sigma_2$, there exists $\Theta \in \mathbf{PN}(\mathcal{P})$ such that $\Sigma_1 \to^* \Theta$ and $\Sigma_2 \to^* \Theta$.
- Let $\equiv$ denotes the smallest equivalence relation containing $\to^*$. A subsystem $\mathcal{P}$ is UN if, whenever $\Pi, \Sigma \in \mathbf{PN}(\mathcal{P})$ are normal forms, and $\Pi \equiv \Sigma$, then $\Pi = \Sigma$.

Recall that WCR is equivalent to CR (cfr. Corollary 3.4.3), and they both imply UN.

The following is a first, naïve, result:

Lemma 5.2.1 (Confluent Subsystems)
Let $\mathcal{P} \subseteq \mathfrak{B}_{\mathbb{X}}$ sensible. $\mathcal{P}$ is CR iff it satisfies all the following conditions:

1. If $\mathcal{P}$ has rules $\mathbf{Y}_{\mathfrak{q}}(\mathfrak{m},\mathfrak{n})$ and $P_{\mathfrak{q}}(\mathfrak{r},\vec{\mathfrak{s}})$ and the cut of these two rules is reducible, then there is only one way to reduce it.

2. If $\mathcal{P}$ has rules $\mathbf{Y}_{\mathfrak{q}}(\mathfrak{m},\mathfrak{n})$, $P_{\mathfrak{q}}(\mathfrak{p},\vec{\mathfrak{r}})$, $P_{\mathfrak{p}}(\vec{\mathfrak{s}})$, and both cuts among the rules are reducible, then there exists a common normal form for $\mathbf{Y}_{\mathfrak{q}}(\mathfrak{m},\mathfrak{n}) \bowtie P_{\mathfrak{q}}(\mathfrak{p},\vec{\mathfrak{r}}) \bowtie P_{\mathfrak{p}}(\vec{\mathfrak{s}})$ in 1 step.

3. If $\mathcal{P}$ has rules $P_{\mathfrak{m}}(\mathfrak{n},\vec{\mathfrak{r}})$, $P_{\mathfrak{n}}(\mathfrak{p},\vec{\mathfrak{s}})$, $P_{\mathfrak{p}}(\vec{\mathfrak{q}})$, and both cuts among the rules are reducible, then there exists a common normal form for $P_{\mathfrak{m}}(\mathfrak{n},\vec{\mathfrak{r}}) \bowtie P_{\mathfrak{n}}(\mathfrak{p},\vec{\mathfrak{s}}) \bowtie P_{\mathfrak{p}}(\vec{\mathfrak{q}})$ in 1 step.

4. If $\mathcal{P}$ has rules $P_{\mathfrak{m}}(\mathfrak{n},\mathfrak{p},\vec{\mathfrak{r}})$, $P_{\mathfrak{n}}(\vec{\mathfrak{s}})$, $P_{\mathfrak{p}}(\vec{\mathfrak{q}})$, and both cuts among the rules are reducible, then there exists a common normal form for $P_{\mathfrak{n}}(\vec{\mathfrak{s}}) \bowtie P_{\mathfrak{m}}(\mathfrak{n},\mathfrak{p},\vec{\mathfrak{r}}) \bowtie P_{\mathfrak{p}}(\vec{\mathfrak{q}})$ in 1 step.

**Proof.** If $\mathcal{P}$ does not satisfy these conditions, than it is easy to build some non-confluent proof nets. For 3. and 4. use the two suggested proof nets: resp. $P_{\mathfrak{m}}(\mathfrak{n},\vec{\mathfrak{r}}) \bowtie P_{\mathfrak{n}}(\mathfrak{p},\vec{\mathfrak{s}}) \bowtie P_{\mathfrak{p}}(\vec{\mathfrak{q}})$ and $P_{\mathfrak{n}}(\vec{\mathfrak{s}}) \bowtie P_{\mathfrak{m}}(\mathfrak{n},\mathfrak{p},\vec{\mathfrak{r}}) \bowtie P_{\mathfrak{p}}(\vec{\mathfrak{q}})$. For 1. and 2. the proceeding is similar, but it is necessary at least a pair of *identities* and a *tensor right* in order to close the two dangling conclusions of the contraction. Sensibility gives the required node.

Conversely, let us assume $\mathcal{P}$ has the given properties. Thanks to Corollary 3.4.3, we just need to check the subsystem is WCR. Let us fix proof nets $\Sigma_1 \leftarrow \Pi \to \Sigma_2$ in $\mathbf{PN}(\mathfrak{B}_{\mathbb{X}})$, obtained eliminating two different cuts $c, c'$. If $c, c'$ do not affect each other, we just apply both them and we obtain a common normal form $\Theta$. Otherwise, the only problematic cases involve $[P/P]$ and $[\mathbf{Y}/P]$ *ns*; thanks to the 4 hypothesis, $\mathcal{P}$ has rules enough to obtain the desired common result $\Theta$. □

The same Lemma 5.2.1 can be rewritten in Corollary 5.2.2 in a more *computational* fashion. Such a new corollary is quite ugly, but we shall see that it implies the important Corollary 5.2.3.

Corollary 5.2.2 (Confluent Systems)
Let $\mathcal{P} \subseteq \mathfrak{B}_{\mathbb{X}}$ sensible. $\mathcal{P}$ is CR iff it satisfies all the following conditions:

1. If $\mathcal{P}$ has rules $\mathbf{Y}_{\mathfrak{q}}(\mathfrak{m},\mathfrak{n})$ and $P_{\mathfrak{q}}(\mathfrak{r},\vec{\mathfrak{s}})$ and the cut of these two rules is reducible, then there is only one way to reduce it.

2. If $\mathcal{P}$ has rules $\mathbf{Y}_{\mathfrak{q}}(\mathfrak{m},\mathfrak{n})$, $P_{\mathfrak{q}}(\mathfrak{p},\vec{\mathfrak{r}})$, $P_{\mathfrak{p}}(\vec{\mathfrak{s}})$, $P_{\mathfrak{m}}(a,\vec{u})$, $P_{\mathfrak{n}}(b,\vec{v})$, $\mathbf{Y}_{\mathfrak{p}}(a,b)$, $\mathbf{Y}_{\mathfrak{r}_\alpha}(u_\alpha,v_\alpha)$ for every $\alpha$ (where $\vec{u} = \langle u_0 \dots u_A \rangle$), and $P_{\mathfrak{q}}(\vec{\mathfrak{r}},\vec{\mathfrak{s}})$, then $\mathcal{P}$ has also, for some $\vec{w},\vec{z}$: $P_a(\vec{w})$, $P_b(\vec{z})$, $\mathbf{Y}_{\mathfrak{s}_\beta}(w_\beta,z_\beta)$ for every $\beta$ (where $\vec{\mathfrak{s}} = \langle \mathfrak{s}_0 \dots \mathfrak{s}_B \rangle$), and $P_{\mathfrak{m}}(\vec{w},\vec{u})$, $P_{\mathfrak{n}}(\vec{z},\vec{v})$.

3. If $\mathcal{P}$ has rules $P_{\mathfrak{m}}(\mathfrak{n},\vec{\mathfrak{r}})$, $P_{\mathfrak{n}}(\mathfrak{p},\vec{\mathfrak{s}})$, $P_{\mathfrak{p}}(\vec{\mathfrak{q}})$, $P_{\mathfrak{m}}(\mathfrak{p},\vec{\mathfrak{r}},\vec{\mathfrak{s}})$, $P_{\mathfrak{n}}(\vec{\mathfrak{q}},\vec{\mathfrak{s}})$, then it has also the rule $P_{\mathfrak{m}}(\vec{\mathfrak{r}},\vec{\mathfrak{s}},\vec{\mathfrak{q}})$.
4. If $\mathcal{P}$ has rules $P_{\mathfrak{m}}(\mathfrak{n},\mathfrak{p},\vec{\mathfrak{r}})$, $P_{\mathfrak{n}}(\vec{\mathfrak{s}})$, $P_{\mathfrak{p}}(\vec{\mathfrak{q}})$, $P_{\mathfrak{m}}(\mathfrak{p},\vec{\mathfrak{r}},\vec{\mathfrak{s}})$, $P_{\mathfrak{m}}(\mathfrak{n},\vec{\mathfrak{r}},\vec{\mathfrak{q}})$, then it has also the rule $P_{\mathfrak{m}}(\vec{\mathfrak{r}},\vec{\mathfrak{s}},\vec{\mathfrak{q}})$.

If $\mathcal{P}$ is saturated (and so in particular if it is maximal), then the two first conditions are enough.

**Proof.** The first part of the statement is a trivial consequence of Lemma 5.2.1. For the last statement, observe that the last two conditions are satisfied thanks to saturation. □

Corollary 5.2.3 (Confluence is Decidable)
Let $\mathcal{P}$ be a sensible subsystem, and let us assume that the promotion rules of $\mathcal{P}$ are described by means of regular expressions. Then, the problem "Is $\mathcal{P}$ CR?" is decidable.

However, in most of the cases the hypothesis of Corollary 5.2.2 cannot be satisfied, as the following, gloomy, Lemma shows:

Lemma 5.2.4 (Maximal Subsystems are not CR)
Let $\mathbb{X}$ have at least three elements, $\mathcal{P} \subseteq \mathfrak{B}_{\mathbb{X}}$ be sensible and maximal, whose $\preceq$ relation is a linear order. Then, $\mathcal{P}$ is not CR.

**Proof.** Let $\mathfrak{r},\mathfrak{q},\mathfrak{m} \in \mathbb{X}$ three different modalities such that $\mathfrak{r} \prec \mathfrak{q} \prec \mathfrak{m}$. We consider the following proof net of **PN**($\mathcal{P}$) (it exists by maximality):

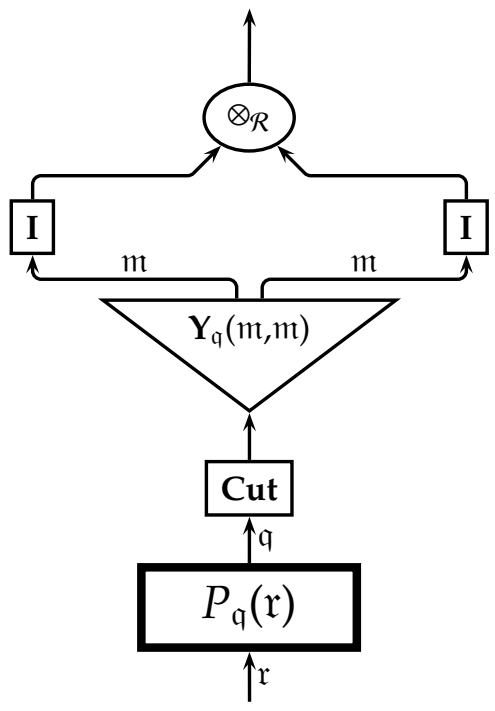

This cut is reducible in at least two different ways (again, they both exist by maximality):

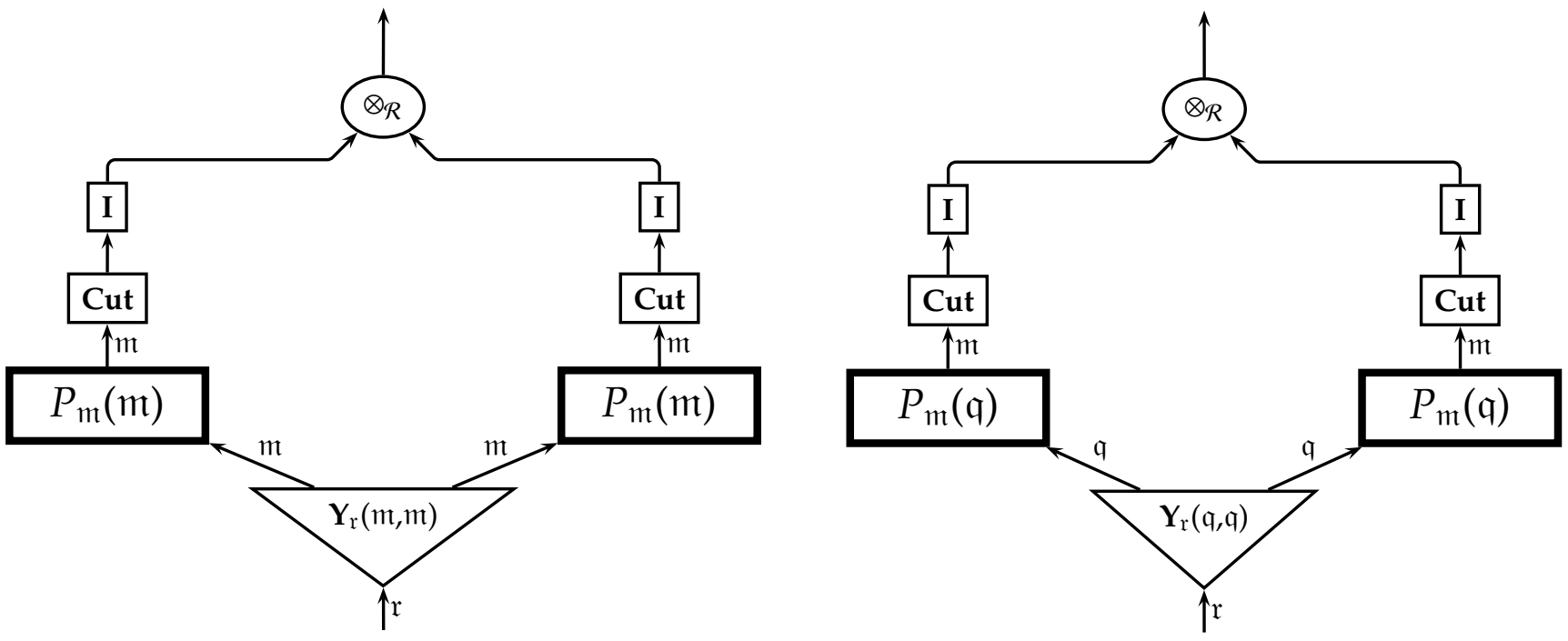

At this point, we can still eliminate the [**I**/_] cuts, and we reach two different normal forms. □

REMARK 5.2.5 In analogy with Proposition 4.4.4, we had erroneously conjectured that a subsystem $\mathcal{P}$ is CR iff its augmented $\mathcal{S}(\mathcal{P})$ is CR, too. This would simplify a lot the study of the confluence of the subsystems. However both the implications of the conjecture are false. We provide two counterexamples. (i) Letting $K$ be the linear kernel, $\mathcal{P} = K \cup \{P_a(b), P_b(c), P_c(d), P_a(c), P_b(d)\}$ is not Church-Rosser because of the lack of rule $P_a(d)$; but this rule is in $\mathcal{S}(\mathcal{P})$, so that this latter is CR. (ii) $\mathcal{P} = K \cup \{\mathbf{Y}_b(a,a), P_c(b), \mathbf{Y}_c(b,b)\}$ is Church-Rosser while $\mathcal{S}(\mathcal{P})$ is not, as it contains $\mathbf{Y}_c(b,b)$ so that every cut $[P_c(b)/\mathbf{Y}_b(a,a)]$ can reduce in two different ways.

The following Lemma describes a class of subsystems enjoying CR:

LEMMA 5.2.6 (SOME CR SUBSYSTEMS)
Let $\mathcal{P} \subseteq \mathfrak{B}_{\mathbb{X}}$ such that:

- $\mathcal{P}$ has only standard contractions, i.e. $\mathbf{Y}^{\mathfrak{m}} = \mathbf{Y}_{\mathfrak{m}}(\mathfrak{m}, \mathfrak{m})$;
- $\mathcal{P}$ enjoys the cut elimination property.

Then, $\mathcal{P}$ is CR.

**Proof.** We have to check the four conditions of Lemma 5.2.1. In this case, all the *ns* are deterministic: there is exactly one way to reduce a cut $[\mathbf{Y}^{\mathfrak{m}}/P_{\mathfrak{m}}(\vec{\mathfrak{r}})]$, i.e. it reduces to

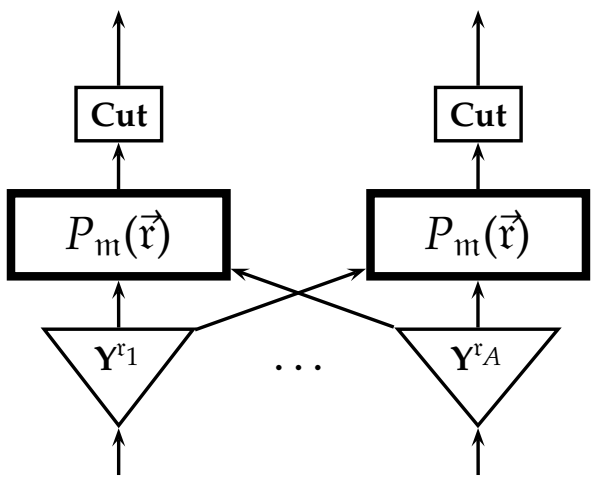

As a consequence, the first condition of Lemma 5.2.1 is satisfied. Let us turn to the cases described by the conditions 2-3-4. In all these cases, the cut elimination property assures that all the reductions can go on, in two different directions. Case by case, it is possible to see that both the reductions arrive to a common proof net. □

We thank the referee for the following conjecture. It seems to hold, but we have not completed the proof, yet.

CONJECTURE 5.2.7 (CUT-ELIMINATION IMPLIES A WEAK FORM OF CR) *Let $\mathcal{P} \subseteq \mathfrak{B}_{\mathbb{X}}$ be a sensible subsystem enjoying the cut elimination property. Let $\Pi \in$ **PN**$(\mathcal{P})$ reduce to two different normal forms $\Sigma_1, \Sigma_2$. Then, $\Sigma_1$ and $\Sigma_2$ have the same underlying graph, differing at most for the labels on their edges.*

# Chapter 6

# Interesting Subsystems of MS

The framework MS is useful only if we are able to identify some *interesting* subsystems. In this Chapter we will attain this target.

In Section 6.1 we will find a subsystem $\mathcal{P}^{M}_{\mathbb{LTS}}$ in which it is possible to represent only functions that execute in quadratic time and linear space (cfr. [RV08]).

In Section 6.2 we present a polytime subsystem soLAL, in which we are able to extend the set of SRN programs that we can encode in ILAL. soLAL is made up of many copies of ILAL, each one with its own modalities "$i!$" and "$i\S$". The natural number $i$ is the *sort* of the modality, and it tells how the modality can be used inside a proof net. We will define a fragment SRN$^{-}$ of functions, strictly grater than BC$^{-}$ but strictly smaller than SRN, that can be encoded in soLAL. We shall give two different encodings of SRN$^{-}$ into soLAL. We shall see that, even if they are very different, they lead to the same result. Indeed, whichever encoding we choose, every time we simulate the safe recursion in soLAL, we are forced to change the type of the functions involved; as a consequence, the obtained function cannot be iterated any more. This is why soLAL is not able to encode the whole SRN, but only a fragment SRN$^{-}$. However, this change of type in soLAL is not as "drastic" as in ILAL, as we shall see. We underline that just a few of the modalities of soLAL are really used in our encoding from SRN$^{-}$ into soLAL. So, probably we are not using all its potential expressive power.

**A soft introduction to the use of our Criteria.** The purpose of this paragraph is to show how to use the criteria proved in the previous chapters. In order to do that, we present some subsystems and we study their polynomiality. All these subsystems are *sensible*. Let $\mathbb{X} = \{!, \S\}$ and $K$ be the *linear kernel* defined on page 29: $K = \{\text{cut}, \mathbf{I}, \mathbf{W}, \mathbf{h}, \otimes_{\mathcal{R}}, \otimes_{\mathcal{L}}, \forall_{\mathcal{R}}, \forall_{\mathcal{L}}, \multimap_{\mathcal{R}}, \multimap_{\mathcal{L}}\}$.

- $K \cup \{\mathbf{Y}^{!}, P_{!}(!^{?}), P_{\S}(!^{*}, \S^{*})\}$ is ILAL, the Intuitionistic Light Affine Logic [AR02]. We can prove that it is polytime: indeed no chains of spindles at all can be created in ILAL, so that Lemma 4.3.31 can be applied. Notice however that some spindle can be built, e.g. $\mathbf{Y}^{!} \bowtie P_{\S}(!, !)$; but they are not dangerous.

- $K \cup \{\mathbf{Y}^{!}, \mathbf{Y}_{!}(!, \S), \mathbf{Y}_{!}(\S, \S), P_{!}(!^{?}), P_{\S}(!^{*}, \S^{*})\}$ is still polytime, again because no chains of spindles can be built. This is a proper extension of ILAL. Referring to Theorem 4.4.25, we see that this system is maximal with ordering $! \prec \S$ among modalities.

- $K \cup \{\mathbf{Y}^{!}, P_{!}(!^{?}, \S^{*}), P_{\S}(\S^{*})\}$ is polytime, again because no chains of spindles can be built. Or,

as an alternative proof, also Lemma 4.3.33 can be applied. This system is different from ILAL: the !-boxes can have some § premises, while the §-boxes cannot have ! premises.

Another way to show that this system is polytime is using Theorem 4.4.25. It says that this system is contained in a maximal system with linear order $\S \prec !$, which has also the rule $\mathbf{Y}_{\S}(!, !)$. And a system smaller than a polytime system is still polytime.

- $K \cup \{\mathbf{Y}^!, P_!(!^*)\}$ is IEAL, the Intuitionistic Elementary Affine Logic. And indeed it is not polytime. To prove this, Lemma 4.3.35 can be used taking the dangerous spindle $\Sigma = \mathbf{Y}^! \bowtie P_1(!, !)$ and the tautology $A = \gamma \multimap \gamma$.

Now we will show how to apply the results of Section 4.4.1.

- The subsystem $K \cup \{\mathbf{Y}^!, \mathbf{Y}_!(!, \S), \mathbf{Y}_!(\S, \S), P_!(!^?), P_{\S}(!^*, \S^*)\}$, already presented, is in fact the augmented subsystem of ILAL. So it is saturated. We already said that it is maximal, and indeed recall that every maximal subsystem is saturated.

- If $\mathcal{P} = K \cup \{\mathbf{Y}^!+, P_!(\S^?), P_{\S}(!^?, \S^?)\}$, then the admissible rules in $\mathcal{S}(\mathcal{P})$ are:

$$\{P_!(\S^?), P_!(!, \S^?), P_!(!, !, \S^?), \ldots, P_!(!^*, \S^?), P_{\S}(!^*, \S^?, \S^?)\}.$$

  The couple $\{\mathbf{Y}^!, P_!(!, !)\}$ tells us that both $\mathcal{P}$ and $\mathcal{S}(\mathcal{P})$ are not polytime, for Proposition 4.4.5.

- If $\mathcal{P} = K \cup \{\mathbf{Y}^!, P_!(!^?, \S^*), P_{\S}(\S^*)\}$, then $\mathcal{P}$ is saturated. It does not contain any spindle (for Proposition 4.4.5 it is enough to check the spindles made up of exactly two rules), so it is polytime.

## 6.1 Quasi-linear Space

In this Section we define a family of subsystems $\left\{\mathcal{P}^M_{\mathbb{LTS}} \mid M \in \mathbb{N}\right\}$ of MS, with $\mathbb{LTS}$ a suitable partial order to be defined. For every fixed $M$, we shall see that $\mathcal{P}^M_{\mathbb{LTS}}$ enjoys the following properties: (i) every of its proof nets normalizes in linear time and space, and (ii) the first $M$ Church numerals together with a successor, a sum and a predecessor exist as proof nets of $\mathcal{P}^M_{\mathbb{LTS}}$. As far as we know, this is the first Light System as asymptotically complex as MLL, that simultaneously shows both (i) and (ii).

As a consequence, we shall prove that $\mathcal{P}^M_{\mathbb{LTS}}$ encodes only functions which execute in quasi-linear space and time $O(n^4)$ (Corollary 6.1.2).

We underline that a partial version of this system has already been presented in [RV08]; however, in that paper we used only usual *constrained* weakening, so that we were not able to encode a "real" predecessor, but only a "moral" predecessor. This latter proof net calculates the result, but it is not able erasing all the *garbage* produced during the computation. The free weakening is needed for that.

**Ramified partial orders.** A partial order $(\mathbb{LTS}, \preceq_{\mathbb{LTS}})$ has $\mathbb{LTS}$ as carrier and $\preceq_{\mathbb{LTS}}$ as order relation. We write $\mathfrak{m} \not\gtrless \mathfrak{n}$ whenever $\mathfrak{m}, \mathfrak{n}$ cannot be compared under $\preceq_{\mathbb{LTS}}$. The anti-reflexive restriction of $\preceq_{\mathbb{LTS}}$ is $\prec_{\mathbb{LTS}}$. Let $\mathfrak{m} \sqcup \mathfrak{n}$ denote the *least upper bound* (lub) and $\mathfrak{m} \sqcap \mathfrak{n}$ the *greatest lower bound* (glb) of $\mathfrak{m}, \mathfrak{n}$ under $\preceq_{\mathbb{LTS}}$. $\preceq_{\mathbb{LTS}}$ is a **ramified partial order** if it satisfies the following conditions: (i) There is $\S \in \mathbb{LTS}$ such that $\S \preceq_{\mathbb{LTS}} \mathfrak{m}$ for every $\mathfrak{m} \in \mathbb{LTS}$, and (ii) For every $\mathfrak{m}, \mathfrak{n} \in \mathbb{LTS}$, if both $\mathfrak{m} \not\gtreqless \mathfrak{n}$ and exists $\mathfrak{r}$ such $\mathfrak{m}, \mathfrak{n} \leq \mathfrak{r}$, then $\mathfrak{m} \sqcap \mathfrak{n} = \S$.

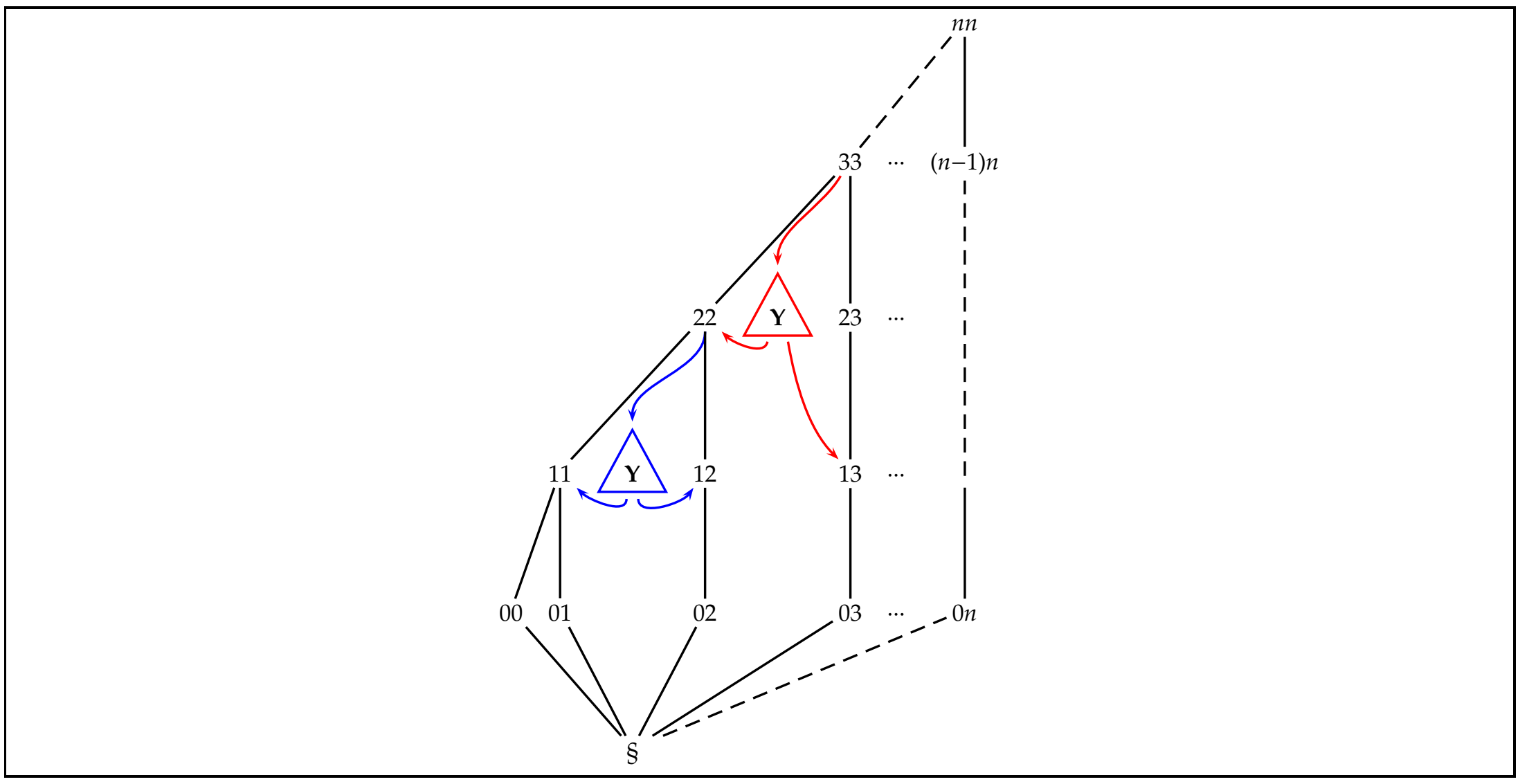

Figure 6.1: An example of ramified partial order. The two contraction nodes show how the modalities influence the construction of the trees of contractions in $\mathcal{P}^{M}_{\mathbb{LTS}}$.

| | |
|---|---|
| $P_{\S}(\S^{*})$ | $P_{\mathfrak{q}}(\mathfrak{m}_1, \ldots, \mathfrak{m}_k)$ **iff** $\S \neq \mathfrak{q} \preceq_{\mathbb{LTS}} \mathfrak{m}_1, \ldots, \mathfrak{m}_k \in \mathbb{LTS}$ |
| $\mathbf{Y}_{\mathfrak{m}\sqcup\mathfrak{n}}(\mathfrak{m}, \mathfrak{n})$ **iff** $\mathfrak{m} \sqcup \mathfrak{n} \in \mathbb{LTS}$ **and** $\mathfrak{m} \not\gtreqless \mathfrak{n}$ | $\mathbf{Y}_{\mathfrak{m}}(\S, \mathfrak{n})$ **iff** $\mathfrak{n} \prec_{\mathbb{LTS}} \mathfrak{m}$ |

Figure 6.2: The modal rules of $\mathcal{P}^{M}_{\mathbb{LTS}}$.

An example of (degenerate) ramified partial order is $\mathbb{N}$. A more interesting example is in Figure 6.1.

**The family $\{\mathcal{P}^{M}_{\mathbb{LTS}} \mid M \in \mathbb{N}\}$.** Let $\mathbb{LTS}$ be a fixed ramified partial order, and $\mathbb{X}_M \subseteq \mathbb{LTS}$ be the set of elements $x$ such that *there are at most $M$ elements smaller than $x$ according to* $\preceq_{\mathbb{LTS}}$. Every $\mathbb{X}_M$ contains §. Every $\mathcal{P}^{M}_{\mathbb{LTS}}$ is a subsystem of $\mathsf{MS}_{\mathbb{X}_M}$ whose modal rules are in Figure 6.2.

The structure of the partial order $(\mathbb{LTS}, \preceq_{\mathbb{LTS}})$ forces the structure of the trees of contractions and boxes to rule out spindles longer that one ring. This is the essential reason why the polytime soundness of $\mathcal{P}^{M}_{\mathbb{LTS}}$ holds. Formally:

PROPOSITION 6.1.1 (POLYTIME SOUNDNESS OF $\mathcal{P}^{M}_{\mathbb{LTS}}$.)
Let $\mathcal{P}^{M}_{\mathbb{LTS}}$ be given. There exist $k_1, k_2$ such that, for every $\Pi \in \mathcal{P}^{M}_{\mathbb{LTS}}$, $\|\Pi\| \leq k_1 \cdot |\Pi|$ and $[\Pi] \leq k_2 \cdot |\Pi|$.

**Proof.** We adapt the proof of Lemma 4.3.31 at page 61.

Let us fix a box $b$ at depth $d$ in $\Pi$. Let us divide the nodes of $\Pi$ in *class-0* and *class-1* as in the previous proof, according to how many spindles there are below them. So, we know that $R_\Pi(b)$ is bounded by the number $m_0$ of the contractions and $\otimes_\mathcal{R}$ nodes of class 0, multiplied by the number $m_1 + 1$ of the same nodes of class 1. However, this bound is still large. Firstly, we notice that, in fact, only the contractions are really needed: if a path passes trough a $\otimes_\mathcal{R}$ node $v$, then it necessarily arrives from a $\otimes_\mathcal{L}$ node $u$; and so *only one* of the two paths outgoing $v$ are effectively used, depending on the path entering $u$. Secondly, $m_1 = 0$, because the modality § can never be the premise of a contraction. Third, at each contraction the number of the modalities labelling the edges decrease, according to $\preceq_{\mathbb{LTS}}$. So, if $M$ is the number of modalities, the tree of contractions has depth at most $M$. And so it has at most $2^M$ leaves. $\forall b\ R_\Pi(b) \leq 2^M$.

This is not enough yet. If we apply directly the previous lemmas we get a quadratic bound, that we do not want. First key observation. The modified weight is defined by

$$\mathrm{T}_d(\Pi, u) \stackrel{\text{def}}{=} \begin{cases} 1 & \alpha(u) \in \{\mathbf{I}, \mathrm{cut}, \mathbf{W}, \mathbf{h}, Pi, \mathbf{i}, \mathbf{o}\} \\ 3 & \alpha(u) \in \{\otimes, \forall, \multimap, \mathbf{Y}\} \\ 2 \cdot (P_\Pi(b) + 1) \cdot R_\Pi(b)^2 & \alpha(u) = Po \end{cases}$$

$$\mathrm{T}_d(\Pi) = \sum_{u \in V_\Pi^d} \mathrm{T}_d(\Pi, u) \leq 3|\Pi|_d + 2 \sum_{b \in B_\Pi^d} (P_\Pi(b) + 1) \cdot R_\Pi(b)^2$$
$$\leq 3|\Pi|_d + 2 \cdot 2^{2M} \sum_{b \in B_\Pi^d} (P_\Pi(b) + 1) \leq |\Pi|_d \left(3 + 2^{2M+1}\right).$$

So the nets reduce in linear time, because $[\Pi]_d \leq \mathrm{T}_d(\Pi)$.

However, for $i > d$, the bound $\|\Pi\|_i^d \leq |\Pi|_i \cdot T_d(\Pi)$ implied by Lemma 4.3.24 is too high. To lower it, recall that if $u$ is a link at level $i > d$, inside a box $b$ at level $d$, $u$ can be copied at most $R_\Pi(b) \leq 2^M$ times: $\|\Pi\|_i^d \leq 2^M \cdot |\Pi|_i$ is linear, as well as $\|\Pi\|^d$. The final statement follows from composing, level by level, the obtained bounds. We obtain the constants $k_1$, $k_2$ that depend on $M$ and $\partial(\Pi)$. □

**Normalization.** We want understand which cuts can be eliminated in $\mathcal{P}^M_{\mathbb{LTS}}$, and which ones cannot. First of all, let $c$ be a cut $[P_\mathfrak{m}(\vec{\mathfrak{r}}, \mathfrak{n})/P_\mathfrak{n}(\vec{\mathfrak{s}})]$. Here, $\mathfrak{m} \preceq_{\mathbb{LTS}} \mathfrak{n} \preceq_{\mathbb{LTS}} \vec{\mathfrak{s}}$. This reduction can always be performed, and the reduct is $P_\mathfrak{m}(\vec{\mathfrak{r}}, \vec{\mathfrak{s}})$. Similarly if the cut involves two §-boxes.

Now, let $c$ be a cut $[\mathbf{Y}_{\mathfrak{m}\sqcup\mathfrak{n}}(\mathfrak{m}, \mathfrak{n})/P_{\mathfrak{m}\sqcup\mathfrak{n}}(\vec{\mathfrak{r}})]$. Of course, if $\vec{\mathfrak{r}}$ is empty, $c$ can be reduced. Otherwise, we assume for simplicity that $\vec{\mathfrak{r}} = \mathfrak{r}$ is a single premise. Here, $\mathfrak{m}, \mathfrak{n} \prec_{\mathbb{LTS}} \mathfrak{m} \sqcup \mathfrak{n} \preceq_{\mathbb{LTS}} \mathfrak{r}$. If $\mathfrak{r} = \mathfrak{m} \sqcup \mathfrak{n}$, $c$ reduces in a unique way to $P_\mathfrak{m}(\mathfrak{m}) \bowtie \mathbf{Y}_{\mathfrak{m}\sqcup\mathfrak{n}}(\mathfrak{m}, \mathfrak{n}) \bowtie P_\mathfrak{n}(\mathfrak{n})$. For every other $\mathfrak{r}$, $c$ does not reduce: indeed, the proof net should reduce to $P_\mathfrak{m}(a) \bowtie \mathbf{Y}_\mathfrak{r}(a, b) \bowtie P_\mathfrak{n}(b)$ for some $a, b$ such that $\mathfrak{r} = a \sqcup b$, but these $a, b$, do not exist thanks to the definition of ramified p.o..

At last, let $c$ be a cut $[\mathbf{Y}_\mathfrak{n}(\S, \mathfrak{m})/P_\mathfrak{n}(\mathfrak{r})]$, where $\mathfrak{m} \prec_{\mathbb{LTS}} \mathfrak{n} \preceq_{\mathbb{LTS}} \mathfrak{r}$. In this case, the cut can always be reduced, possibly in many ways, to $P_\S(\S) \bowtie \mathbf{Y}_\mathfrak{r}(\S, b) \bowtie P_\mathfrak{m}(b)$ for every $\mathfrak{m} \preceq_{\mathbb{LTS}} b \prec_{\mathbb{LTS}} \mathfrak{r}$.

**Encoding functions in $\mathcal{P}^M_{\mathbb{LTS}}$.** Now, we show that every $\mathcal{P}^M_{\mathbb{LTS}}$ is strictly extends MLL, even though they belong to the same time and space complexity class. A type for Church numerals is $\mathfrak{m}\mathbb{CN} = \forall\alpha.\mathfrak{m}(\alpha \multimap \alpha) \multimap \S(\alpha \multimap \alpha)$, for every $\mathfrak{m} \in \mathbb{LTS}$. Figure 6.3a shows the net that corresponds to the Church numeral $\overline{n}$ for some $n \leq M$. Namely, once fixed $M$, we can expect to represent at most $M$ different Church numerals.

Among the functions that we are able to represent there are: (1) a successor proof net with type $\mathfrak{n}\mathbb{CN} \multimap \mathfrak{m}\mathbb{CN}$ for every $\mathfrak{n} <_{\mathbb{LTS}} \mathfrak{m}$ (Figure 6.3b); (2) a sum proof net with type $\mathfrak{n}\mathbb{CN} \multimap \mathfrak{m}\mathbb{CN} \multimap (\mathfrak{n} \sqcup \mathfrak{m})\mathbb{CN}$, whenever $\mathfrak{n} \ngtrless \mathfrak{m}$ and $\mathfrak{n} \sqcup \mathfrak{m} \in \mathbb{LTS}$ (Figure 6.4); (3) a predecessor proof net with type $\mathfrak{n}\mathbb{CN} \multimap \mathfrak{m}\mathbb{CN}$ for every $\mathfrak{n} \leq_{\mathbb{LTS}} \mathfrak{m}$ (Figure 6.5).

It is easy to show that all these proof nets do reduce: for example, they reduce using the **closed normalization strategy** defined on page 35. The result is not necessaily unique, due to the non-deterministic behaviour of the cuts $[\mathbf{Y}_{\mathfrak{n}}(\S, \mathfrak{m})/P_{\mathfrak{n}}(\mathfrak{r})]$; anyway, the possibly different results may differ only for the labels on their edges, so they all represent the same Church numeral.

If the reader has on hand a *coloured* copy of this thesis, he will see that we have distinguished two kinds of boxes: in blue, the §-boxes, that are not duplicable; in red, all the other ones, that can be duplicated.

Corollary 6.1.2 (Space and Time)
Let $W_1, W_2$ be types that represent the binary words in $\mathcal{P}^{M}_{\mathbb{LTS}}$. For each $w \in \mathbb{W}$, we call $\overline{w} \triangleright \vdash W_i$, $i \in \{0, 1\}$, its representative proof net. Suppose that the size of $\overline{w}$, as well as the number of edges in $\overline{w}$, is $O(|w|)$. At last, let $\Pi \triangleright W_1 \vdash W_2$ be a proof net of $\mathcal{P}^{M}_{\mathbb{LTS}}$ representing a function $f : \mathbb{W} \to \mathbb{W}$. Then, $f$ is computable in quasi-linear space and time $O(n^4)$.

The hypothesis of the previous Corollary is usually satisfied for Church binary words.
**Proof.** Let us call $M$ the Turing machine that
1. takes an input $w \in \mathbb{W}$,
2. writes on his inner tape the binary representation of the proof net $\Pi \bowtie \overline{w}$,
3. reduces step by step the proof net on his inner tape,
4. translates the result $\overline{r}$ into the word $r \in \mathbb{W}$.

We shall call $n = |w|$. We want to show that $M$ executes in quadratic time and space, i.e. $O(n^2)$. The reader can refer to Section 3.4.2 for the representation of a proof net on the tape of a Turing machine. We decide to represent the graph as a list of edges, and each edge is couple of nodes, because (a) this representation is the most compact, but also (b) duplications are easier. Recall that the boxes are kept separately on the tape. Moreover, we assume that erasures can be performed without phisically moving anything, just marking an edge or a box as "erased".

Step 2 takes time and space $O(n \log n)$, as explained in Section 3.4.2. Notice that the size of $\Pi$ does not influence the execution speed nor the used space, as it does not depend on $n$.

Considering step 3, the inner tape of $M$ contains at each time a reduct of $\Pi \bowtie \overline{w}$; the size of of $\Pi \bowtie \overline{w}$ is always $O(n)$ because of Proposition 6.1.1, so that the space used on the inner tape is always $O(n \log n)$.

The translation of Step 4 do not require any additional space, and is performed just looking at the whole tape, in time $O(n \log n)$.

So, $M$ executes using space $O(n \log n)$, i.e. quasi-linear. It remains to understand how much time it is needed by step 3. We shall give a very *rough* extimation. A single *ns* may annihilate some edges or boxes. In this case, the *ns* is simulated by $M$ just moving along the tape, in time $O(n \log n)$. Or, a single *ns* may duplicate a box. The box has size $O(n \log n)$, so the duplication requires going forth and back for a total of $O(n^2 \log^2 n)$ steps. Multiplying by the number of *ns*, $M$ executes in time $O(n^3 \log^3 n)$, which is certainly smaller than $O(n^4)$.[1] □

[1] To be precise, the function $n^3 \log^3 n$ is asyntotically greater than $n^3$, but smaller than any $n^{3+\epsilon}$, for any $\epsilon > 0$.

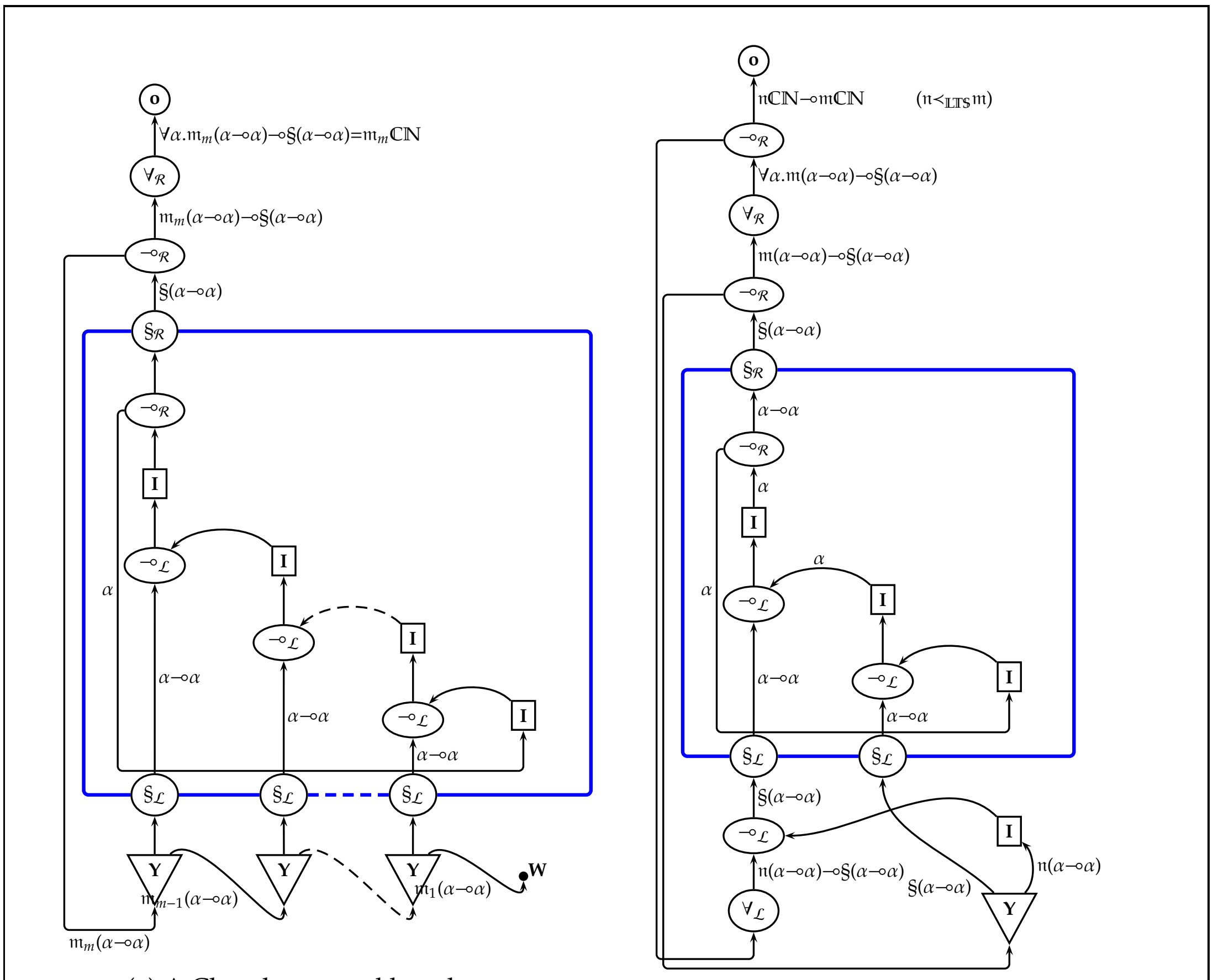

(a) A Church numeral less than $m$.

(b) Successor on Church numerals

Figure 6.3: $\mathcal{P}^{M}_{\mathbb{LTS}}$: encoding Church numerals and successor.

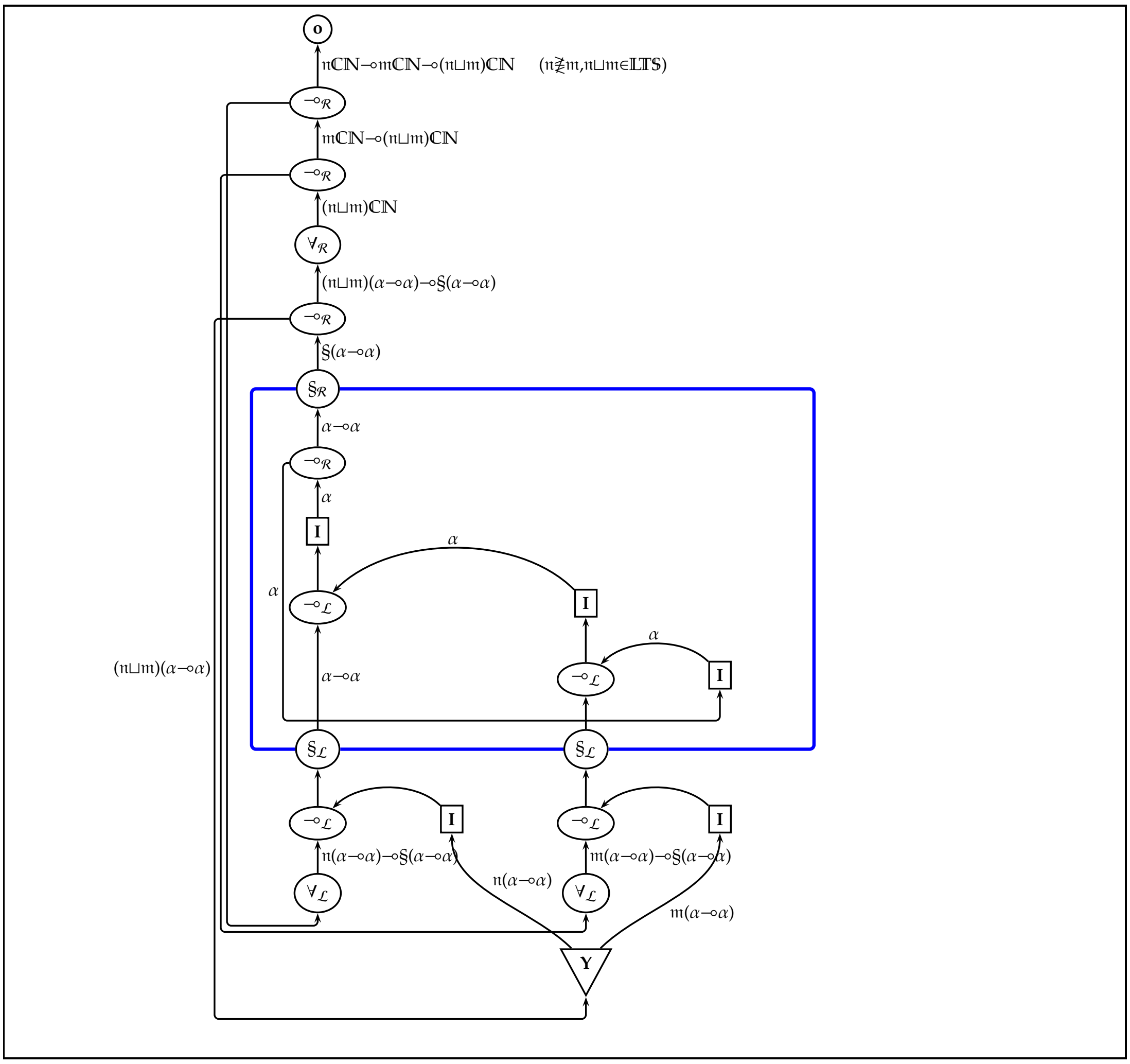

Figure 6.4: $\mathcal{P}^{M}_{\mathbb{LTS}}$: the sum of two Church numerals of type $\mathfrak{n}\mathbb{CN}, \mathfrak{m}\mathbb{CN}$ with $\mathfrak{n} \ngtrsim \mathfrak{m}$ and $\mathfrak{n} \sqcup \mathfrak{m} \in \mathbb{LTS}$.

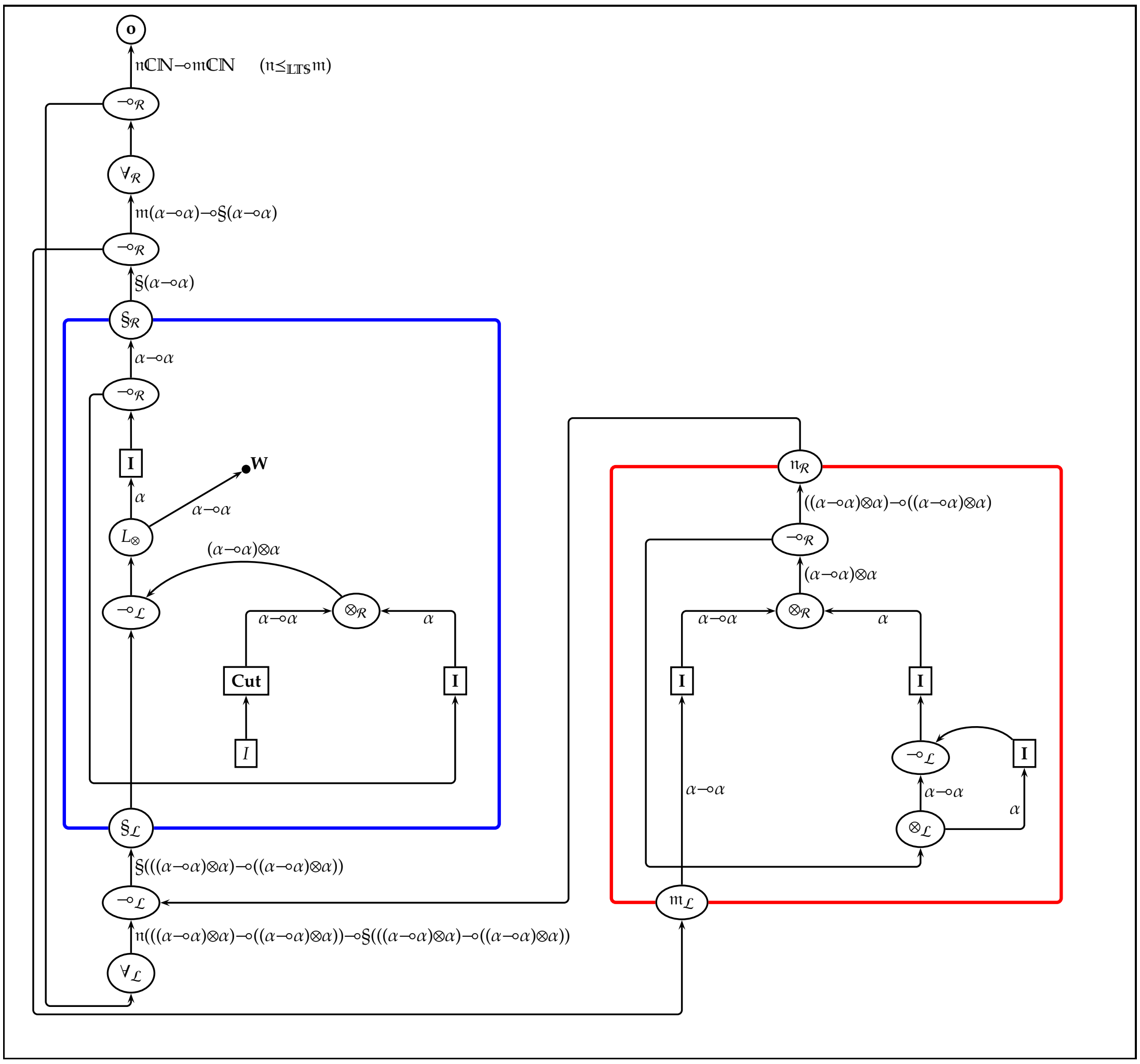

Figure 6.5: $\mathcal{P}^{M}_{\mathbb{LTS}}$: the Predecessor on Church numerals, with type $\mathfrak{n}\mathbb{CN} \multimap \mathfrak{m}\mathbb{CN}$ for every $\mathfrak{n} \preceq_{\mathbb{LTS}} \mathfrak{m}$.

$$P_{i!}(i!^?, j_1\S^*, \ldots, j_m\S^*) \text{ for every } j_1, \ldots, j_m < i \leq M$$
$$P_{i\S}(i!^*, i\S^*, j_1\S^*, \ldots, j_m\S^*) \text{ for every } j_1, \ldots, j_m < i \leq M$$
$$\mathbf{Y}_{i!}(i!, i!) \text{ for every } i \leq M$$

Figure 6.6: The modal rules of $\mathsf{soLAL}_M$.

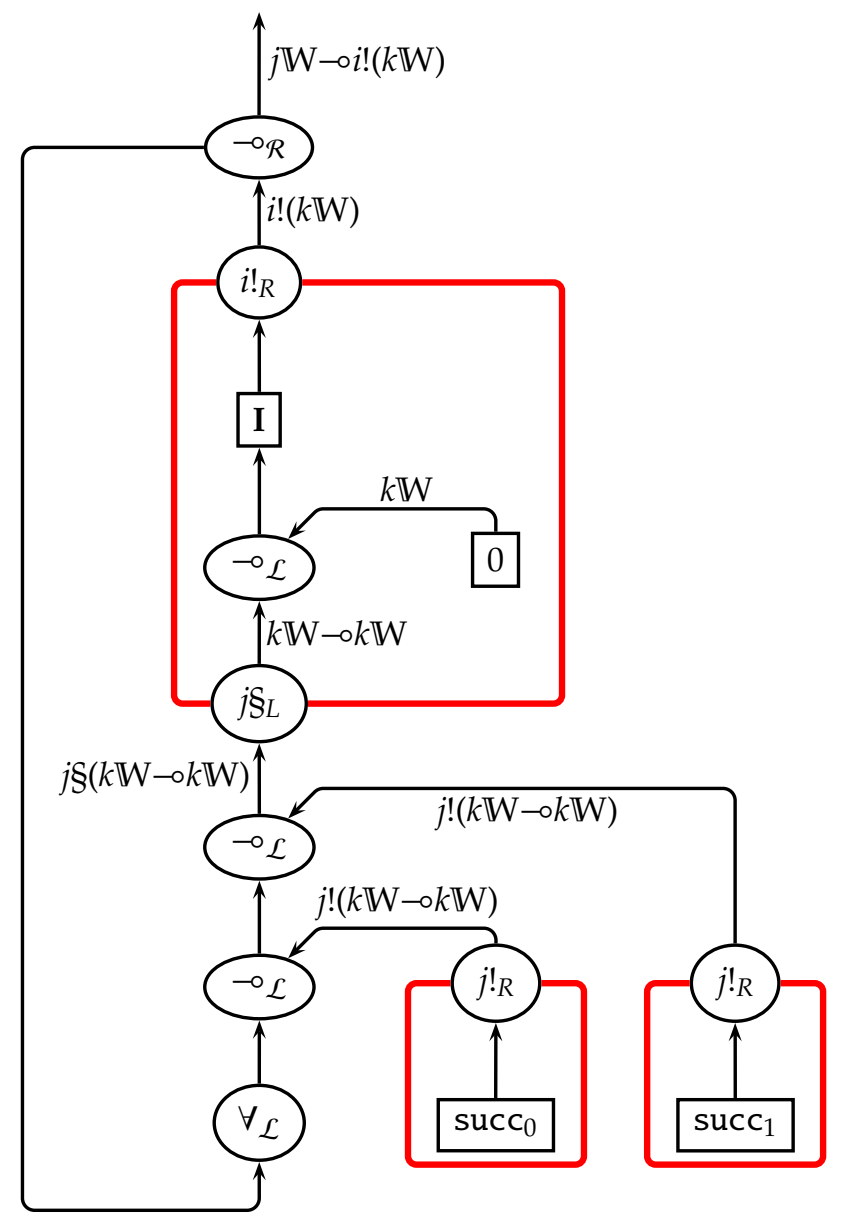

Figure 6.7: The proof nets $\mathtt{Crc}_{j<i,k}$ in soLAL, with $j < i$, for any $k$.

## 6.2 soLAL

We have pointed out, many times, that we are interested in finding some subsystem $\mathcal{P} \subseteq \mathsf{MS}$ able to represent as many SRN programs as possible. soLAL is designed precisely with this target in mind. See Section 2.7 for a definition of SRN. Firstly, we will define a family of subsystems $\mathsf{soLAL}_M \subseteq \mathfrak{B}_\Upsilon$, with $\Upsilon = \{i! \mid i \leq M\} \cup \{i\S \mid i \leq M\}$ for some fixed $M \geq 2$. Every one of such subsystems is a proper extension of ILAL. Later we encode a fragment $\mathsf{SRN}^-$ of SRN into $\mathsf{soLAL}_M$ (Proposition 6.2.5). The more, we will come to the same result through two different encodings (Lemma 6.2.7 and 6.2.8); this, together with some concluding observations in Section 6.3, suggests that the complexity of soLAL is *really* the one of $\mathsf{SRN}^-$, whatever encoding we choose. The details of the encoding will be postponed to Appendix A, for the *very* interested reader only, because of its length.

**Definitions.** $\mathsf{soLAL}_M$ (soLAL for short) contains the linear kernel of MS, plus the modal rules in Figure 6.6. Figure 6.7 shows an example of proof net we can build in soLAL but not in ILAL. The meaning of such a proof net will be explained later on. Exactly as in the previous Section 6.1, if the reader has on hand a *coloured* copy of this thesis, he will see that we have distinguished two kinds of boxes: in blue, the §-boxes, that are not (directly) duplicable; in red, the !-boxes, that can be duplicated. The name soLAL stands for *sorted* ILAL, because soLAL extends ILAL with the introduction of *sorted modalities*. For example, every $i!(\alpha \multimap \alpha)$ is an *of-course* (modal) formula of sort $i$, and $i\S(\alpha \multimap \alpha)$ is a *paragraph* one of sort $i$. $i\S^n A$ abbreviates $i\S \cdots i\S A$, with $n \geq 0$ occurrences of $i\S$. An analogous meaning holds for $i!^n A$. We observe that restricting $P_{i!}(i!^?, j_1\S^*, \ldots, j_m\S^*)$ to $P_{i!}(i!^?)$, and $P_{i\S}(i!^*, i\S^*, j_1\S^*, \ldots, j_m\S^*)$ to $P_{i\S}(i!^*, i\S^*)$ we get a subsystem $i$- soLAL of soLAL. Every $i$-soLAL is ILAL identifying $!A \equiv i!A, i\S B \equiv \S B$, for every $A, B$.

**Basic properties.** Let $\Re_\uparrow$ the preordering among the modalities of $\text{soLAL}_M$ defined in Section 4.4.2. Observing that the preordering in this case is in fact a linear ordering, we understand that in $\text{soLAL}_M$ it is not possible to build dangerous spindles, so:

Fact 6.2.1
For every fixed $M$, $\text{soLAL}_M$ is polytime.

Remark 6.2.2 soLAL does not enjoy cut-elimination, because e.g. a cut $P_{i!}(j\S) \bowtie \mathbf{Y}^{i!}$ cannot reduce.

soLAL is not confluent as well, because the two cuts $P_{i!}(j\S) \bowtie P_{i!}(i!) \bowtie \mathbf{Y}^{i!}$ can both reduce to different normal forms. (By the way, we have used Lemma 5.2.1 to find out such an example). Here we want to remark why *the above two negative aspects, which relate to* soLAL *taken in its generality, are harmless in the case we consider.* Indeed, we can certainly adopt the **closed normalization strategy**, already defined on page 35: reduce one instance of the cuts $[P/P]$, $[P/\mathbf{Y}]$ only if the lowermost box they involve is closed. All the reductions that we need may be performed using such a strategy. Moreover, the normal form we end up with is unique.

Anyway, we *conjecture* that such a normalization strategy can in fact be avoided, for we focus on a very a specific set of proof nets that soLAL can contain.

soLAL can prove $j\S A \multimap i!A$, for $j < i$ and for every $A$. This is its most important feature. This is the reason why soLAL generalizes ILAL: ILAL can prove $!A \multimap \S A$, but not $\S A \multimap !A$.

In soLAL we can define $i\mathbb{W} = \forall\alpha.i!(\alpha \multimap \alpha) \multimap i!(\alpha \multimap \alpha) \multimap i\S(\alpha \multimap \alpha)$ as the type of *(Church) (binary) words of sort* $i$, and $i\mathbb{S} = \forall\alpha.i!(\alpha \multimap \alpha) \multimap i\S(\alpha \multimap \alpha)$ as the $i$-sort type of *(unary) strings*. Also, from ILAL we recall that a *word-coercion* proof net, of type $\mathbb{W} \multimap \S \dagger \mathbb{W}$ for any sequence of modalities $\dagger$, takes a word as argument to rebuild it inside some boxes, the outermost being a §-box. In ILAL, the outermost box can only be §. Instead, the type of the word-coercion proof nets of soLAL can be also $j\mathbb{W} \multimap i!(\dagger k\mathbb{W})$, for every $j < i, k \in \mathbb{N}$, as in Figure 6.7. Please, notice $i!$ in front of $k\mathbb{W}$. soLAL is a kind of hierarchy of systems $i$-soLAL that can "communicate" at least by word-coercion proof nets that map words from one level to the other.

**The restrictions $\text{SRN}^=$ and $\text{SRN}^-$ of SRN.** The reader can find the definitions of recursion, linear recursion and linear composition in Figure 2.5.

Definition 6.2.3 $\text{SRN}^=$ *is the set containing* $\text{BC}^-$*, plus all the functions that can be obtained from* $\text{BC}^-$ *with 1 application of* (not necessarily linear) *Safe Recursion.*

Namely, like in $\mathsf{BC}^-$, we allow in $\mathsf{SRN}^=$ arbitrarily nested linear recursions; moreover, we allow the outermost recursion to be non-linear. Moreover, the functions obtained using an outermost recursion can be further linearly composed. As an example, the following program $g$ is in $\mathsf{SRN}^-$ but not in $\mathsf{BC}^-$. We omit the projections, to lighten the presentation:

$$h(w; y, z, t) = \mathtt{cond}(\,; w, y, t)$$

$$\begin{cases} g(0; y, z) & = z \\ g(s_i w; y, z) & = h(w; y, z, g(w; y, z)) \end{cases} \qquad (*)$$

The program $g(w; y, z)$ returns $y$ if $w \neq 0$ contains a digit '0' that is not the lowermost digit, and $z$ otherwise.

Definition 6.2.4 $\mathsf{SRN}^-$ *is the least set containing* $\mathsf{BC}^-$*, closed under Safe Linear Composition, and closed under Safe Recursion limited to any* $g, h_0, h_1 \in \mathsf{BC}^-$.

For example, if $g(w; y, z)$ is as in (*), then $h(w; y, z) = \mathsf{s}_i(\,; g(w; y, z))$ is in $\mathsf{SRN}^-$.

It holds $\mathsf{BC}^- \subsetneq \mathsf{SRN}^= \subsetneq \mathsf{SRN}^- \subsetneq \mathsf{SRN}$. Our goal is to focus on $\mathsf{SRN}^-$; $\mathsf{SRN}^=$ is only useful to identify some functions of $\mathsf{SRN}^-$ crucial for our proofs. The theorem we want to prove is the following:

Proposition 6.2.5 (From $\mathsf{SRN}^-$ to soLAL.)
There is a compositional map $\ulcorner \cdot \urcorner$ from $\mathsf{SRN}^-$ to soLAL, such that for every $f \in \mathsf{SRN}$ the proof net $\ulcorner f \urcorner$ represents $f$.

In the next paragraphs we provide two different proofs for it.

**Following Murawski-Ong's encoding.** We will prove Proposition 6.2.5 using an encoding perfectly analogous to the one of [MO04]. We shall see that it is possible, in that way, to encode some non-linear recursion; however their type is not further iterable, so that we are not able representing the whole SRN but only its fragment $\mathsf{SRN}^-$.

Lemma 6.2.6 ($\mathsf{BC}^-$ in soLAL)
Every function $f \in (\mathsf{BC}^-)^{\mathsf{n;s}}$ can be represented in soLAL with type:

$$\ulcorner f \urcorner \triangleright j_1\mathbb{W}, \ldots, j_n\mathbb{W}, \underbrace{j\S^m(k\mathbb{W}), \ldots, j\S^m(k\mathbb{W})}_{s} \vdash j\S^m(k\mathbb{W})$$

for every $j, k$, every $j_1, \ldots, j_n, j \leq j$ and every $m$ big enough.

**Proof.** If we follow exactly [MO04], we get an encoding in which all the sorts are equal: $j_1 = \ldots = j_n = j = k$. Here we can be more flexible on the sorts, because soLAL has, in particular, the promotion rule $j\S^m(k\mathbb{W}) \multimap k\S^m(k\mathbb{W})$ and the coercion $j_1\mathbb{W} \multimap k\S(k\mathbb{W})$. □

Lemma 6.2.7 (Non-linear Recursion in soLAL, #1)
Let us assume that $g \in \mathsf{SRN}^{\mathsf{n;s}}$ and $h_0, h_1 \in \mathsf{SRN}^{1+\mathsf{n;s}+1}$ can all be represented in soLAL with proof nets:

$$\ulcorner h_0 \urcorner, \ulcorner h_1 \urcorner \triangleright (\multimap_{r=0}^{\mathsf{n}} j_r\mathbb{W}) \multimap (\multimap_{r=1}^{\mathsf{s}} j\S^m(k\mathbb{W})) \multimap j\S^m(k\mathbb{W}) \multimap j\S^m(k\mathbb{W})$$
$$\ulcorner g \urcorner \triangleright (\multimap_{r=1}^{\mathsf{n}} j_r\mathbb{W}) \multimap (\multimap_{r=1}^{\mathsf{s}} j\S^m(k\mathbb{W})) \multimap j\S^m(k\mathbb{W}) \ .$$

for some $j_0, \ldots, j_n, j, k, m$. Then, it is also possible to encode the function $f \in \mathsf{SRN}^{1+n;s}$ obtained by Recursion of $g, h_0, h_1$ as a proof net $\ulcorner f \urcorner$ of soLAL, whose type is:

$$\ulcorner f \urcorner \triangleright (\multimap^{n}_{r=0} j_r\mathbb{W}) \multimap (\multimap^{s}_{r=1} i\S^{m+2}(k\mathbb{W})) \multimap j\S^{m+2}(k\mathbb{W}) \ ,$$

for every $i < j$.

**Proof.** The construction that proves this Lemma is in Figure 6.8. Please notice that such a proof net contains some irreducible cuts; anyway, the proof net obtained cutting the *non-linear recursion* with some numerals is closed, and thus it can be reduced starting from the closed boxes. □

**Proof of Proposition 6.2.5.** Lemma 6.2.6 shows how to represent the functions of $\mathsf{BC}^{-}$ in soLAL. Lemma 6.2.7 shows that it is then possible to apply a non-linear recursion, thus representing in soLAL all $\mathsf{SRN}^{=}$. Now, we want to linearly compose the $\mathsf{SRN}^{=}$ functions, thus getting the $\mathsf{SRN}^{-}$ functions. This is possible thanks to the type provided by Lemma 6.2.7. Indeed, imagine we want to cut a function $f$ with output type $j\S^m(k\mathbb{W})$ with a function $g$ with safe input type $i\S^m(k\mathbb{W})$. Lemma 6.2.7 assures that, in this situation, $f$ can also be written as another proof net $f'$ with output type $i\S^m(k\mathbb{W})$, changing also its input types. Hence we can compose $g$ with $f'$. □

**A different encoding.** As we anticipated, we have proved the same Proposition 6.2.5 following two completely different approaches. Here, we show the second one. Instead of using Lemma 6.2.7, we use the forthcoming Lemma 6.2.8. All the remaining parts of the proof of Proposition 6.2.5 are identical.

Notice that the only difference between Lemma 6.2.7 and Lemma 6.2.8 is in the type of the got proof net. The encoding, that here we just sketch, exploits the use of the boxes of type $j\S A \multimap i!B$, so we would expect to reach a result better than the one of Lemma 6.2.7. However, this is false. The functions that we got still are non-iterable, so that we again encode the same set of functions $\mathsf{SRN}^{-}$.

LEMMA 6.2.8 (NON-LINEAR RECURSION IN SOLAL, #2)
Let us assume that $g \in \mathsf{SRN}^{n;s}$ and $h_0, h_1 \in \mathsf{SRN}^{1+n;s+1}$ can all be represented in soLAL with proof nets:

$$\ulcorner h_0 \urcorner, \ulcorner h_1 \urcorner \triangleright (\multimap^{n}_{r=0} j_r\mathbb{W}) \multimap (\multimap^{s}_{r=1} j\S^{m}(k\mathbb{W})) \multimap j\S^{m}(k\mathbb{W}) \multimap j\S^{m}(k\mathbb{W})$$
$$\ulcorner g \urcorner \triangleright (\multimap^{n}_{r=1} j_r\mathbb{W}) \multimap (\multimap^{s}_{r=1} j\S^{m}(k\mathbb{W})) \multimap j\S^{m}(k\mathbb{W}) \ .$$

for some $j, m$. Then, it is also possible to encode the function $f \in \mathsf{SRN}^{1+n;s}$ as a proof net $\ulcorner f \urcorner$ of soLAL, whose type is:

$$\ulcorner f \urcorner \triangleright (\multimap^{n}_{r=0} j_r\mathbb{W}) \multimap (\multimap^{s}_{r=1} i\S^{3} j\S^{m+1}(k\mathbb{W})) \multimap i\S^{4} j\S^{m}(k\mathbb{W}) \ ,$$

for every $i > j$.

The encoding that allows the proof of Lemma 6.2.8 is quite heavy. Here we just sketch the main ideas. Details can be found in Appendix A.

The embedding $\ulcorner \cdot \urcorner$ is inductively defined on the structure of its argument. We will describe how to build $\ulcorner f^{1+n;s} \urcorner$ in the case that $f^{1+n;s}$ is (not necessarily linear) recursive, i.e.

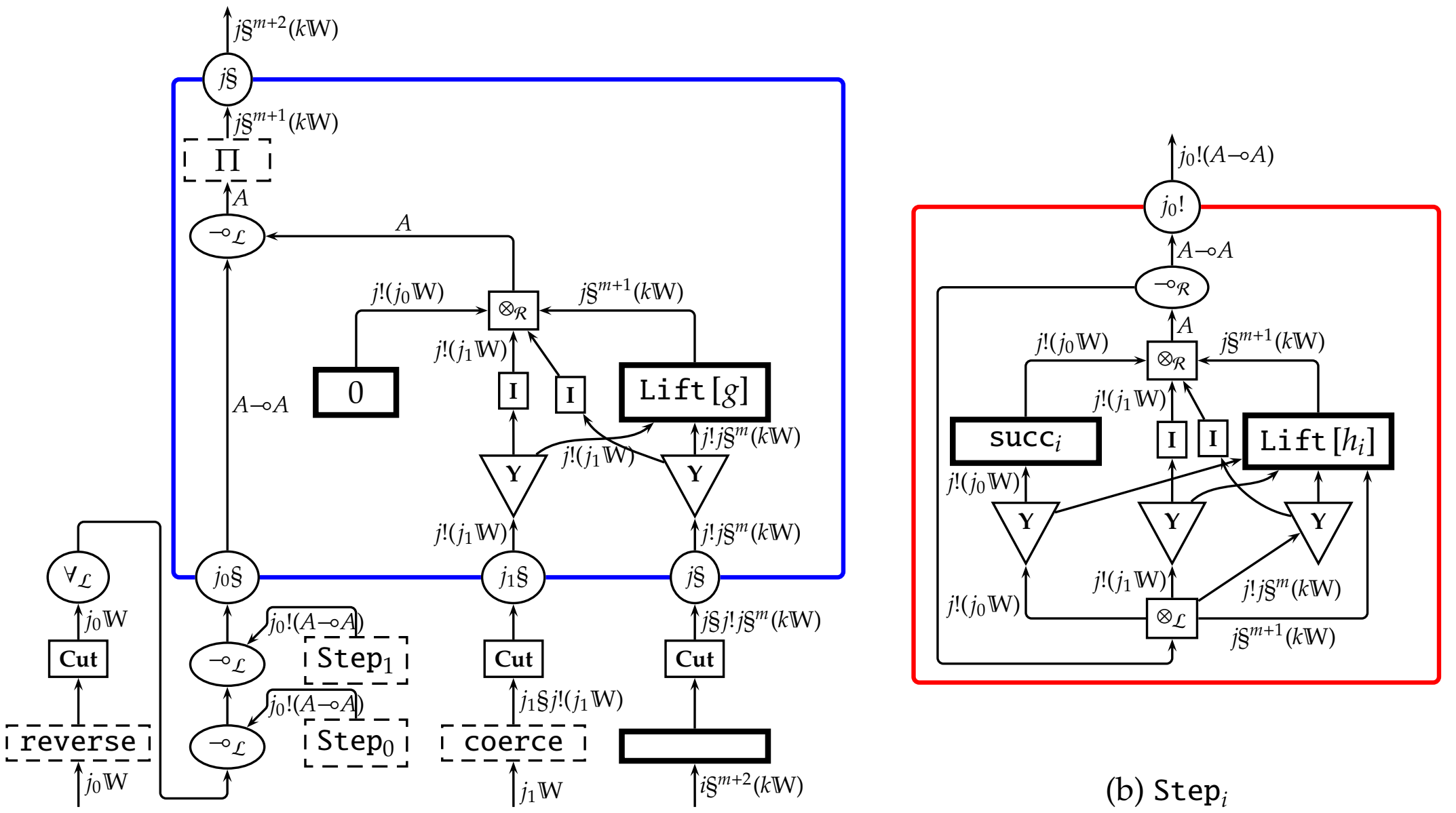

(a) Safe and Linear Recursion

(b) $\mathtt{Step}_i$

where we assume $\mathsf{n} = \mathsf{s} = 1$, $\Pi$ is projection of its last argument, $\boxed{\otimes_{\mathcal{R}}}$ and $\boxed{\otimes_{\mathcal{L}}}$ are sequences of $\otimes_{\mathcal{R}}$ and $\otimes_{\mathcal{L}}$ nodes resp., $\mathtt{Step}_i$, with $i \in 0,1$, is on the right, and

$$A = j!(j_0\mathbb{W}) \otimes \ldots \otimes j!(j_{\mathsf{n}}\mathbb{W}) \otimes \underbrace{j!j\S^m(k\mathbb{W}) \otimes \ldots \otimes j!j\S^m(k\mathbb{W})}_{\mathsf{s}} \otimes j\S^{m+1}(k\mathbb{W}).$$

Figure 6.8: The safe *not necessarily linear* recursion proof net in SOLAL. We assume that $\mathsf{n} = \mathsf{s} = 1$.

of type (2.10), built on $g, h_0, h_1$. Instead, if $f$ is basic or it is a linear composition, it is enough to follow [MO04] that defines the map from $\mathsf{BC}^-$ to $\mathsf{ILAL}$.

$\ulcorner f(w, \overrightarrow{x}^{\mathsf{n}}; \overrightarrow{y}^{\mathsf{s}})\urcorner$ iterates $\ulcorner h_0 \urcorner, \ulcorner h_1 \urcorner$ as many times as the length of $w$, from an initial $\ulcorner g(\overrightarrow{x}^{\mathsf{n}}; \overrightarrow{y}^{\mathsf{s}})\urcorner$. In order to do this, $\ulcorner f(w, \overrightarrow{x}^{\mathsf{n}}; \overrightarrow{y}^{\mathsf{s}})\urcorner$ maintains a *configuration*, that changes at each step; a configuration can be ideally represented as a tuple

$$\langle \overline{\overline{r}}, [\overline{\overline{a_1}}, \ldots, \overline{\overline{a_{\mathsf{r}}}}], [\overline{\overline{x_{11}}}, \ldots, \overline{\overline{x_{1\mathsf{r}}}}], \ldots, [\overline{\overline{x_{\mathsf{n}1}}}, \ldots, \overline{\overline{x_{\mathsf{nr}}}}], [\overline{\overline{y_{11}}}, \ldots, \overline{\overline{y_{1\mathsf{r}}}}], \ldots, [\overline{\overline{y_{\mathsf{s}1}}}, \ldots, \overline{\overline{y_{\mathsf{sr}}}}]\rangle$$

The word $\overline{\overline{r}}$ is initially $\ulcorner g(\overrightarrow{x}^{\mathsf{n}}; \overrightarrow{y}^{\mathsf{s}})\urcorner$, then step-by-step accumulates the result of the recursion unfolding. The remaining elements of the configuration are lists of words, all with length $\mathsf{r}$; at the beginning, $\mathsf{r} = |w| + 1$, then the lists are consumed along the computation.

**An example.** We give a simple example of the rôle of the configurations to represent $\ulcorner f^{1+\mathsf{n};\mathsf{s}}\urcorner$. Let $\mathsf{n} = 0$, $\mathsf{s} = 1$, $G = \ulcorner g \urcorner$, $H_i = \ulcorner h_i \urcorner$, with $i \in \{0,1\}$, $\overline{\overline{0}} = \ulcorner \mathtt{0} \urcorner$, and $\overline{\overline{a}} = \ulcorner a \urcorner$. Let a possible

unfolding of $f$ be:

$$
\begin{aligned}
f(\mathsf{s}_{i_1}(;\mathsf{s}_{i_2}(;\mathbb{0}));a) &= h_{i_1}(\mathsf{s}_{i_2}(;\mathbb{0});a, f(\mathsf{s}_{i_2}(;\mathbb{0});a)) \\
&= h_{i_1}(\mathsf{s}_{i_2}(;\mathbb{0});a, h_{i_2}(\mathbb{0};a, f(\mathbb{0};a))) \\
&= h_{i_1}(\mathsf{s}_{i_2}(;\mathbb{0});a, h_{i_2}(\mathbb{0};a, g(;a)))
\end{aligned} \tag{6.1}
$$

with $i_1, i_2 \in \{0,1\}$. Let us assume we can someway produce the initial configuration $\langle G\overline{\overline{a}}, [\overline{\overline{\mathbb{0}}}, \overline{\overline{\mathbb{0}}}], [\overline{\overline{a}}, \overline{\overline{a}}]\rangle$, where every list is as long as $\mathsf{s}_{i_1}(;\mathsf{s}_{i_2}(;\mathbb{0}))$, *i.e.* 2 elements. The iterative application of a suitable step function from the initial configuration produces the following main steps:

$$
\begin{aligned}
&\langle G\,\overline{\overline{a}}, [\overline{\overline{0}}, \overline{\overline{0}}], [\overline{\overline{a}}, \overline{\overline{a}}]\rangle \to^* \\
&\langle G\,\overline{\overline{a}}, \langle \overline{\overline{0}}, [\overline{\overline{\mathsf{s}_{i_2}(;\mathbb{0})}}]\rangle, \langle \overline{\overline{a}}, [\overline{\overline{a}}]\rangle\rangle \to^* \\
&\langle H_{i_2}\,\overline{\overline{0}}\,\overline{\overline{a}}\,(G\,\overline{\overline{a}}), [\overline{\overline{\mathsf{s}_{i_2}(;\mathbb{0})}}], [\overline{\overline{a}}]\rangle \to^* \\
&\langle H_{i_2}\,\overline{\overline{0}}\,\overline{\overline{a}}\,(G\overline{\overline{a}}), \langle \overline{\overline{\mathsf{s}_{i_2}(;\mathbb{0})}}, [\cdot]\rangle, \langle \overline{\overline{a}}, [\cdot]\rangle\rangle \to^* \\
&\langle H_{i_1}\,\overline{\overline{\mathsf{s}_{i_2}(;\mathbb{0})}}\,\overline{\overline{a}}\,(H_{i_2}\,\overline{\overline{0}}\,\overline{\overline{a}}\,(G\,\overline{\overline{a}})), [\cdot], [\cdot]\rangle
\end{aligned} \tag{6.2}
$$

The odd lines of (6.2) are *configurations* and the even ones *pre-configurations*. A pre-configuration comes from its preceding configuration by: (i) preserving the first argument, (ii) separating head and tail of every list, and storing them as the two components of a same pair, (iii) mapping the appropriate successor between $\ulcorner \mathsf{s}_0 \urcorner, \ulcorner \mathsf{s}_1 \urcorner$ along the tail, only on the leftmost list.

The *suitable step function* we mentioned above will be called $\mathtt{C2C}_{1;1}$ in Appendix A.

## 6.3 Conclusions about soLAL and SRN

We have concentrated our attention on soLAL because of a precise reason. We were searching for a subsystem $\mathcal{P}$ that would be able to encode all the SRN programs. The more, we expected that such a $\mathcal{P}$ would generalize the principles of ILAL, and would respect all the *natural* assumptions that are present in the encoding from $\mathsf{BC}^-$ into ILAL of [MO04]. soLAL is a subsystem with precisely these features. soLAL is an extension of ILAL; more precisely, soLAL is the union of *many different copies* of ILAL, one for each sort of the modalities. The natural numbers can be represented in soLAL as binary Church numerals, exactly as in ILAL, even if there is one different type for each different sort. This is certainly reminiscent of the *tiering* of [Lei93], of which SRN is an example. The application of a Church numeral to any function $f$, i.e. the iteration of $f$, *changes the type* of $f$ leading to a new function $g$ that involves a higher number of paragraphs than $f$. The number of paragraphs is a measure of *how many recursions and compositions* are used to build $g$; exactly as in ILAL. Under these assumptions, soLAL appears the best candidate subsystem we can use to accomplish our goal.

On the other side, ILAL too can represent some functions of $\mathsf{SRN}^-$ that are not in $\mathsf{BC}^-$ (maybe all?). The real difference is *how* these functions are represented. The only way that we see, in ILAL, to represent a $\mathsf{SRN}^-$ function $f$ is to *flatten* the distinctions among normal and safe arguments, thus having all the arguments of type $\mathbb{W}$. On the contrary, soLAL is able to distinguish the normal arguments of $f$ (of type $i\mathbb{W}$ for some $i$) and its safe arguments (of type $j\S^m(k\mathbb{W})$ for some $j,k$). This distinction is crucial: it means that the type of the inputs and of the output of $f$ is (almost) preserved. This is in accordance with the intuition: *the number m of paragraphs in the type of the output of $f$, that is equal to number of paragraphs in the type of the safe arguments, is a measure of the complexity of $f$. Namely, it is a bound on the number of nested recursion and composition schemes occurring in $f$.*

Now, soLAL is anyway not able to represent the whole SRN. This suggests some considerations. Let us focus for a moment on the theorems $j\S A \multimap i!A$, for $j < i$, that are provable in soLAL. These implications are a kind of *inverse* of the theorems $!A \multimap \S A$, typical of ILAL. But, of course, ILAL and soLAL cannot prove $\S A \multimap !A$ because this theorem would break their polynomial time soundness. We like to imagine all the theorems $j\S A \multimap i!A$ as *approximations* of $\S A \multimap !A$. Every such approximation is compatible with the polynomial time soundness, while its *upper bound* $\S A \multimap !A$ is not. This approach suggests that soLAL is the best subsystem of MS that extends ILAL under the natural assumptions we mentioned above.

A pair of words about other, possible, subsystems of MS encoding SRN. We have tried *lots* of them, without getting any result better than the one of soLAL. Just as an example, we can imagine a subsystem based on $\mathcal{P}^{M}_{\mathbb{LTS}}$ of Section 6.1, where the numerals are encoded using *asymmetric* contractions. Of course this approach — as all the other ones we can imagine — is quite different from ILAL, thus getting far from the assumptions we underlined at the beginning of this Section. This is why we preferred to omit from this thesis all these attempts. Only a few of them — the ones more strictly related with soLAL — will be discussed in Section 8.1.

# Chapter 7

# An alternative approach: LALL

We have already seen that it is an interesting challenge to represent SRN programs inside some subsystem of MS. This was, in fact, the main motivation that led us to the study of MS. However, we did not reach our goal. We identified some subsystems that slightly extend the expressiveness of ILAL, but none of them is really able to capture the whole SRN. It appears like that the *stratification property*, that is fundamental for ILAL, is too strong w.r.t. SRN.

In this last Chapter we will partially solve the problem using a different approach. We will present a polytime logic LALL that extends ILAL using both the levels of $L^4$, and some recursive types. We will be able to compositionally encode into LALL all the *finite initial fragments* of all the SRN programs (Proposition 7.5.2). LALL is not a subsystem of MS. LALL is not stratified, but anyway it is *weakly stratified*, i.e. it satisfies a weaker notion of stratification. When we say *weakly stratified*, we do not refer to the property of Weak Stratification as defined in [BM10], that strictly depends on the syntax of proof nets they use. More simply, we refer to the fact that, during a reduction, the residues of any node $u$ are either at the same depth or at the same level of $u$ (Fact 7.1.4).

We strongly suggest the reader to read the summary of this Chapter at page 9, before this Chapter itself, where we have underlined the most important features of LALL.

One possible objection is the following. Apparently, in this way, *it is possible to represent in* LALL *every total function, even the non-computable ones*. Indeed, if $f$ is a non-computable total function, and $f \upharpoonright l$ is a finite initial fragment of $f$, then $f \upharpoonright l$ is also an initial fragment of some SRN program $t_l$, so $f \upharpoonright l$ can be represented in LALL, and at last we have represented $f$ in LALL. However, we underline two facts. (i) The representation just given for $f$ is non constructive, as we do not know which $t_l$ can represent $f$. (ii) We are not interested in the functions, but in the algorithms. It is not possible to represent *every algorithm* in LALL, but at least it is possible to represent the SRN ones.

The previous observations are supported by the forthcoming Lemma 7.4.2. It shows that, in fact, all the functions $g : \mathbb{W} \to \mathbb{W}$ *eventually constant* can be easily represented in LALL. Eventual constant means constant for all but finitely many inputs. Every such a function is computable in constant time, so it is not surprising that they can be encoded in whatever logic. Anyway, $g$ is encoded using a naïve *look-up* algorithm, so that (a) the embedding is all but compositional, (b) the proof net required to encode $g$ has really large size, and finally (c) Lemma 7.4.2 is useless in a *complexity* perspective, and we cannot use it for proving Proposition 7.5.2.

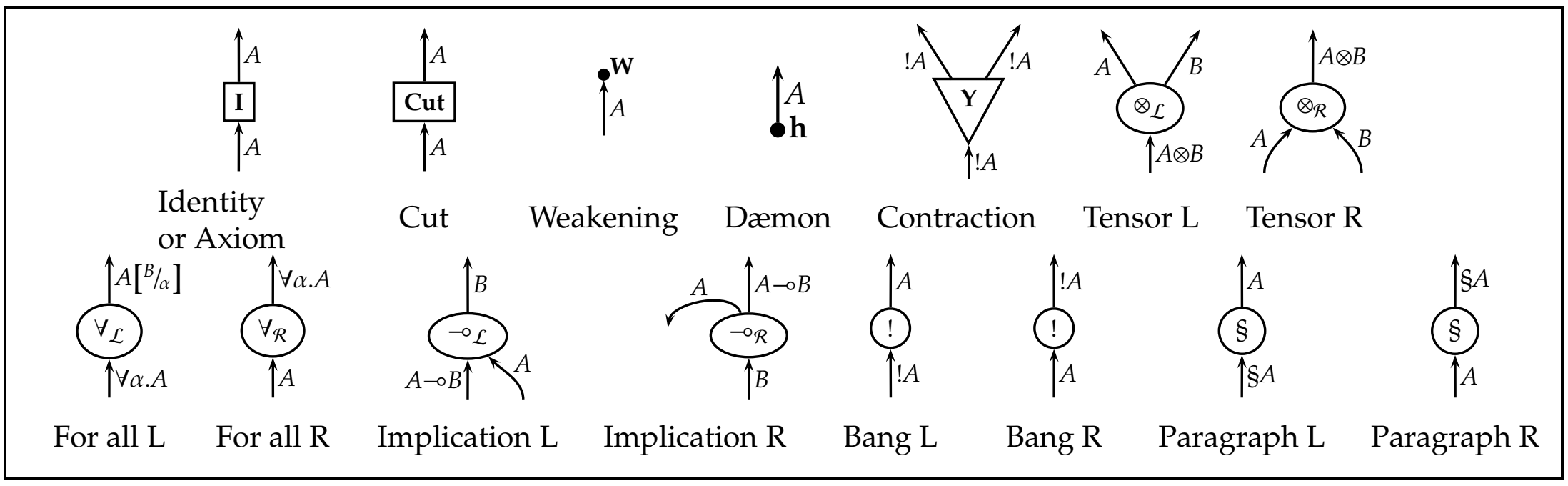

Figure 7.1: The nodes in the proof nets of LALL.

## 7.1 Light Affine Logic by Levels

**The language of formulæ.** First, for any fixed countable set $\mathcal{V}$ of propositional variables, the set $\mathcal{F}$ of formulæ is generated by the following grammar:

$$F ::= \mathscr{S} \mid \alpha \mid F \otimes F \mid F \multimap F \mid \forall\alpha.F \mid !F \mid \S F \qquad \alpha \in \mathcal{V}$$

where $\mathscr{S}$ is a constant. Second, we define the quotient $\mathcal{F}_{\mathscr{S}}$ of $\mathcal{F}$ by assuming:

$$\mathscr{S} = \forall\alpha.(\alpha \multimap (\mathbb{B} \multimap \mathscr{S} \multimap \alpha) \multimap \alpha) \tag{7.1}$$

among the elements of $\mathcal{F}$. Namely, (7.1) gives to $\mathscr{S}$ the meaning of Scott numerals [ACP93]. The formulæ we shall effectively use are the equivalence classes in $\mathcal{F}_{\mathscr{S}}$. Every time we label an edge of a proof net of LALL by $\mathscr{S}$, we can also label that edge by any "unfolding" of $\mathscr{S}$ that obeys (7.1). $B\left[{}^{C}/_{\alpha}\right]$ is the substitution of every free occurrence of $\alpha$ in $B$ with $C$. Similarly, $(\forall\alpha.A)\{C\}$ means $A\left[{}^{C}/_{\alpha}\right]$.

**Proof structures and proof nets.** LALL is a language of proof nets. Proof nets are defined, essentially, as for MS. For sake of simplicity, we will recall them here. Given the nodes in Figure 7.1, we say that *an Axiom node and a Dæmon nodes are proof structures*. Moreover, given two proof structures $\Pi$ and $\Sigma$:

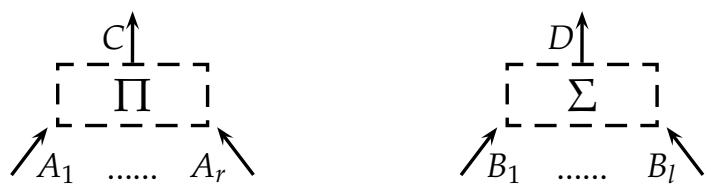

denoted as $\Pi \triangleright A_1, \ldots, A_r \vdash C$ and $\Sigma \triangleright B_1, \ldots, B_l \vdash D$, respectively, with $r, l \geq 0$, then all the graphs inductively built from $\Pi$ and $\Sigma$ by the rule schemes in Figure 7.3 are proof structures.

Now, indexing is introduced, exactly as in Section 8.2.2.

Definition 7.1.1 (Indexing and Proof nets, adapted from [BM10]) *Let $\Pi$ be a proof structure.*

1. *An* indexing *for $\Pi$ is a function $I$ from the edges of $\Pi$ to $\mathbb{Z}$ that satisfies the constraints in Figure 7.2 and such that $I(e) = I(e')$, for every pair $e, e'$ of inputs and output of $\Pi$.*
2. *A* proof net *is a proof structure that admits an indexing.*

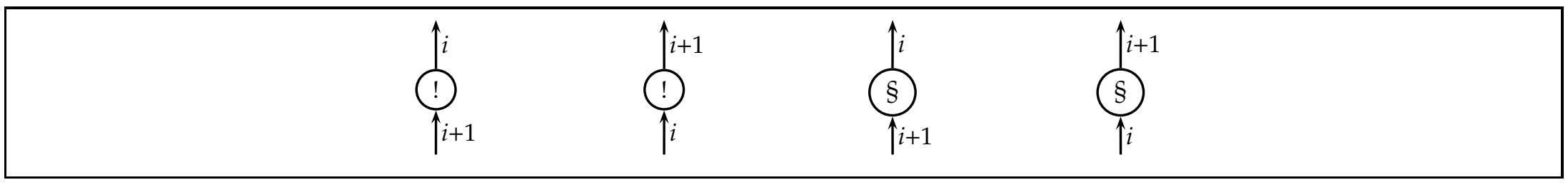

Figure 7.2: Constraints on the indexing. The nodes we omit have the same index on all of their incident edges.

3. *An indexing $I$ of $\Pi$ is* canonical *if $\Pi$ has an edge $e$ such that $I(e) = 0$, and $I(e') \geq 0$ for all edges $e'$ of $\Pi$.*

Fact 7.1.2 (Existence and Uniqueness of Canonical Indexing)
Every proof net of LALL admits a unique canonical indexing.

The indexing tells that the nodes $!_{\mathcal{L}}$ and $\S_{\mathcal{L}}$ are not standard *dereliction* nodes. Instead, they are auxiliary ports of §-boxes whose border is somewhat *fuzzy*. We mean that a §-boxes is not necessarily either contained into or disjoint from another box. Instead, it can "overlap" a !-box, and it can have more than one conclusion Paragraph R. To distinguish §-boxes from the ! ones we adopt a dotted border (and also a different colour, if the reader is reading a coloured version of this thesis). Please notice that these boxes are completely useless, and they can be avoided as in [BM10]; anyway, we think they are useful to underline that § and ! nodes are not dereliction nodes.

Let $I_0$ be the canonical indexing of $\Pi$ and $e \in E_{\Pi}$. The *level of* $e$ is $l(e)$. It is defined as $I_0(e)$. The *level of* $\Pi$ is $l(\Pi)$. It is defined as the greatest value assumed by $I_0$ on the edges of $\Pi$. We denote as usual as $B_{\Pi}$ the set of the !-boxes in $\Pi$, and it is naturally in bijection with the set of the $!_{\mathcal{R}}$ nodes in $\Pi$.

**Normalization.** Most of the normalization steps (*ns*) are analogous to the namesake of MS, already described in Section 3.3. We just recall what they do. The *linear* normalization steps annihilate in the natural way a pair of linear nodes (Identity/Cut, $\multimap_{\mathcal{L}}$ / $\multimap_{\mathcal{R}}$, $\otimes_{\mathcal{L}}/\otimes_{\mathcal{R}}$, $\S_{\mathcal{L}}/\S_{\mathcal{R}}$, $\forall_{\mathcal{L}}/\forall_{\mathcal{R}}$). The *modal* normalization steps are of two kinds: $!_{\mathcal{L}}/!_{\mathcal{R}}$ is reduced merging the two involved boxes which can be !-boxes as well as §-boxes fuzzy borders. Instead, *contraction*/$!_{\mathcal{R}}$ duplicates the whole !-box cut with the contraction, as in ILAL. The *garbage collection* normalization steps involve the weakening or the dæmon nodes, cut with any other node. It is always possible to reduce such a cut with the help of some more weakening and dæmon nodes, as done in ILAL [AR02]. The set of cut nodes of $\Pi$ is cuts($\Pi$).

Fact 7.1.3 (Cut-elimination)
Every LALL proof net reduces to some cut-free proof net.

We recall that the definition of *stratification* is at page 39.

Fact 7.1.4 (Weak Stratification)
Let $\Pi$ be a LALL proof net such that $\Pi \to \Pi'$ by reducing a cut node at level $i$ and depth $d$. Let $e$ be any edge in $\Pi$, and $e'$ any one among its residues in $\Pi'$. Then, either $\partial(e) = \partial(e')$, or $l(e) = l(e')$.

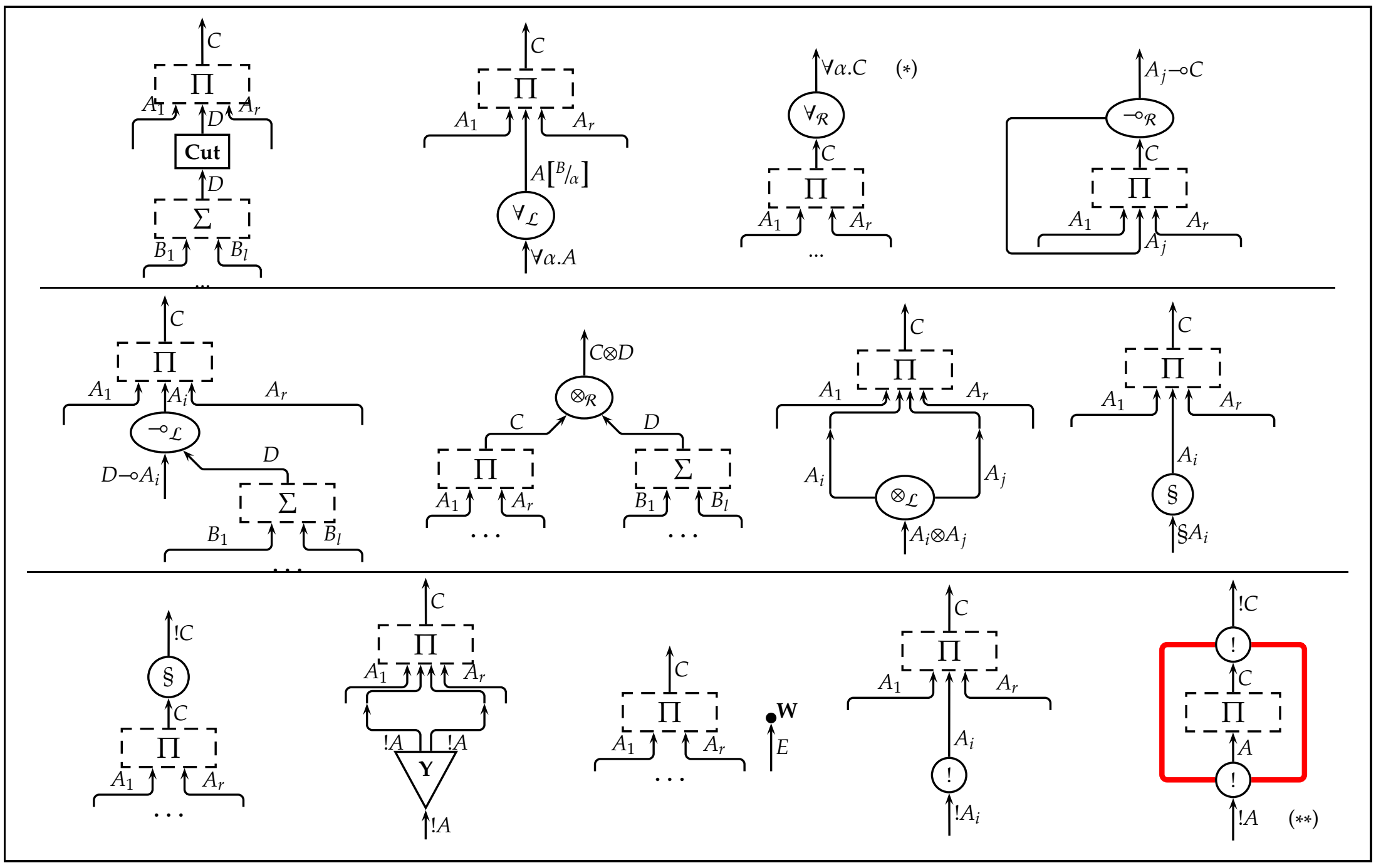

Figure 7.3: Inductive rule schemes to build proof structures of LALL. (*) $\alpha$ does not occur free in $A_1, \dots, A_r$. (**) A !-box, which has *at most* a single assumption.

## 7.2 LALL is polytime

We adapt the definitions of [BM10], with some minor modifications to handle the free weakening nodes. Let us fix a proof net $\Pi$ to reduce. We define an ordering over cuts($\Pi$) that determines which cuts to reduce first.

A graph theoretic path in any proof net $\Pi$ is *exponential* if it contains a, possibly empty, sequence of consecutive contractions and stops at a $!_{\mathcal{L}}$ node.

Let $B, C \in \mathcal{B}_{\Pi}$. Let $B \prec_1^L C$ if the roots of $B$ and $C$ lie at the same level, and the root of $B$ is in cut with an exponential path that enters a left port of $C$. $\preceq^L$ is the reflexive and transitive closure of $\prec_1^L$. One can show that $\preceq^L$ is a partial order, *upward arborescent*: for every $C$ there is at most one $B$ such that $B \prec_1^L C$.

Let $c, c' \in$ cuts($\Pi$). We write $c \leq c'$ iff one of the following conditions holds. (i) $c'$ is connected to a weakening or a dæmon, and $c$ is not. (ii) The condition (i) is false but $l(c) < l(c')$ holds. (iii) The conditions (i) and (ii) are false, so $l(c) = l(c')$. In this case, $c \leq c'$ iff: (a) either $c'$ is connected to a contraction, and $c$ is not, or (b) $c, c'$ are connected to a contraction on one side, to the boxes $B, B'$, respectively, on the other, and $B \preceq^L B'$.

Definition 7.2.1 (Canonical normalization) *A sequence of normalization steps that starts from a given proof net $\Pi$ is* **canonical** *whenever cuts that are smaller under $\leq$ are eliminated before higher ones.*

Theorem 7.2.2 (Polynomial bound for LALL)
Let $\Pi$ be a LALL proof net of size $s$, level $l$, and depth $d$. Then, every canonical reduction is long at most $(l+1)s^{(d+2)^l}$ steps.

The proof strategy coincides with the one in [BM10], where the reduction of the garbage collection steps is delayed till the end.

## 7.3 Preliminary notions about SRN

We recall from Section 2.7 that $\mathsf{SRN}^{\mathsf{n};\mathsf{s}}$ is the subset of SRN whose terms have normal arity n, and safe arity s. If not otherwise stated, $\vec{t^m} = t_1, \ldots, t_m$ will always denote sequences of $m \geq 0$ terms of SRN. Moreover, we write $|\vec{t^m}| \leq l$, for some $l > 0$, meaning that the size of every term $t_i$ is not greater than $l$. Now, we recall that, for every $t$ in $\mathsf{SRN}^{\mathsf{n};\mathsf{s}}$, and $\vec{x} = x_1, \ldots, x_{\mathsf{n}}$, $\vec{y} = y_1, \ldots, y_{\mathsf{s}}$, it holds

$$|t(\vec{x};\vec{y})| \leq p_t\left(|x_1|, \ldots, |x_{\mathsf{n}}|\right) + \max\left\{|y_1|, \ldots, |y_{\mathsf{s}}|\right\} \tag{7.2}$$

where $p_t$ is the *characteristic polynomial of* $t$ which is non-decreasing and depends on $t$. We notice that if $u$ is a subterm of $t$, then $\deg(p_u) \leq \deg(p_t)$, $\deg(p)$ denoting the degree of the polynomial $p$. At last, we define the *composition degree* $\partial_{\mathsf{C}}(t)$ and the *recursion degree* $\partial_{\mathsf{R}}(t)$ of $t$, as the functions that count resp. the number of safe composition and of recursion schemes inside $t$.

Definition 7.3.1 (Output Bounding Function $ob.(\cdot)$) *Let $t$ in $\mathsf{SRN}^{\mathsf{n};\mathsf{s}}$ and $l \geq 0$. We define $ob.(\cdot)$, that takes $t$ and $l$ as arguments, as $ob_t(l) = p_t(l, \ldots, l) + l$.*

Fact 7.3.2 ($ob.(\cdot)$ Bounds the Output Length of $t \in$ SRN)
For every $t$ in $\mathsf{SRN}^{\mathsf{n};\mathsf{s}}$, $l \geq 0$, and sequences $\vec{x}, \vec{y}$ such that $|\vec{x}|, |\vec{y}| \leq l$, we have $|t(\vec{x};\vec{y})| \leq ob_t(l)$.

The following definition will be used only in Proposition 7.5.2. Notice that $nb_t(l)$ is a polynomial, for every fixed $t$.

Definition 7.3.3 (Net Bounding Function $nb.(\cdot)$) *Let $t$ in $\mathsf{SRN}^{\mathsf{n};\mathsf{s}}$ and $l \geq 0$. We define $nb.(\cdot)$, that takes $t$ and $l$ as arguments, as $nb_t(l) = \underbrace{ob_t(ob_t(\ldots ob_t(l)\ldots))}_{\partial_{\mathsf{R}}(t)+\partial_{\mathsf{C}}(t)\ \textit{times}}$.*

We have to deal with various kinds of numbers, so the following correspondence is convenient:

Fact 7.3.4 (Relation between Naturals, Scott words, and Words-as-Terms)
Every sequence $(d_1, \ldots, d_l)$ with $d_1, \ldots, d_l \in \{0,1\}$ and $l \geq 0$, identifies uniquely a number $n = 2^{l-1} \cdot d_l + \cdots + 2^0 \cdot d_1 \in \mathbb{N}$. So, both the term of SRN $\mathsf{s}_{d_l}(;\ldots \mathsf{s}_{d_1}(;\mathbb{0})\ldots)$ and the Scott word $[n]$ identify $n$, too. We say that the sequence, as well as the Scott number and the SRN term, *represent* $n$.

We underline that an infinite number of sequences, and of terms, represent the same $n$.

## 7.4 Preliminary useful proof nets in LALL

**Booleans.** The type of booleans is $\mathbb{B} = \forall\gamma.\gamma \multimap \gamma \multimap \gamma$ whose representative proof nets are in Figure 7.4. The proof net $\nabla\mathsf{B}[b]$ duplicates any boolean we may plug into $b$ by a Cut node.

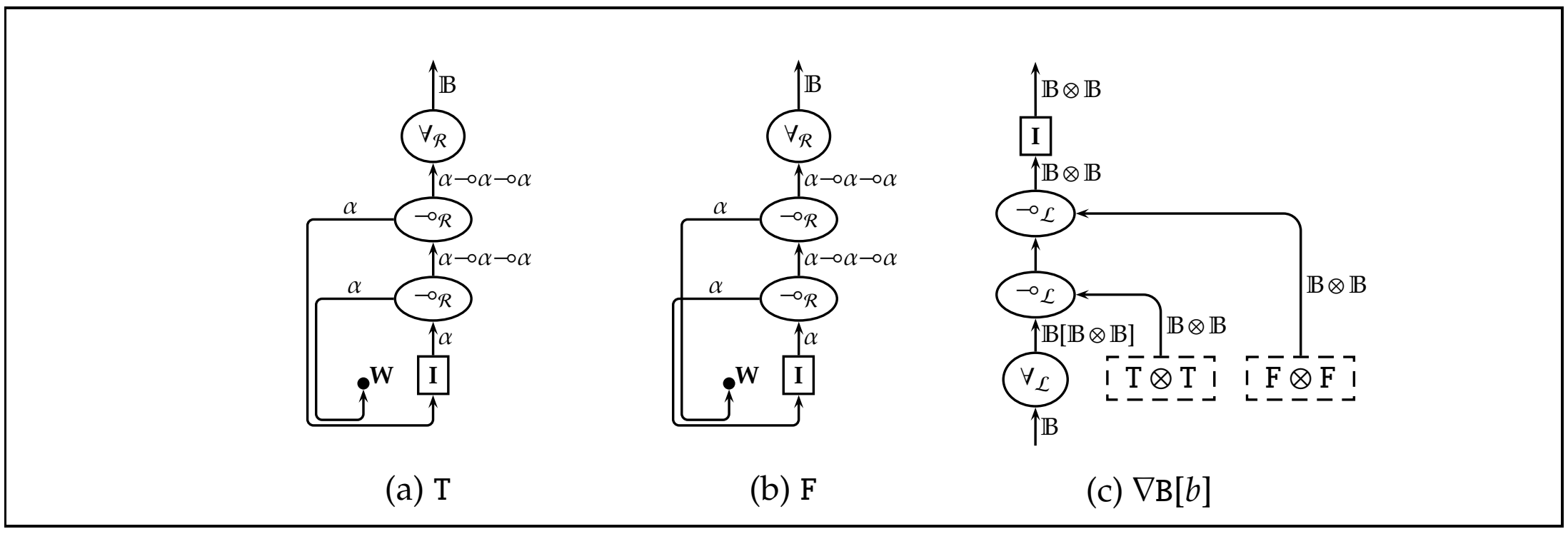

Figure 7.4: Booleans.

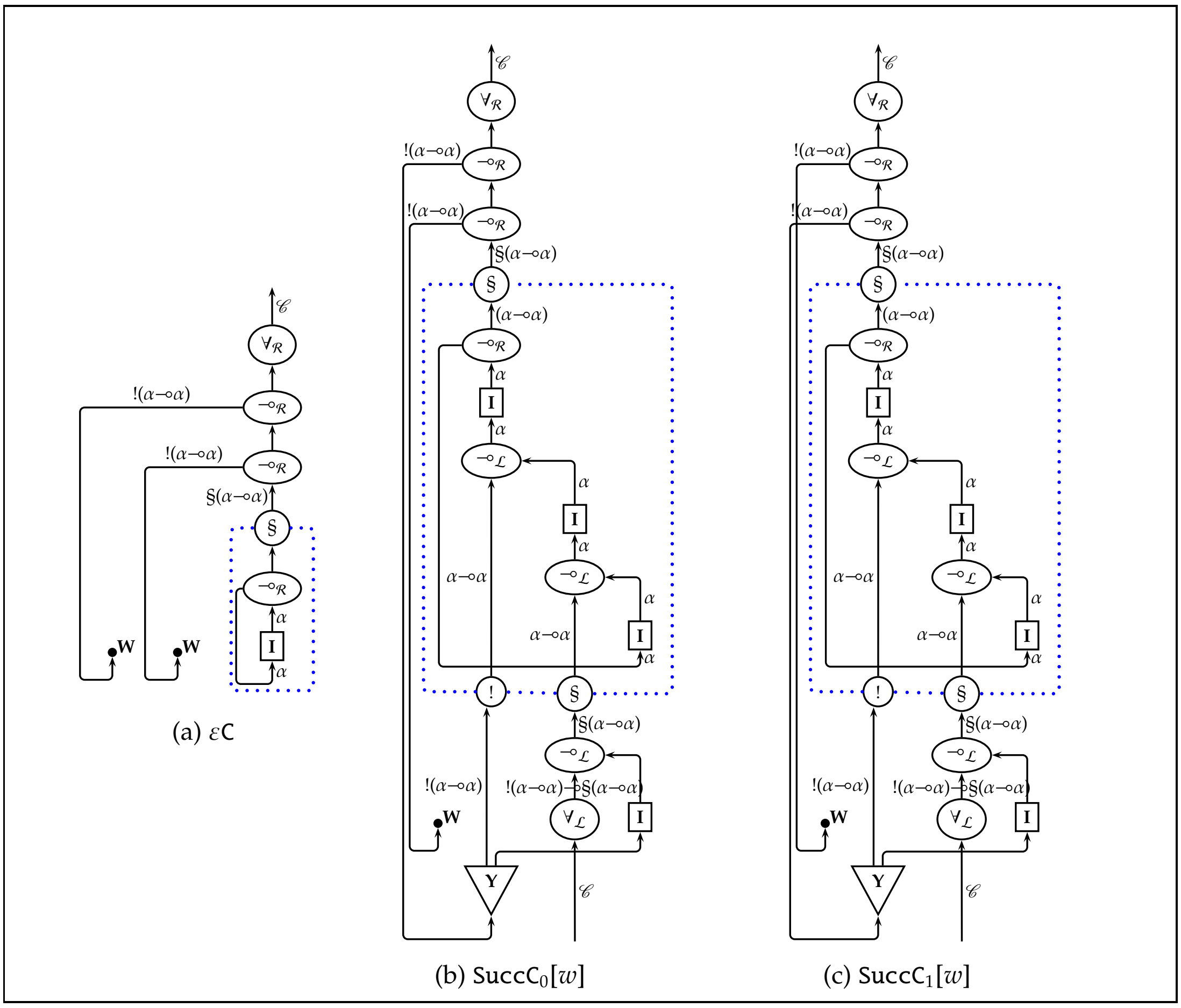

Figure 7.5: (Church) Words.

**Church words, or simply words.** The type of words is $\mathscr{C} = \forall\alpha.!(\alpha \multimap \alpha) \multimap !(\alpha \multimap \alpha) \multimap \S(\alpha \multimap \alpha)$. Figure 7.5 introduces the zero $\varepsilon$C and the two successors. If $w$ is a natural number in binary notation, $\overline{w}$ is its usual representation by a proof net. Figure 7.6 introduce a proof net that invert the bits inside any $\overline{w}$, plugged by Cut into the dangling input of RevC[$w$].

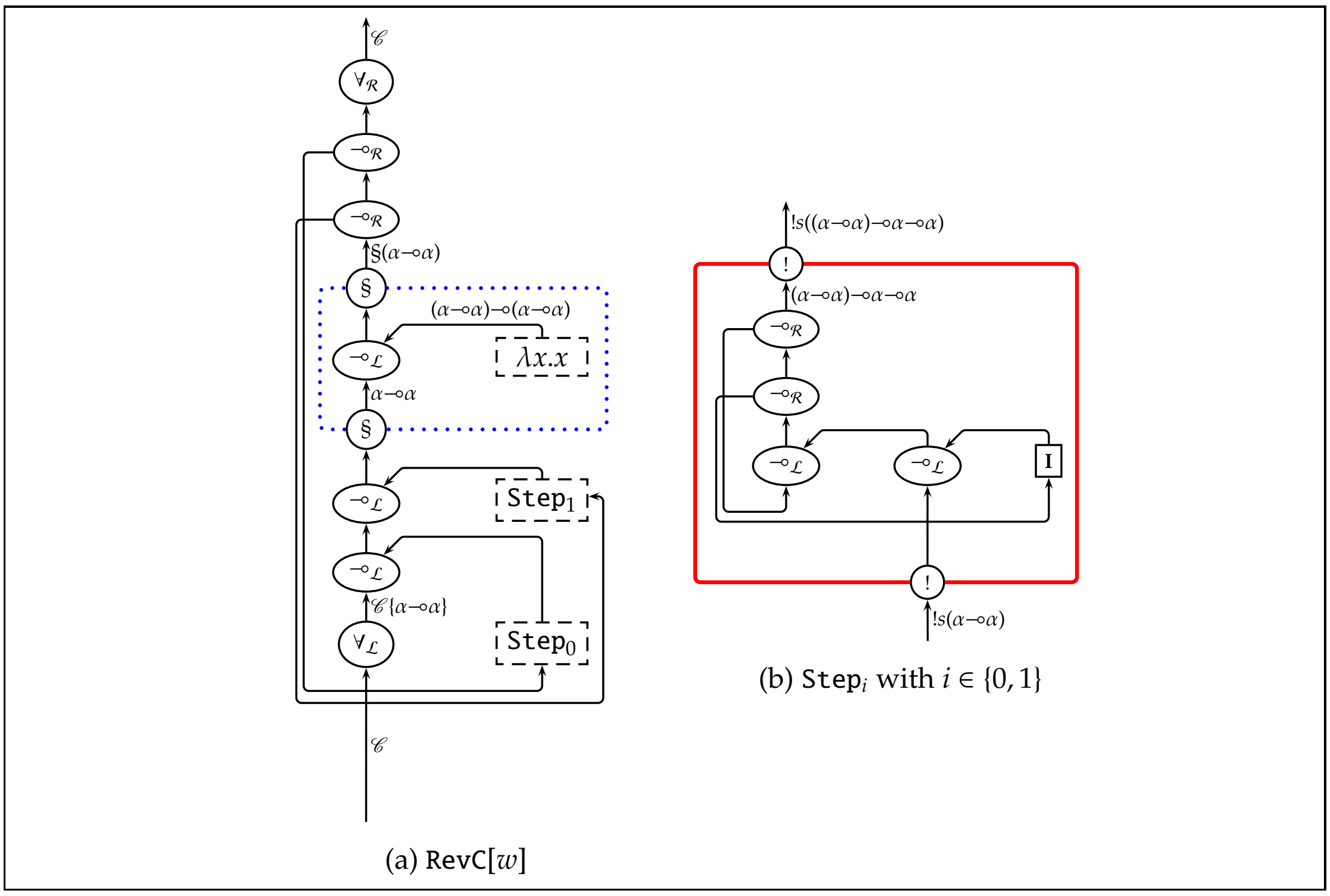

Figure 7.6: Bit reversion over (Church) Words.

**Scott words.** Intuitively, the type $\mathscr{S}$ of Scott words describes a tuple of booleans. They are defined in Figure 7.7. On Scott words we have the proof nets in Figure 7.8. We remark that SuccS$_0$[$s$] adds to $s$ the least significant bit T, which stands for the digit 0, and SuccS$_1$[$s$] adds F, instead, which stands for 1. PredS[$s$] shifts $s$ to its right deleting the *least significant* bit. So:

REMARK 7.4.1 A Scott word is in fact a stack of bits, the least significant bit being on the top of the stack.

Moreover, CondS[$s, x, y$] branches a computation, depending on the value of $s$. It yields $x$ if the least significant bit of $s$ is 0, *or* if $s = \varepsilon$S, while it yields $y$ if the least significant bit of $s$ is 1. The preprocessing avoids to return $\varepsilon$S: if $s = \varepsilon$S, then $s$ becomes SuccS$_0$[$\varepsilon$S]. Also, the three assumptions of type $\mathscr{S}$ in $\mathscr{S}, \mathscr{S}, \mathscr{S} \vdash \mathscr{S}$ specify the type of $s, x$, and $y$, respectively.

**Finite functions over Scott words.** A function $f : \mathbb{W} \to A$ is **eventually constant** if there exists a constant $C \in A$ such that $f(w) = C$ for all but finitely many inputs $w$. In this case, the **support** of $f$ is the subset $\{w \in \mathbb{W} \mid f(w) \neq C\}$, which is finite. Then:

LEMMA 7.4.2 (FINITE FUNCTIONS IN LALL)
Every finite function $f : \mathbb{W} \to A$, where $A$ is a type representable in LALL, can be represented as a proof net in LALL.

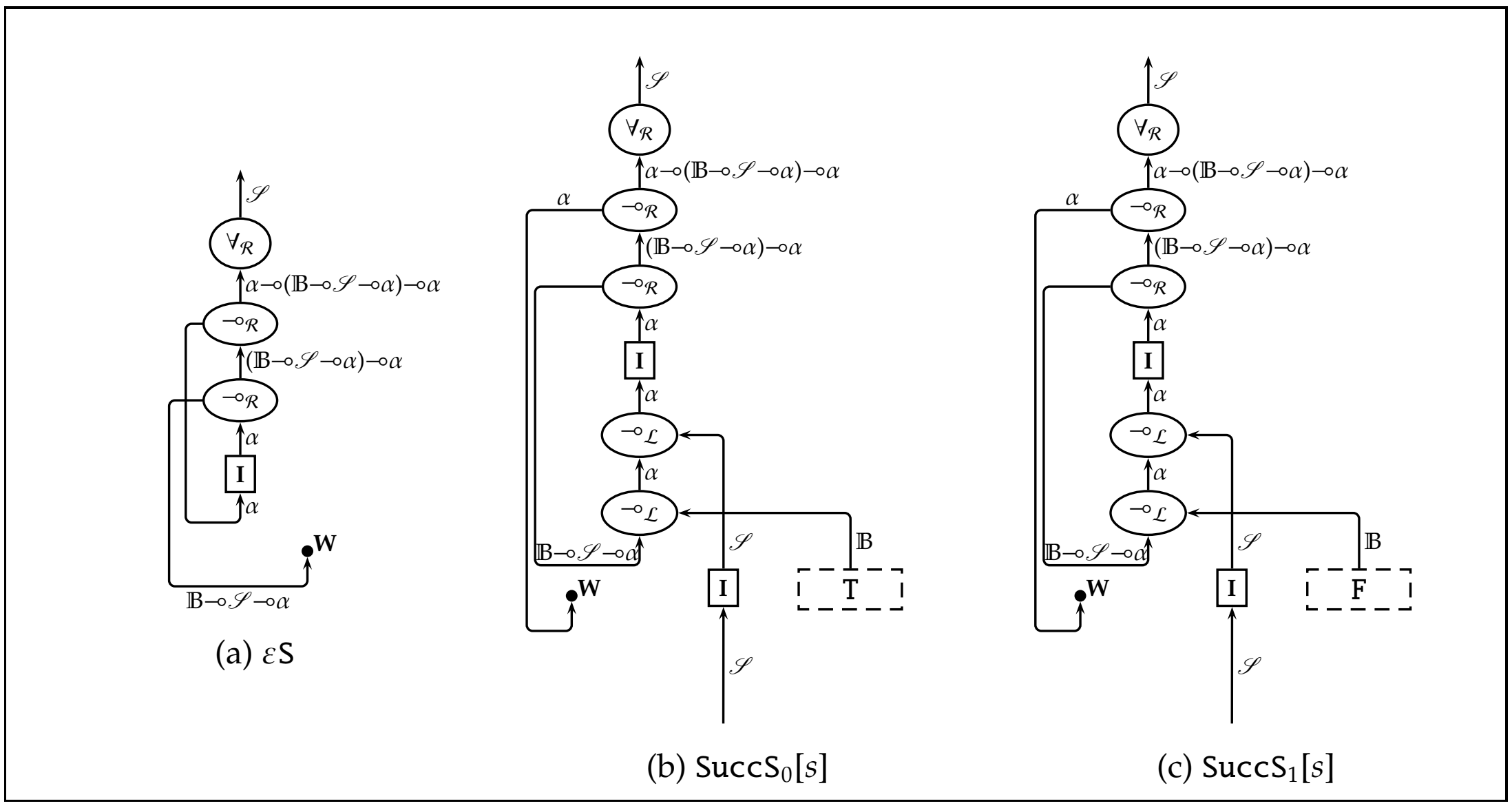

Figure 7.7: Binary Scott numerals.

A finite function $f$ is completely described by its values on the inputs of the support. This description is finite. If the elements in the support are at most $l$ bits long, $f$ has at most $2^l$ *interesting* couples (input,output). So, $f$ is represented using a *look-up* table with $2^l$ entries, that contains every possible output for all the $2^l$ *interesting* inputs (Figure 7.9).

This fact is theoretically interesting, but it will not be needed when proving Proposition 7.5.2. Indeed, the size of the proof net representing $f$ is $O\left(2^l\right)$, that is too high for us. In the next two paragraphs we will show that, in some interesting cases, this bound can be reduced to $O(l)$.

**Duplicating Scott words.** For any $l \geq 0$, the proof net $\nabla\mathtt{S}_l[s] \triangleright \mathscr{S} \vdash \mathscr{S}^2$, where $\mathscr{S}^2 = \mathscr{S} \otimes \mathscr{S}$, is inductively defined on $l$. It is in Figure 7.10. $\nabla\mathtt{S}_l[s]$ builds two copies of any Scott word *at most as long as* $l$. The generalization $\nabla\mathtt{S}^k_l[s] \triangleright \S^k\mathscr{S} \vdash \S^k\mathscr{S} \otimes \S^k\mathscr{S}$ of $\nabla\mathtt{S}_l[s]$ duplicates a given Scott word *at most as long as* $l$ which lies inside $k \geq 0$ paragraph boxes. Specifically, $\nabla\mathtt{S}^0_l[s]$ is $\nabla\mathtt{S}_l[s]$, while $\nabla\mathtt{S}^k_l[s]$ is in Figure 7.11, with $k > 0$, which is the only proof net that exploits the *fuzzy borders* of paragraph boxes. A boring exercise shows that: $|\nabla\mathtt{S}^k_l[s]| = 19 + 89l + 3k = O(l)$.

**Coercing Scott words.** For any $k, l \geq 0$, we define $\mathtt{CoerS}^k_l[s] \triangleright \mathscr{S} \vdash \S^k\mathscr{S}$ by cases on $k$, and by induction on $l$. If $k = 0$, then $\mathtt{CoerS}^0_l[s]$ is the node Axiom. Otherwise, the proof net is in Figure 7.12, where, for $i \in \{0, 1\}$, $\lambda s.\S^k(\mathtt{SuccS}_i[s]) \triangleright\vdash \S^k\mathscr{S} \multimap \S^k\mathscr{S}$ *denotes* the proof net that we build by: (i) enclosing $\mathtt{SuccS}_i[s]$ into $k$ paragraph boxes to get $\S^k(\mathtt{SuccS}_i[s]) \triangleright \S^k\mathscr{S} \vdash \S^k\mathscr{S}$, and (ii) adding an Implication R to $\S^k(\mathtt{SuccS}_i[s])$ so to close it and get its type $\S^k\mathscr{S} \multimap \S^k\mathscr{S}$. The proof net $\mathtt{CoerS}^k_l[s]$ reconstructs a given Scott word *at most as long as* $l$ into an identical Scott word inside $k$ paragraph boxes. We can show that $|\mathtt{CoerS}^k_l[s]| = 43l + 3kl = O(l)$.

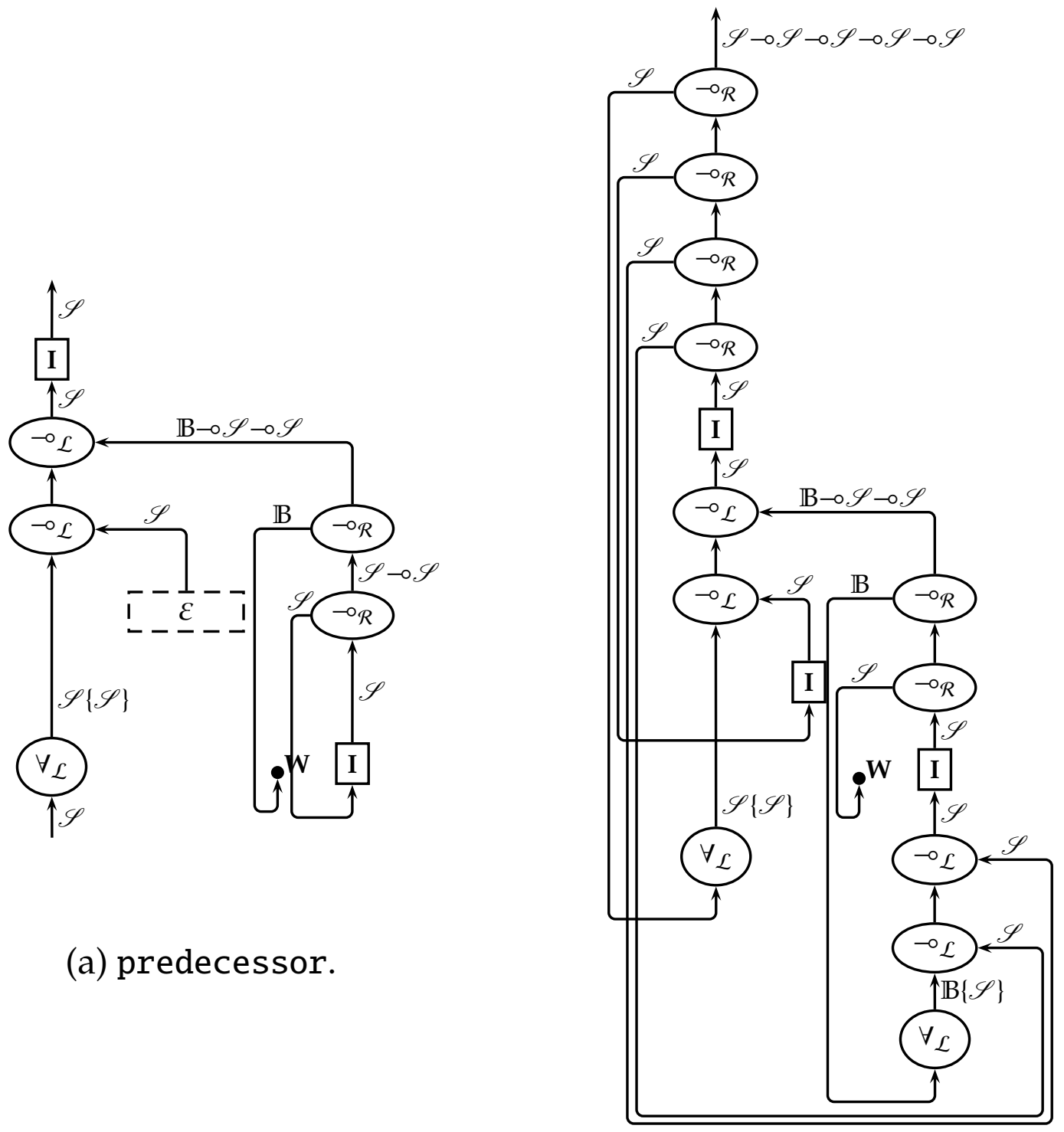

(a) `predecessor`.

(b) `conditional`.

Figure 7.8: Basic operations over binary Scott numerals.

**Scott words to words.** The proof net $\mathtt{StoC}_l[s]$ normalizes to the word $\overline{w}$, for any Scott word *at most as long as l*. We do not describe it, as it has a structure analogous to the proof net $\mathtt{CoerS}_l^0[s]$ in Figure 7.12.

**Lifting.** Let $\Pi \triangleright \vec{\mathscr{S}}^{\mathsf{n}}, \S^k \vec{\mathscr{S}}^{\mathsf{s}} \vdash \S^k \mathscr{S}$ for some $\mathsf{n}, \mathsf{s}, k \geq 0$. For every $k' \geq 0$ we can build $\mathtt{Lift}[\Pi] \triangleright \vec{\mathscr{S}}^{\mathsf{n}}, \S^{k+k'} \vec{\mathscr{S}}^{\mathsf{s}} \vdash \S^{k+k'} \mathscr{S}$ by: (i) enclosing $\Pi$ into $k'$ paragraph boxes, getting $\Pi'$, and (ii) plugging the conclusion of $\mathtt{CoerS}_l^{k'}[s]$, using Cut, into every of the $\mathsf{n}$ premises with type $\S^{k'} \mathscr{S}$ of $\Pi'$. The final proof net is in Figure 7.13. The proof net $\mathtt{Lift}[\Pi]$ is $\Pi$ deepened inside $k'$ paragraph boxes. Notice that $|\mathtt{Lift}[\Pi]| = |\Pi| + k'(\mathsf{n} + \mathsf{s} + 1) + \mathsf{n}|\mathtt{CoerS}_l^{k'}[s]| = O(|\Pi| + l)$.

**Contracting the premises of a proof net.** Let $\Pi \triangleright \vec{A}, \S^k \mathscr{S}, \S^k \mathscr{S}, \vec{A'} \vdash A$ for some $l, k \geq 0$. We can build $\nabla_l^k[\Pi] \triangleright \vec{A}, \S^k \mathscr{S}, \vec{A'} \vdash A$ by: (i) writing $\Pi'$ which is $\Pi$ with a new Tensor L between the two outlined premises of type $\S^k \mathscr{S}$, and (ii) plugging the conclusion of $\nabla \mathtt{S}_l^k[s] \triangleright \S^k \mathscr{S} \vdash \S^k \mathscr{S} \otimes \S^k \mathscr{S}$, by a Cut, into the premise of the new Tensor L in $\Pi'$. Notice that $|\nabla_l^k[\Pi]| = |\Pi| + |\nabla \mathtt{S}_l^k[s]| + 2 = O(l)$.

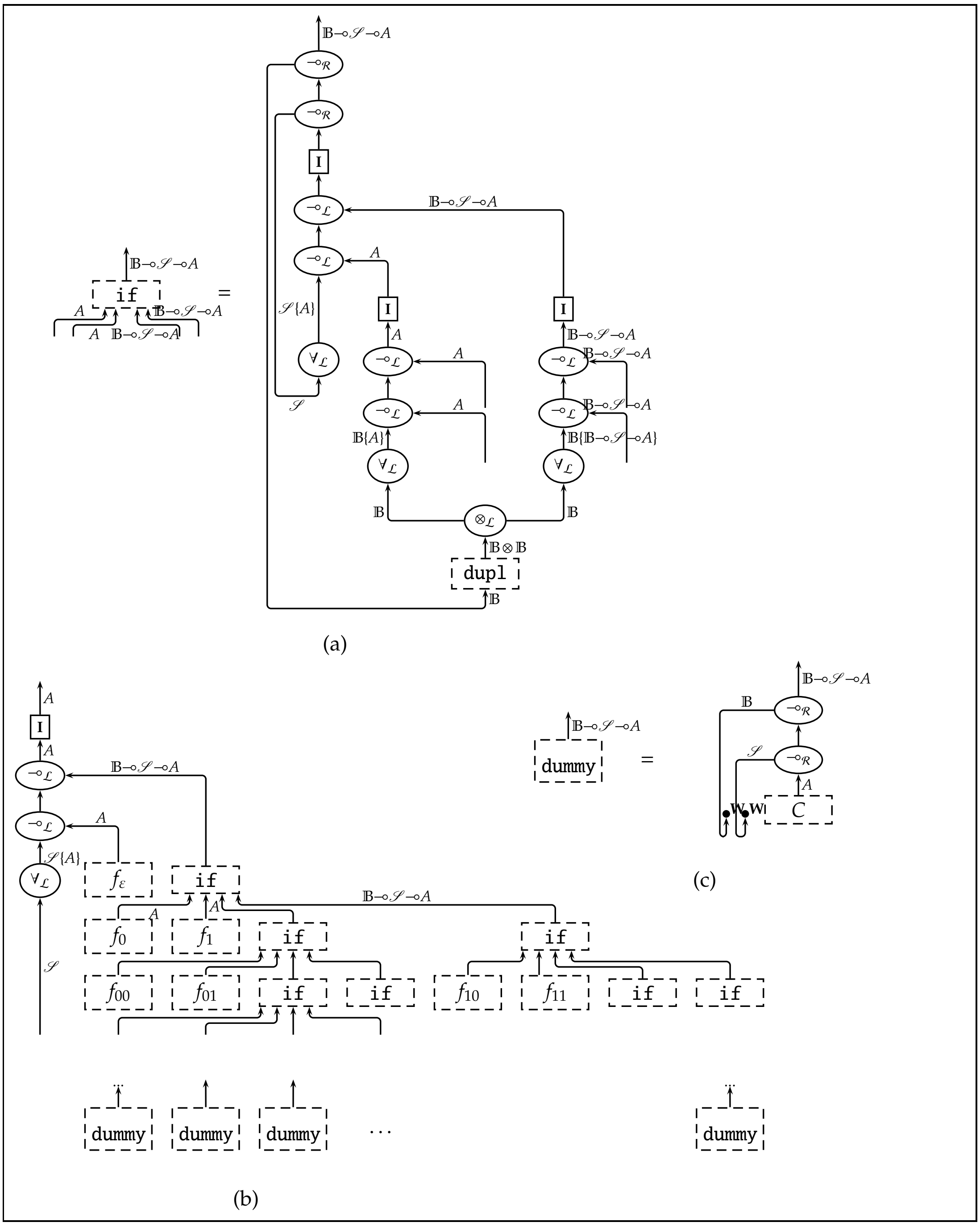

Figure 7.9: Proof nets that represent a finite function $f : \mathbb{W} \to A$.

## 7.5 Embedding SRN into LALL

The goal is to compositionally embed SRN into LALL, with a map as much analogous as possible to the natural one from $\mathsf{BC}^-$ into ILAL of [MO04]. For any fixed n, and s the map $\lceil \cdot \rceil^{\cdot}$ takes a term $t$ of $\mathsf{SRN}^{n;s}$ as first and $l \geq 0$ as second argument, and yields a proof net

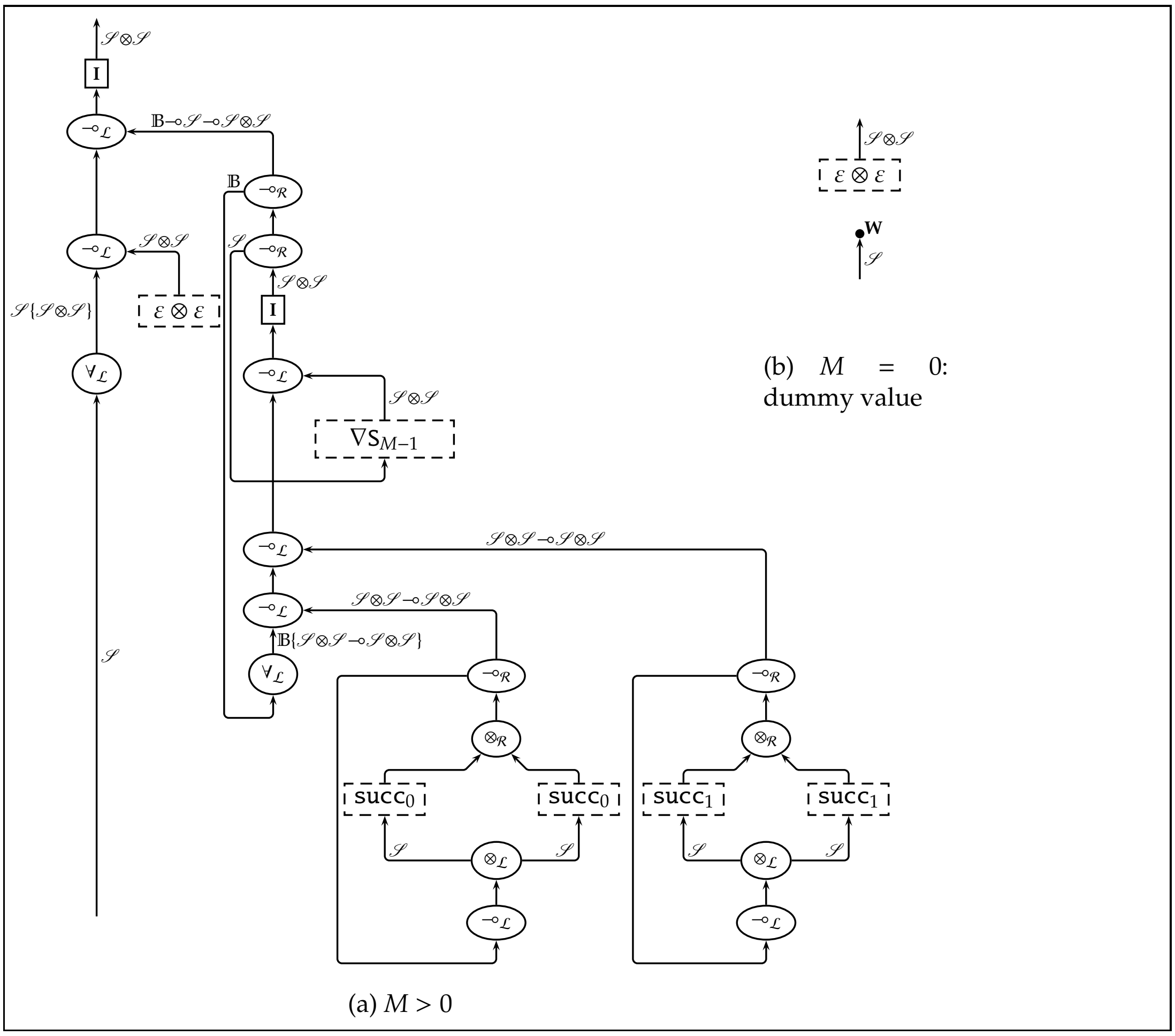

Figure 7.10: The *Duplication* proof net $\nabla \mathsf{s}_M : \mathscr{S} \multimap \mathscr{S} \otimes \mathscr{S}$.

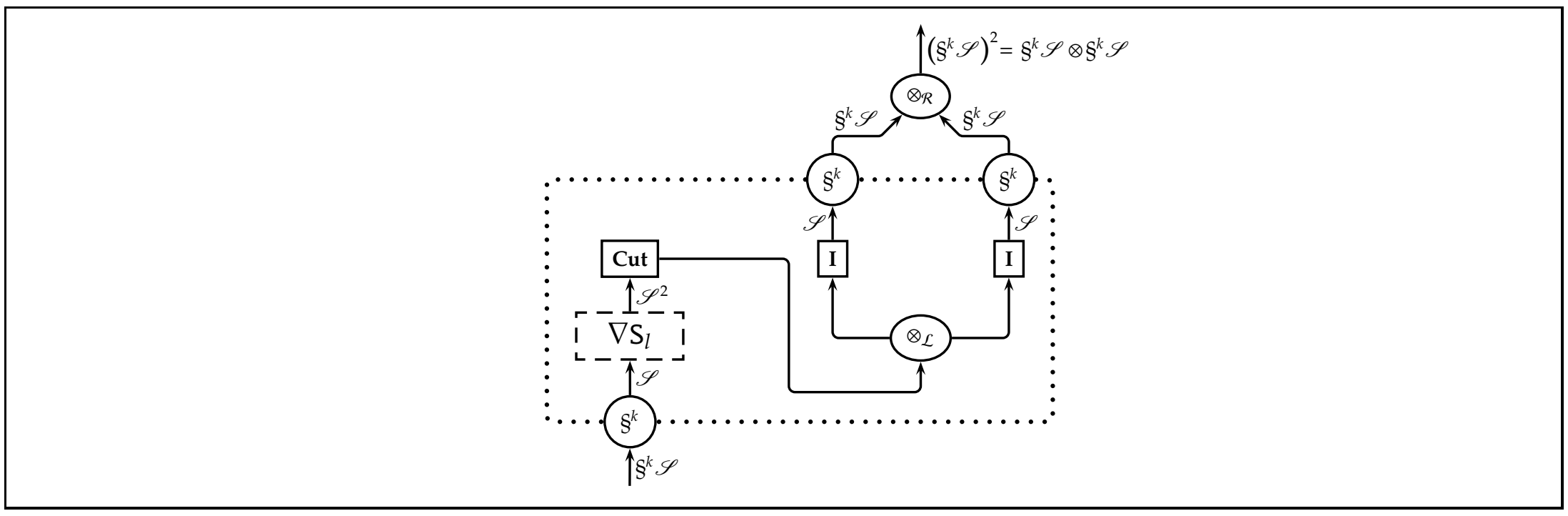

Figure 7.11: The generalized duplication $\nabla \mathsf{s}_l^k[s]$ of Scott words.

$\lceil t \rceil^l \triangleright \overrightarrow{\mathscr{S}}^{\mathsf{n}}, \overrightarrow{\S^k \mathscr{S}}^{\mathsf{s}} \vdash \S^k \mathscr{S}$, for some $k$. We define the map inductively on the first argument.

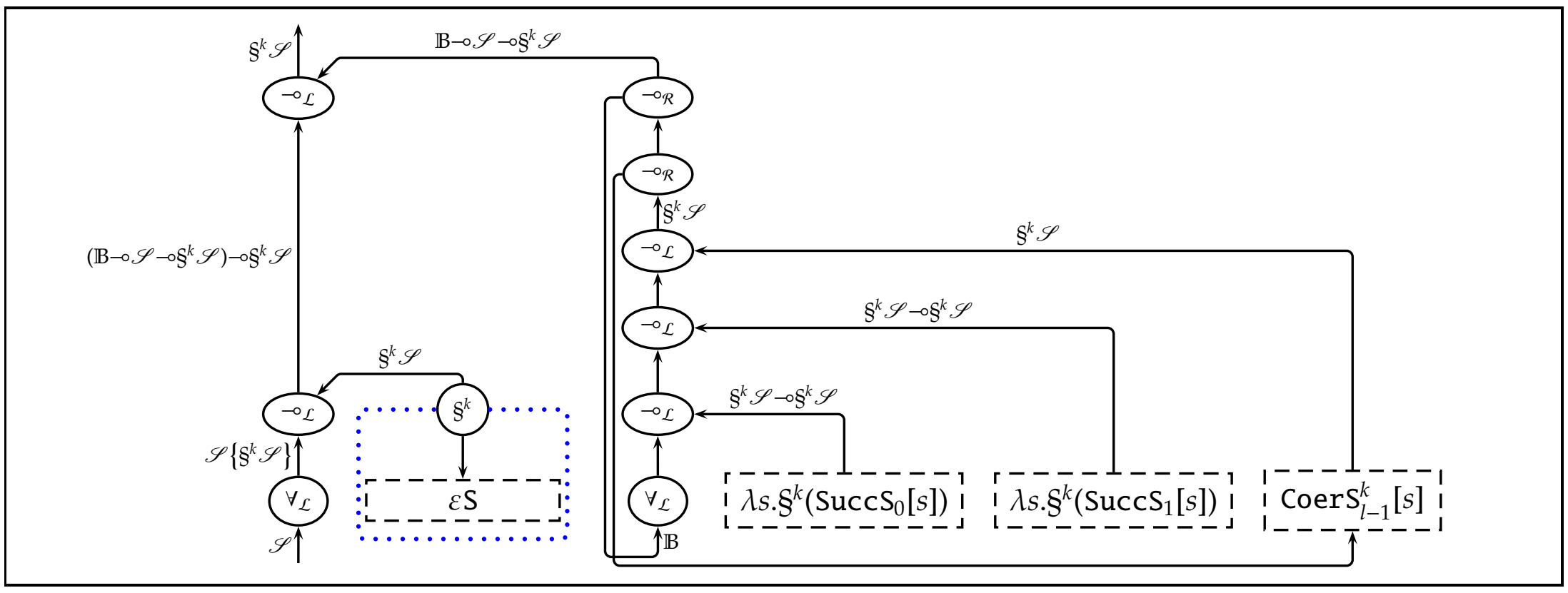

Figure 7.12: The coerce proof net $\mathtt{CoerS}^k_l[s]$ on Scott words.

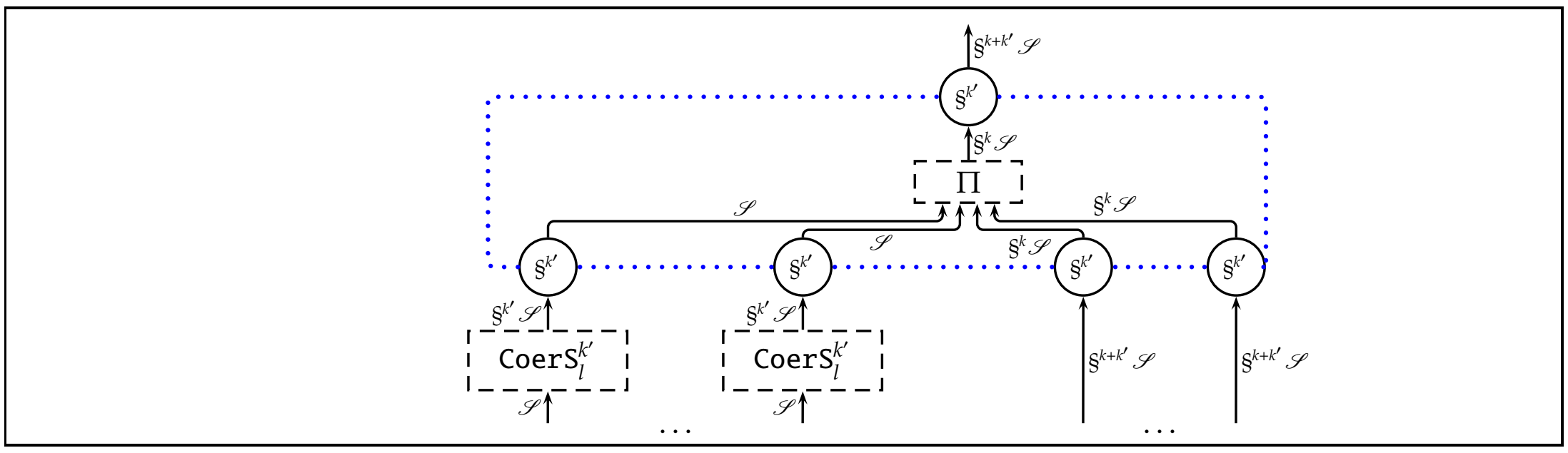

Figure 7.13: The *Lifting* of a proof net $\Pi$ representing $f$ up to $M$ with $k$ paragraphs.

**The base cases of $\lceil\cdot\rceil^{\cdot}$.** Some of them are straightforward:

$$\lceil \mathtt{0} \rceil^l = \varepsilon\mathrm{S} \triangleright\ \vdash \mathscr{S} \qquad \lceil \mathsf{s}_i \rceil^l = \mathtt{SuccS}_i[s] \triangleright \mathscr{S} \vdash \mathscr{S} \qquad (i \in \{0,1\})$$
$$\lceil \mathtt{P} \rceil^l = \mathtt{PredS}[s] \triangleright \mathscr{S} \vdash \mathscr{S} \qquad \lceil \mathtt{B} \rceil^l = \mathtt{CondS}[s,x,y] \triangleright \mathscr{S},\mathscr{S},\mathscr{S} \vdash \mathscr{S}$$

where, $s, x, y$ jute denote the inputs of the proof nets they appear into. Concerning the projection, $\lceil \pi_i^{\mathsf{n};\mathsf{s}} \rceil^l$ is an Axiom that connects the $i$-th input to the conclusion, erasing all the other inputs by Weakenings. An example with $1 \le i \le \mathsf{n}$ is:

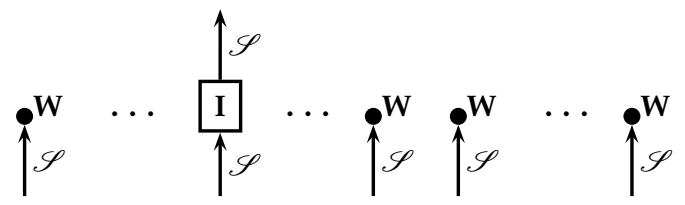

Notice that $k = 0$ here.

**The case of $\lceil\cdot\rceil^{\cdot}$ on the composition.** We now focus on $t = \circ[t', u_1, \ldots, u_{\mathsf{n}}, v_1, \ldots, v_{\mathsf{s}}]$ such that $t'$ be in $\mathsf{SRN}^{\mathsf{m};\mathsf{r}}$. Because of space constraints, but without loss of generality, we show how to build $\lceil t \rceil^l$ by assuming $\mathsf{m} = \mathsf{n} = \mathsf{s} = 1$, and $\mathsf{r} = 2$. By induction we have:

$$\lceil t' \rceil^{ob_l(t)} \triangleright \mathscr{S}, \S^{k'}\mathscr{S}, \S^{k'}\mathscr{S} \vdash \S^{k'}\mathscr{S} \qquad \lceil u_1 \rceil^l \triangleright \mathscr{S} \vdash \S^{k_u}\mathscr{S}$$
$$\lceil v_i \rceil^l \triangleright \mathscr{S}, \S^{k_i}\mathscr{S} \vdash \S^{k_i}\mathscr{S} \qquad (i \in \{1,2\})$$

By letting $k = \max\{k', k_u, k_1, k_2\}$, we get:

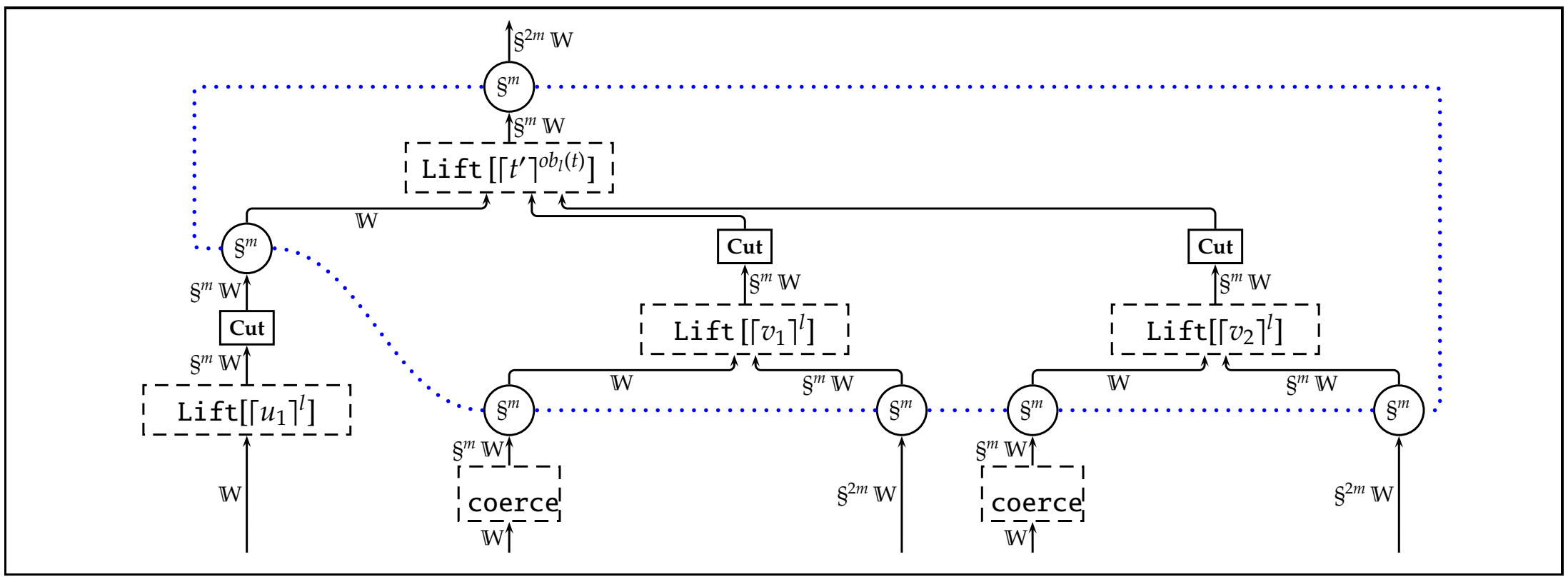

Figure 7.14: The (partial) translation of $\circ[t', u_1, v_1, v_2]$ with missing contractions.

$$\mathtt{Lift}[\lceil t'\rceil^{ob_l(t)}] \triangleright \mathscr{S}, \S^k\mathscr{S}, \S^k\mathscr{S} \vdash \S^k\mathscr{S} \qquad \mathtt{Lift}[\lceil u_1\rceil^l] \triangleright \mathscr{S} \vdash \S^k\mathscr{S}$$
$$\mathtt{Lift}[\lceil v_i\rceil^l] \triangleright \mathscr{S}, \S^k\mathscr{S} \vdash \S^k\mathscr{S} \qquad (i \in \{1,2\})$$

Next, if we build $\Pi'$ in Figure 7.14, then $\lceil t\rceil^l$ is $\nabla_l^{2k}[\nabla_l^0[\nabla_l^0[\Pi']]]$. The two occurrences of $\nabla_l^0$ contract three "normal" premises. One is from $\lceil u_1\rceil^l$. The other two are from $\lceil v_1\rceil^l, \lceil v_2\rceil^l$. The occurrence of $\nabla_l^{2k}$ contracts the single "safe" premise of $\lceil v_1\rceil^l$ and $\lceil v_2\rceil^l$. We insist remarking the existence of $\lceil t\rceil^l$ for any $\mathsf{m}, \mathsf{n}, \mathsf{r}, \mathsf{s}$. One can count: $|\lceil t\rceil^l| \leq |\mathtt{Lift}[\lceil t'\rceil^{ob_l(t)}]| + \sum_i^{\mathsf{m}} |\mathtt{Lift}[\lceil u_i\rceil^l]| + \sum_j^{\mathsf{r}} |\mathtt{Lift}[\lceil u_j\rceil^l]| + k(1+\mathsf{n}+\mathsf{s}'\mathsf{n}+\mathsf{s}'\mathsf{s}) + \mathsf{s}'\mathsf{n}|\mathtt{CoerS}_l^k[\mathsf{s}]|$. So $|\lceil t\rceil^l| = O\left(|\lceil t'\rceil^{ob_l(t)}| + \sum_i^{\mathsf{m}} |\lceil u_i\rceil^l| + \sum_j^{\mathsf{r}} |\lceil v_j\rceil^l| + ob_l(t)\right)$.

**The case of $\lceil\cdot\rceil^{\cdot}$ on the recursion.** Let $t = \mathtt{r}[u_\varepsilon, u_0, u_1]$ with $u_\varepsilon \in \mathsf{SRN}^{\mathsf{n};\mathsf{s}}, u_0, u_1 \in \mathsf{SRN}^{\mathsf{n}+1;\mathsf{s}+1}$. Again, because of space constraints, we set $\mathsf{n} = \mathsf{s} = 1$ which is general enough to show the key point of the embedding. In the course of the iteration unfolding that $\lceil t\rceil^l$ carries out, the safe argument gets duplicated, so we must contract them. By induction:

$$\lceil u_\varepsilon\rceil^l \triangleright \mathscr{S}, \S^{k_\varepsilon}\mathscr{S} \vdash \S^{k_\varepsilon}\mathscr{S} \qquad \lceil u_i\rceil^{ob_t(l)} \triangleright \mathscr{S}, \mathscr{S}, \S^{k_i}\mathscr{S}, \S^{k_i}\mathscr{S} \vdash \S^{k_i}\mathscr{S} \qquad (i \in \{0,1\})$$

By letting $k = \max\{k_\varepsilon, k_0, k_1\}$, and using $\mathtt{Lift}[\cdot]$ in analogy to the translation of the composition, $\lceil t\rceil^l$ is in Figure 7.15. The first argument, *i.e.* the Scott word that drives the recursion unfolding, becomes a word, and, then, it is necessary to reverse it by $\mathtt{RevC}$. Otherwise we would unfold the iteration according to a wrong bit order, as implied by Remark 7.4.1. Moreover, (i) $\Pi$ projects the rightmost $\mathsf{n}+\mathsf{s}+1$-th element of type $A$ it gets in input and which contains the result, (ii) The two proof nets $\boxed{\otimes_{\mathcal{R}}}$, $\boxed{\otimes_{\mathcal{L}}}$ are two obvious trees of Tensor R and L nodes. Calculating the size, we get $|\lceil t\rceil^l| = O\left(|\lceil u_\varepsilon\rceil^l| + |\lceil u_0\rceil^{ob_t(l)}| + |\lceil u_1\rceil^{ob_t(l)}| + ob_t(l)\right)$.

Definition 7.5.1 (Simulating a Term by a Proof net) *Let $t$ be in $\mathsf{SRN}^{\mathsf{n};\mathsf{s}}$, $l \in \mathbb{N}$, and $\Pi \triangleright \vec{\mathscr{S}}^{\mathsf{n}}\ (\overrightarrow{\S^k\mathscr{S}})^{\mathsf{s}} \vdash \S^k\mathscr{S}$, for some $k \in \mathbb{N}$. Then, $\Pi$ $k$-*simulates *$t$* with *$l$*-bounded inputs *if, for every pair of vectors of natural numbers $\vec{x}^{\mathsf{n}}, \vec{y}^{\mathsf{s}}$, such that $|\vec{x}^{\mathsf{n}}|, |\vec{y}^{\mathsf{s}}| \leq l$, the proof net we get by plugging $\lceil x_1\rceil^l, \ldots, \lceil x_{\mathsf{n}}\rceil^l, \S^k\lceil y_1\rceil^l, \ldots, \S^k\lceil y_{\mathsf{s}}\rceil^l$ into the inputs of $\Pi$, in the natural way, normalizes to $\S^k\lceil z\rceil^l$, whenever $z$ is the result of $t(\vec{x}^{\mathsf{n}}; \vec{y}^{\mathsf{s}})$.*

Proposition 7.5.2 (LALL simulates SRN)
Let $l \geq 0$, and $t \in \mathsf{SRN}^{\mathsf{n};\mathsf{s}}$. Then, $\lceil t\rceil^l$ $k$-simulates $t$ with $l$-bounded input. Moreover, (i) $k \leq \partial_{\mathsf{R}}(t) \cdot 2^{\partial_{\mathsf{C}}(t)}$, and (ii) for every fixed $t$, $|\lceil t\rceil^l|$ is $O(nb_t(l))$.

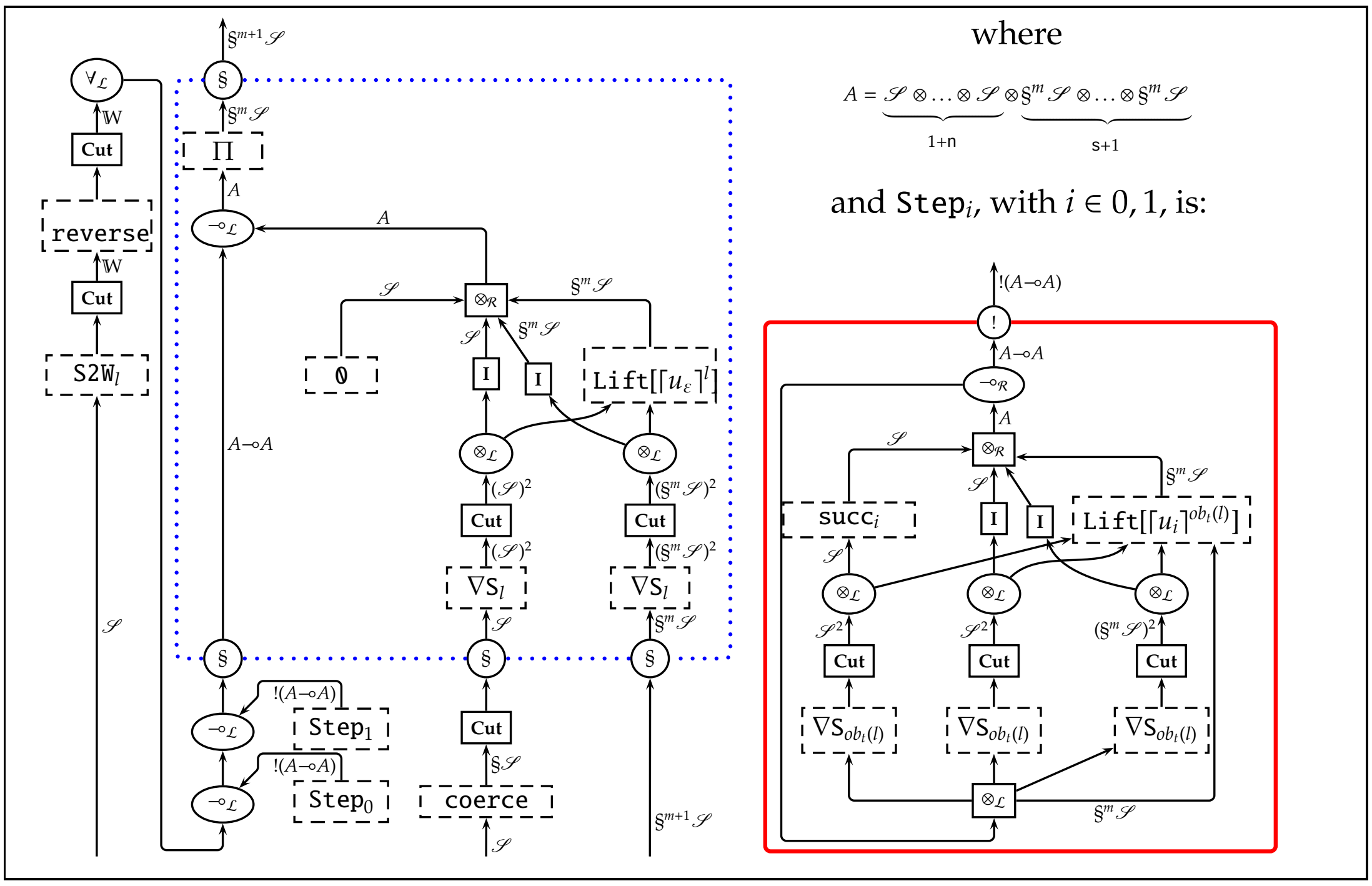

Figure 7.15: Safe Recursion.

The statement holds by induction on $t$, using Fact 7.3.2, Fact 7.3.4, the definition of $\lceil\cdot\rceil^{\cdot}$ and of the size of every proof net used by $\lceil\cdot\rceil^{\cdot}$, and recalling that $ob()$ is defined in terms of $p_t(\cdot)$.

# Chapter 8

# Further directions

This last Chapter considers some possible directions for the future research. We have already begun the exploration of some of these directions; some others, on the contrary, are just ideas for future work.

First of all, in Section 8.1 we consider some subsystems that may possibly encode some more functions of SRN than the ones that soLAL can represent. Two of them, co-soLAL and $\text{soLAL}^{\infty}$, are natural extensions of soLAL. The third one is IEAL, Intuitionistic Elementary Affine Logic, that is known to be not polytime. We conjecture anyway that it could be polytime, under some reduction strategy.

In Section 8.2 we consider some possible generalizations of the framework MS. MS lacks several of the rules that one can find in literature. As we already said, we were interested only in *stratified* logics, because they are easier to handle than the other ones, and because of their philosophical similarity with *tiering*. We show that it is possible to extend MS without losing stratification. $\text{MS}^{+}$ is the extension of MS with *generalized contractions*, i.e. contractions with an arbitrary number of premises, on the style of [BM10, Laf04]. Then, $\text{MS}^{\dagger}$ extends MS by the use of the *levels* of [BM10]. At last we shall consider *untyped proof nets*, and we shall allow some *recursive types* on the labels, as [BM07]. The reason for all these generalizations is twofold. On one side, this potentially allows to capture inside MS other characterizations of FPTIME different from ILAL: for example, $\text{L}^{4}$ [BM10]. On the other side, as we said, MS revealed not as expressive as we wanted: we were not able to encode all the SRN programs inside any subsystem of MS. So, maybe, a *generalized* subsystem of MS could be used for this purpose. In Chapter 7 we have seen that this sometimes is true: the system LALL is not a subsystem of MS, but it is a generalized subsystem, and it has revealed useful for our research. LALL simulates all the SRN programs.

## 8.1 Beyond soLAL

As we pointed out, soLAL strictly extends ILAL, but we can't embed the whole SRN into soLAL. Here we propose two variation co-soLAL and $\text{soLAL}^{\infty}$ of soLAL that strictly extend soLAL, and that could attain the target. IEAL will be considered in Section 8.1.3.

### 8.1.1 co-soLAL

co-soLAL is the extension of soLAL with *level-coercions*, i.e. axioms extraneous to MS that prove $j\S A \multimap i\S A$ for every $i < j$. Essentially, these nodes make all the modalities equivalent, so that co-soLAL is not strongly polytime; it is as expressive as IEAL (Intuitionistic Elementary Affine Logic). "Equivalent" here is used with a technical meaning: the modalities are all equivalent under $\Re_{\uparrow}$. On the other side it is easy to encode SRN into co-soLAL. We show that there exists a polytime reduction strategy, but using it not all the SRN programs reduce. So, at the moment, co-soLAL does not reach the prefixed target. Anyway, we conjecture that some strategy exists such that co-soLAL is polytime and all the SRN programs can reduce.

Definition 8.1.1 (co-soLAL) *co-soLAL is the system soLAL plus the LC nodes (*level-coercions*) for every $i < j$:*

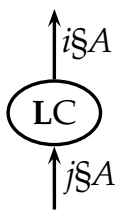

Notice that co-soLAL is not a subsystem of MS, because of the presence of the external nodes LC.

Lemma 8.1.2 (About Equivalence of Modalities)
In co-soLAL, for every pair of modalities $(\mathfrak{m}, \mathfrak{n})$, where $\mathfrak{n} \neq 1!$, it is possible to prove $\vdash \mathfrak{m}\alpha \multimap \mathfrak{n}\alpha$, using the "Flat" proof nets in Figure 8.1. On the other hand, it is not possible to prove $\mathfrak{m}\alpha \multimap 1!\alpha$, for any modality $\mathfrak{m} \neq 1!$.

Lemma 8.1.3 (co-soLAL is Not Strongly Polytime)
There exists a proof net in co-soLAL that takes an integer $n$ and calculates $2^n$. Such a proof net is in Figure 8.1.

Nevertheless, the system co-soLAL is polynomial-time in the following, weaker, sense.

Definition 8.1.4 (Fast Strategy) *A reduction $\Pi \to \Sigma$ is* **fast** *if the* ns $[\mathbf{Y}/P]$ *is applied only when the box does not contain any* LC *node.*

Lemma 8.1.5 (Weak Polynomiality of co-soLAL)
There exists a family of polynomials $\{p_{\partial,L}(x) \mid \partial, L \in \mathbb{N}\}$ such that for every proof net $\Pi$ of co-soLAL, whose depth is at most $\partial$, contains at most $L$ LC nodes, and for every *fast* reduction $\Pi \to_k \Sigma$, $p_{\partial,L}(|\Pi|)$ bounds both $k$ and $|\Sigma|$.

**Proof.** Let $\Pi$ be as in the hypothesis, and $b$ be a fixed box at depth $d$ in $\Pi$. We want to show a bound on the number of consecutive spindles in $\Pi$, so we will able to adapt Lemma 4.3.31 to this situation. Let us consider a chain of $N$ consecutive spindles. The spindles are connected through modal formulas $i_1!A_1, \dots, i_{n-1}!A_{n-1}$ whose sorts $i_j$ strictly increase, unless some LC box is present. Without LC's, at most $M$ consecutive spindles may exist. Each LC let to add at most $M$ new spindles to the chain; so $M \cdot L$ is a bound for the number of consecutive spindles. Now, thanks to (the proof of) Lemma 4.3.31, for each box $b$ it holds $R_{\Pi}(b) \leq |\Pi|^{M \cdot L+1}$. During the reduction, $L$ does not increase because we are using the fast strategy. Now, we can look

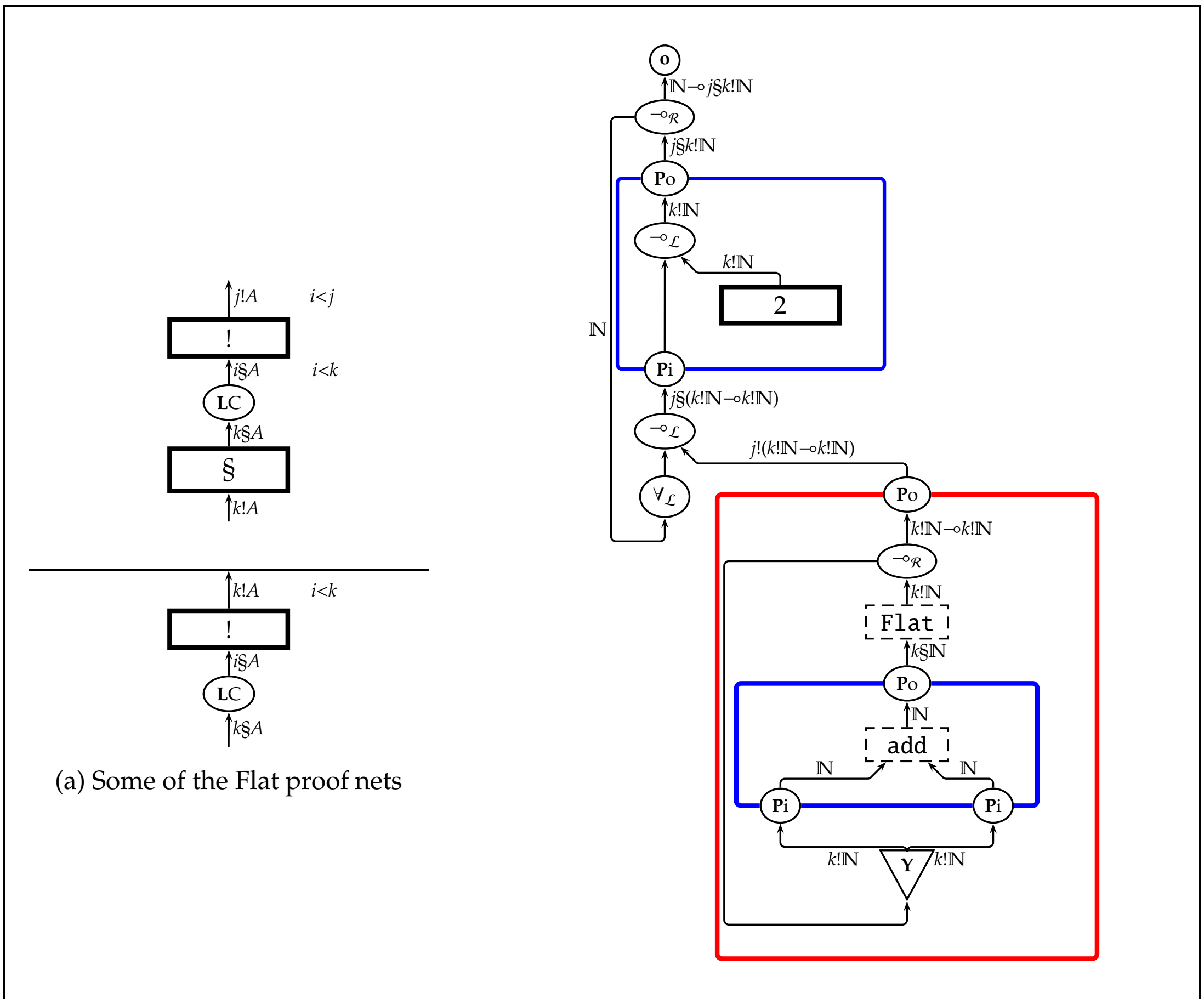

(a) Some of the Flat proof nets

(b) The exponential proof net, with $k \geq 2$

Figure 8.1: This proof net, in co-soLAL, if reduced, reduces in exponential time. It corresponds to the $\lambda$-term $\lambda n.(\lambda y.\mathtt{add}\ \ yy)\overline{2}$. The integers are $\mathbb{N} = \forall\alpha.j!(\alpha \multimap \alpha) \multimap j\S(\alpha \multimap \alpha)$ for some $j$.

back at the proof of Lemma 4.3.24 and we understand that it still holds: our bound on $R_\Pi$ implies a bound on the normalization time and size. □

Notice that we cannot deduce that $\mathcal{P}$ is polynomial time, according to Definition 4.1.4, but just that there is a polynomial that bounds the reduction as described in the statement of this Lemma.

The *fast reduction* is not enough to represent SRN in co-soLAL. Indeed, let us think again to the recursive scheme. It takes a function $g : j\mathbb{W} \multimap (j\mathbb{W}^n) \multimap (j\mathbb{W})^s \multimap j\mathbb{W}$, encloses it in a box, duplicates it as many times as needed, and at last applies the functions to the argument. Then, the recursion scheme returns a function $f : j\mathbb{W} \multimap (j\mathbb{W}^n) \multimap \left(i\S^3 j\S j\mathbb{W}\right)^s \multimap i\S^4 j\mathbb{W}$; it is necessary at least one LC node to transform it to the right type $f' : j\mathbb{W} \multimap (j\mathbb{W}^n) \multimap (j\mathbb{W})^s \multimap j\mathbb{W}$. To iterate $f'$, we put it inside a box, thus we would need to duplicate a box containing an LC node, and the fast strategy does not allow this.

### 8.1.2 soLAL$^\infty$

In this Section, again, we extend soLAL. soLAL$^\infty$ is the generalization of soLAL to infinite modalities. We show that the system soLAL$^\infty$ is not polytime, and nevertheless all the functions that can be represented in it are polytime.

Please notice that soLAL$^\infty$ is not sensible, because every sensible subsystem has only finite modalities. So, we will not be able to use the general theorems. soLAL$^\infty$ fits, on the contrary, in the setting described in Section 4.5.

soLAL$^\infty$ is the union of all the soLAL$_M$, for every number of modalities $M$. That is, soLAL$^\infty$ is a system with infinite modalities: $\mathbb{X} = \{i! \mid i \in \mathbb{N}\} \cup \{i\S \mid i \in \mathbb{N}\}$. soLAL$^\infty$ is not polytime, because it can build chains of spindles arbitrarily long. However, we can show that:

Lemma 8.1.6 (Polynomiality of soLAL$^\infty$)
There exists a family of polynomials $\{p_{\partial,M}(x) \mid \partial, M \in \mathbb{N}\}$ such that for every proof net $\Pi$ of soLAL$^\infty$, whose depth is at most $\partial$ and containing at most $M$ modalities, $p_{\partial,M}(|\Pi|)$ bounds both $[\Pi]$ and $||\Pi||$.

**Proof.** Every such proof net is a proof net of soLAL$_M$. So, $p_{\partial,M}(x)$ is the polynomial provided by the polynomiality of soLAL$_M$ (Fact 6.2.1). Notice moreover that the number $M$ cannot increase during the reduction, this due to the peculiar form of the rewriting rules of soLAL. □

Corollary 8.1.7 (Polynomiality of soLAL$^\infty$, #2)
Let $\Pi : i_1\mathbb{W}, \ldots, i_k\mathbb{W} \vdash j\mathbb{W}$ be a proof net of soLAL$^\infty$ that represents a function $f$. Then, $f$ is a polytime function.

**Proof.** The thesis means that: there exists a polynomial $p_\Pi(x_1, \ldots, x_k)$ such that for every word $\overline{w}_1 : i_1\mathbb{W}, \ldots, \overline{w}_k : i_1\mathbb{W}$, the proof net obtained cutting $\Pi$ with $\overline{w}_1, \ldots, \overline{w}_k$ reduces in time and size bounded by $p_\Pi(|\overline{w}_1|, \ldots, |\overline{w}_k|)$.

Let us call $\Sigma = \Pi \bowtie \overline{w}_1 \bowtie \ldots \bowtie \overline{w}_k$. $\Sigma$ is in soLAL$_M$, for some $M$. Let $d = \partial(\Sigma)$. Thus, according to the previous Lemma, we can find a polynomial $p(x)$ such that the reduction time and size of $\Sigma$ is bounded by $p(|\Sigma|)$. Now, notice that the parameters $|\Pi|$, $M$ and $d$ do not depend on the particular values of the arguments $\overline{w}_i$'s. The size of $\Sigma$ is essentially the size of the arguments of $\Pi$: $|\Sigma| = |\Pi| + |\overline{w}_1| + \ldots + |\overline{w}_k|$, so we may choose $p_\Pi(|\overline{w}_1|, \ldots, |\overline{w}_k|) = p(|\Pi| + |\overline{w}_1| + \ldots + |\overline{w}_k|)$. □

We conjecture that it is possible to represent into soLAL$^\infty$ some functions not in SRN$^-$. However, the encoding is not completely trivial.The work in progress under this line is based on the following preliminary definitions:

Definition 8.1.8 SRN$^\pm$ *is the set containing* SRN$^-$*, plus all the functions that can be obtained from* SRN$^-$ *with 1 application of* (not necessarily linear) *safe recursion.*

Definition 8.1.9 *Let $f : \mathbb{N}^k \to \mathbb{N}$ be a computable function, and $\Pi$ be a proof net of* soLAL$^\infty$*. For every $l \in \mathbb{N}$, we say that $\Pi$* **simulates** *$f$* **with $l$-bounded inputs** *if, whenever $f(x_1, \ldots, x_k) = x$ and $|x_1|, \ldots, |x_k| \leq l$, the proof net obtained cutting $\Pi$ with the corresponding proof nets $\hat{x}_1, \ldots, \hat{x}_k$ reduces to $\hat{x}$.*

Then, our conjecture is:

Conjecture 8.1.10 (soLAL$^\infty$ represents SRN$^\pm$ for $l$-bounded inputs) *There exists a compositional map* $\lceil\cdot\rceil^{\cdot}$ *that translates every* SRN$^\pm$ *program* $f(\vec{x};\vec{y})$ *into a proof net* $\lceil f(\vec{x};\vec{y})\rceil^l$ *of* soLAL$^\infty$, *such that* $\lceil f(\vec{x};\vec{y})\rceil^l$ *represents* $f(\vec{x};\vec{y})$ *with l-bounded inputs.*

**Proof idea.** Let us imagine we want implement a proof net $\Pi$ calculating the recursive function starting from functions $g, h_0, h_1 \in$ SRN$^-$. Instead of really implementing a recursion, we just implement a composition of a sufficient number of copies of the step functions $h_0, h_1$. The greater is $l$, the greater is the tier of the output of $\Pi$; so it is necessary to work in soLAL$^\infty$, and not just insoLAL. □

### 8.1.3 IEAL

We have already pointed out (page 89) that IEAL is a non-polytime subsystem of MS. Here we describe how to map every SRN program into IEAL in a compositional way. Of course, this is not very interesting because IEAL is not polytime; anyway, we hope that some day IEAL will be proved weakly polytime, under some reduction strategy $\sigma$ that we don't know yet.

The most natural translation is an adaptation of the Murawski-Ong coding for ILAL. $f(\vec{x};\vec{y})$ is mapped into $\Pi : \mathbb{W}, \ldots, \mathbb{W}, !^m\, \mathbb{W}, \ldots, !^m\, \mathbb{W} \vdash !^m\, \mathbb{W}$, with $\mathbb{W} = !(\alpha \multimap \alpha) \multimap !(\alpha \multimap \alpha)$ and for some $m \geq 0$. Of course all the functions in BC$^-$ can be encoded in this way; but, this time, it is possible to contract the safe variables, thus getting full SRN. The number $k$ is connected to the number of safe recursions used to build $f$.

This result is not interesting, **not** only because IEAL is elementary-time, but because it is possible to represent in IEAL some elementary functions. Nevertheless, notice that some of the restrictions on the safe/normal variables typical of SRN still hold in IEAL. For example, IEAL proves $\mathbb{W} \multimap !\,\mathbb{W}$ but $!\,\mathbb{W} \not\multimap \mathbb{W}$.

Now, the real problem lies in juxtaposition: in IEAL it can be encoded as `concat` : $\mathbb{W} \multimap \mathbb{W} \multimap \mathbb{W}$, that is with only safe variables. As we said in Section 2.9, this allows the construction of some non-polytime function. So, the unknown strategy $\sigma$ that we mentioned before, necessarily must not reduce such a proof net `concat`.

Another related observation.

Proposition 8.1.11
Let $\mathcal{P} \subseteq$ MS be a subsystem that enjoys the strong normalization property. Let us suppose that $\mathcal{P}$ is weakly polytime. Then, for every $\Pi \in \mathbf{PN}(\mathcal{P})$, $|\mathrm{nf}(\Pi)|$ is polynomial in $|\Pi|$.

**Proof.** Almost immediate. By hypotheses there is a normalization strategy $\sigma$ such that $\Pi$ reduces to $\mathrm{nf}(\Pi)$ in polynomial time and polynomial size. So, $\mathrm{nf}(\Pi)$ has polynomial size. □

The latter Proposition is quite naïve, and nevertheless captures a behaviour that is quite uncommon in the practice of subsystems of MS: firstly, there exists a normal form; secondly, the size of $\mathrm{nf}(\Pi)$ is polynomial, independently on the chosen (possibly elementary-time) normalization strategy. This is exactly the behaviour of SRN.

## 8.2 Generalizations

Let us return to the original problem: embedding SRN into some stratified logic. Instead of going on about soLAL, we could imagine several different light logic for this purpose. In particular, we can imagine to extend the framework MS with several different constructions, without giving up to the *stratification*.

### 8.2.1 Generalized Contractions

The *multiplexor* of [Laf04], as well as the *contraction* of [BM10], have an arbitrary number of premises. Let us consider a contraction with $n \geq 1$ conclusions:

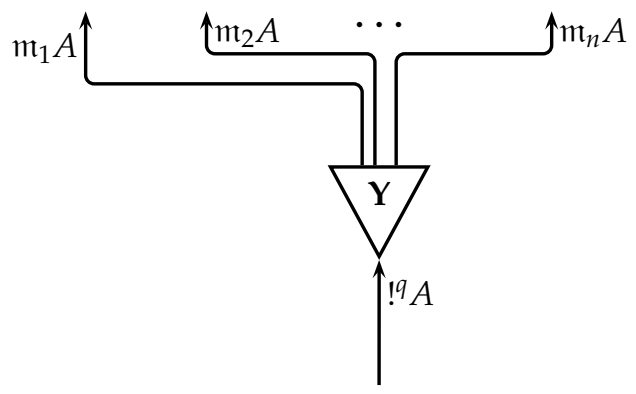

$n \geq 1$

We denote $\mathbf{Y}_{\mathfrak{q}}(\mathfrak{m}_1, \ldots, \mathfrak{m}_n)$ such a rule. $n$ is called the *rank* of the contraction. The *rank* $\text{rk}(\Pi)$ of a proof net $\Pi$ is the maximum of the ranks of its contractions; and we will call $\text{rk}_d(\Pi)$, for $d \leq \partial(\Pi)$, the maximum of the ranks of the contractions at depth $d$. $\mathsf{MS}^+ \supseteq \mathsf{MS}$ is the framework including all these new rules. We want to show that all the theorems for MS still hold in $\mathsf{MS}^+$.

Lemma 8.2.1 (Generalization of Corollary 4.3.25)
Let $\mathcal{P} \subseteq \mathsf{MS}^+$. Let us assume there exists a polynomial $r(x)$ such that for every proof net $\Pi$ of $\mathcal{P}$ and for every $d$ $\mathrm{W}_d(\Pi) \leq r(|\Pi|)$. Then $\mathcal{P}$ is strongly polytime.

**Proof.** The proof is long, but similar to the one of Corollary 4.3.25. The only difference is the definition of the modified weight $T_d$ (defined on page 55):

$$\mathrm{T}_d(\Pi) = \sum_{u \in V_\Pi^d} \mathrm{T}_d(\Pi, u)$$

$$\mathrm{T}_d(\Pi, u) \stackrel{\text{def}}{=} \begin{cases} 1 & u \in \{\mathbf{I}, \text{cut}, \mathbf{W}, \mathbf{h}, Pi(\mathfrak{m}), \mathbf{i}, \mathbf{o}\} \\ 3 & u \in \{\otimes_{\mathcal{L}}, \otimes_{\mathcal{R}}, \forall_{\mathcal{L}}, \forall_{\mathcal{R}}, \multimap_{\mathcal{L}}, \multimap_{\mathcal{R}}, \mathbf{Y}_{\mathfrak{q}}(\mathfrak{m}, \mathfrak{n})\} \\ 3 \cdot (P_\Pi(u) + 1) \cdot \sum_{\tau \in O(b)} \text{len}(\tau) & u \in \{Po(\mathfrak{m})\}. \end{cases}$$

Notice that the only difference is the case for $Po(\mathfrak{m})$. The modification is due to the $[P/\mathbf{Y}]$ reduction rule: here, we do not know *a priori* the number of premises of the contraction, so a more complicated expression is needed. The new expression involves the *length* $\text{len}(\tau)$ of the paths starting from the $Po$ node, exactly as in [DL08]. □

Lemma 8.2.2 (Generalization of Lemma 4.3.31: Short Chains and Number of Paths)
Let $\mathcal{P} \subseteq \mathsf{MS}^+$. Let us assume that every chain of spindles that can be built in $\mathcal{P}$ cannot have more than $L$ spindles, for some constant $L \in \mathbb{N}$. Then there exists a polynomial $p(x)$ such that for every $\Pi \in \mathbf{PN}(\mathcal{P})$, for every $d \leq \partial(\Pi)$ and for every $b \in B_\Pi^d$,

$$R_\Pi(b) \leq p(|\Pi|_d).$$

**Proof.** Let us fix a box $b$ at depth $d$; we want to count how many copies of $d$ will be created during the reduction. We call $n = \mathrm{rk}_d(\Pi)$. Then, we can proceed as in the proof of Lemma 4.3.31 at page 61: we divide the nodes of $\Pi$ at depth $d$ in (at most) $L+1$ classes (we *do not* write again the classification, because it can be found in the cited proof). $\Pi_i$ is the module (modules were defined at page 32) containing all the nodes of class at most $i$, and $m_i$ is the number of (generalized) contractions of class-$i$. $R_i$ is the number of paths starting from $b$ and maximal relatively to $\Pi_i$, so that $R_\Pi(b) = R_L$. Every contraction may split a path into at most $n$ new paths, so

$$
\begin{aligned}
R_0 &\leq (n-1)\cdot m_0 + 1 \\
R_{i+1} &\leq R_i \cdot ((n-1)\cdot m_i + 1) \\
R_\Pi(b) &= R_L \leq \prod_{j=0}^{L}((n-1)\cdot m_i + 1) \leq n^{L+1}\cdot |\Pi|_d{}^{L+1} \leq |\Pi|_d{}^{2(L+1)}.
\end{aligned}
$$

□

Proposition 8.2.3 (Generalization of Proposition 4.3.40: A Polynomiality Criterion)
Let $\mathcal{P} \subseteq \mathsf{MS}^+$ sensible. The following are equivalent:
1. $\mathcal{P}$ is polytime.
2. $\mathcal{P}$ cannot build any dangerous spindle.

**Proof.** 2. ⇒ 1. Follows from Lemma 8.2.2. 1. ⇒ 2. The presence of a dangerous spindle leads to the construction of a family of proof nets that reduces in exponential time, exactly as in Lemma 4.3.39. □

Theorem 8.2.4 (Generalization of Theorem 4.4.25: Structure of Maximal Sensible Subsystems)
Let $\mathcal{P} \subseteq \mathsf{MS}^+$ be a sensible subsystem with a linear $\preceq$ and with modalities in $\mathbb{X}$. $\mathcal{P}$ is maximal iff $\mathcal{P}$ contains exactly the following rules:
1. all the rules $\mathbf{Y}_{\mathfrak{q}}(\mathfrak{m},\mathfrak{n})$ for every choice of $\mathfrak{q} \prec \mathfrak{m},\mathfrak{n}$ in $\mathbb{X}$;
2. all the rules $\mathbf{Y}_{\mathfrak{q}}(\mathfrak{q},\mathfrak{n})$ for every choice of $\mathfrak{q} \prec \mathfrak{n}$ in $\mathbb{X}$;
3. all the rules $P_{\mathfrak{q}}(\mathfrak{q}^?,\mathfrak{m}_1^*,\dots,\mathfrak{m}_k^*)$ for every choice of $\mathfrak{m}_1,\dots,\mathfrak{m}_k \prec \mathfrak{q}$ in $\mathbb{X}$;
4. only one among $\mathbf{Y}_{\mathfrak{q}}(\mathfrak{q},\mathfrak{q})$ and $P_{\mathfrak{q}}(\mathfrak{q}^*,\mathfrak{m}_1^*,\dots,\mathfrak{m}_k^*)$, for every $\mathfrak{m}_1\dots,\mathfrak{m}_k \prec \mathfrak{q}$ in $\mathbb{X}$.

The proof is analogous to the one of Theorem 4.4.25.

### 8.2.2 Levels

The system Light Linear Logic by Levels, or $\mathsf{L}^4$, of [BM10] is an extension of ILAL characterized by the notion of *levels* and *indexings*. Intuitively, the system allows some rules that, *apparently*, seem some *derelictions*:

$$\frac{\Gamma, A \vdash B}{\Gamma, \S A \vdash B}\ \S \qquad \frac{\Gamma \vdash B}{\Gamma, !A \vdash B}\ !$$

However, inside a proof, each occurrence of such rules must be in some way *balanced* with other occurrences, thus forming some *levels* inside the proof net; as a consequence each group of such rules appear more similar to a promotion than to a dereliction. $\mathsf{L}^4$ allows only such *balanced* proofs. The formal definition of proof net of $\mathsf{L}^4$ relies on the notion of *indexing*, a proof net being accepted if and only if it allows an indexing.

Here we will extend MS with *levels*. The new framework is called $\mathsf{MS}^{\dagger}$. There will be two different kinds of boxes: some ones will be *physical*, with a clear border, and containing a proof net inside them; the other ones will be *fuzzy*, their border will not be drawn and they can contain a graph that is not necessarily a proof net. *A fuzzy box can also have more conclusions!* Just to help the readability, sometimes we will draw these fuzzy boxes, with a dashed line. However, we consider only a very simplified situation: there will be no way to duplicate the content of a fuzzy box. This is what happens, e.g., in $\mathsf{L}^4$: the !-boxes are physical and can be duplicated, while the §-boxes are *fuzzy* and cannot be duplicated. This property will be assured by the presence of two disjoint sets $\mathbb{X}, \mathbb{Y}$ of modalities: the modalities belonging to $\mathbb{X}$ will behaves differently from the modalities in $\mathbb{Y}$.

We find quite difficult to handle levels. The only result of this Section is a sufficient condition for polytime soundness (Proposition 8.2.21). To prove it, it will be necessary the whole power of the CS as it appears in [DL08]. We now try to give an intuition of *why* our CS is not sufficient. The subsystems that use indexings are not really stratified, but just *weakly* stratified. For example, a normal box can *intersect* with a fuzzy box. In this case, it can still happen that a CS-path enters into one of such boxes; but the point is, where should it continue? after the *solid* box, or after the *fuzzy* box? or none? The only way to manage the situation is to consider all the possible CS-paths provided by the standard CS, and not only *our* CS-paths.

Formally, let us fix two sets of modalities $\mathbb{X}$ and $\mathbb{Y}$ disjoint. We are going to define the *quasi-proof nets*: only some of them will be called *proof nets*. The **quasi-proof nets** of $\mathsf{MS}^{\dagger}_{\mathbb{X},\mathbb{Y}}$ are defined by induction using all the linear and affine cases of Definition 3.1.3 at page 26, plus the following non linear cases:

**Rule $P_{\mathfrak{q}}(\mathfrak{m}_1, \ldots, \mathfrak{m}_k)$ or Promotion:** If $\mathfrak{q} \in \mathbb{X}$, and $\mathfrak{m}_1, \ldots, \mathfrak{m}_k \in \mathbb{X} \cup \mathbb{Y}$, then the following is a proof net:

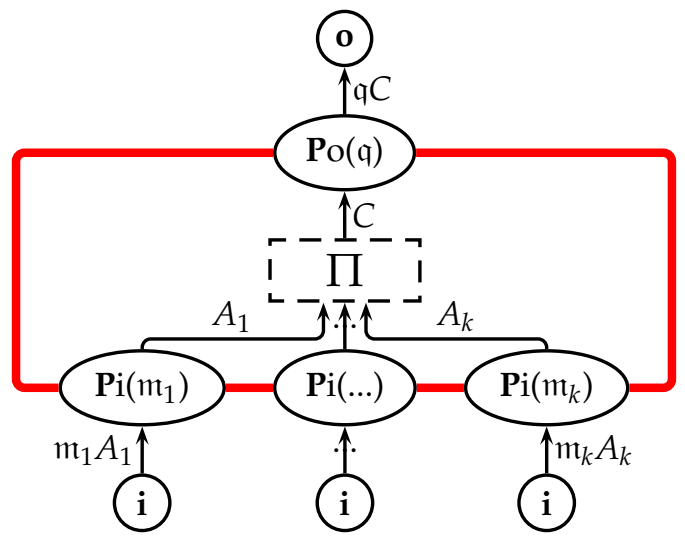

**Rule $\mathbf{Y}_{\mathfrak{p}}(\mathfrak{m}, \mathfrak{n})$ or Contraction:** If $A_i = \mathfrak{m}A$ and $A_j = \mathfrak{n}A$ for some $i, j$, and moreover $\mathfrak{p} \in \mathbb{X}$ and $\mathfrak{m}, \mathfrak{n} \in \mathbb{X} \cup \mathbb{Y}$, then the following is a proof net:

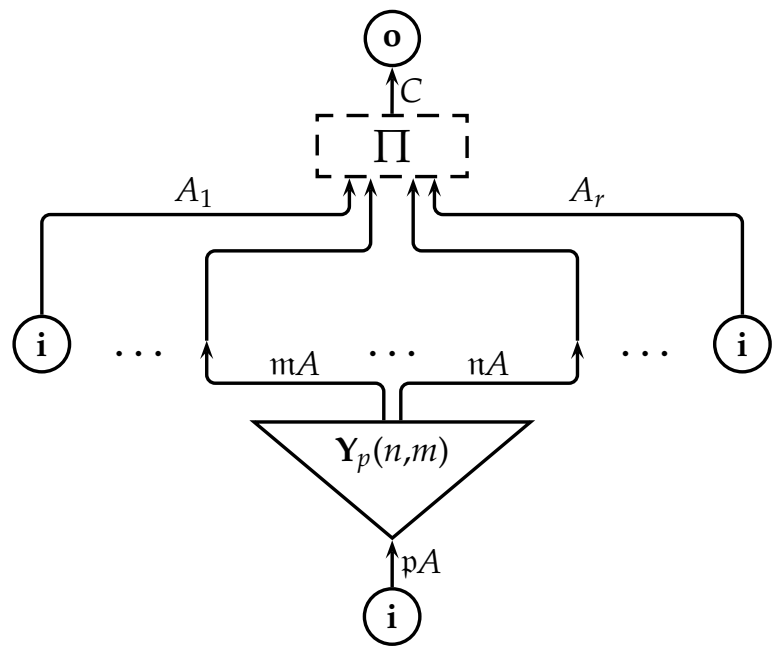

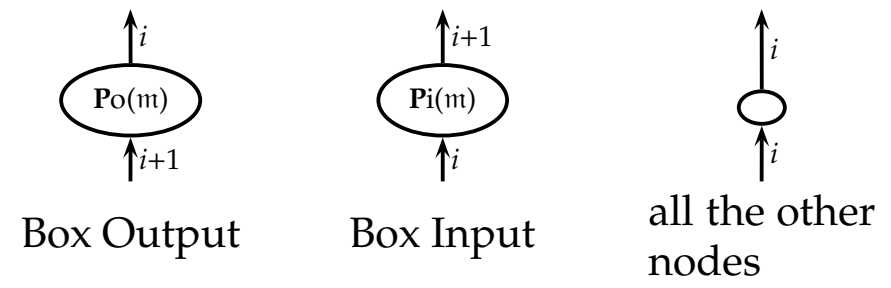

Figure 8.2: Constraints for indexing proof structures.

**Port right:** If $\mathfrak{m} \in \mathbb{Y}$, the following is a proof net:

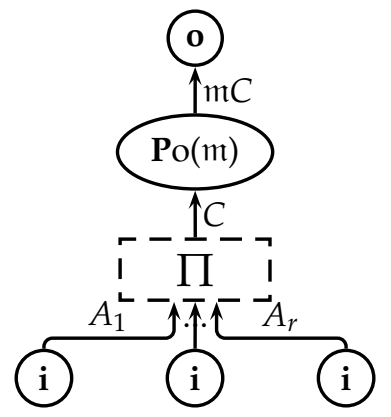

**Port left:** If $\mathfrak{m} \in \mathbb{X} \cup \mathbb{Y}$, the following is a proof net:

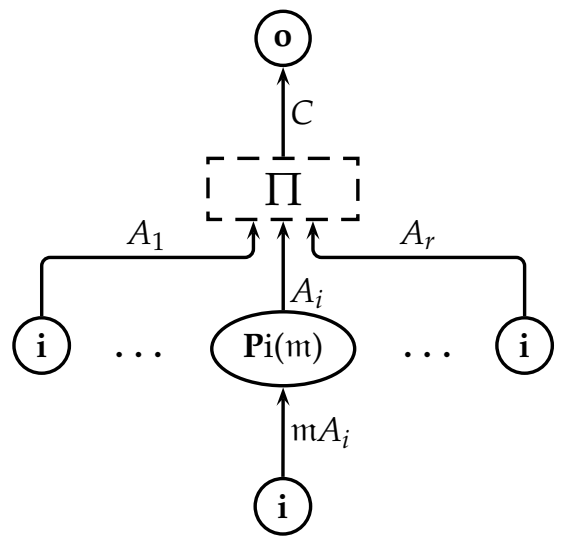

DEFINITION 8.2.5 (INDEXING) *Let $\Pi$ be a quasi-proof net of $\mathsf{MS}_{\mathbb{X},\mathbb{Y}}$. An* **indexing** *for $\Pi$ is a function $I$ from the edges of $\Pi$ to $\mathbb{Z}$ that satisfies the constraints in Figure 8.2 and such that $I(e) = I(e')$, for every premise or conclusion $e, e'$ of $\Pi$. An indexing $I$ is* **canonical** *if $\Pi$ has an edge $e$ such that $I(e) = 0$, and $I(e') \geq 0$ for all edges $e'$ of $\Pi$.*

DEFINITION 8.2.6 (PROOF NETS) *Let $\Pi$ be a quasi-proof net. $\Pi$ is a proof net of $\mathsf{MS}_{\mathbb{X},\mathbb{Y}}$ iff it admits an indexing.*

REMARK 8.2.7 A box input is *not* a dereliction. The indexing forces every box input to be *balanced* with some box output, so that in fact box inputs and outputs form a boundary for a subgraph that we can call a *fuzzy box*: a region with a fuzzy border.

As an example, $\mathsf{L}^4$ is a subsystem of $\mathsf{MS}^{\dagger}_{\mathbb{X},\mathbb{Y}}$ with $\mathbb{X} = \{!\}$ and $\mathbb{Y} = \{\S\}$.

FACT 8.2.8 (RIGIDITY)
If $I$ is an indexing for $\Pi$ and $n \in \mathbb{Z}$, then also $I'(e) \stackrel{\text{def}}{=} I(e) + n$ defines an indexing for $\Pi$.

FACT 8.2.9 (CANONICAL INDEX)
Every proof net of $\mathsf{MS}_{\mathbb{X},\mathbb{Y}}$ has one and only one canonical indexing.

We will write $I_0$ for the canonical indexing and $I(l)$ for the indexing that takes value $l$ on the premises. The worth point is that the $I_0$ can take its 0 value on an edge that is not a premise:

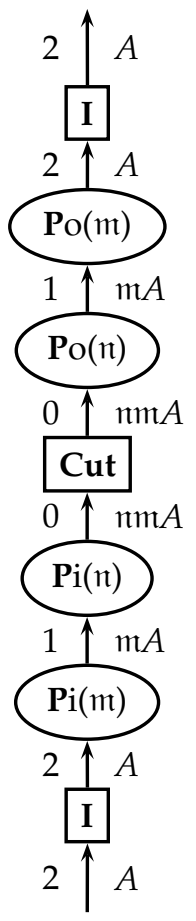

Definition 8.2.10 (Level) *Let $\Pi$ be a proof net with canonical indexing $I_0$. For $e \in E_\Pi$, the level of $e$ is $l(e) = I_0(e)$. For $u \in V_\Pi$, $l(v)$ is the least value taken by $I_0$ on the edges incident in $u$. The level of $\Pi$, $l(\Pi)$, is the greatest value taken by $I_0$ on $E_\Pi$.*

Thanks to our conditions over $\mathbb{X}, \mathbb{Y}$, it follows that:

Fact 8.2.11 (Contractive cuts)
Let $c$ be a cut of kind $[Po(\mathfrak{m})/\mathbf{Y}_\mathfrak{m}(\mathfrak{r}, \mathfrak{s})]$. Then, $\mathfrak{m} \in \mathbb{X}$.

On the other side, if $c'$ is a cut of kind $[Po(\mathfrak{m})/Pi(\mathfrak{m})]$, then $\mathfrak{m}$ can be either in $\mathbb{X}$ or in $\mathbb{Y}$. This means that a box may be in cut with a *fuzzy* box.

**Normalization.** The normalization is extended from MS to $\mathrm{MS}^\dagger$ in the natural way, replacing the $[P/P]$ *ns* with $[Po(\mathfrak{m})/Pi(\mathfrak{m})]$, for every $\mathfrak{m} \in \mathbb{X} \cup \mathbb{Y}$. This *ns* annihilates the two involved nodes. The boxes, both physical and fuzzy, merge in the natural way. 2 of the 4 possible combinations are in Figure 8.3.

Notice that, thanks to Fact 8.2.11, a cut $[Po(\mathfrak{m})/\mathbf{Y}_\mathfrak{m}(\mathfrak{r}, \mathfrak{s})]$ may exist only if the $Po$ node lies on the border of a box; such a cut reduces in the usual way. This is the main reason that motivates our restrictions on $\mathbb{X}$ and $\mathbb{Y}$ in the construction of the proof nets.

Notice also that $\mathrm{MS}^\dagger$ is not stratified, in the usual sense, as Figure 8.3 shows: the *fuzzy* boxes do not influence the depth of the nodes. In that Figure, all the nodes of $\Sigma$ change their depth.

We now present some technical Facts that are useful when studying the reduction of the subsystems of $\mathrm{MS}^\dagger$. Such Facts appear complicated, but, in fact, they are absolutely trivial.

Fact 8.2.12 (Preservation of the Canonical Indexing)
Let $\Pi \to_S \Sigma$. If $S$ is a linear *ns* or a $[P/\mathbf{Y}]$ *ns*, $u \in E_\Pi$ and $f \in E_\Sigma$ is a residual of $e$, then $I_0(e) = I_0(f)$. That is, the canonical indexing is preserved. Else, if $S$ is a garbage collection step or a $[Po/Pi]$ step, then $I_0(e) \geq I_0(f)$. That is, the canonical index may have *lifted down*.

Fact 8.2.13
Let $\Pi \to^* \Sigma$. Then $\partial(\Pi) \geq \partial(\Sigma)$ and $l(\Pi) \geq l(\Sigma)$.

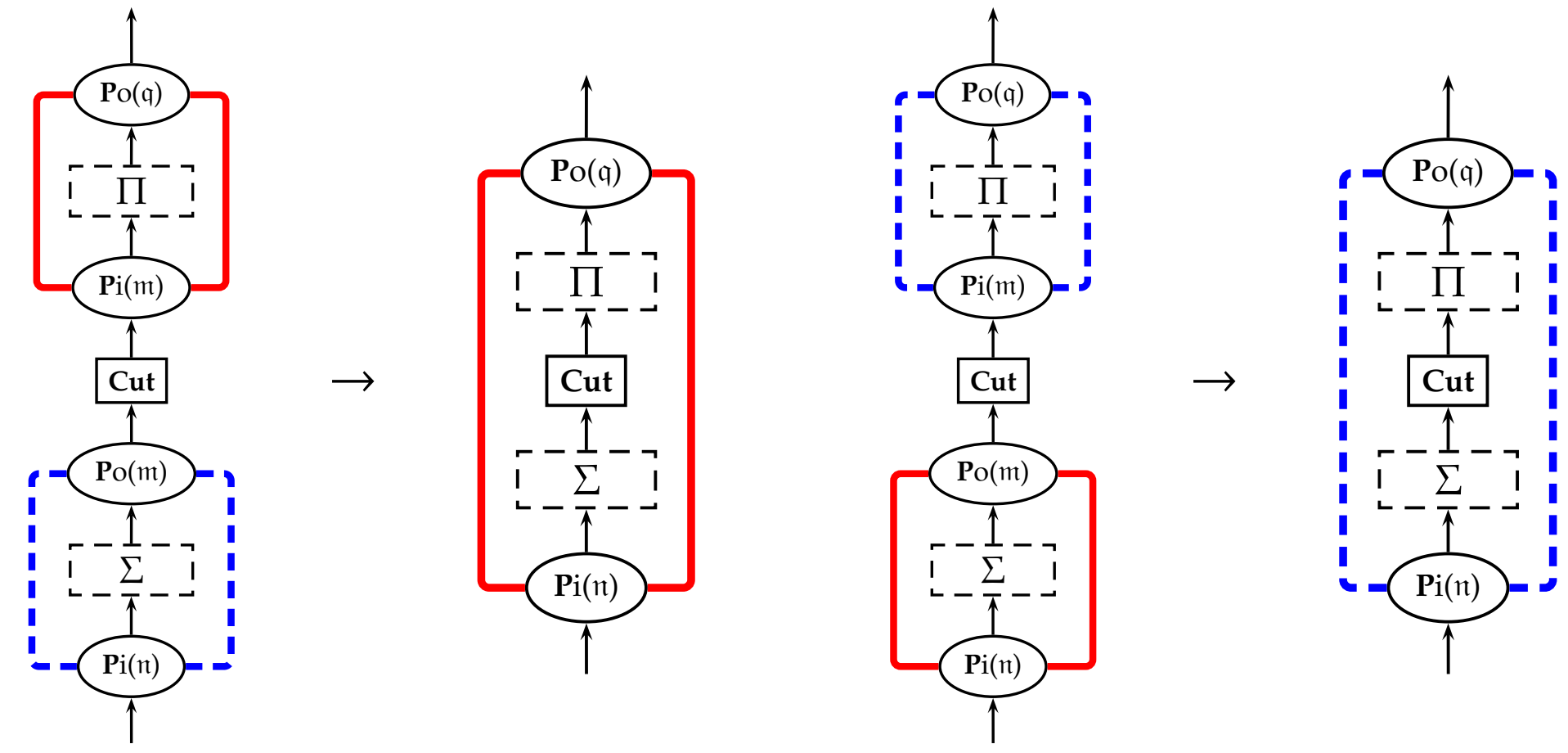

(a) $[Po(\mathfrak{m})/Pi(\mathfrak{m})]$, with $\mathfrak{m} \in \mathbb{Y}, \mathfrak{q} \in \mathbb{X}$ (b) $[Po(\mathfrak{m})/Pi(\mathfrak{m})]$, with $\mathfrak{m} \in \mathbb{X}, \mathfrak{q} \in \mathbb{Y}$

Figure 8.3: Two of the 4 possible cases in the reduction $[Po(\mathfrak{m})/Pi(\mathfrak{m})]$. For clarity reasons, we have drawn all the boxes with one only premise and one only conclusion.

FACT 8.2.14
Let $\Pi \to_S \Sigma$. We consider the same indexing $I(l)$ over $\Pi$ and $\Sigma$. $S$ is a *ns* involving a cut of $\Pi$ at depth $d$ and index $i$. If $S = [Po/\mathbf{Y}]$, then $\Sigma$ can contain new cuts only at depth $\geq d$. If $S = [Po/Pi]$, then $\Sigma$ can contain new cuts only at level $\geq i$. For every other $S$, $\Sigma$ can contain new cuts only at depth $d$ and level $i$.

Our notion of context semantics (Section 4.3.1) appears too weak to handle *levels*. From now on, we will use the full power of context semantics of [DL08]. Context semantics is generally used to study properties of *strong* polynomiality, i.e. independently from the reduction strategy. To simplify the work, on the contrary, we will require that *gc* steps are always performed only when no other *ns* exist. These are the slowest possible reductions: every duplication is always performed, even if the reducts will eventually disappear due to a *gc*.

We now recall the CS as in [DL08]. An **exponential signature** is a formula generated by the following grammar:

$$t \stackrel{\text{def}}{=} \mathtt{e} \mid \mathtt{r}(t) \mid \mathtt{l}(t).$$

$\mathcal{E}$ is the set of exponential signatures. A **stack element** is an element of $S \stackrel{\text{def}}{=} \{\mathtt{a}, \mathtt{o}, \mathtt{f}, \mathtt{s}, \mathtt{x}\} \cup \mathcal{E}$, for every $\mathfrak{m} \in \mathbb{X} \cup \mathbb{Y}$. A **stack** is a finite non empty sequence of stack elements. Let $\Pi$ a proof net. Then a **context** of $\Pi$ is an element of

$$C_\Pi \stackrel{\text{def}}{=} E_\Pi \times \mathcal{E}^* \times S^+ \times \mathcal{B}.$$

The rewriting relation among contexts $C \mapsto_\Pi C'$ is described in Figure 8.4.

Particular kinds of stacks $U \in S^+$ are the final stacks. We distinguish **positive** (denoted

| | | | |
|---|---|---|---|
| 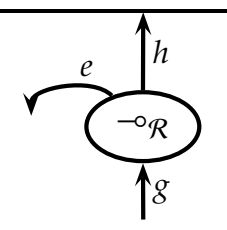 | $(g,U,V,+) \mapsto_\Pi (h,U,V\cdot \mathtt{o},+)$<br>$(e,U,V,-) \mapsto_\Pi (h,U,V\cdot \mathtt{a},+)$<br>$(h,U,V\cdot \mathtt{o},-) \mapsto_\Pi (g,U,V,-)$<br>$(h,U,V\cdot \mathtt{a},-) \mapsto_\Pi (e,U,V,+)$ | 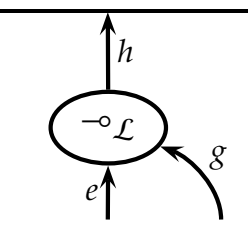 | $(g,U,V,+) \mapsto_\Pi (e,U,V\cdot \mathtt{a},-)$<br>$(h,U,V,-) \mapsto_\Pi (e,U,V\cdot \mathtt{o},-)$<br>$(e,U,V\cdot \mathtt{a},+) \mapsto_\Pi (g,U,V,-)$<br>$(e,U,V\cdot \mathtt{o},+) \mapsto_\Pi (h,U,V,+)$ |
| 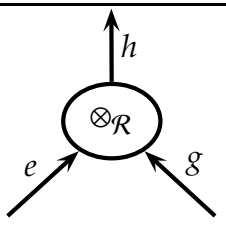 | $(e,U,V,+) \mapsto_\Pi (h,U,V\cdot \mathtt{f},+)$<br>$(g,U,V,+) \mapsto_\Pi (h,U,V\cdot \mathtt{x},+)$<br>$(h,U,V\cdot \mathtt{f},-) \mapsto_\Pi (e,U,V,-)$<br>$(h,U,V\cdot \mathtt{x},-) \mapsto_\Pi (g,U,V,-)$ | 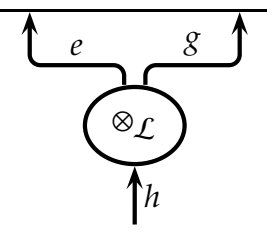 | $(h,U,V\cdot \mathtt{f},+) \mapsto_\Pi (e,U,V,+)$<br>$(h,U,V\cdot \mathtt{x},+) \mapsto_\Pi (g,U,V,+)$<br>$(e,U,V,-) \mapsto_\Pi (h,U,V\cdot \mathtt{f},-)$<br>$(g,U,V,-) \mapsto_\Pi (h,U,V\cdot \mathtt{x},-)$ |
| 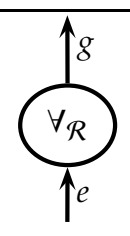 | $(e,U,V,+) \mapsto_\Pi (g,U,V\cdot \mathtt{s},+)$<br>$(g,U,V\cdot \mathtt{s},-) \mapsto_\Pi (e,U,V,-)$ | 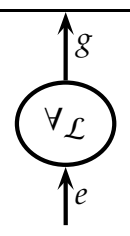 | $(e,U,V\cdot \mathtt{s},+) \mapsto_\Pi (g,U,V,+)$<br>$(g,U,V,-) \mapsto_\Pi (e,U,V\cdot \mathtt{s},-)$ |
| 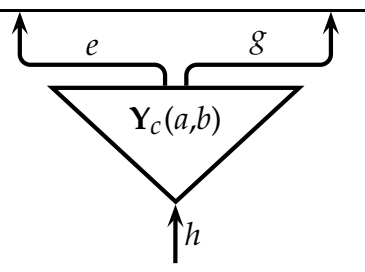 | $(h,U,V\cdot \mathtt{l}(t),+) \mapsto_\Pi (e,U,V\cdot t,+)$<br>$(h,U,V\cdot \mathtt{r}(t),+) \mapsto_\Pi (g,U,V\cdot t,+)$<br>$(e,U,V\cdot t,-) \mapsto_\Pi (h,U,V\cdot \mathtt{l}(t),-)$<br>$(g,U,V\cdot t,-) \mapsto_\Pi (h,U,V\cdot \mathtt{r}(t),-)$ | 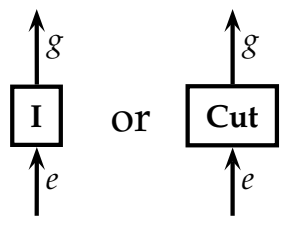 | $(e,U,V,+) \mapsto_\Pi (g,U,V,+)$<br>$(g,U,V,-) \mapsto_\Pi (e,U,V,-)$ |
| 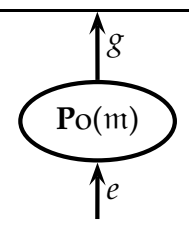 | $(e,U\cdot \mathtt{e},V,+) \mapsto_\Pi (g,U,V\cdot \mathtt{e},+)$<br>$(g,U,V\cdot \mathtt{e},-) \mapsto_\Pi (e,U\cdot \mathtt{e},V,-)$ | 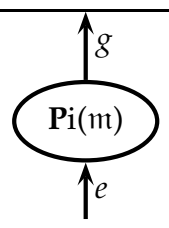 | $(e,U,V\cdot \mathtt{e},+) \mapsto_\Pi (g,U\cdot \mathtt{e},V,+)$<br>$(g,U\cdot \mathtt{e},V,-) \mapsto_\Pi (e,U,V\cdot \mathtt{e},-)$ |
| 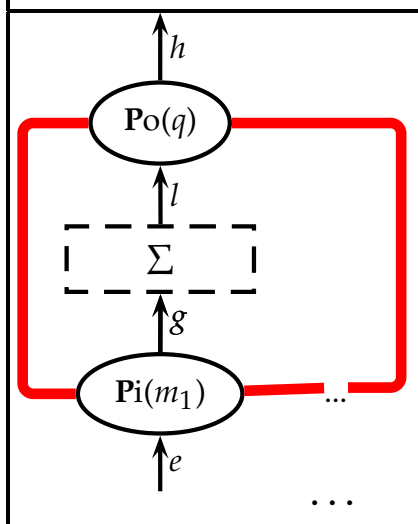 | $(e,U,V,+) \mapsto_\Pi (h,U,V,+)$<br>$(e,U,V\cdot t,+) \mapsto_\Pi (g,U\cdot t,V,+)$<br>$(h,U,V,-) \mapsto_\Pi (e,U,V,-)$<br>$(g,U\cdot t,V,-) \mapsto_\Pi (e,U,V\cdot t,-)$ | | |

Figure 8.4: Rewriting Relation among contexts. Notice that if $(e,U,V,b) \mapsto_\Pi (e',V',U',b')$ then also $(e',U',V',b'\downarrow) \mapsto_\Pi (e,U,V,b\downarrow)$.

with $P$) and **negative** (denoted with $N$) **final stacks** and define them mutually recursively:

$$P \ ::= \ \mathtt{e} \mid P\cdot\mathtt{e} \mid P\cdot o \mid P\cdot\mathtt{f} \mid P\cdot\mathtt{x} \mid P\cdot\mathtt{s} \mid N\cdot\mathtt{a};$$
$$N \ ::= \ N\cdot t \mid N\cdot\mathtt{o} \mid N\cdot\mathtt{f} \mid N\cdot\mathtt{x} \mid N\cdot\mathtt{s} \mid P\cdot\mathtt{a}.$$

For example, the stack $\mathtt{e}\cdot\mathtt{a}\cdot\mathtt{l}(\mathtt{e})$ is negative final, while $\mathtt{e}\cdot\mathtt{a}\cdot\mathtt{f}\cdot\mathtt{a}$ is positive final. A context $C \in \mathsf{C}_\Pi$ is final iff one of the following four cases hold:

- If $C = ((u,v),U,V,+)$, $\alpha(v) = \mathbf{o}$ and $V$ is a positive final stack;
- If $C = ((u,v),U,V,+)$, $\alpha(v) = \mathbf{W}$ and $V$ is a positive final stack;
- If $C = ((u,v),U,V,-)$, $\alpha(v) = \mathbf{i}$ and $V$ is a negative final stack.

A context is said **canonical** if $|U| = l(e)$.

A **CS-path** over $\Pi \in \mathbf{PN}(\mathsf{MS}\,\mathbb{X},\mathbb{Y})$ is a sequence of canonical contexts connected through the rewriting relation. A CS-path is **maximal** if it begins from a *Po* and terminates in a final context.

In Section 4.3.1 we defined the *weight* $\mathrm{W}_d$ and the *modified weight* $\mathrm{T}_d$ for each depth $d$; here,

we cannot distinguish the various depths:

$$\begin{aligned}
\mathrm{W}(\Pi) &= \sum_{b \in B_\Pi} R_\Pi(b) \qquad \text{where} \\
R_\Pi(b) &= \text{number of maximal CS-paths starting from } b. \\
\mathrm{T}(\Pi) &= \sum_{u \in V_\Pi} \mathrm{T}_\Pi(u) \qquad \text{where} \\
\mathrm{T}_\Pi(u) &= \begin{cases} 1 & \alpha(u) \notin \{Po, Pi\} \\ 0 & \alpha(u) = Pi \\ (P_\Pi(u)+1)\sum_{t \in O(u)} \operatorname{len} t & \alpha(u) = Po \end{cases}
\end{aligned}$$

The following three Lemmas are essentially proved in [DL08].

**Lemma 8.2.15 (Properties of the Modified Weight)**
Let $\Pi \to \Sigma$. (a) $\mathrm{T}(\Pi) > \mathrm{T}(\Sigma)$, so that $[\Pi] \leq \mathrm{T}(\Pi)$. (b) $|\Pi| \leq \mathrm{T}(\Pi)$, so that $||\Pi|| \leq \mathrm{T}(\Pi)$.

**Lemma 8.2.16 (Paths, Reduction Time, Used Space)**
Let $\Pi \to \Sigma$. (a) $\mathrm{W}(\Pi) \geq \mathrm{W}(\Sigma)$. (b) There exists a polynomial $p(x, y)$ such that $\mathrm{T}(\Pi) \leq p(\mathrm{W}(\Pi), |\Pi|)$, so that also $[\Pi], ||\Pi|| \leq p(\mathrm{W}(\Pi), |\Pi|)$. (c) $\mathrm{W}(\Pi) \leq [\Pi] + ||\Pi||$.

Letting $||U||$ the number of exponential signatures in the stack $U$, we can prove that:

**Lemma 8.2.17 (Stratification)**
Let $\Pi \in \mathbf{PN}(\mathsf{MS}_{\mathbb{X},\mathbb{Y}})$, and $(e, U, V, b) \mapsto^*_\Pi (e', U', V', b')$. Then $||U|| + ||V|| = ||U'|| + ||V'||$.

The Definition 4.3.26 of *spindle*, as well as the Definition 4.3.27 of *chain of spindles* are untouched, just be careful that they now relies on a different definition of CS-paths:

Definition 8.2.18 (Spindles and Dangerous Spindles) *Let $\mathcal{P} \subseteq \mathfrak{B}^\dagger_{\mathbb{X},\mathbb{Y}}$, $\Pi \in \mathbf{PN}(\mathcal{P})$; $e \in E_\Pi$ an edge labelled $\mathfrak{m}A$ entering a contraction $u$; $f \in E_\Pi$ an edge labelled $\mathfrak{n}B$ outgoing a $Po$ node $b$; $\partial(e) = \partial(f)$. A* **spindle** *$\mathfrak{m}A : \Sigma : \mathfrak{n}B$ between $e$ and $f$ (or also between $u$ and $b$) is a pair of* CS*-paths: $\tau$ from $e$ to $f$ passing through the left conclusion of $u$; $\rho$ from $e$ to $f$ passing through the right conclusion of $u$; and such that $\tau$ and $\rho$ are the only* CS*-paths connecting $e$ with $f$. $e$, $f$, $u$ are resp. the* **principal premise**, **principal conclusion** *and* **principal contraction** *of $\Sigma$. An edge that is premise (resp. conclusion) of a node of $\Sigma$, but that is not part of $\Sigma$, is said* **non-principal** *premise (resp. conclusion). We shorten $\mathfrak{m}A : \Sigma : \mathfrak{n}B$ with $\mathfrak{m} : \Sigma : \mathfrak{n}$. Finally, every $\mathfrak{m}A : \Sigma : \mathfrak{m}B$ is* dangerous.

Definition 8.2.19 (Chains of Spindles) *Let $\mathcal{P} \subseteq \mathfrak{B}_{\mathbb{X}}$, $\Pi \in \mathbf{PN}(\mathcal{P})$, $\Pi$ containing $r \geq 1$ spindles $modm_1 A_1 : \Sigma_1 : \mathfrak{n}_1 B_1, \ldots, \mathfrak{m}_r A_r : \Sigma_r : \mathfrak{n}_r B_r$. Let us suppose that each spindle $\Sigma_i$ is between nodes $u_i$ and $b_i$, and for each $i < r$ $\Pi$ contains a* CS*-path between $b_i$ and $u_{i+1}$. The graph composed of the $r$ spindles and the $r-1$* CS*-paths is called a* **chain of $r$ spindles**, *and we shall write $\mathfrak{m}_1 A_1 : \Theta_r : \mathfrak{n}_r B_r$ for it. $|\Theta_r| = r$ denotes the number of spindles it contain.*

*Finally, every $\mathfrak{m}A : \Theta_r : \mathfrak{m}B$ is* dangerous.

Let $\mathcal{P} \subseteq \mathsf{MS}_{\mathbb{X},\mathbb{Y}}$. Let us assume there is a constant $L \in \mathbb{N}$ such that no more than $L$ consecutive spindles may exist in a proof net $\Pi \in \mathbf{PN}(\mathcal{P})$. We say that $\mathcal{P}$ has **determinacy degree** $L$. The **strong determinacy** of [DL08] is in fact a particular case: the determinacy degree is 1.

LEMMA 8.2.20 (DETERMINACY)
Let $\mathcal{P} \subseteq \mathsf{MS}^{\dagger}_{\mathbb{X},\mathbb{Y}}$ with determinacy degree $L \in \mathbb{N}$. Then, there exists a polynomial $p(x)$ such that for every $u \in V_{\Pi}$, $R_{\Pi}(u) \leq p(|\Pi|)$. The degree of $p(x)$ is $L+1$.

**Proof.** This proof is analogous to the one of Lemma 4.3.31, this time classifying *all* the contractions in $\Pi$ instead of just contractions at level $d$.

Let $b$ be a fixed box at depth $d$ in $\Pi$.

1. We want to classify the links $u \in V_{\Pi}$ in classes, class-0, class-1, class-2, . . . according to *how many consecutive spindles there are between $b$ and $u$*. Let us consider all the possible paths $\tau$ between $b$ and $u$. Let us call $\Sigma^u_b$ the module made up of all the nodes of all the $\tau$'s. Let us consider all the possible chains of spindles $\Phi^u_b, \Psi^u_b, \ldots, \Omega^u_b$ whose nodes are among the nodes of $\Sigma^u_b$. $u$ is said *class-$i$* if $i = \max\left\{\left|\Phi^u_b\right|, \left|\Psi^u_b\right|, \ldots, \left|\Omega^u_b\right|\right\}$. Notice that there are at most $L+1$ classes $(0, \ldots, L)$ by hypothesis.
2. Let $m_i$ be the *number of Contractions and Po nodes* of class-$i$, and $m = m_0 + \ldots + m_L$. We observe that $m \leq |\Pi|$.
3. $P$ will denote the greatest number of premises of the (non-fuzzy) boxes in $\Pi$, if this is greater or equal than 2; otherwise, we state $P = 2$.
4. For $0 \leq i \leq L$, let $\Pi_i$ be the *module containing all and only the class-$j$ nodes*, for every $j \leq i$.
5. A path $\tau$ starting from $b$ is *maximal relatively to* $\Pi_i$ if there exists a maximal path $\tau'$ starting from $b$ whose intersection with $\Pi_i$ is exactly $\tau$.
6. We call $R_i$, for $0 \leq i \leq L$, the number of paths starting from $b$ and maximal relatively to $\Pi_i$. The definitions imply $R_{\Pi}(b) = R_L$.

We prove that $R_i \leq \prod_{j=0}^{i}\left(P \cdot m_j + 1\right)$ by induction on $i$. Two paths separate only in a contraction node or in a (non-fuzzy) promotion box, but *not* in a $\otimes_{\mathcal{L}}$ node, which, instead, forces to go in one specific direction. This time, differently from Lemma 4.3.31, we have to deal also with paths that arrive to a *Po* link with negative context, and also in that point the paths can separate, in at most $P$ different paths.

In $\Pi_0$ there are no spindles, meaning $R_0 \leq m_0+1$. By induction, let $R_{i-1} \leq \prod_{j=0}^{i-1}\left(P \cdot m_j + 1\right)$. Then, the paths maximal relatively to $\Pi_i$ are at most $R_i \leq R_{i-1} \cdot (P \cdot m_i + 1) \leq \prod_{j=0}^{i}\left(P \cdot m_j + 1\right)$. So, $R_{\Pi}(b) = R_L \leq \prod_{j=0}^{L}(P \cdot m_i + 1) \leq (P \cdot |\Pi|)^{L+1}$. □

PROPOSITION 8.2.21 (DETERMINACY AND POLYTIME SOUNDNESS)
Let $\mathcal{P} \subseteq \mathsf{MS}^{\dagger}_{\mathbb{X},\mathbb{Y}}$ with determinacy degree $L \in \mathbb{N}$. Then, $\mathcal{P}$ is polytime.

**Proof.** Consequence of Lemma 8.2.20 and Lemma 8.2.16. □

### 8.2.3 Untyped Proof nets, Recursive Types

Here we consider two classical generalizations of proof nets, related to each other. Both such generalizations may lead to some new subsystems stratified, and nevertheless much more expressive than the actual subsystems of MS. This is why we are interested in a new framework allowing them. We shall see that it is reasonable to have MS proof nets with recursive types; less reasonable to have untyped MS proof nets.

**Untyped proof nets.** It is in the tradition of Light Logic that proof nets can be defined with or without *types*, that is *formulæ*, labelling their edges. The results of polynomial time soundness for LLL, ILAL, SLL, hold for all the proof nets, typed and untyped ones. Untyped proof nets, on the logical point of view, are *wrong proofs*; and anyway there are two good reasons to consider them. The first reason is technical: considering the formulæ, the previous cited logics are *not* polytime (cfr. Section 3.4.2). The second reason is more relevant: untyped proof nets stay to typed proof nets in the same way as untyped $\lambda$-terms stay to $\lambda$-terms; they are programs, even if their behaviour may be sometimes weird (cfr. Section 2.1).

We underline that most of the theorems that we proved for proof nets do not depend on the particular formulas appearing on the edges of the proof nets themselves. This means that such theorems can be easily generalized to the case of untyped proof nets, and to proof nets containing recursive types. **HOWEVER**, notice that the definition of "subsystem" strongly relies on the formulæ. There is hardly any reason to decide if a $[P/\mathbf{Y}]$ or $[P/P]$ cut can reduce or not, unless looking at the modalities labeling their edges.

**Recursive types.** More interesting is the case of *recursive types*. In theory of domains [AC98], a recursive type is a type $T$ that satisfies some recurrence equation $T = A(T)$. In terms of proof nets, this means to consider formulæ up to the equivalence $T \equiv A(T)$. This identifies a class of proof nets larger than the typed ones, but smaller than the untyped ones. This approach has been used, e.g., in Chapter 7, when defining LALL; the Scott numerals/words can be typed only using a recursive type. For example, the following is the recursive type used for Scott numerals in [ACP93]:

$$\mathscr{S} = \forall\alpha.\alpha \multimap (\mathscr{S} \multimap \alpha) \multimap \alpha$$

$$0 = \lambda x \lambda y.x$$

$$\text{succ} = \lambda n.\lambda x \lambda y.yn$$

(no need for modalities). In this numerical system it is easy to write predecessor and conditional, but we can hardly encode anything else. Recursive types can be used to give great expressiveness to logical systems. For example, [DLB06] shows that ILAL without second order quantification but with some recursive types is complete w.r.t. polytime functions.

There exists a large literature about recursive types. We recall a second approach we found in literature. We call $T = \mu\alpha.A(\alpha)$ a particular solution of $T = A(T)$; then, we add a new kind of node $\mu$ (i.e. logical rule), left or right, whose only effect is to change the type along an edge:

$$\frac{\Gamma \vdash A\left[{}^{\mu\alpha.A}/_{\alpha}\right]}{\Gamma \vdash \mu\alpha.A}\ \mu_{\mathcal{R}} \qquad\qquad \frac{\Gamma, A\left[{}^{\mu\alpha.A}/_{\alpha}\right] \vdash B}{\Gamma, \mu\alpha.A \vdash B}\ \mu_{\mathcal{L}}$$

This approach has been used e.g. in [BM07, DLB06]. The advantage of this approach is that it is not necessary any quotient over formulæ. Also Scott numerals are a least fixpoint:

$$\mathscr{S} = \mu\beta.\forall\alpha.\alpha \multimap (\beta \multimap \alpha) \multimap \alpha.$$

Of course, this second approach has also some disadvantages, first of all a heavier syntax for proof nets.

In MS, as we said, most of the proofs of the theorems do not depend on the formulæ involved in the reductions. So, it should be easy to generalize such proofs to proof nets with recursive types, either quotienting formulæ, or using $\mu$ nodes, without running into the problems that we saw about untyped proof nets.

# Appendix A

# Encoding from $\mathsf{SRN}^{=}$ to $\mathsf{soLAL}$

Here we give most of the details of the proof of Lemma 6.2.8 on page 100.

The proof nets described in this Appendix represent the height of a work that traces back at least to [Rov07].

**Briefly.** The proof net that simulates $\ulcorner f(w, \overrightarrow{x}^{\mathsf{n}}; \overrightarrow{y}^{\mathsf{s}}) \urcorner$ is $\mathtt{It}_{1+\mathsf{n},\mathsf{s}}[H_0, H_1, G]$ in Figure A.13 (but its main components are in Figure A.15). It generates two copies $W_1, W_2$ of $\ulcorner w \urcorner$. The first copy $W_1$ is used to build the initial configuration; this is done by the proof net $\mathtt{W2C}_{1+\mathsf{n};\mathsf{s}}$ (*Word to Configurations*) in Figure A.17. Then, the second copy $W_2$ iterates a transition function $\mathtt{C2C}_{1+\mathsf{n};\mathsf{s}}$ (*Configurations to Configurations*) in Figure A.4 and ff.. At last, the result is stored as the first element of the last configuration, and it is necessary some work to extract it. The proof nets $\mathtt{C2FC}_{1+\mathsf{n};\mathsf{s}}$ (*Configurations to Final Configurations*) in Figure A.23 and $\mathtt{FC2W}_{1+\mathsf{n};\mathsf{s}}$ (*Final Configurations to Word*) in Figure A.24 do the job.

## A.1 The details on the map from $\mathsf{SRN}^{=}$ to $\mathsf{soLAL}$

We proceed as follows. Section A.1.1 is about the basic data types we shall be using, like words, strings, higher-order tuples, *etc.*. Section A.1.2 introduces the basic nets, namely combinators for basic operation, like, for example, the coercions, that embed a word inside a certain number of boxes. Section A.1.3 defines configurations, pre-configurations, and the transition function $\mathtt{C2C}_{1+\mathsf{n};\mathsf{s}}$ at the heart of the virtual machine that implements the recursion unfolding inside $\mathsf{soLAL}$. Finally, Section A.1.4, gives the iteration of the transition function, starting from an initial configuration.

### A.1.1 Basic datatypes

We define structure and operations relative to *(sorted unary) naturals*, *(sorted binary) words*, *(sorted) lists*, *pairs* and *tuples* exploiting the higher-order features of $\mathsf{soLAL}$. We shall use them extensively in the course of the definition of $\mathtt{It}_{1+\mathsf{n},\mathsf{s}}[H_0, H_1, G]$.

**Sorted unary numerals.** The type $i\mathbb{S}$ of *sorted unary numerals*, or simply *numerals*, is $i\mathbb{S} = \forall\alpha.i!(\alpha \multimap \alpha) \multimap i\S(\alpha \multimap \alpha)$, with $i$ its *sort*. Each natural naturally corresponds to a closed proof net $\overline{n} \triangleright \vdash i\mathbb{S}$ in $\mathsf{soLAL}$. In $\mathsf{soLAL}$ we can easily program the closed proof net $\mathtt{Ss} \triangleright j\mathbb{S} \multimap j\mathbb{S}$ that yields the *successor* of argument.

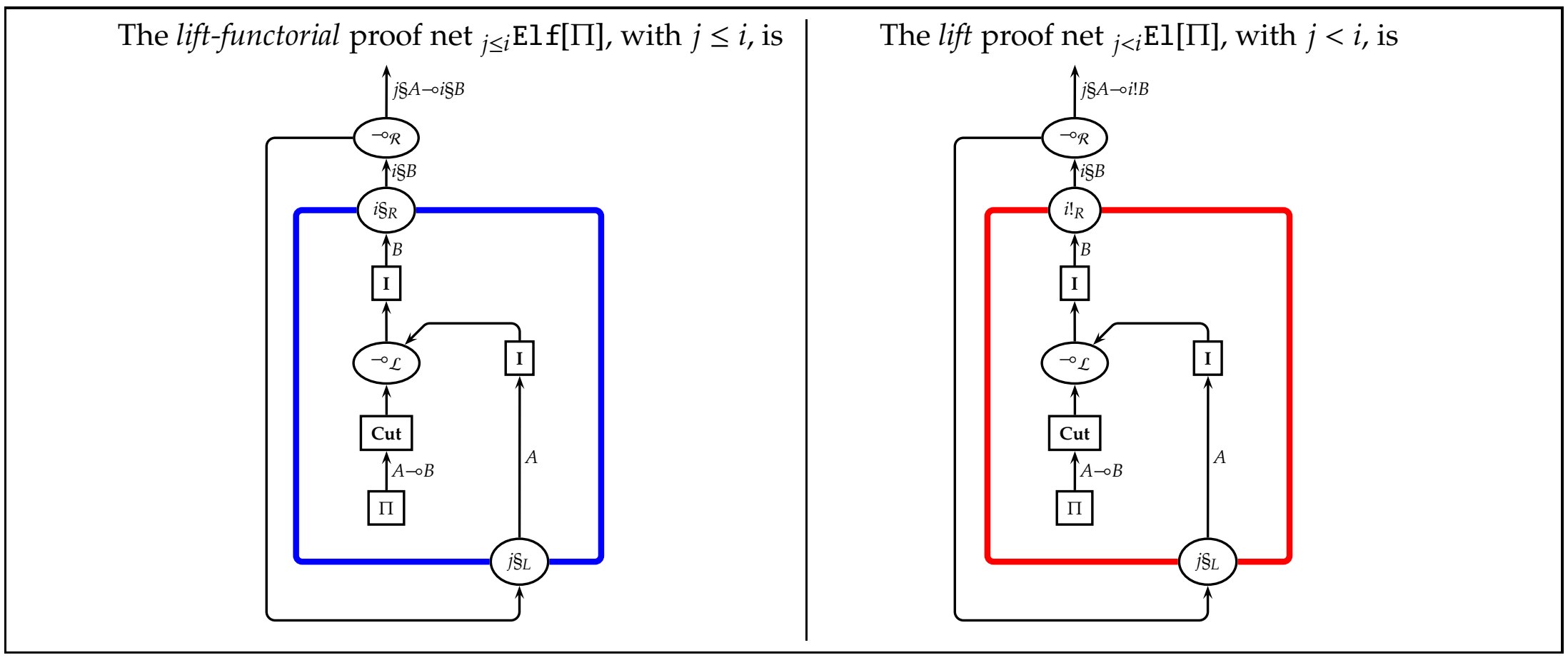

Figure A.1: Two of the four embedding proof nets we use in SOLAL.

**Sorted binary words.** The type $i\mathbb{W}$ of *sorted binary words*, or simply *words*, is $i\mathbb{W} = \forall\alpha.i!(\alpha \multimap \alpha) \multimap i!(\alpha \multimap \alpha) \multimap i\S(\alpha \multimap \alpha)$, with $i$ its *sort*. Each word naturally corresponds to a closed proof net $\overline{\overline{w}} \triangleright \vdash i\mathbb{W}$ in SOLAL. In SOLAL we can program standard combinators on words. `Ws0` $\triangleright \vdash j\mathbb{W} \multimap j\mathbb{W}$, `Ws1` $\triangleright \vdash j\mathbb{W} \multimap j\mathbb{W}$ are two successors that we can use to set $\ulcorner s_0 \urcorner$ = `Ws0` and $\ulcorner s_1 \urcorner$ = `Ws1`. `Ws1` obviously generalizes the successor on strings. `Ws0` must be defined to erase every occurrence of the symbol 0 to the right hand side of the most significant bit of a word, as in [MO04]. This forces every $\overline{\overline{w}}$, with $w \neq 0$, to have 1 as its most significant bit. `Wp` $\triangleright \vdash j\mathbb{W} \multimap j\mathbb{W}$ is the predecessor which erases the least significant bit of its argument. We can use it to set $\ulcorner$P$\urcorner$ = `Wp`. Finally, `W2S` $\triangleright \vdash j\mathbb{W} \multimap j\mathbb{S}$ maps a word $\overline{\overline{w}}$ to a numeral $\overline{n}$ so that $n$ counts the number of bits of $w$.

**Sorted lists.** The type $i\mathbf{L}(A)$ of *sorted lists* whose elements are of type $A$, or simply *lists of type* $A$, is $i\mathbf{L}(A) = \forall\alpha.i!(A \multimap \alpha \multimap \alpha) \multimap i\S(\alpha \multimap \alpha)$, with $i$ its *sort*. `nil` is the name of the *empty list* [ ].

**Higher-order pairs and tuples.** We choose to exploit the expressive power of higher and second-order formulæ in SOLAL to represent pairs and tuples. The type $A \odot B$ of *pairs* is $A \odot B = \forall\gamma.(A \multimap B \multimap \gamma) \multimap \gamma$, a standard second-order encoding. Its generalization to tuples of $n \geq 1$ elements is $A_1 \odot \ldots \odot A_n = \forall\gamma.((\multimap_{k=1}^{n} A_k) \multimap \gamma) \multimap \gamma$. $\odot^n A$ will abbreviate $A \odot \ldots \odot A$, with $n$ occurrences of $A$.

### A.1.2 Basic proof nets

We call basic the proof nets we are going to introduce because, after some programming experience inside SOLAL, they seem unavoidable tools to use proof nets-as-programs. They realize embedding and coercion functions that enclose proof nets inside boxes, and the diagonal functions that replicate the same word into tuples.

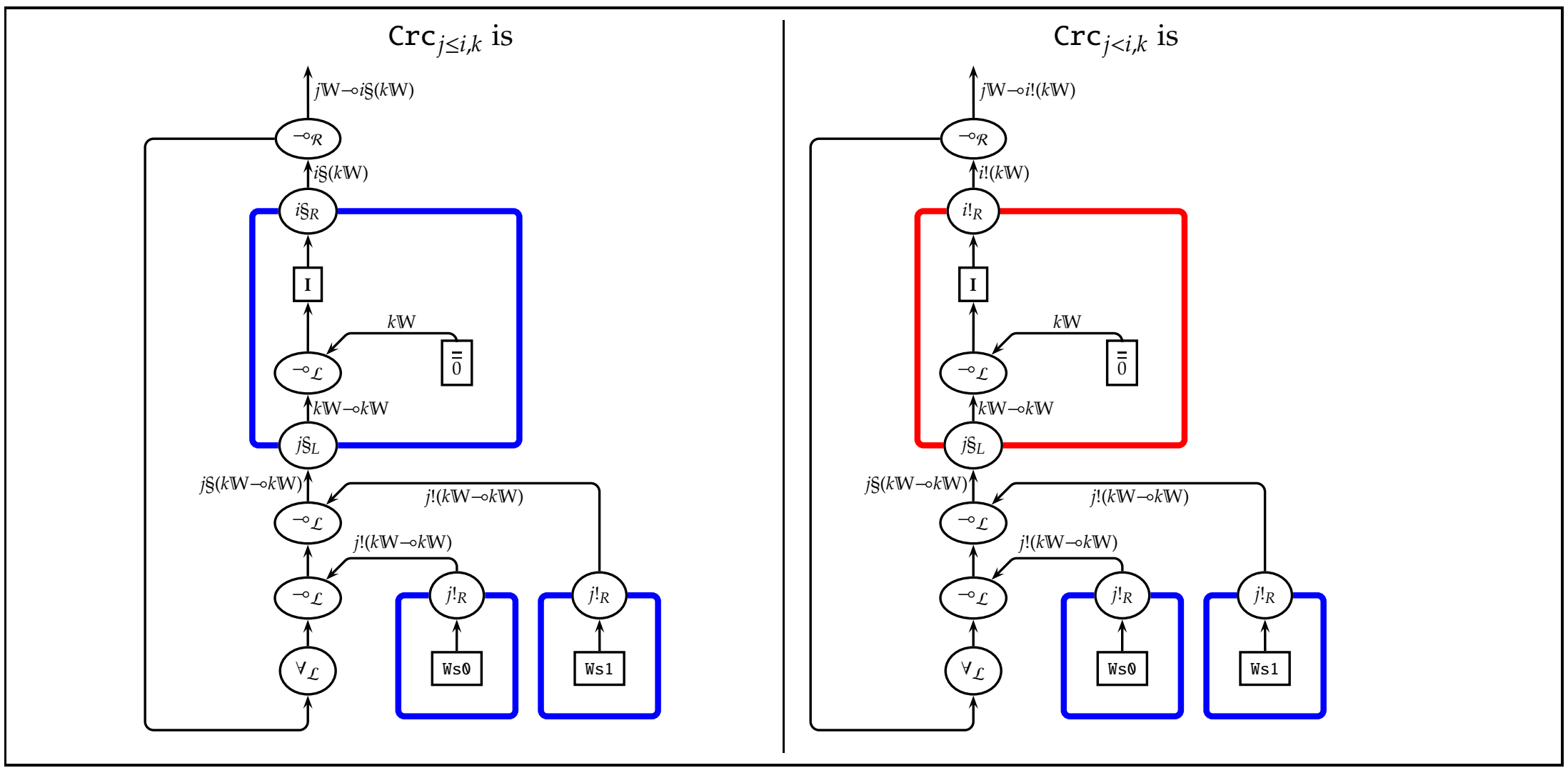

Figure A.2: The proof nets $\texttt{Crc}_{j\leq i,k}$ with $j \leq i$, and $\texttt{Crc}_{j<i,k}$, with $j < i$, for any $k$.

**Embedding.** There are four kinds of embedding proof nets, every of them taking a proof net $\Pi \triangleright \vdash A \multimap B$ as parameter. Every embedding places the parameter into boxes as the boxes were functors. Thee proof nets simply change the *type* of $\Pi$, but do not change its intended computational behaviour. Two examples are in Figure A.1

**Coercions.** There are two kinds of coercion proof nets, as in Figure A.2. Every of them takes a word of type $j\mathbb{W}$ to reconstruct it inside a box. It is worth remarking that the analogous of $\texttt{Crc}_{j<i,k} \triangleright \vdash j\mathbb{W} \multimap i!(k\mathbb{W})$, with $j < i$, for any $k$, cannot exist in ILAL where the result of a coerce proof net can only be inside at least a paragraph box.

**Diagonals.** There are two kinds $\nabla^2_{j\leq i,k}, \nabla^2_{j<i,k}$ of *diagonal* proof nets. Figure A.3 introduces $\nabla^2_{j\leq i,k}$. The other can be obtained from $\nabla^2_{j\leq i,k}$ by transforming the paragraph box into an of-course box. $\nabla^2_{j\leq i,k}$ takes an argument of type $j\mathbb{W}$ to replicate it twice inside a box. The replication comes by iterating $H_0, H_1$ in Figure A.3 with the argument, starting from a pair of occurrences of $\overline{\overline{0}}$. Both diagonals can be generalized to $\nabla^m_{j\leq i,k}$, and $\nabla^m_{j<i,k}$, with type $j\mathbb{W} \multimap i\S(\odot^m k\mathbb{W})$, and $j\mathbb{W} \multimap i!(\odot^m k\mathbb{W})$, respectively, for any $m \geq 2$.

### A.1.3 Configurations and transition function between them

**Configurations.** For every $\mathsf{k} \geq 1, i \in \mathbb{N}$, the type of the *configurations of sort* $i$ is:

$$\mathbf{C}[C_1 \ldots C_{\mathsf{k}}; C] = \forall \alpha_1 \ldots \alpha_{\mathsf{k}}.(\multimap^{\mathsf{k}}_{i=1} i!(C_i \multimap \alpha_i \multimap \alpha_i)) \multimap i\S((\multimap^{\mathsf{k}}_{i=1} \alpha_i) \multimap \forall \gamma.((C \multimap (\multimap^{\mathsf{k}}_{i=1} \alpha_i) \multimap \gamma) \multimap \gamma))$$

such that $\{\alpha_1, \ldots, \alpha_{\mathsf{k}}, \gamma\} \cap \mathrm{FV}(C) = \emptyset$. The structure of $\mathbf{C}[C_1 \ldots C_{\mathsf{k}}; C]$ implies that every configuration is a tuple of lists whose element type is $C_i$, with $1 \leq i \leq k$, but the first element, which is not a list, but a single element of type $C$. We shall use the canonical instance:

$$\mathbf{C}[B_1 \ldots B_{\mathsf{s}}; B] = \mathbf{C}[\underbrace{j\mathbb{W} \ldots j\mathbb{W}}_{1+\mathsf{n}}\ B_1 \ldots B_{\mathsf{s}}; B]$$

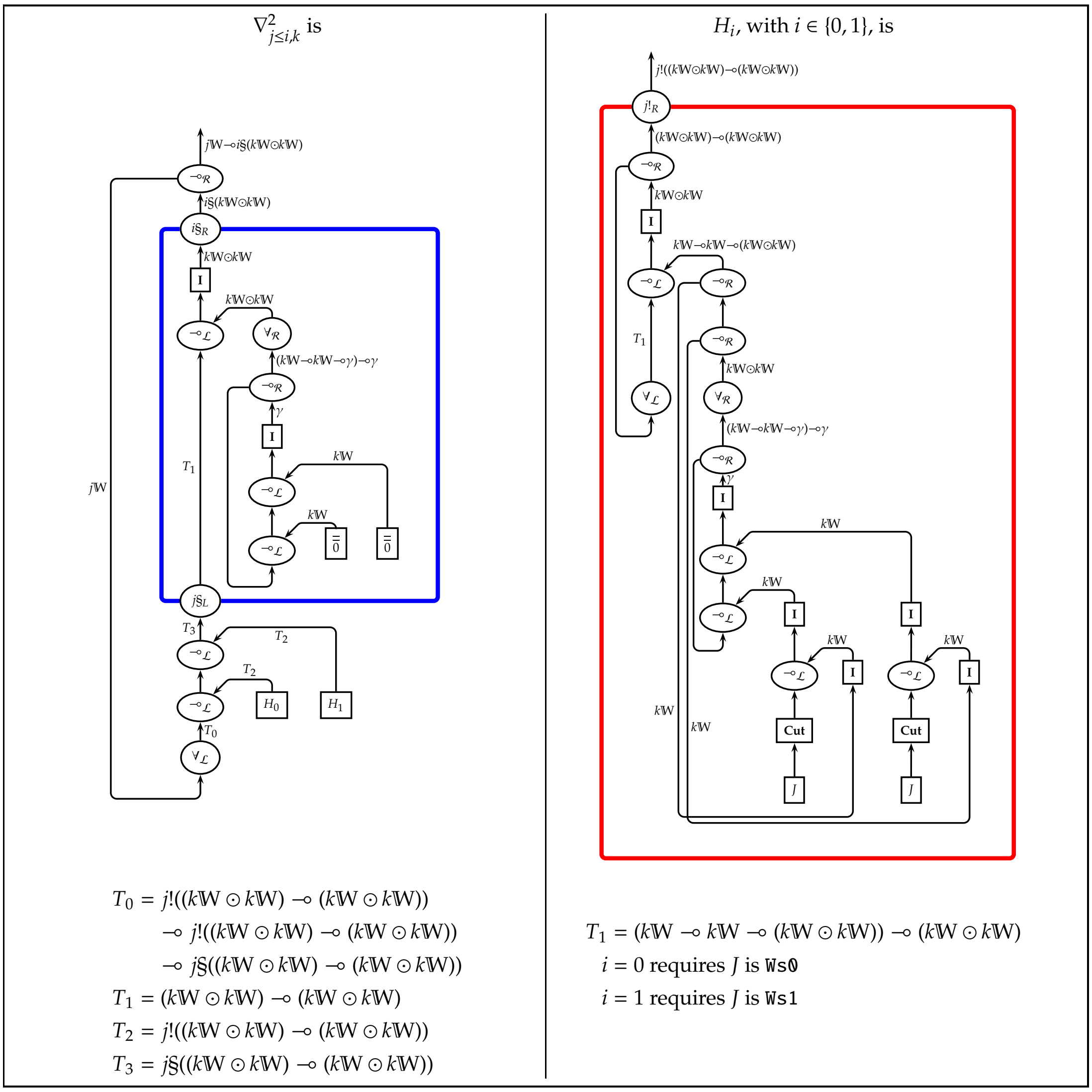

Figure A.3: The proof net $\nabla^2_{j\leq i,k}$ with $j \leq i$, for any $k$.

of the type of the configurations, where $\mathsf{k} = 1 + \mathsf{n} + \mathsf{s}$, for some given $\mathsf{n}, \mathsf{s} \geq 0$. Every configuration, as we have seen, is a tuple:

$$\langle \overline{\overline{r}}, [\overline{\overline{a_1}}, \ldots, \overline{\overline{a_\mathsf{r}}}], [\overline{\overline{n_{11}}}, \ldots, \overline{\overline{n_{1\mathsf{r}}}}], \ldots, [\overline{\overline{n_{\mathsf{n}1}}}, \ldots, \overline{\overline{n_{\mathsf{nr}}}}], [\Pi_{11}, \ldots, \Pi_{1\mathsf{r}}], \ldots, [\Pi_{\mathsf{s}1}, \ldots, \Pi_{\mathsf{sr}}] \rangle$$

It follows that (i) every configuration is a tuple whose first element is a word $\overline{\overline{r}}$, and all the remaining elements are lists, all with the same length $\mathsf{r}$, (ii) Proposition 6.2.5 implies that every $\Pi_{pq}$ is, in fact, a word inside a certain number of boxes, and (iii) the elements of the lists are all at the same level, which, as first noticed in [Rov99], is basic to hope getting an iterable transition function in a stratified system like SOLAL.

**Pre-Configurations.** For every $\mathsf{k} \geq 1$, the type of the *pre-configurations* is:

$$\mathbb{PC}[\alpha_1 \dots \alpha_{\mathsf{k}}; C_1 \dots C_{\mathsf{k}}; C] = \forall \gamma.((C \multimap (\multimap_{i=1}^{\mathsf{k}} \mathbf{T}[\alpha_i; C_i]) \multimap \gamma) \multimap \gamma)$$
$$\mathbf{T}[\alpha; C] = \forall \beta.((\mathbf{U}[\alpha; C] \multimap \beta) \multimap \beta)$$
$$\mathbf{U}[\alpha; C] = (C \multimap \alpha \multimap \alpha) \multimap (C \multimap C) \multimap C \multimap \alpha \qquad \text{(tuple of couples of } (C \multimap C) \text{ and } C)$$

such that $\{\alpha_1, \dots, \alpha_{\mathsf{k}}, \gamma\} \cap \mathrm{FV}(C) = \emptyset$. We shall use the following canonical instance of the type of the pre-configurations:

$$\mathbb{PC}[\alpha_0 \dots \alpha_{\mathsf{n+s}}; B_1 \dots B_{\mathsf{s}}; B] = \mathbb{PC}[\alpha_0 \dots \alpha_{\mathsf{n+s}}; \underbrace{i\mathbb{W} \dots i\mathbb{W}}_{1+\mathsf{n}}\ B_1 \dots B_{\mathsf{s}}; B]$$

where $\mathsf{k} = 1 + \mathsf{n} + \mathsf{s}$, for some $\mathsf{n}, \mathsf{s} \geq 0, m \geq 1$. Every pre-configuration is in fact a tuple of tuples:

$$\langle \bar{\bar{r}}, \langle \overline{\overline{a_1}}, [\overline{\overline{a_2}}, \dots, \overline{\overline{a_{\mathsf{r}}}}] \rangle \;\; , \langle \overline{\overline{n_{11}}}, [\overline{\overline{n_{12}}}, \dots, \overline{\overline{n_{1\mathsf{r}}}}] \rangle, \dots, \langle \overline{\overline{n_{\mathsf{n}1}}}, [\overline{\overline{n_{\mathsf{n}2}}}, \dots, \overline{\overline{n_{\mathsf{nr}}}}] \rangle \;\; , \langle \Pi_{11}, [\Pi_{12}, \dots, \Pi_{1\mathsf{r}}] \rangle, \dots, \langle \Pi_{\mathsf{s}1}, [\Pi_{\mathsf{s}2}, \dots, \Pi_{\mathsf{sr}}] \rangle \rangle$$

**The proof net $\mathtt{C2C}_{1+\mathsf{n};\mathsf{s}}[F, F']$.** The root of its definition is in Figure A.4. Its main components are the proof nets $\mathtt{C2PC}_{1+\mathsf{n};\mathsf{s}}$ and $\mathtt{PC2C}_{1+\mathsf{n};\mathsf{s}}$, which split the move "configuration to configuration" in two simpler steps: "configuration to pre-configuration" and "pre-configuration to configuration".

$\mathtt{C2C}_{1+\mathsf{n};\mathsf{s}}$ requires two parameters $F$ and $F'$. $F$ must be a closed proof net of type $C \multimap C$, for some $C$. It will be either a successor on words, mapped on the first list of the configuration in input to get the next value of the step function, or a function identity, mapped along all the remaining lists of the configuration in input. $F'$ is a closed proof net that is used as a step function by $\mathtt{PC2C}_{1+\mathsf{n};\mathsf{s}}$. Looking back to Proposition 6.2.5, instances of $F'$ must be $\ulcorner h_0 \urcorner$ and $\ulcorner h_1 \urcorner$, relative the recursive function $f^{1+\mathsf{n};\mathsf{s}}$ of SRN we want to represent. Both $\ulcorner h_0 \urcorner$ and $\ulcorner h_1 \urcorner$ have type $(\multimap_{k=0}^{\mathsf{n}} j\mathbb{W}) \multimap (\multimap_{k=1}^{\mathsf{s}} B_k) \multimap B \multimap B$, with $B_1 \dots B_{\mathsf{s}}, B$ instantiated by $j\mathbb{W}$.

**The proof net $\mathtt{C2PC}_{1+\mathsf{n};\mathsf{s}}[F]$.** It is defined by the set of proof nets in Figure A.5, A.6, A.7, and A.8. The proof net takes a configuration in input, separates the first component of every list in the configuration from its tail, and maps $F$ along the tail. In particular, if $F$ is a successor on words, all the words in the tail get increased.

Specifically, Figure A.8 contains a step function $\mathtt{StC2PC}[F]$ that takes a list and applies $F : C \multimap C$ ($F$ is the successor or the identity) to each element of the list but the head.

**The proof net $\mathtt{PC2C}_{1+\mathsf{n};\mathsf{s}}[F']$.** It is defined by the set of proof nets in Figure A.9, A.10, A.11, A.12. $F'$ has type $(\multimap_{k=0}^{\mathsf{n}} j\mathbb{W}) \multimap (\multimap_{k=1}^{\mathsf{s}} B_k) \multimap B \multimap B$. The graph with name "$1 + \mathsf{n} + \mathsf{s}$ **W** nodes" in Figure A.11 contains only as many as $1 + \mathsf{n} + \mathsf{s}$ nodes **W**.

### A.1.4 The iteration

Finally, we can define $\mathtt{It}_{1+\mathsf{n},\mathsf{s}}[H_0, H_1, G]$ that iterates $\mathtt{C2C}_{1+\mathsf{n};\mathsf{s}}[F, F']$ from an initial configuration, for a sufficient number of times. From the last configuration the iteration produces we can extract the word that represents the result of $\ulcorner f^{1+\mathsf{n};\mathsf{s}} \urcorner$. We have already described (at the beginning of this Appendix) the interaction among the proof nets $\mathtt{It}_{1+\mathsf{n},\mathsf{s}}[H_0, H_1, G]$, $\mathtt{W2C}_{1+\mathsf{n};\mathsf{s}}$, $\mathtt{C2C}_{1+\mathsf{n};\mathsf{s}}$, $\mathtt{C2FC}_{1+\mathsf{n};\mathsf{s}}$, and $\mathtt{FC2W}_{1+\mathsf{n};\mathsf{s}}$. The last two proof nets use *final configurations*, that we have not defined yet. We proceed by first setting the final configurations, and, then, by bottom up introducing $\mathtt{It}_{1+\mathsf{n},\mathsf{s}}[H_0, H_1, G]$.

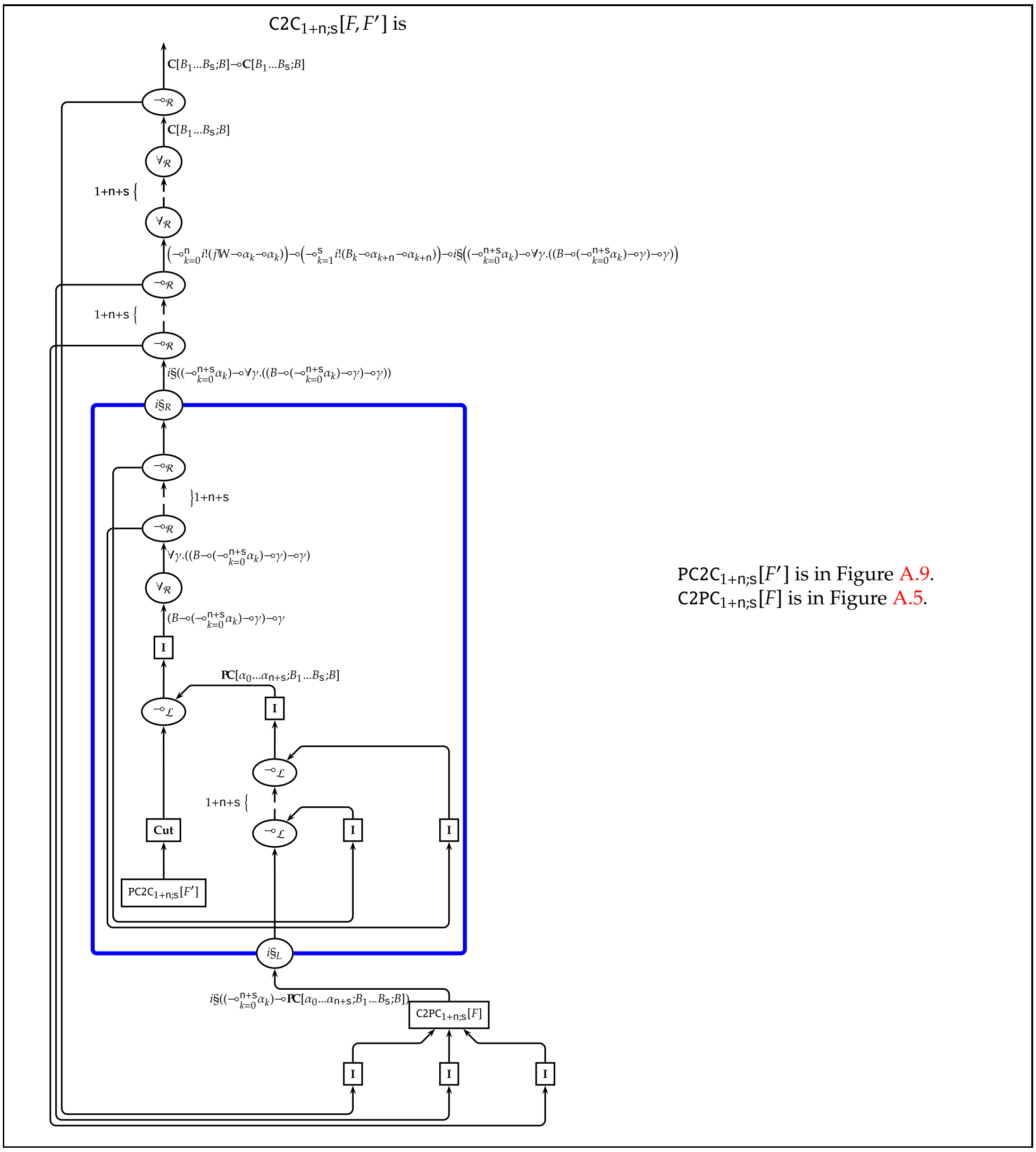

Figure A.4: The proof net $\mathsf{C2C}_{1+n;s}[F, F']$

**Final configurations.** For every $k \geq 1$, the type of the final configurations is:

$$\mathbf{FC}[C_1 \ldots C_k; C] = \forall \alpha_1 \ldots \alpha_k \gamma.(\multimap_{i=1}^{k} i!(C_i \multimap \alpha_i \multimap \alpha_i)) \multimap i\S((\multimap_{i=1}^{k} \alpha_i) \multimap (C \multimap (\multimap_{i=1}^{k} \alpha_i) \multimap \gamma) \multimap \gamma)$$

such that $\{\alpha_1, \ldots, \alpha_k, \gamma\} \cap \mathrm{FV}(C) =$. The difference with the type of the configurations is the extrusion of the universal quantifier on $\gamma$. We shall use the following *canonical instance* of the type of the final configurations, for some given $n, s \geq 0$, and $m \geq 1$:

$$\mathbf{FC}[B_1 \ldots B_s; B] = \mathbf{FC}[\underbrace{i\mathbb{W} \ldots i\mathbb{W}}_{1+n}\ B_1 \ldots B_s; B]$$

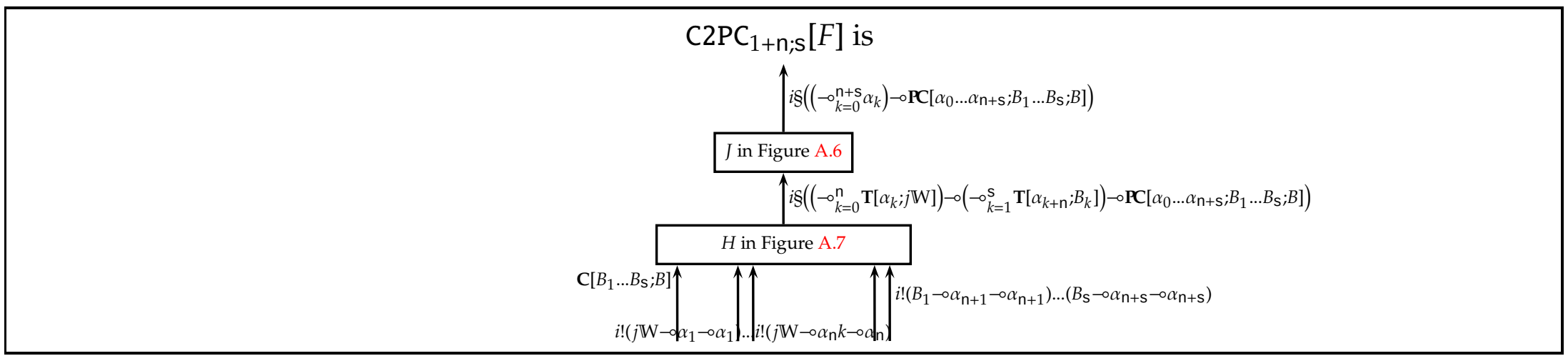

Figure A.5: The proof net $\mathtt{C2PC}_{1+n;s}[F]$

The final configurations are a necessary step to extract, with the correct typing, the first component of a configuration, that will represent the result of an iteration.

**The proof net $\mathtt{It}_{1+n,s}[H_0, H_1, G]$.** Assuming $H_0, H_1, G$ be proof nets such that:

$$H_i \rhd \vdash j\mathbb{W} \multimap (\multimap_{k=1}^{n} j\mathbb{W}) \multimap (\multimap_{k=1}^{s} B_k) \multimap B \multimap B \qquad (i \in \{0,1\})$$
$$G \rhd \vdash (\multimap_{k=1}^{n} j\mathbb{W}) \multimap (\multimap_{k=1}^{s} B_k) \multimap B$$

$\mathtt{It}_{1+n,s}[H_0, H_1, G]$ is in Figure A.13, A.14, A.15, and A.16. We have already described the computational behaviour of this proof net.

**The proof net $\mathtt{W2C}_{1+n;s}$.** It is in Figure A.17 and A.18.

We conclude with some comments on the structure of the proof nets in Figure A.18. For every $1 \leq q \leq n$, $\mathtt{Ss}(\mathtt{W2S}\,k_q)$ is obtained by applying $\mathtt{Ss} \rhd i\mathbb{S} \multimap i\mathbb{S}$ to the result of the application of $\mathtt{W2S} \rhd j\mathbb{W} \multimap i\mathbb{S}$ to an argument of type $j\mathbb{W}$, represented by $k_q$ in the name of the proof net. An analogous definition holds for every $\mathtt{Ss}(\mathtt{W2S}\,h_p)$, with $1 \leq p \leq s$.

**The proof net $\mathtt{L2C}_{1+n;s}[F]$.** Assuming that $F$ be a proof net such that $F \rhd \vdash B$, $\mathtt{L2C}_{1+n;s}[F]$ is in Figure A.19, A.20, and A.21. It takes $1 + n + s$ lists, each with as many elements as the number of digits of the main argument of $f^{1+n;s}$, augmented by 1, and yields the expected configurations.

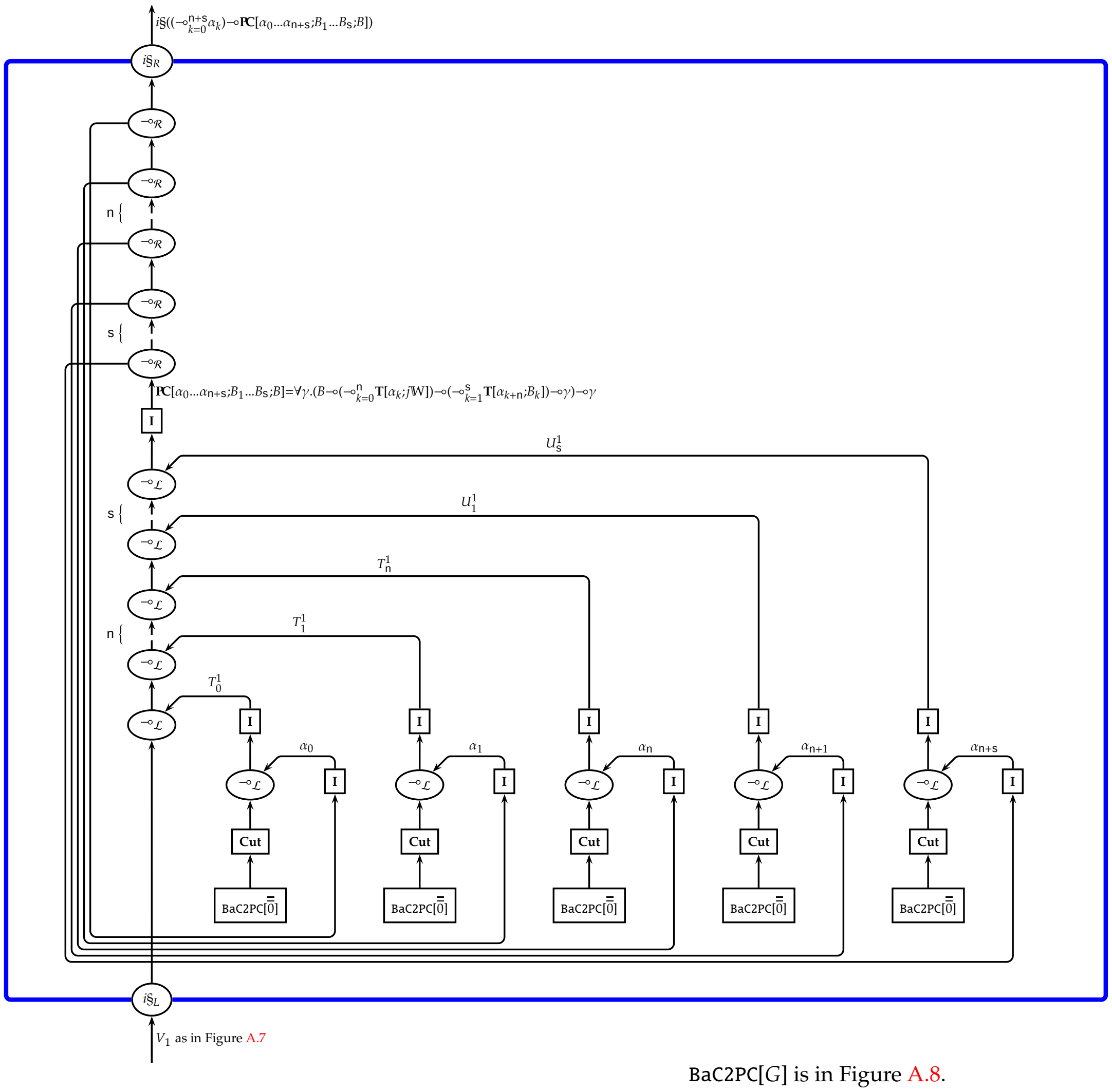

BaC2PC[$G$] is in Figure A.8.

$$T_k^1 = \mathbf{T}[\alpha_k; j\mathbb{W}] \qquad (0 \le k \le \mathsf{n})$$

$$U_k^1 = \mathbf{T}[\alpha_{k+\mathsf{n}}; B_k] \qquad (1 \le k \le \mathsf{s})$$

Figure A.6: The proof net $J$ for $\mathsf{C2PC}_{1+\mathsf{n};\mathsf{s}}[F]$

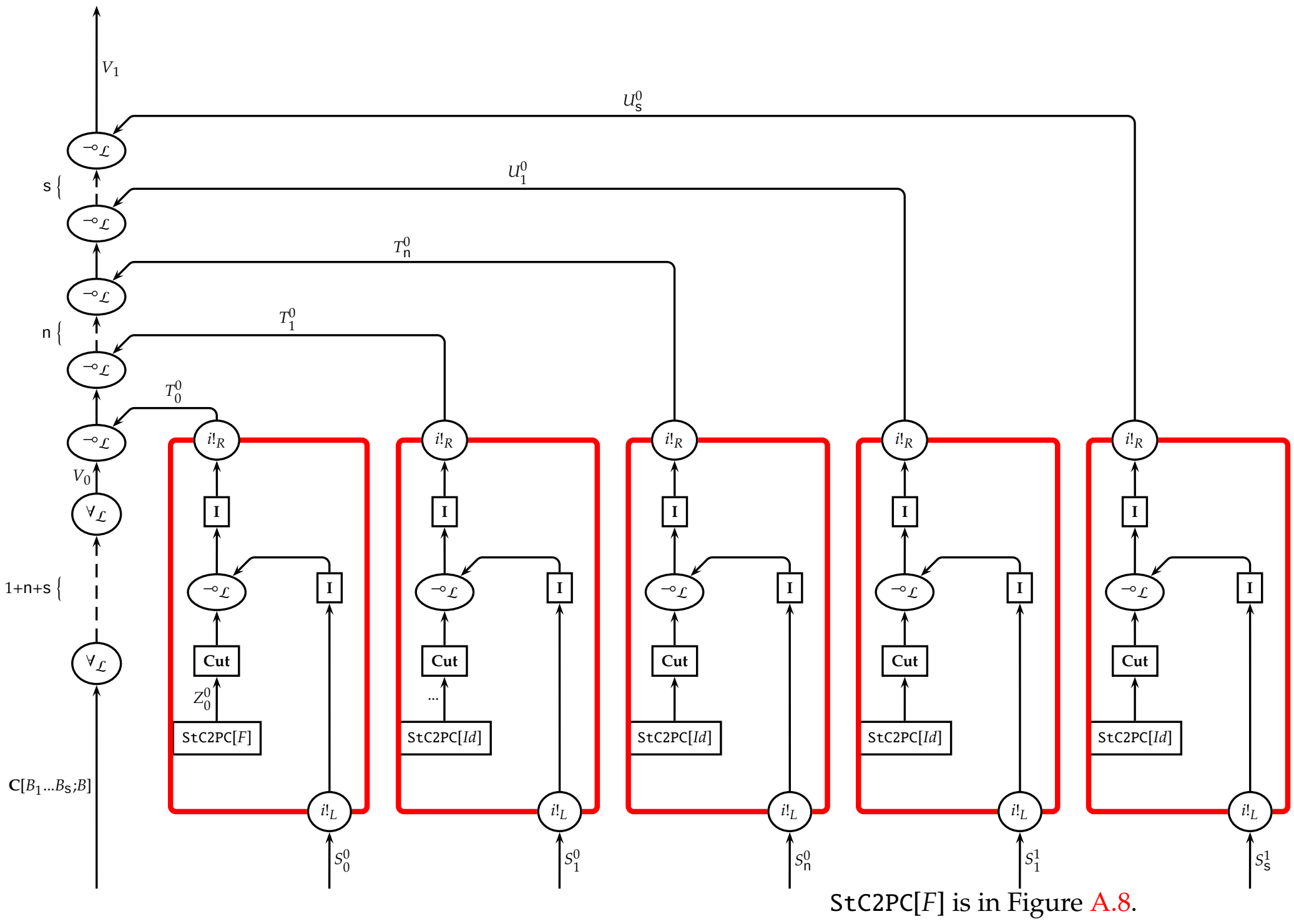

StC2PC[$F$] is in Figure A.8.

$$
\begin{aligned}
\mathbf{C}[B_1 \ldots B_s; B] &= (\multimap^{n}_{k=0} i!(j\mathbb{W} \multimap \alpha_k \multimap \alpha_k)) \multimap (\multimap^{s}_{k=1} i!(B_k \multimap \alpha_{k+n} \multimap \alpha_{k+n})) \multimap \\
&\quad i\S((\multimap^{n+s}_{k=0} \alpha_k) \multimap \forall\gamma.((B \multimap (\multimap^{n+s}_{k=0} \alpha_k) \multimap \gamma) \multimap \gamma)) \\
V_0 &= (\multimap^{n}_{k=0} i!(j\mathbb{W} \multimap \mathbf{T}[\alpha_k; j\mathbb{W}] \multimap \mathbf{T}[\alpha_k; j\mathbb{W}])) \multimap (\multimap^{s}_{k=1} i!(B_k \multimap \mathbf{T}[\alpha_{k+n}; B_k] \multimap \mathbf{T}[\alpha_{k+n}; B_k])) \multimap i\S V_1 \\
T^0_k &= i!(j\mathbb{W} \multimap \mathbf{T}[\alpha_k; j\mathbb{W}] \multimap \mathbf{T}[\alpha_k; j\mathbb{W}]) \qquad (0 \le k \le n) \\
U^0_k &= i!(B_k \multimap \mathbf{T}[\alpha_{k+n}; B_k] \multimap \mathbf{T}[\alpha_{k+n}; B_k]) \qquad (1 \le k \le s) \\
S^0_k &= i!(j\mathbb{W} \multimap \alpha_k \multimap \alpha_k) \qquad (0 \le k \le n) \\
S^1_k &= i!(B_k \multimap \alpha_{k+s} \multimap \alpha_{k+s}) \qquad (1 \le k \le s) \\
V_1 &= i\S\big((\multimap^{n}_{k=0} \mathbf{T}[\alpha_k; j\mathbb{W}]) \multimap (\multimap^{s}_{k=1} \mathbf{T}[\alpha_{k+n}; B_k]) \multimap \mathbf{PC}[\alpha_0 \ldots \alpha_{n+s}; B_1 \ldots B_s; B]\big) \\
Z^0_k &= S^0_k \multimap (j\mathbb{W} \multimap \mathbf{T}[\alpha_k; j\mathbb{W}] \multimap \mathbf{T}[\alpha_k; j\mathbb{W}])
\end{aligned}
$$

Figure A.7: The proof net $H$ for C2PC$_{1+n;s}[F]$

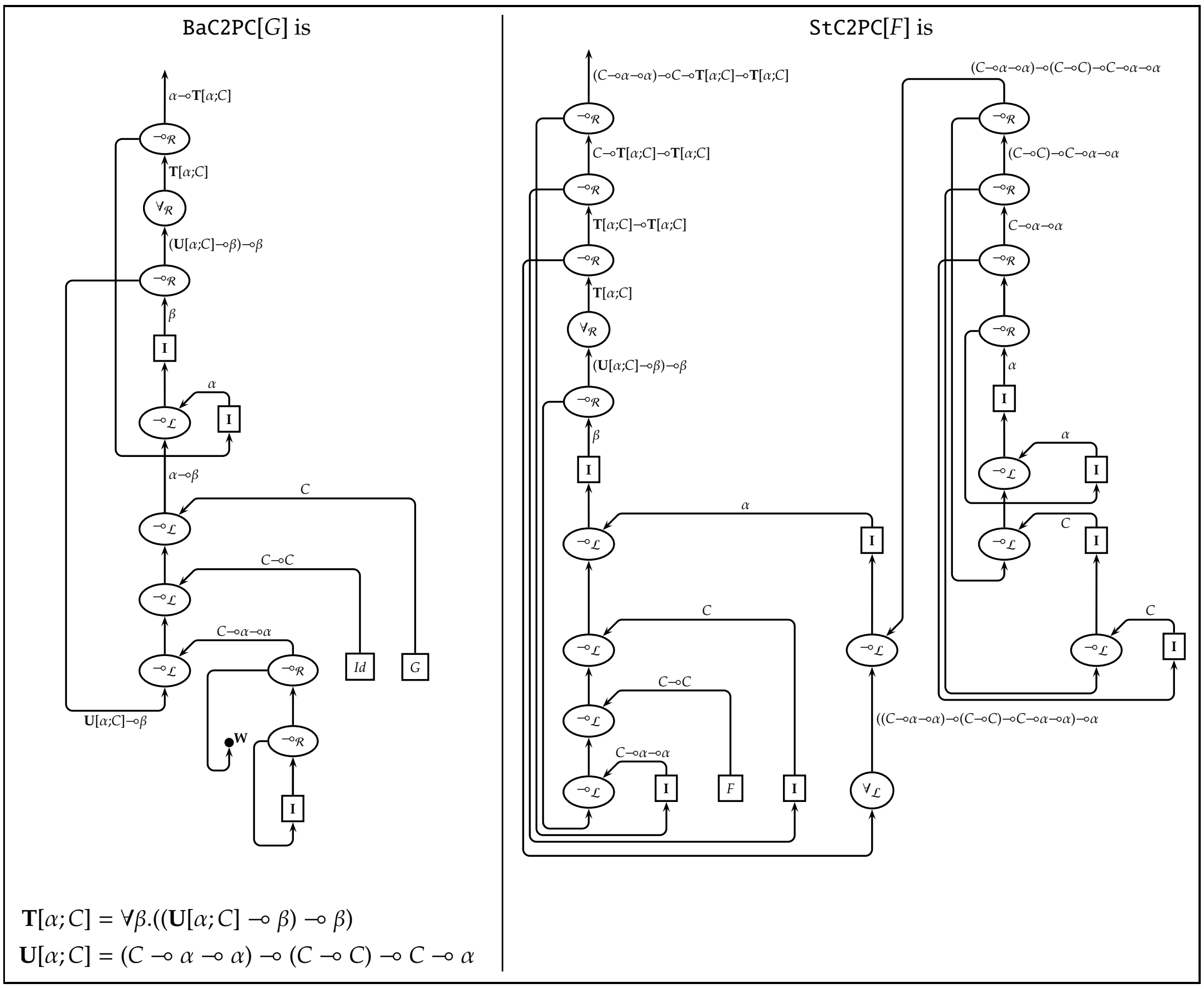

Figure A.8: The proof nets BaC2PC[$G$] and StC2PC[$F$]

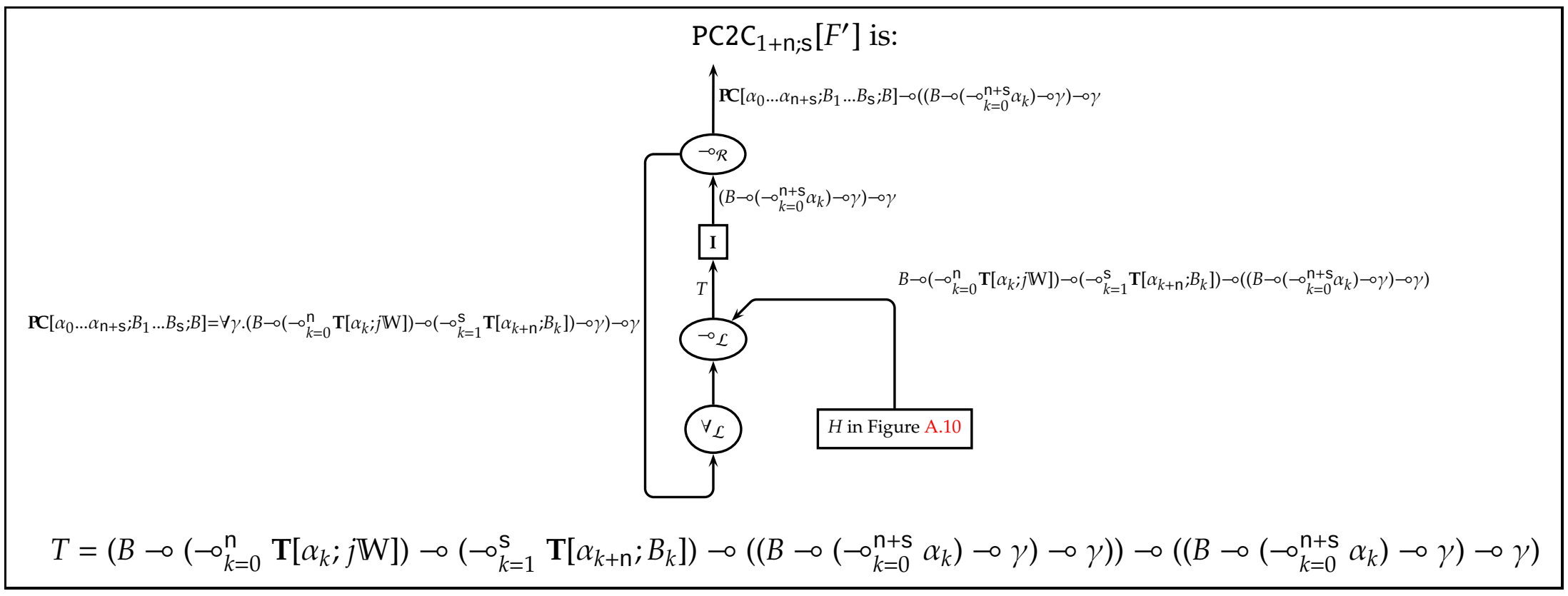

Figure A.9: The proof net PC2C$_{1+n;s}$[$F'$]

$H$ for $\mathsf{PC2C}_{1+\mathsf{n};\mathsf{s}}[F']$ is:

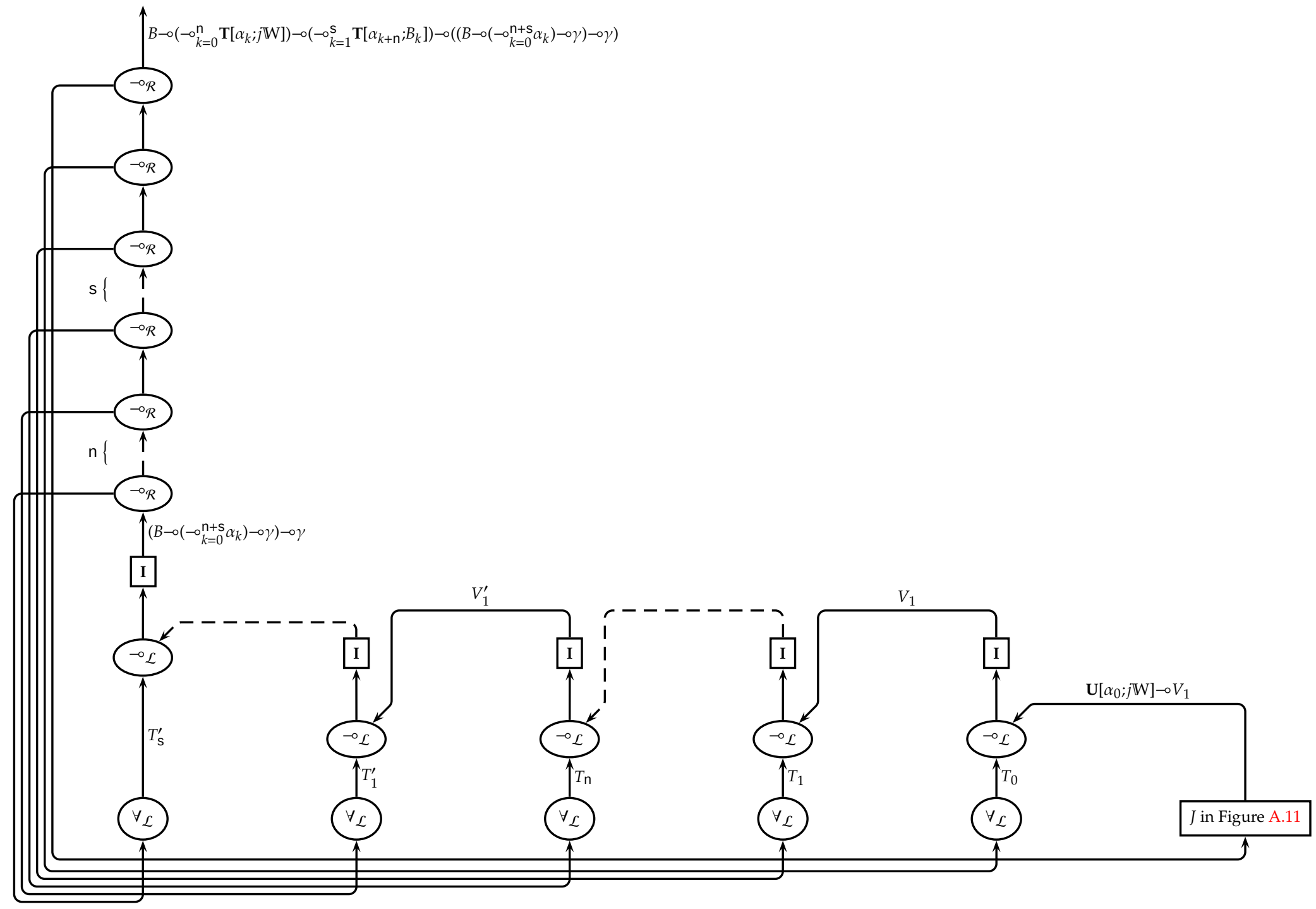

$T_k = (\mathbf{U}[\alpha_k; j\mathbb{W}] \multimap V_{k+1}) \multimap V_{k+1}$ $(0 \le k \le \mathsf{n})$

$T'_k = (\mathbf{U}[\alpha_{\mathsf{n}+k}; B_k] \multimap V'_{k+1}) \multimap V'_{k+1}$ $(1 \le k \le \mathsf{s})$

$V_k = \mathbf{U}[\alpha_k; j\mathbb{W}] \multimap (\multimap_{m=k+1}^{\mathsf{n}} \mathbf{U}[\alpha_m; j\mathbb{W}]) \multimap (\multimap_{m=1}^{\mathsf{s}} \mathbf{U}[\alpha_{\mathsf{n}+m}; B_m]) \multimap ((B \multimap (\multimap_{m=0}^{\mathsf{n}+\mathsf{s}} \alpha_m) \multimap \gamma) \multimap \gamma)$ $(0 \le k \le \mathsf{n})$

$V'_k = \mathbf{U}[\alpha_{\mathsf{n}+k}; B_k] \multimap (\multimap_{m=k+1}^{\mathsf{s}} \mathbf{U}[\alpha_{\mathsf{n}+m}; B_m]) \multimap ((B \multimap (\multimap_{m=0}^{\mathsf{n}+\mathsf{s}} \alpha_m) \multimap \gamma) \multimap \gamma)$ $(1 \le m \le \mathsf{s})$

Figure A.10: The proof net $H$ for $\mathsf{PC2C}_{1+\mathsf{n};\mathsf{s}}[F']$

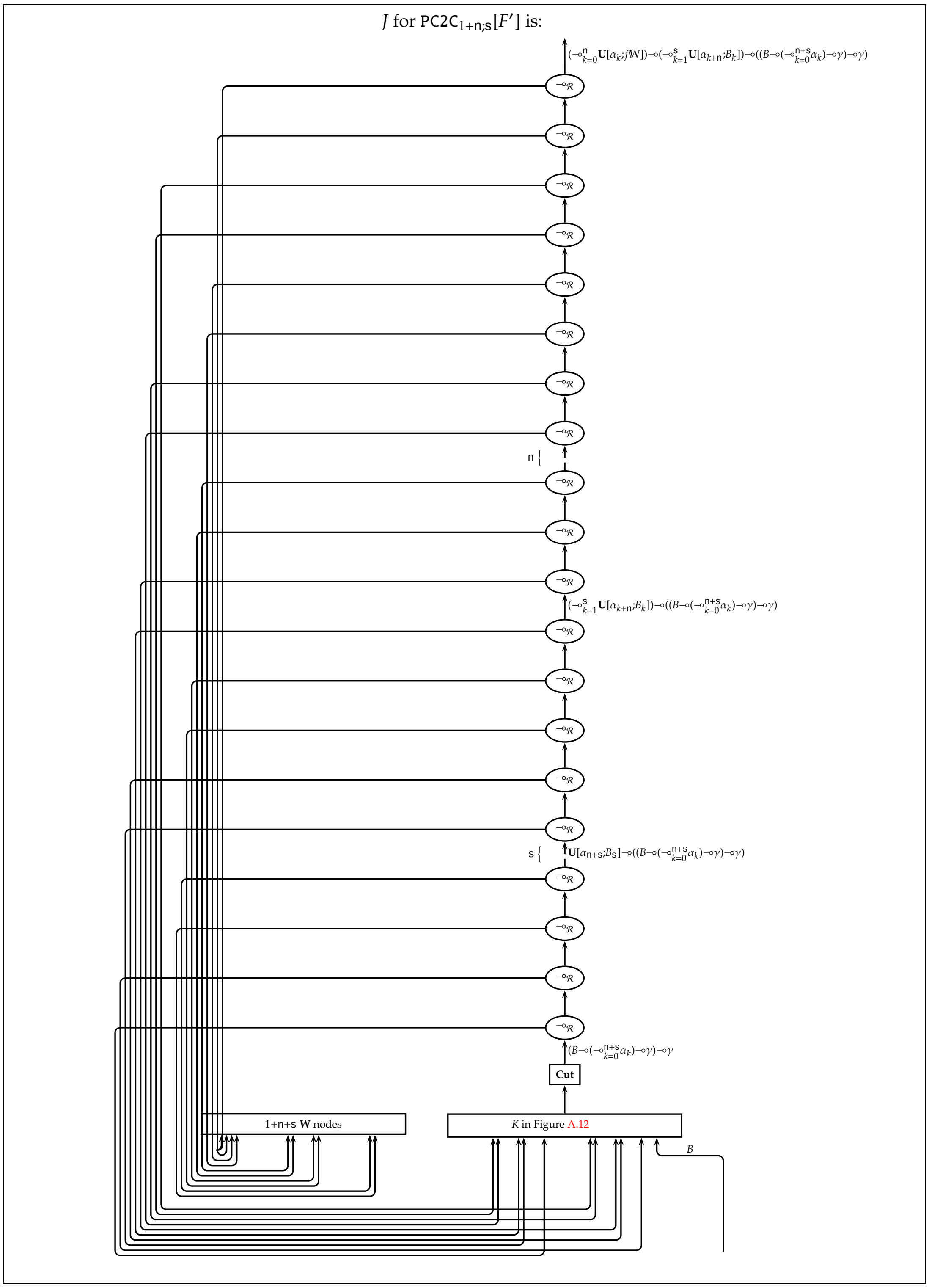

Figure A.11: The proof net $J$ for PC2C$_{1+n;s}[F']$

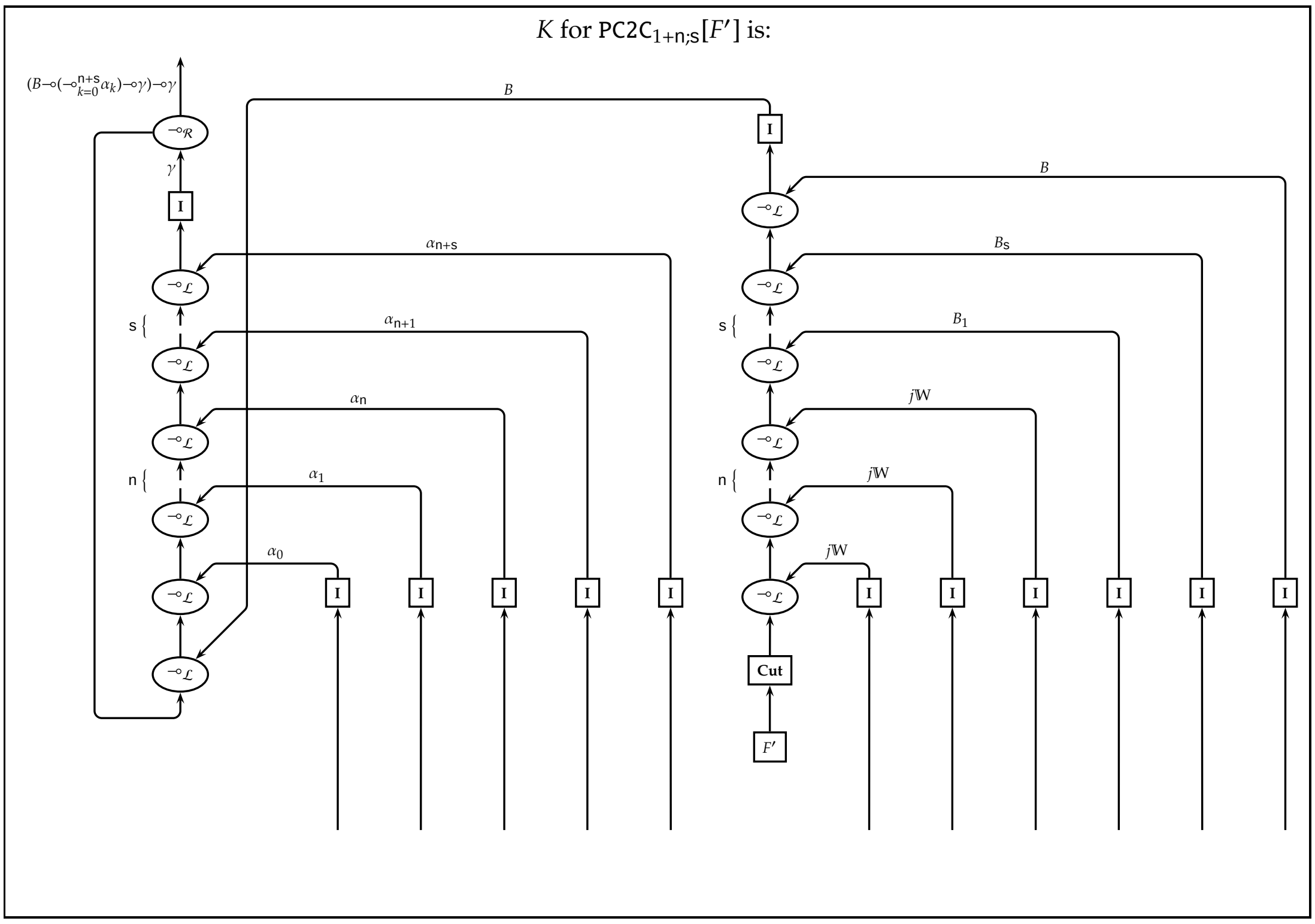

Figure A.12: The proof net $K$ for $\mathtt{PC2C}_{1+n;s}[F']$

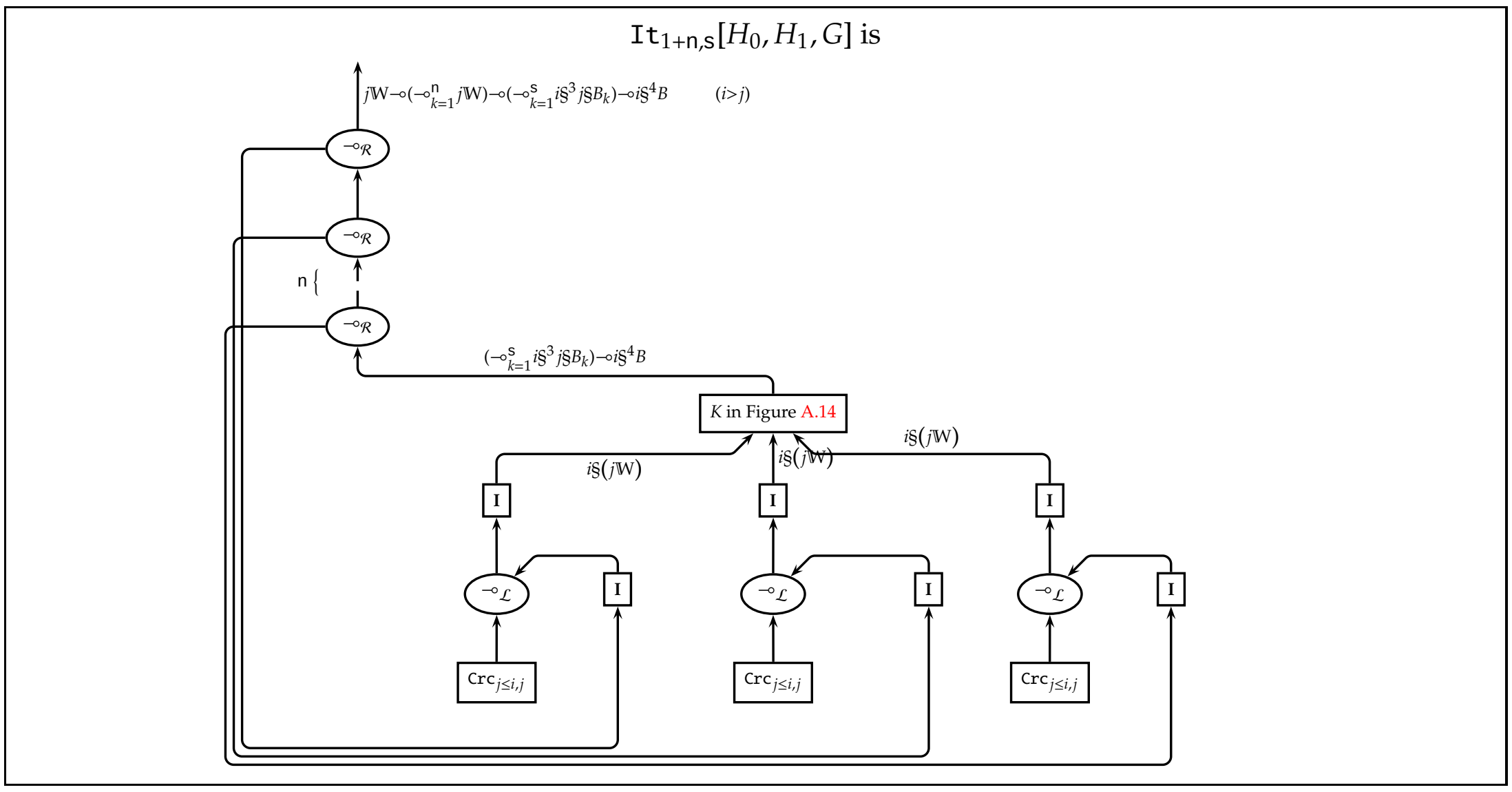

Figure A.13: The proof net $\mathtt{It}_{1+n,s}[H_0, H_1, G]$.

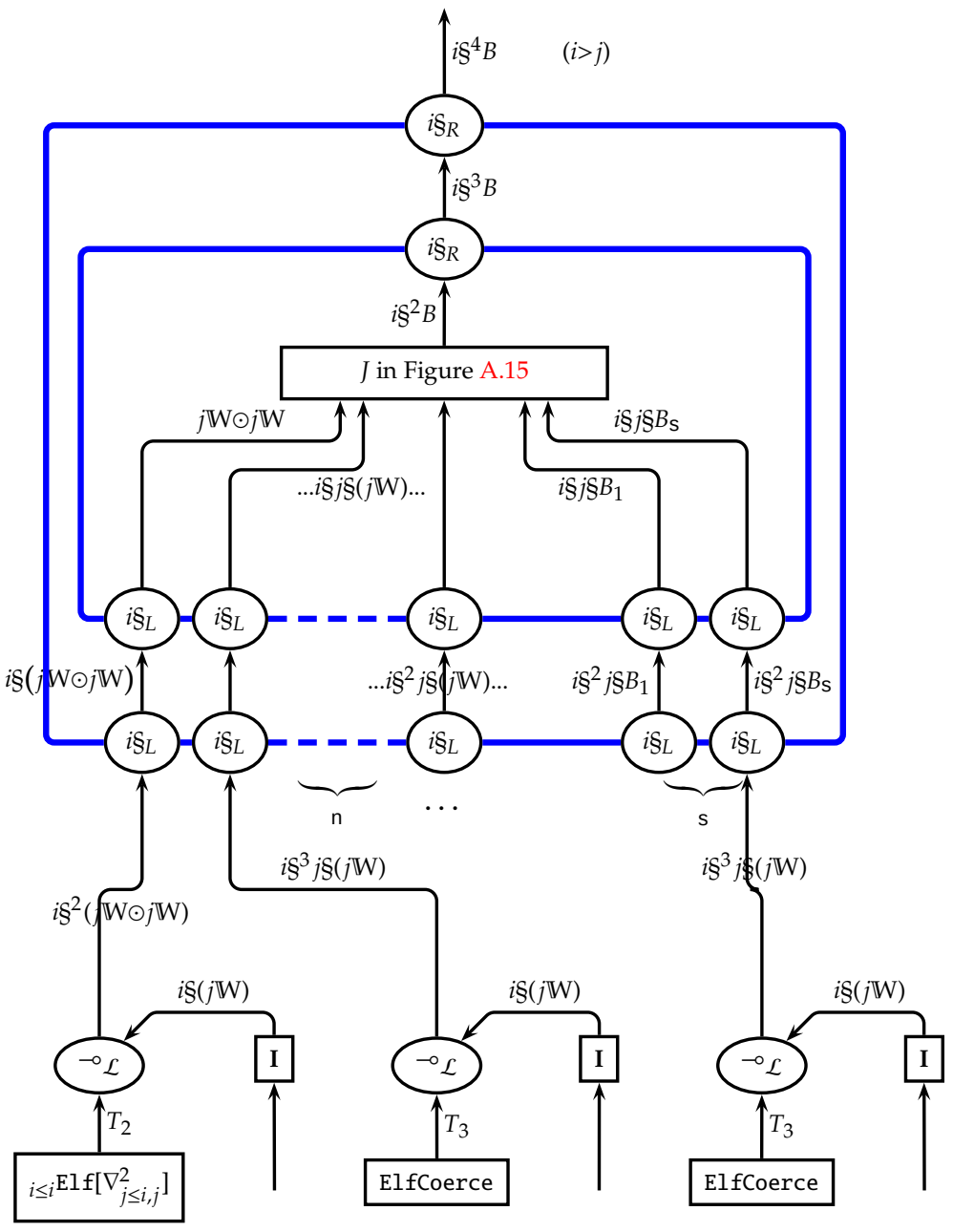

$$T_2 = i\S(j\mathbb{W}) \multimap i\S^2(j\mathbb{W} \odot j\mathbb{W})$$

$$T_3 = i\S(j\mathbb{W}) \multimap i\S^3\, j\S(j\mathbb{W})$$

Figure A.14: The proof net $K$ for $\mathtt{It}_{1+n,s}[H_0, H_1, G]$.

C2C$_{1+n;s}[F,G]$ is in Figure A.4
W2C$_{1+n;s}$ is in Figure A.17
FC2W$_{1+n;s}$ is in Figure A.24
C2FC$_{1+n;s}$ is in Figure A.23
Ba2St$[G]$ is in Figure A.16

$J$ for It$_{1+n,s}[H_0, H_1, G]$ is

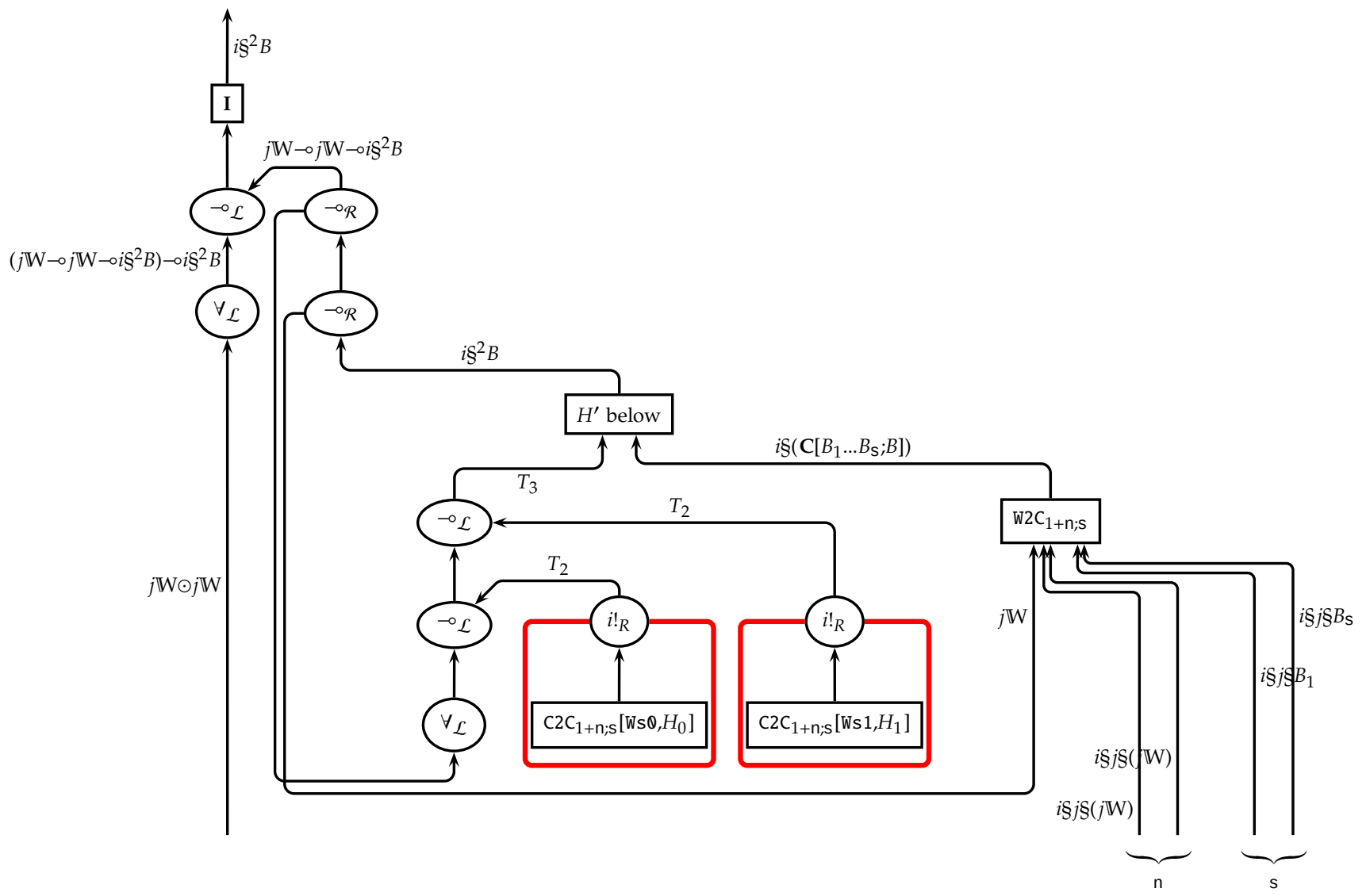

$$T_2 = i!(\mathbf{C}[B_1 \ldots B_s; B] \multimap \mathbf{C}[B_1 \ldots B_s; B])$$
$$T_3 = i\S(\mathbf{C}[B_1 \ldots B_s; B] \multimap \mathbf{C}[B_1 \ldots B_s; B])$$
$$j\mathbb{W} \odot j\mathbb{W} = \forall\gamma.(j\mathbb{W} \multimap j\mathbb{W} \multimap \gamma) \multimap \gamma$$

---

$H'$ for It$_{1+n,s}[H_0, H_1, G]$ is

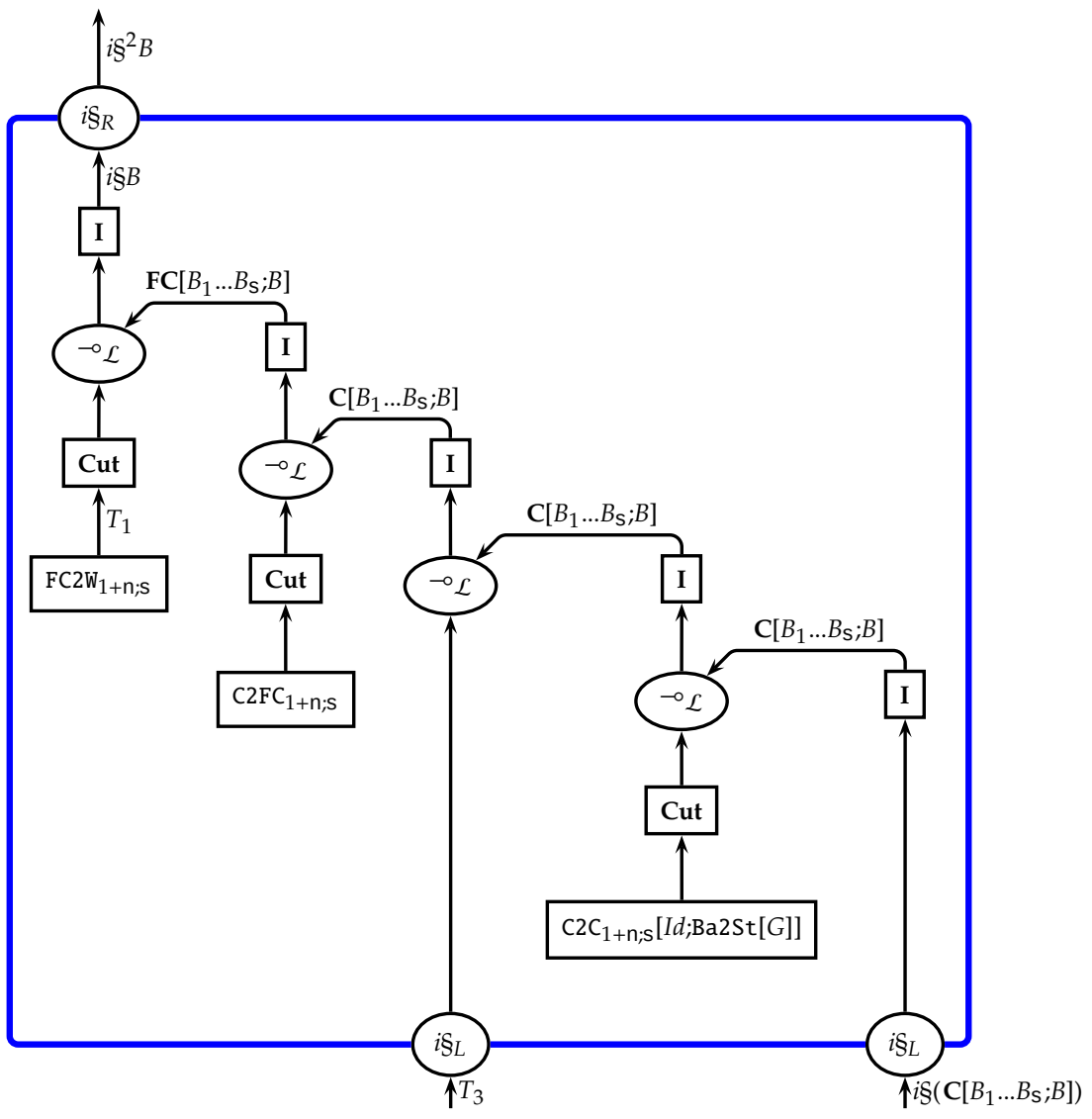

$$T_1 = \mathbf{FC}[B_1 \ldots B_s; B] \multimap i\S B$$

Figure A.15: The proof nets $J$ and $H'$ for It$_{1+n,s}[H_0, H_1, G]$.

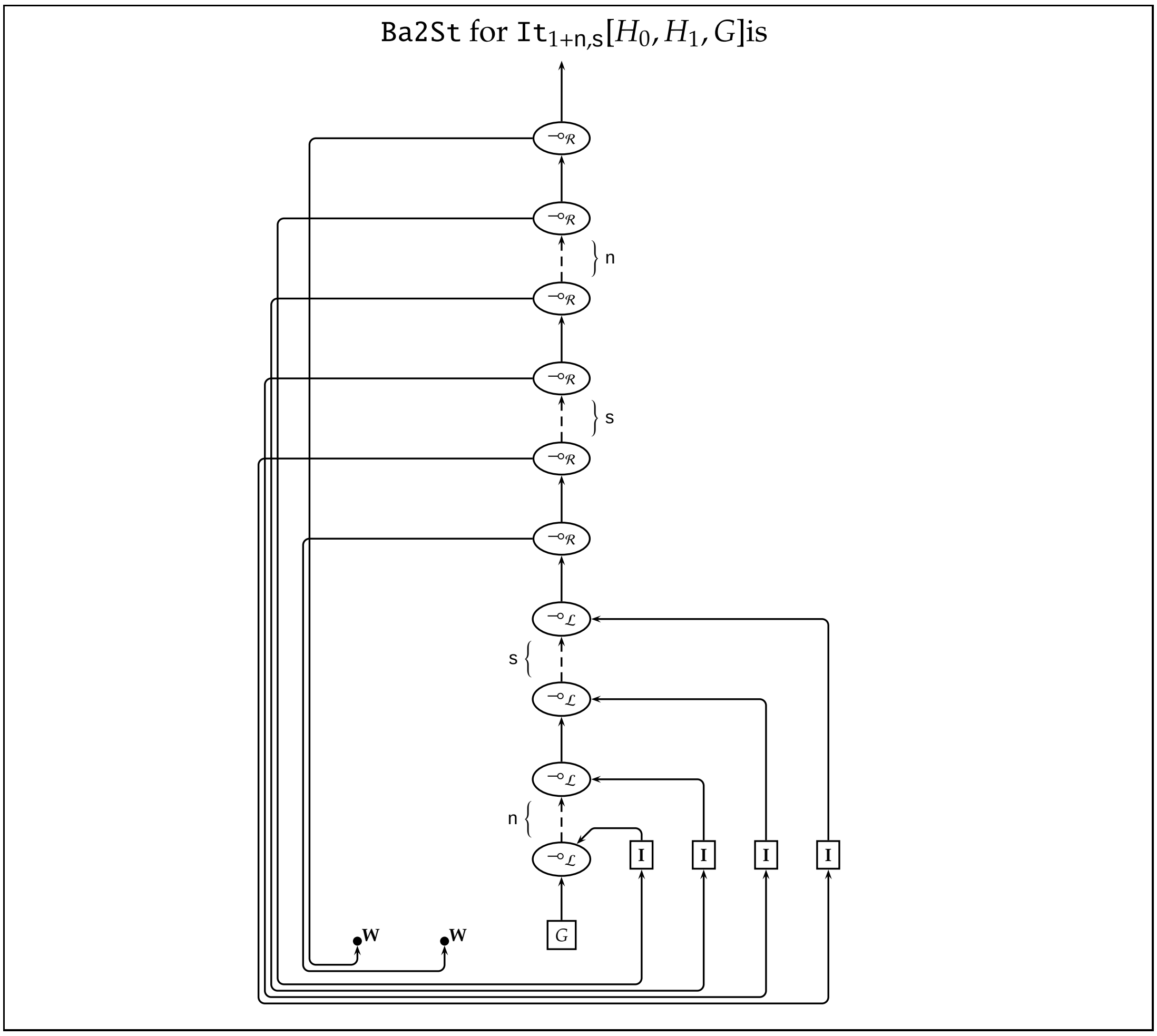

Figure A.16: The proof net Ba2St for $\mathtt{It}_{1+n,s}[H_0, H_1, G]$

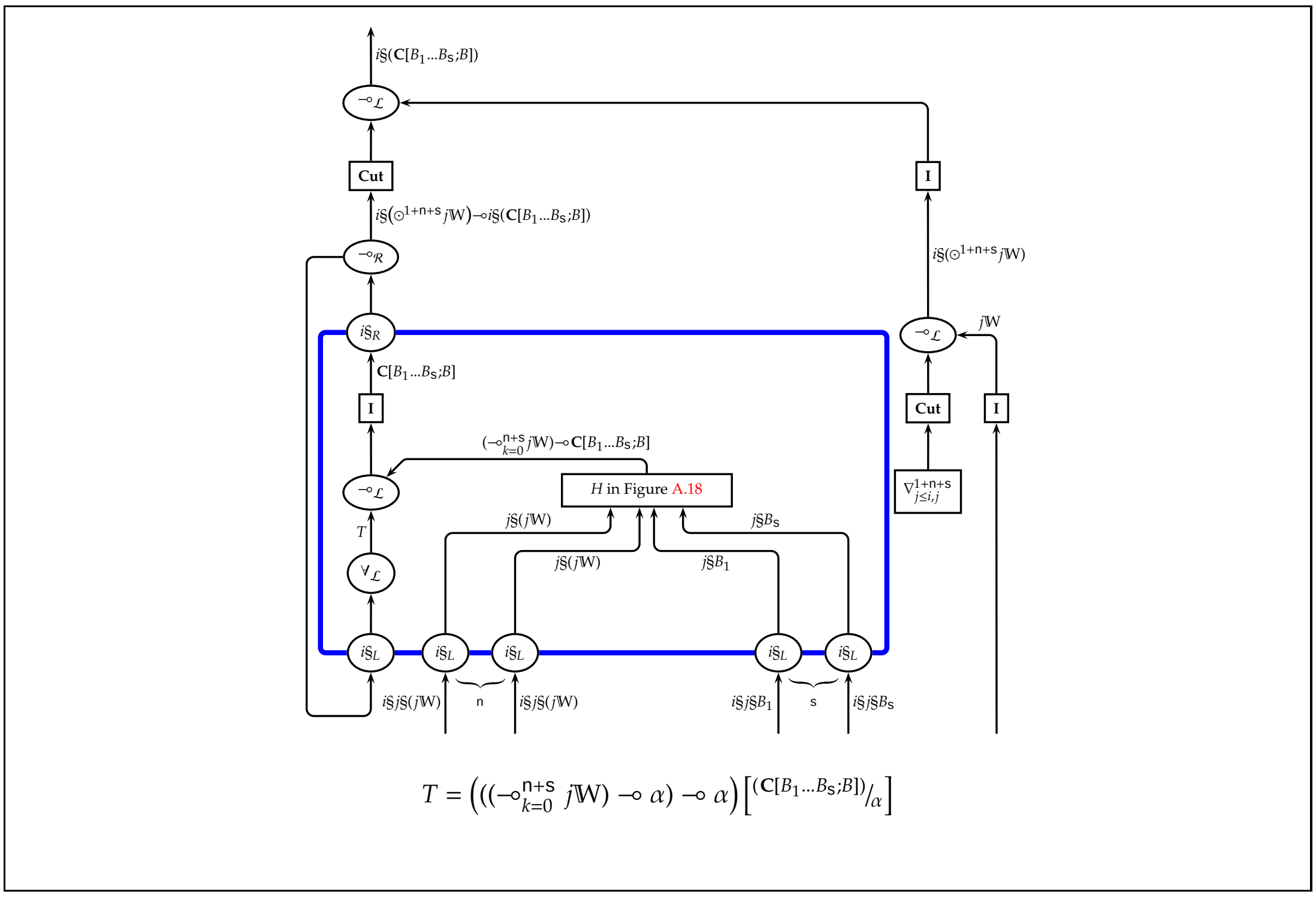

Figure A.17: The proof net $\mathtt{W2C}_{1+n;s}$

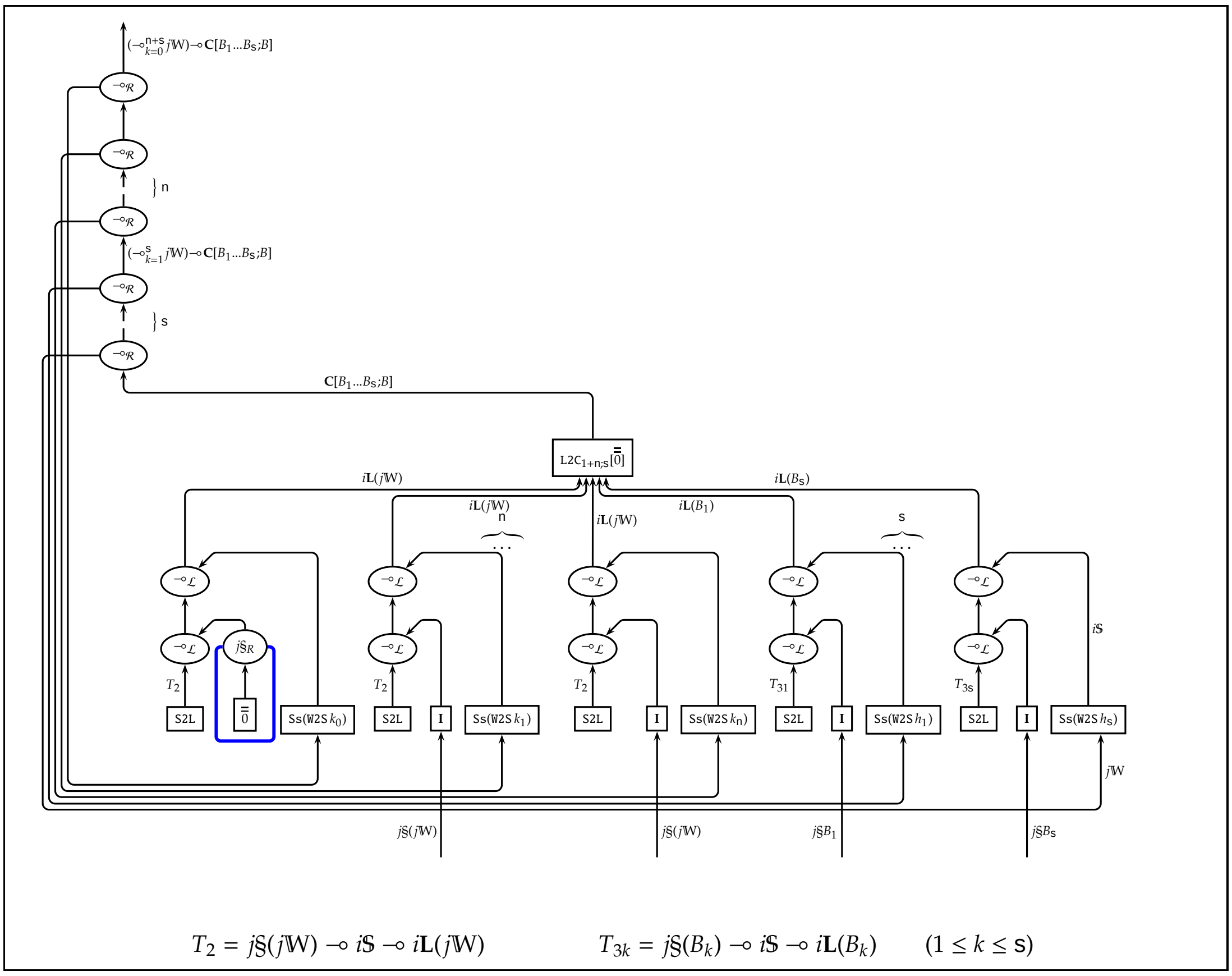

Figure A.18: The proof net $H$ for W2C$_{1+n;s}$

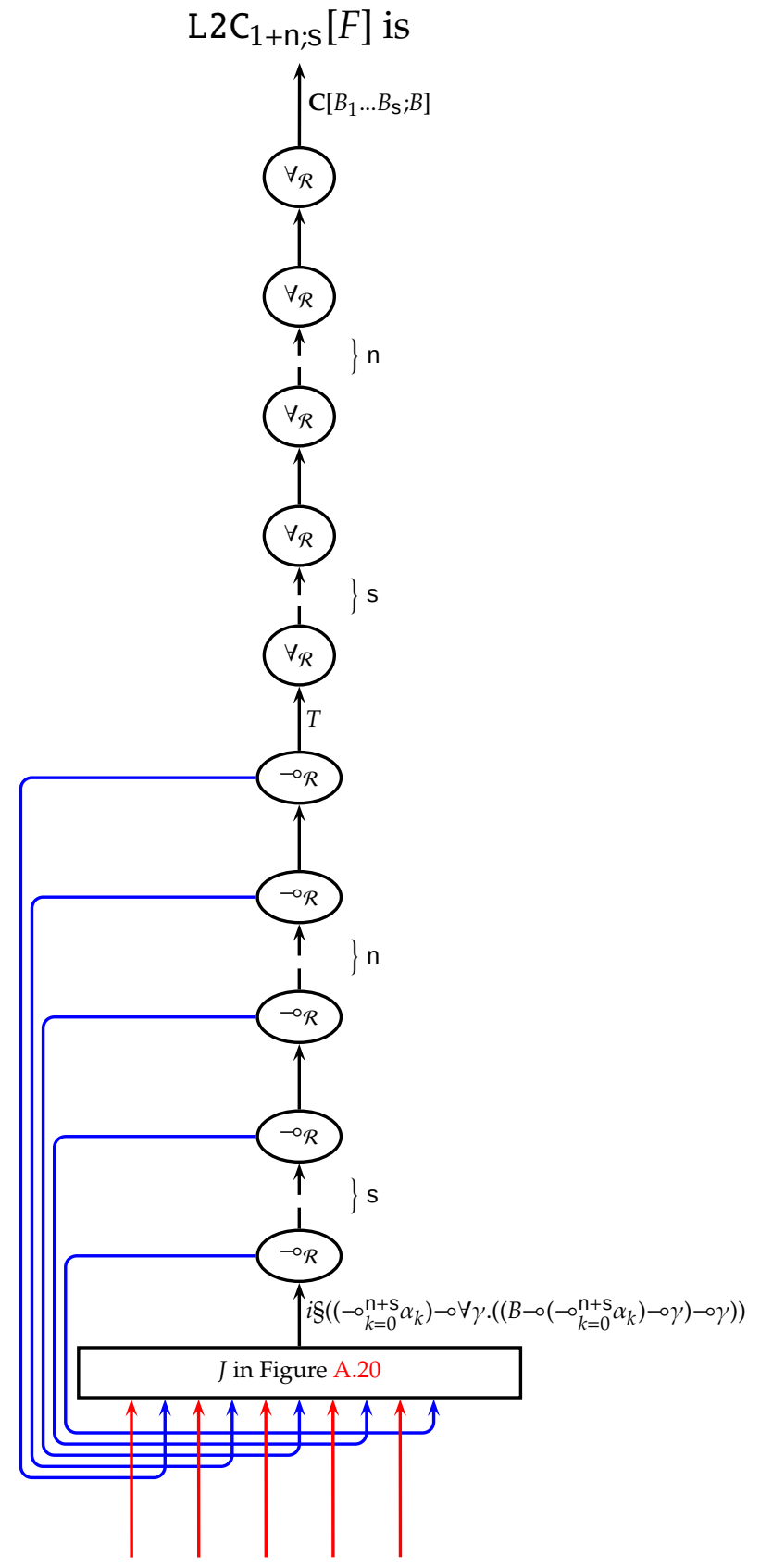

$$T = (\multimap_{k=0}^{n} i!(j\mathbb{W} \multimap \alpha_k \multimap \alpha_k)) \multimap (\multimap_{k=1}^{s} i!(B_k \multimap \alpha_{n+k} \multimap \alpha_{n+k})) \multimap i\S((\multimap_{k=0}^{n+s} \alpha_k) \multimap \forall\gamma.((B \multimap (\multimap_{k=0}^{n+s} \alpha_k) \multimap \gamma) \multimap \gamma))$$

Figure A.19: The proof net L2C$_{1+n;s}[F]$ with $i > j$.

**The proof net S2L.** It is in Figure A.22. It remarkably exploits the more general structure of the bang boxes of SOLAL, as compared to those ones of ILAL, to generate a list of copies of a word. Using S2L we shall also duplicate the safe arguments of the recursive function $f^{1+n;s}$ we want to represent. This is the basic step to represents recursive functions whose safe arguments must not necessarily be linear, as, instead, is required for the recursive functions of BC$^{-}$ representable inside ILAL [MO04]. Namely, this is the basic step to extend BC$^{-}$.

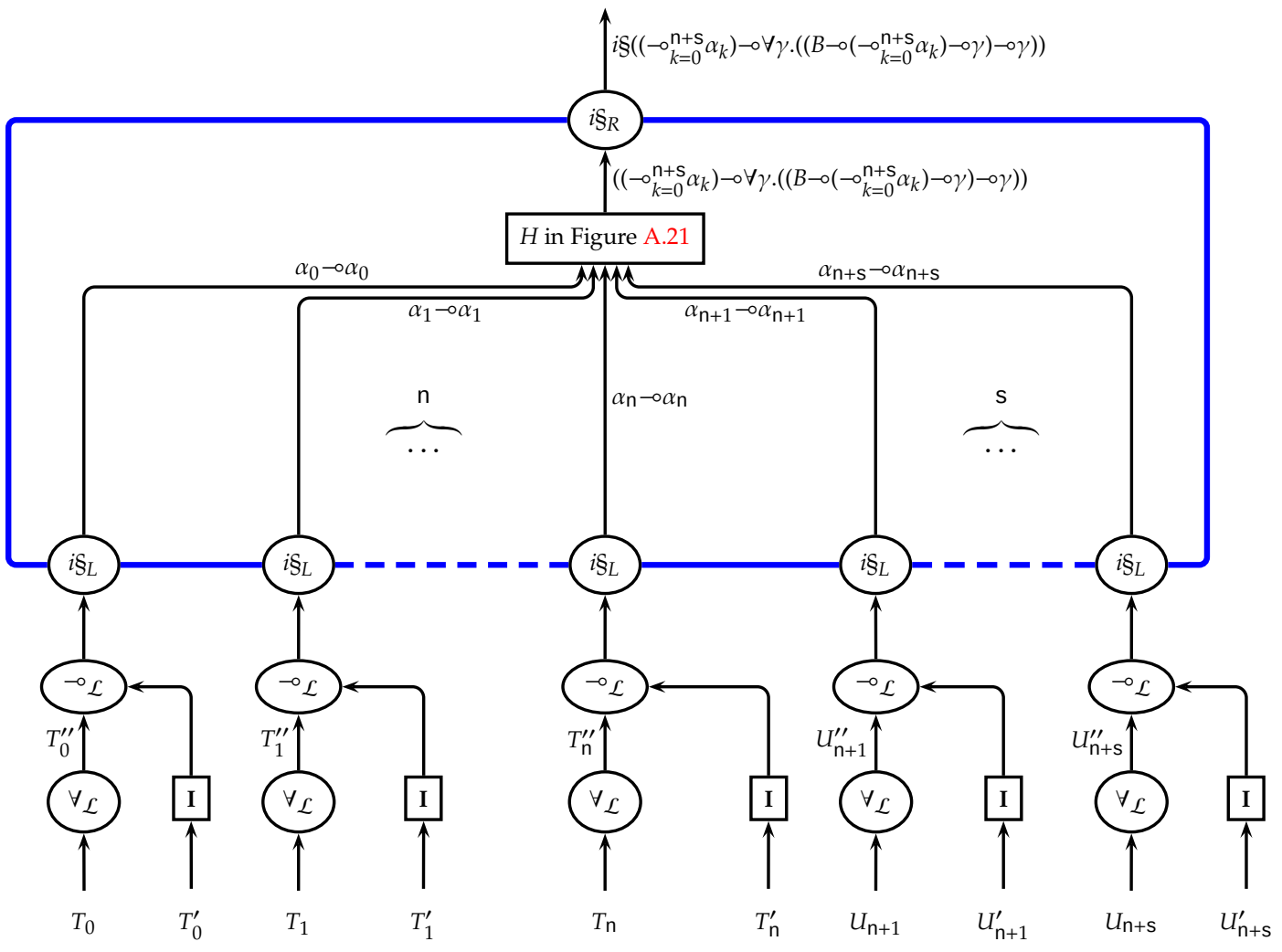

$$
\begin{aligned}
T_k &= \forall\alpha_k.i!(j\mathbb{W} \multimap \alpha_k \multimap \alpha_k) \multimap i\S(\alpha_k \multimap \alpha_k) && (0 \le k \le \mathsf{n})\\
T'_k &= i!(j\mathbb{W} \multimap \alpha_k \multimap \alpha_k) && (0 \le j \le \mathsf{n})\\
T''_k &= i!(j\mathbb{W} \multimap \alpha_k \multimap \alpha_k) \multimap i\S(\alpha_k \multimap \alpha_k) && (0 \le k \le \mathsf{n})\\
U_k &= \forall\alpha_k.i!(B_k \multimap \alpha_k \multimap \alpha_k) \multimap i\S(\alpha_k \multimap \alpha_k) && (1 \le k \le \mathsf{s})\\
U'_k &= i!(B_k \multimap \alpha_k \multimap \alpha_k) && (1 \le k \le \mathsf{s})\\
U''_k &= i!(B_k \multimap \alpha_k \multimap \alpha_k) \multimap i\S(\alpha_k \multimap \alpha_k) && (1 \le k \le \mathsf{s})
\end{aligned}
$$

Figure A.20: The proof net $J$ for $\mathtt{L2C}_{1+\mathsf{n};\mathsf{s}}[F]$.

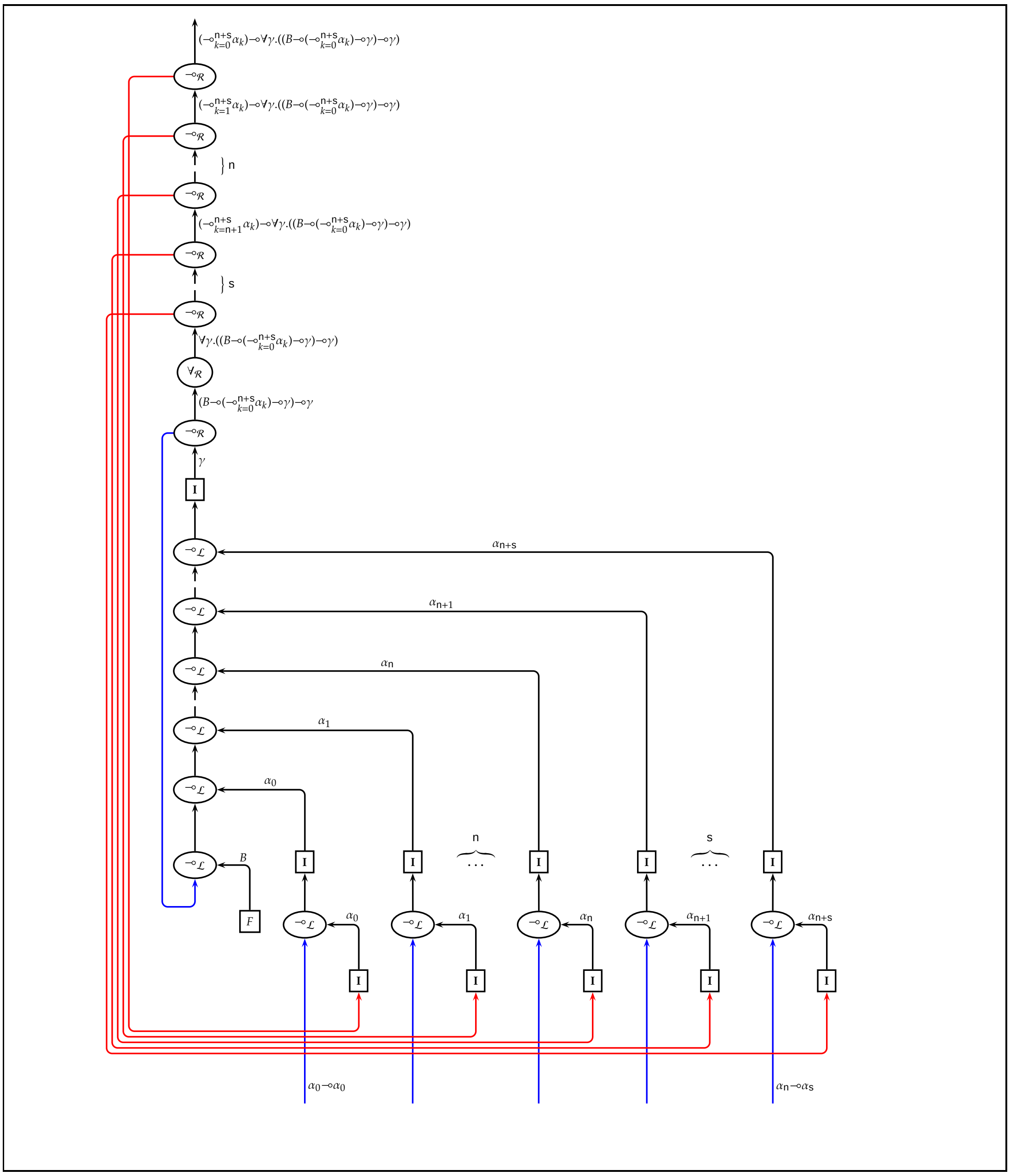

Figure A.21: The proof net $H$ for $\mathsf{L2C}_{1+n;s}[F]$.

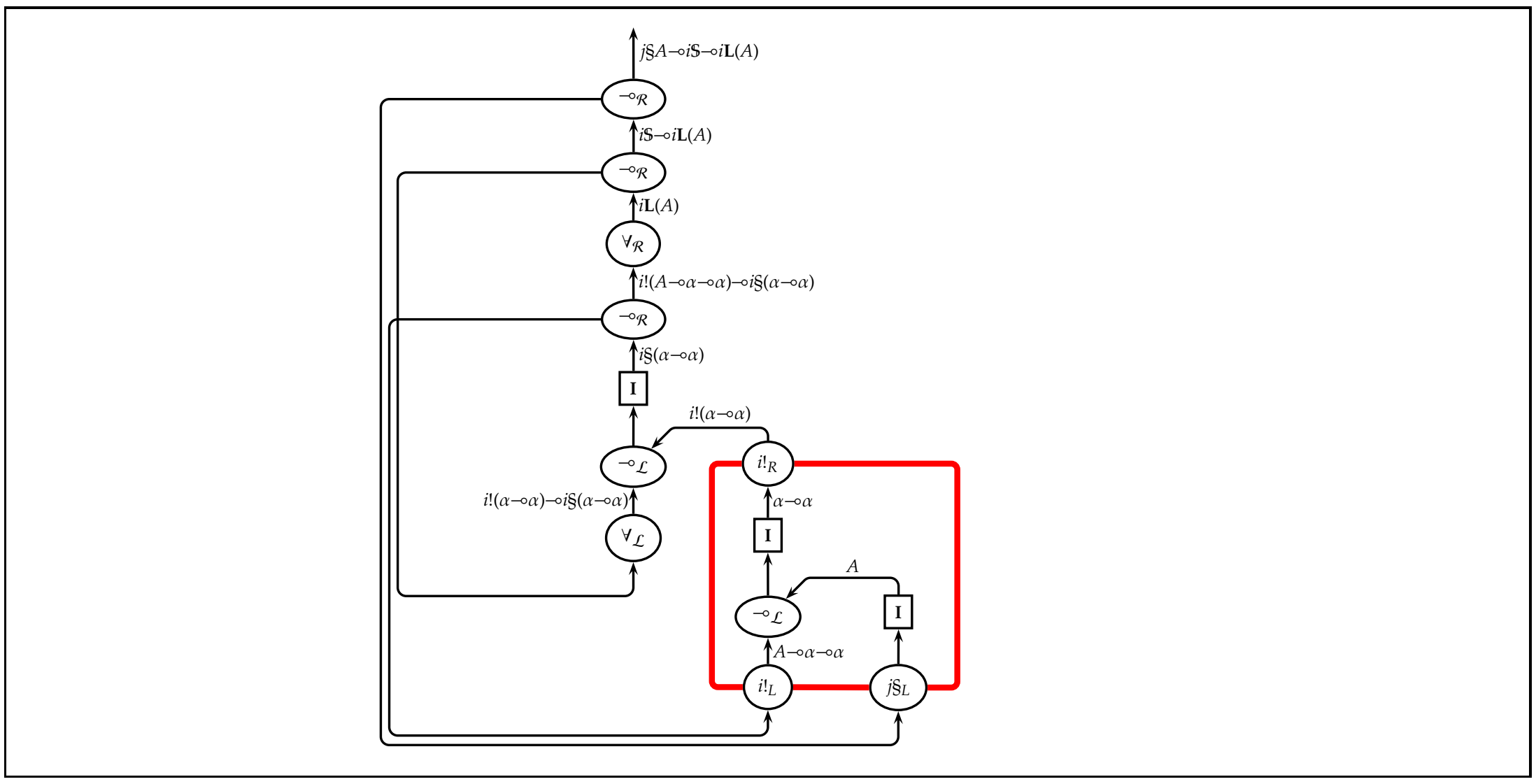

Figure A.22: The proof net S2L of soLAL, with $i > j$.

**The proof net** $\mathtt{C2FC}_{1+\mathsf{n};\mathsf{s}}$**.** It is in Figure A.23. It maps a configuration to the same configuration, but with a slightly different type that allows to extract the first component of the resulting configuration, namely, the representation of $f^{1+\mathsf{n};\mathsf{s}}$, applied to some normal and safe argument instances.

**The proof net** $\mathtt{FC2W}_{1+\mathsf{n};\mathsf{s}}$**.** It is in Figure A.24, where we assume (i) $\Pi_k \triangleright \vdash B_k$, for every $1 \leq k \leq \mathsf{s}$, (ii) $\Pi'$ calculates the first projection of two elements, and (iii) $\Pi''$ calculates the first projection of $\mathsf{n} + \mathsf{s}$ elements. which erases all the lists inside the final configuration, keeping only its first element.

C2FC$_{1+n;s}$ is

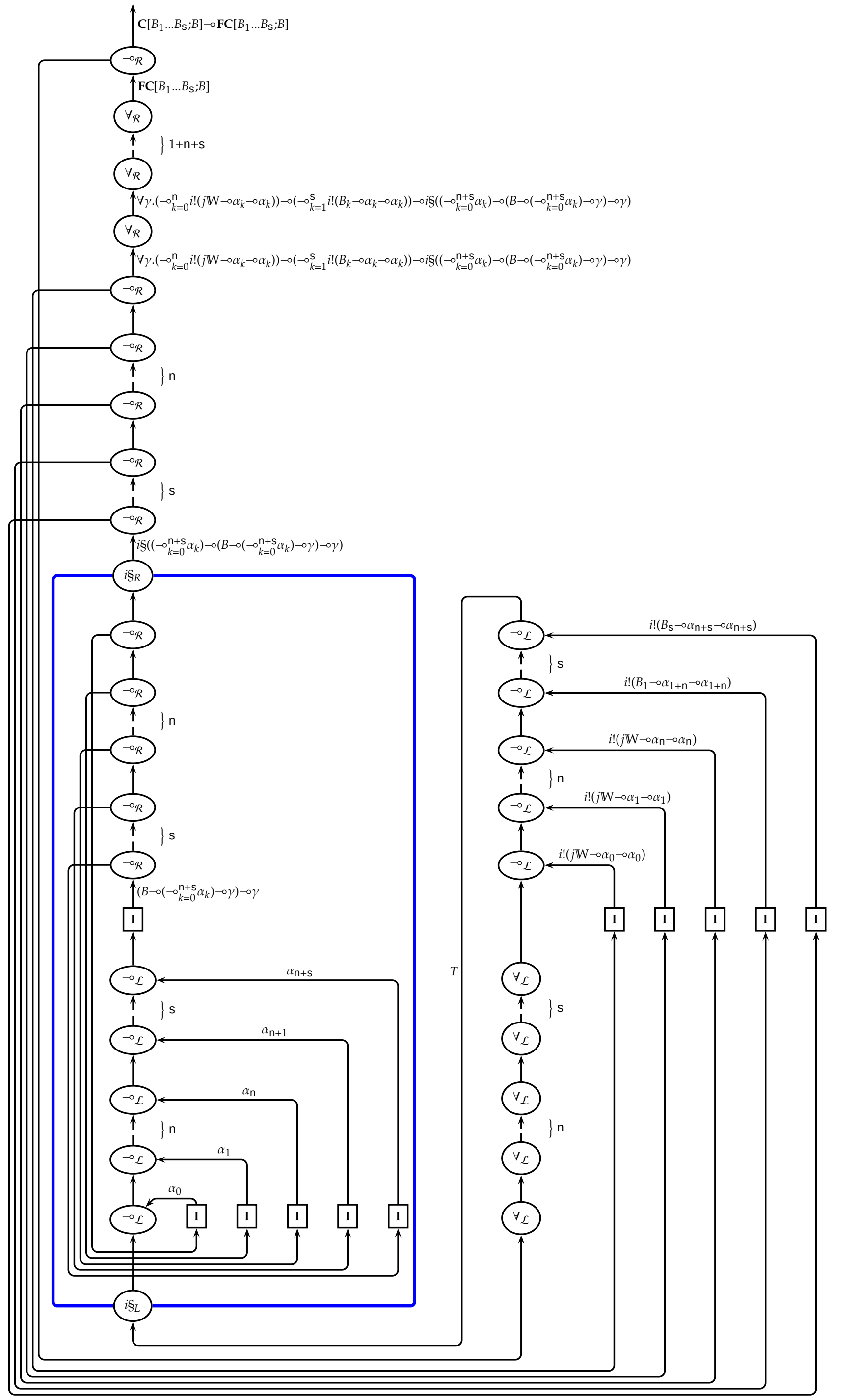

$T = i\S((\multimap_{k=0}^{n+s} \alpha_k) \multimap \forall\gamma.(B \multimap (\multimap_{k=0}^{n+s} \alpha_k) \multimap \gamma) \multimap \gamma)$

Figure A.23: The proof net C2FC$_{1+n;s}$.

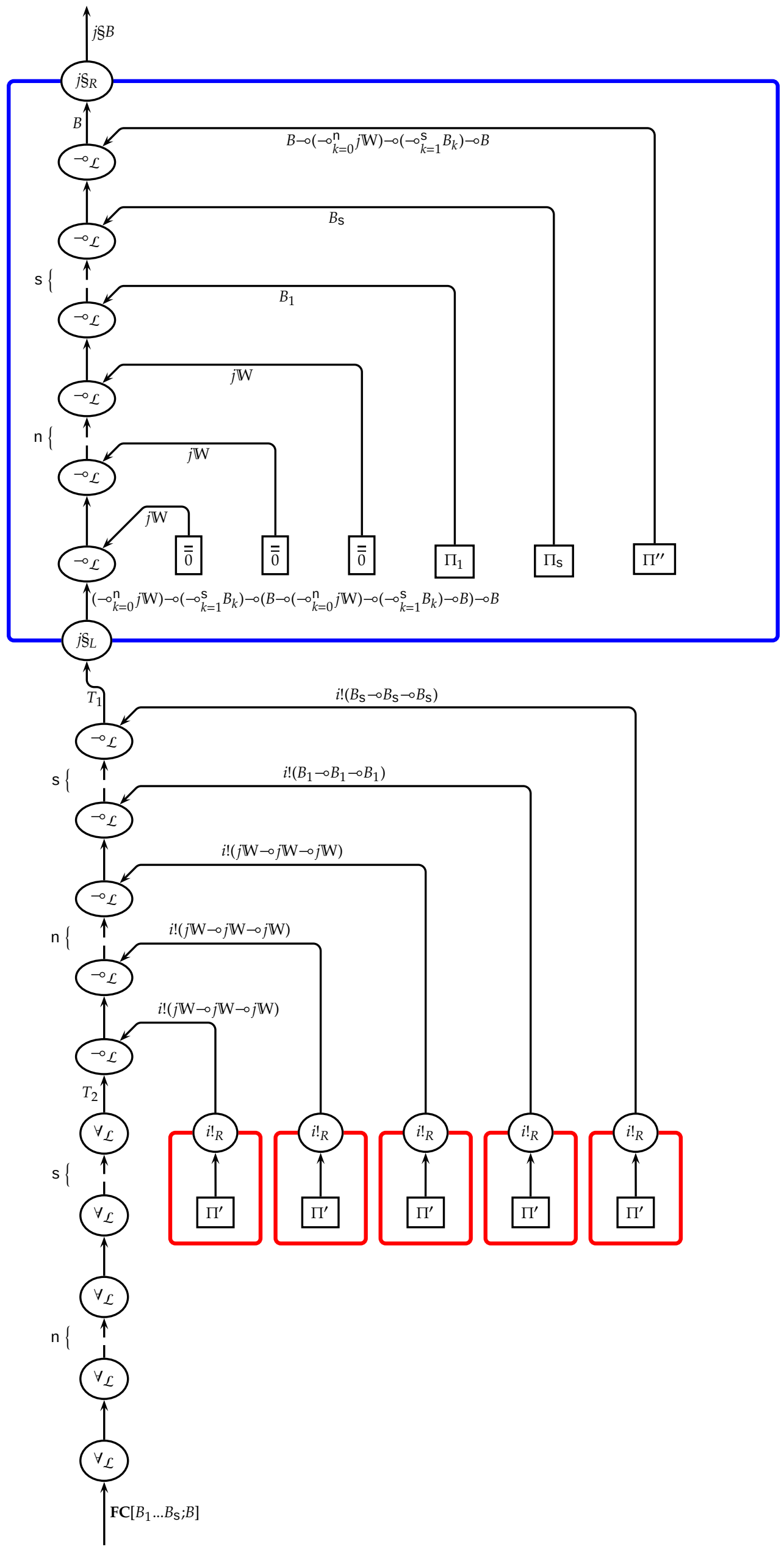

$$T_1 = \S((\multimap_{k=0}^{\mathsf{n}} j\mathbb{W}) \multimap (\multimap_{k=1}^{\mathsf{s}} B_k) \multimap (B \multimap (\multimap_{k=0}^{\mathsf{n}} j\mathbb{W}) \multimap (\multimap_{k=1}^{\mathsf{s}} B_k) \multimap B) \multimap B)$$

$$\begin{aligned} T_2 = {} & (\multimap_{k=0}^{\mathsf{n}} i!(j\mathbb{W} \multimap j\mathbb{W} \multimap j\mathbb{W})) \multimap \\ & (\multimap_{k=1}^{\mathsf{s}} i!(B_k \multimap B_k \multimap B_k)) \multimap \\ & \S((\multimap_{i=0}^{\mathsf{n}} j\mathbb{W}) \multimap (\multimap_{k=1}^{\mathsf{s}} B_k) \multimap (B \multimap (\multimap_{k=0}^{\mathsf{n}} j\mathbb{W}) \multimap (\multimap_{k=1}^{\mathsf{s}} B_k) \multimap B) \multimap B) \end{aligned}$$

Figure A.24: The proof net $\mathtt{FC2W}_{1+\mathsf{n};\mathsf{s}}$.

# Index

# Table of Symbols